\documentclass[fleqn,usenatbib]{mnras}

\usepackage{newtxtext,newtxmath}

\usepackage[T1]{fontenc}

\DeclareRobustCommand{\VAN}[3]{#2}
\let\VANthebibliography\thebibliography
\def\thebibliography{\DeclareRobustCommand{\VAN}[3]{##3}\VANthebibliography}

\usepackage{graphicx}	
\usepackage{amsmath}	
\usepackage{comment}    
\usepackage{multirow}   

\title[VLBI properties  of compact IPS sources]{VLBI properties of compact interplanetary scintillators detected by the Murchison Widefield Array}

\author[Jaiswal et al.]{
Sumit Jaiswal$^{1}$\thanks{E-mail: sumit@shao.ac.cn}, 
Tao An,$^{1}$\thanks{E-mail: antao@shao.ac.cn}
Ailing Wang$^{1,2}$ and
Steven Tingay$^{3}$
\\
$^{1}$Shanghai Astronomical Observatory, Key Laboratory of Radio Astronomy, CAS, 80 Nandan Road, Shanghai 200030, China \\
$^{2}$ University of Chinese Academy of Sciences, 19A Yuquanlu, Beijing 100049, China \\
$^{3}$International Centre for Radio Astronomy Research, Curtin University, GPO Box U1987, Perth, WA 6845, Australia\\
}

\date{Accepted XXX. Received YYY; in original form ZZZ}

\pubyear{2021}

\begin{document}
\label{firstpage}
\pagerange{\pageref{firstpage}--\pageref{lastpage}}
\maketitle

\begin{abstract}
Interplanetary scintillation (IPS) provides an approach for identifying the presence of sub-arcsec structures in radio sources, and very long baseline interferometry (VLBI) technique can help verify whether the IPS sources have fine structures on milli-arcsec (mas) scales. We searched the available VLBI archive for the 244 IPS sources detected by the Murchison Widefield Array at 162~MHz and found 63 cross-matches.
We analysed the VLBI data of the 63 sources and characterised the compactness index in terms of the ratio $R$ of the VLBI-measured flux density at 4.3~GHz to the flux density estimated using the Very Large Array Sky Survey (VLASS) at 3~GHz and NRAO VLA Sky Survey (NVSS) at 1.4~GHz ($S_{\rm VLBI}/S_{\rm SA}$).
Eleven sources are identified as blazars according to their flat spectra and strong variability. They show high compactness indices with $R>0.4$, compact core-jet structure, and a broad distribution of normalised scintillation index (NSI). 
Other sources show diverse morphologies (compact core, core and one-sided jet, core and two-sided jets), but there is a correlation between $R$ and NSI with a correlation coefficient $r=0.47$.
A similar $R$--NSI correlation is found in sources showing single steep power-law or convex spectra.
After excluding blazars (which are already known to be compact sources) from the VLBI-detected IPS sources, a strong correlation is found between the compactness and scintillation index of the remaining samples, indicating that stronger scintillating sources are more compact. This pilot study shows that IPS offers a convenient method to identify compact radio sources without the need to invoke high-resolution imaging observations, which often require significant observational time.

\end{abstract}




\section{Introduction} \label{sec:intro}
Scintillation is a phenomenon caused by changes in the phase and/or amplitude of radio waves due to the scattering effects of the turbulent neutral and ionised media lying between a radio source and the Earth  \citep{2017isra.book.....T}. 
There are three main types of scintillations, namely ionospheric, interplanetary and interstellar, depending on the distance of the scattering medium from the observer and the size of the plasma structure. It can be readily shown that scintillation is weak at high frequencies and for nearby scattering media. Interplanetary scintillation (IPS) was discovered by \citet{1964Thesis.....000C} who identified compact ($<$2 arcsec) radio sources with rapid ($\sim$1 s time-scale) and random flux density fluctuations caused by turbulent solar wind plasma. IPS has
been used as an astrophysical tool to infer the sub-arcsecond scale structures of radio sources \citep{1964Natur.203.1214H}, without the need for high-resolution imaging observations of each source separately.
Observations of IPS sources can also be used to probe the characteristics of the intervening plasma and thereby the corresponding solar activity \citep[e.g., ][]{1967ApJ...147..433S,1967ApJ...147..449C,1971A&A....10..310E,1978SSRv...21..411C,1998JGR...10312049J}. \citet{1987MNRAS.229..589P} published a catalogue of 1789 IPS sources at 81.5~MHz using the Cambridge IPS array. 

Large synthesis arrays of multiple elements such as the Murchison Widefield Array \citep[MWA; ][]{2013PASA...30....7T} have sufficient instantaneous (u,v) coverage and temporal resolution to allow rapid measurements of a large number of IPS sources throughout the entire field of view. Recently, \citet{2018MNRAS.473.2965M} and \citet{2018MNRAS.474.4937C,2018MNRAS.479.2318C} detected sub-arcsecond components within many hundreds of radio sources using the $30\degr \times 30\degr$ MWA field-of-view at $\sim$160~MHz, with just 5~min of data and a time resolution of 0.5~s. Interestingly, they found that those sources displaying strong scintillation are mainly peaked or inverted spectrum sources, analogous to the GHz-peaked spectrum  sources \citep[GPS,][]{1998PASP..110..493O,2021A&ARv..29....3O}.~
Inverted-spectrum or compact steep-spectrum sources are morphologically associated with radio-loud active galactic nuclei (AGN) with sub-kpc to several kpc scale structures, which are thought to be young radio AGN in their early evolutionary stages or aged AGN frustrated by the surrounding dense  interstellar medium (ISM) \citep[e.g.,][]{2010MNRAS.408.2261K,2012ApJ...760...77A,2016AN....337..105S,2017ApJ...836..174C,2021A&ARv..29....3O}.

In terms of its timescale, mechanism, and frequency dependence, the IPS phenomenon is very different from intra-day variability (IDV), which is due to a combination of intrinsic variability of the radio source itself and extrinsic variability caused by the interstellar medium. IDV is evident at much higher frequencies than IPS. Moreover, the timescale of IDV ranges from hours to days, much larger than the IPS timescale of a few seconds \citep{2015aska.confE..58B}. IDV has been observed in many flat-spectrum AGNs \citep{1992A&A...258..279Q,2003ApJ...585..653B,2012RAA....12..147L} with the associated angular size smaller than a few tens of micro-arcseconds according to the scintillation criteria of the first Fresnel zone at $\gtrsim5$~GHz frequencies for typical scattering screen distances in case of the interstellar scintillation. The IPS sources, on the other hand, are expected to have sub-arcsecond size \citep[e.g., ][]{1964Natur.203.1214H,1967ApJ...147..449C,2019PASA...36....2M}.  IDV has been well-studied in the literature, especially in the Micro-Arcsecond Scintillation-Induced Variability (MASIV) survey using Very Large Array \citep{2002PASA...19...14K,2003AJ....126.1699L,2007A&AT...26..575J,2008ApJ...689..108L,2011AJ....142..108K,2018MNRAS.474.4396K}. For some sources, very long baseline interferometry (VLBI) observational studies \citep[e.g., ][]{2004AJ....128.1570O,2004ApJ...614..607O,2006ApJS..166...37O,2021MNRAS.tmp.2457S} were conducted to confirm their compactness and to understand how the interstellar scintillation depends on the source compactness. However, no such high-resolution imaging studies of IPS sources have been carried out to understand their behaviour on VLBI scales.

In this study, we present an analysis of GHz-frequency VLBI data of a sample of MWA-detected IPS sources in an attempt to understand how their pc-scale properties, such as morphology, radio spectrum, brightness temperature, and variability are related to the scintillation properties inferred from the MWA observations. 

\section{Sample construction} \label{sec:sample}
We used the IPS catalogue compiled by \citet{2018MNRAS.474.4937C,2018MNRAS.479.2318C}. The observation was made on 2016 May 1 at 02:25:20 UTC and lasted 5 minutes. The pointing centre was at RA~=~00$^{\rm h}$49$^{\rm m}$02$^{\rm s}$,  Dec~=~$-$19$\degr$58$\arcmin$54$\arcsec$. 
Of the 2550 continuum sources detected in the 900 deg$^2$ MWA field-of-view, 247 IPS sources showing rapid fluctuations were identified with a signal-to-noise ratio of $\ge 5$.
After removing three gravitational lensing sources (J013212$-$065232, J014013$-$095654 and J015741$-$104345) identified from the literature, the IPS sample contains 244 AGN. We searched these sources in the available VLBI archive, mainly the VLBA Calibrator Surveys\footnote{\url{http://astrogeo.org/} maintained by L. Petrov.}  \citep{2002ApJS..141...13B,2003AJ....126.2562F,2005AJ....129.1163P,2006AJ....131.1872P,2007AJ....133.1236K,2008AJ....136..580P}, 
and found 63 sources after removing two sources (J004839$-$294718 and J005429$-$235127) with poor quality VLBI data within a search radius of $\sim$1~arcmin. The resolution of MWA observation was $\sim$2~arcmin \citep{2015PASA...32...25W}, so the search radius is approximately a half of the MWA beam size. The cross-matching of the sources was further confirmed by examining the corresponding NRAO VLA Sky Survey \citep[NVSS;][]{1998AJ....115.1693C} and Very Large Array Sky Survey \citep[VLASS;][]{2020PASP..132c5001L,2020RNAAS...4..175G} images.
Therefore, the parent sample size of the IPS sources is 244 out of which the VLBI-selected sample size is 63.

Although these VLBI observations were mostly made in snapshot mode at 2, 5 and 8 GHz, they are capable of detecting bright compact sources above $\sim$15~mJy at GHz frequencies and can be used to check whether these sources have compact structure on mas scales.
The VLBI survey campaigns cover the sky zone of the IPS observation made by \citet{2018MNRAS.473.2965M} and \citet{2018MNRAS.474.4937C}.
However, for sources in the southern sky, the sensitivity of VLBA observations decreases due to the high temperature of the antenna and the unavailability of some high-latitude telescopes. Therefore, even if all sources in the parent sample are at $\delta>-40^\circ$, the lower sensitivity of the VLBA snapshot observations may result in some of the sources not being detected.  
A total of 484 sources are found in VLBI archive within the 900 deg$^2$ MWA field-of-view, suggesting that a large fraction of these sources are not detected by the MWA during the time span (i.e., 5 min). Continuous IPS observations on longer time scales can help determine the fraction of VLBI sources in the IPS sample and the physical reasons (e.g., sensitivity, variability) behind this fraction.

These VLBI sources were well distributed over a variety of pc-scale morphological classes (i.e., compact core, core + one sided jet, core + two sided jets, and core + extended structure) and a variety of overall spectrum shapes (i.e., simple power-law spectrum, inverted/convex spectrum, and flat spectrum); see Sections~\ref{sec:results} and \ref{sec:discuss} for details. 
As mentioned above, although the VLBI sample in this paper is not complete and inclined to bright, flat-spectrum sources, the widely distribution of the radio properties still allow for an exploratory study.
The basic properties of the IPS--VLBI sample are presented in Table~\ref{tab:sample}, which include the source coordinates, redshifts (obtained from the NASA/IPAC Extragalactic Database, NED, if available), MWA flux densities at 162~MHz ($S_{\rm MWA}^{\rm 162MHz}$), spectral index $\alpha$ from MWA GLEAM ($\alpha_{\rm GLEAM}$), the NVSS 1.4-GHz flux densities ($S_{\rm NVSS}^{\rm 1.4GHz}$), the VLASS 3-GHz flux densities ($S_{\rm VLASS}^{\rm 3GHz}$), and normalised scintillation indices \citep[NSI;][]{2018MNRAS.474.4937C}.
The scintillation index of an IPS source is defined as the ratio of the root mean square (rms) of the flux density fluctuations of the source to its mean flux density, and varies with the source elongation from the Sun \citep[e.g.][]{1993SoPh..148..153M}. 
The normalised scintillation index is defined as the ratio of the observed scintillation index and the scintillation index predicted by the solar elongation of the source, and is close to unity for sources dominated by a single sub-arcsec compact component \citep{2018MNRAS.474.4937C}.

\begin{figure}
\centering
\includegraphics[scale=0.5]{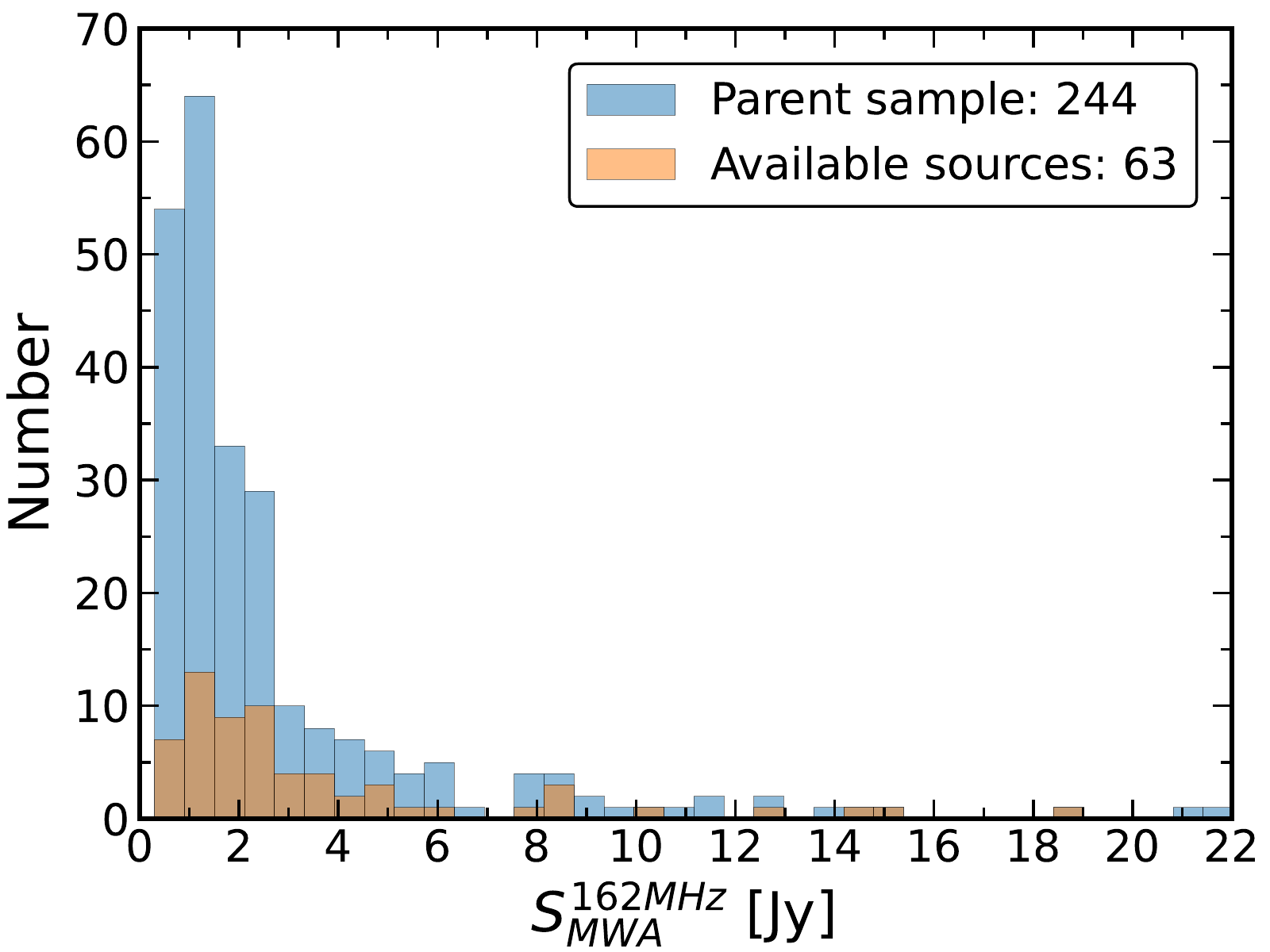}
\includegraphics[scale=0.5]{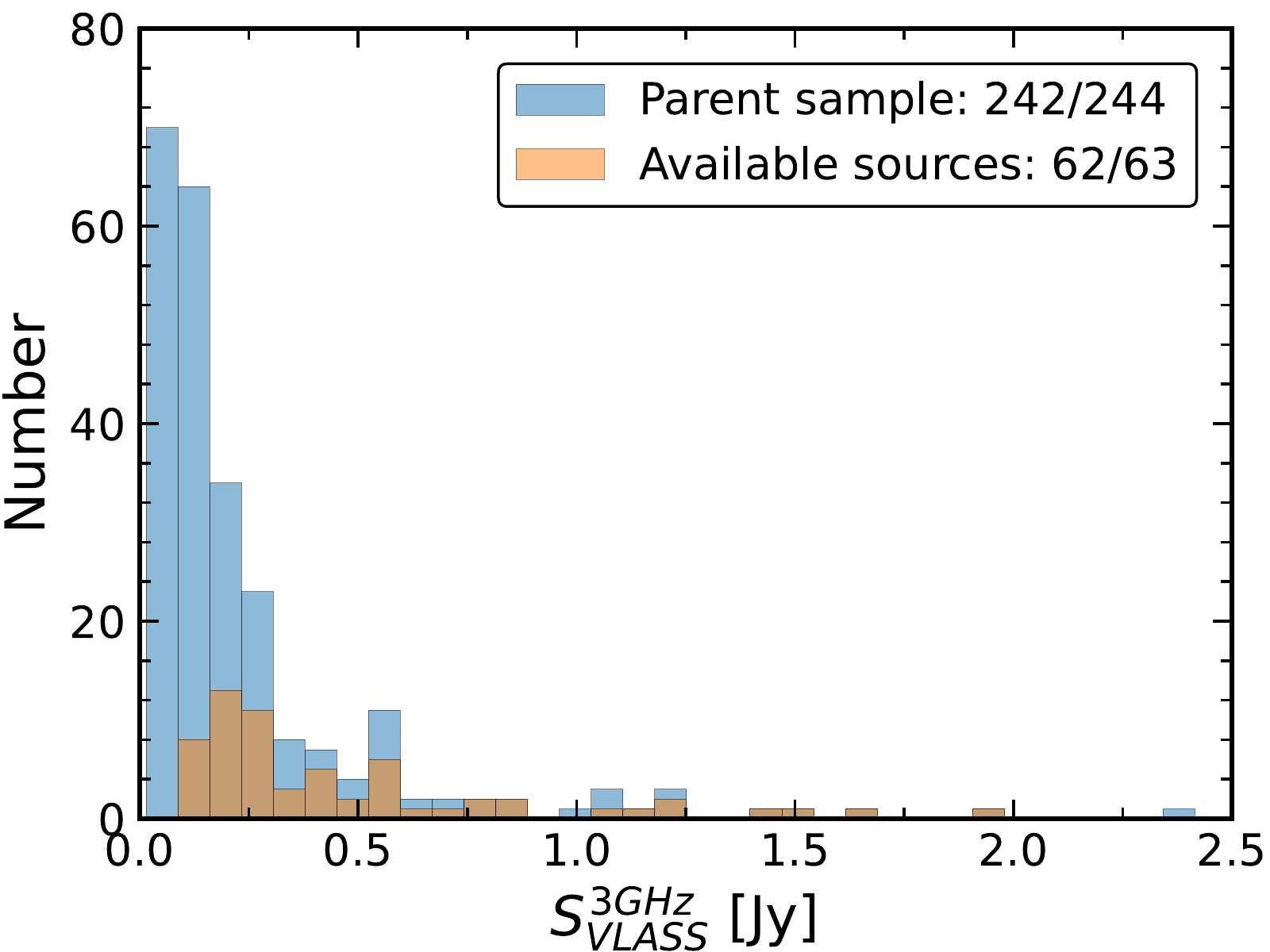}
\caption{The distributions of MWA flux density at 162~MHz and VLASS 3~GHz flux density. }
\label{fig:hist-1}
\end{figure}

Figure~\ref{fig:hist-1} shows the distributions of $S_{\rm MWA}^{\rm 162MHz}$ and $S_{\rm VLASS}^{\rm 3GHz}$ for the 63 sources 
and their comparison with the parent sample data. 
These histograms indicate that for most sources, $S_{\rm MWA}^{\rm 162MHz} < 6$~Jy and $S_{\rm VLASS}^{\rm 3GHz} < 0.6$~Jy. The flux density distribution of the IPS-VLBI sample
at low frequency (162~MHz) is consistent with that of the parent sample. However, the peak of the flux density distribution of the selected sample at 3~GHz is shifted toward higher flux densities compared to the parent sample. This is expected as the VLBA Calibrator Surveys were designed to detect compact sources that are brighter than $\sim$100~mJy at 1.4~GHz at arcsec scales if the spectral index $\ge -0.5$, otherwise brighter than $\sim$200~mJy. However, for the southern sky,  the flat-spectrum sources with 1.4~GHz flux density $\ge$200~mJy at arcsec scales were mainly observed using the VLBA in snapshot mode. In the parent sample, 81 IPS sources are fainter than 100~mJy at 3~GHz, and 152 sources are fainter than 200 mJy. This indicates at least one-third of sources the parent sample are not observed by the VLBA because of this flux density limitation.

\begin{figure*}
\includegraphics[scale=0.36]{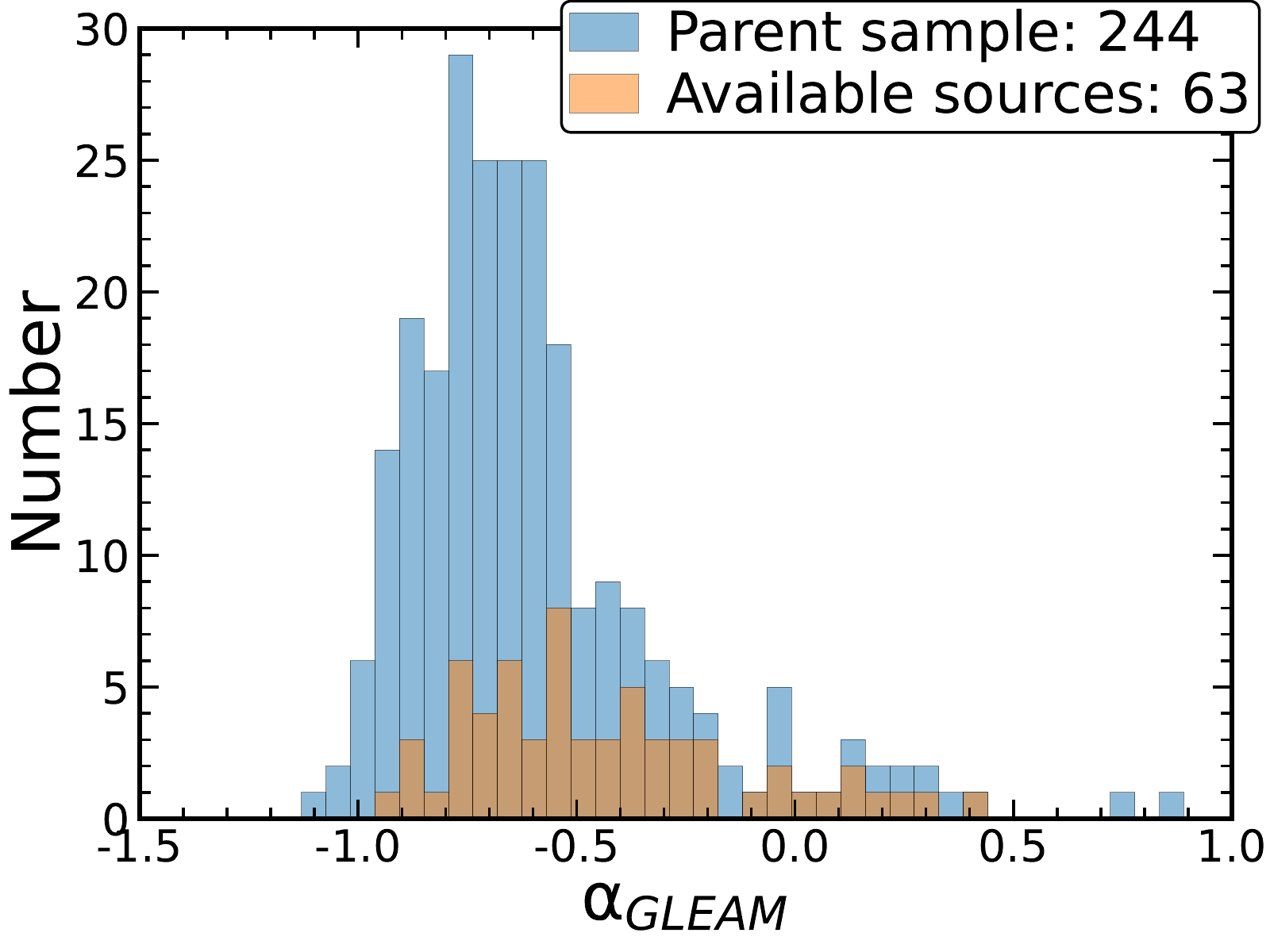}
\includegraphics[scale=0.36]{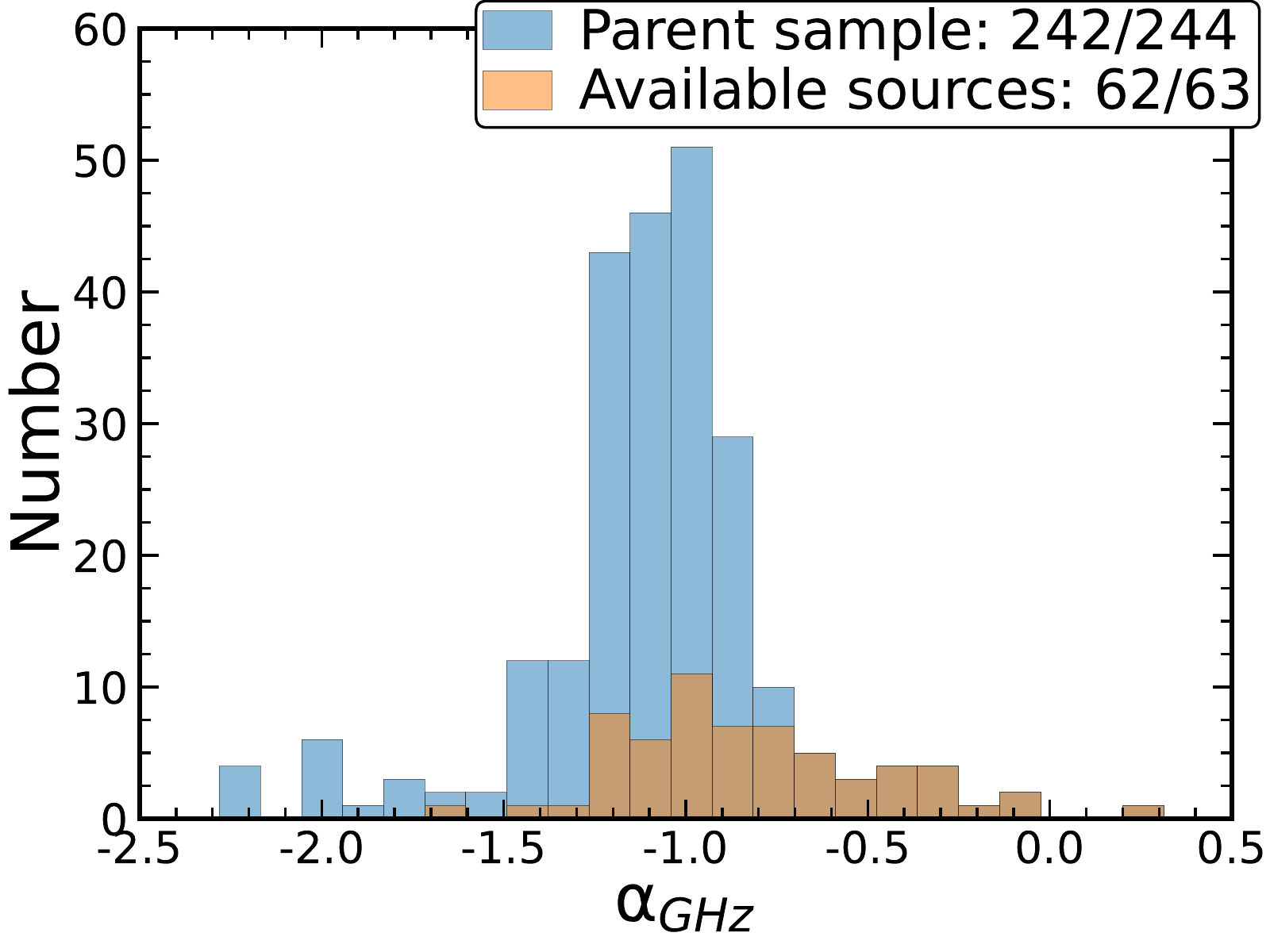}
\includegraphics[scale=0.36]{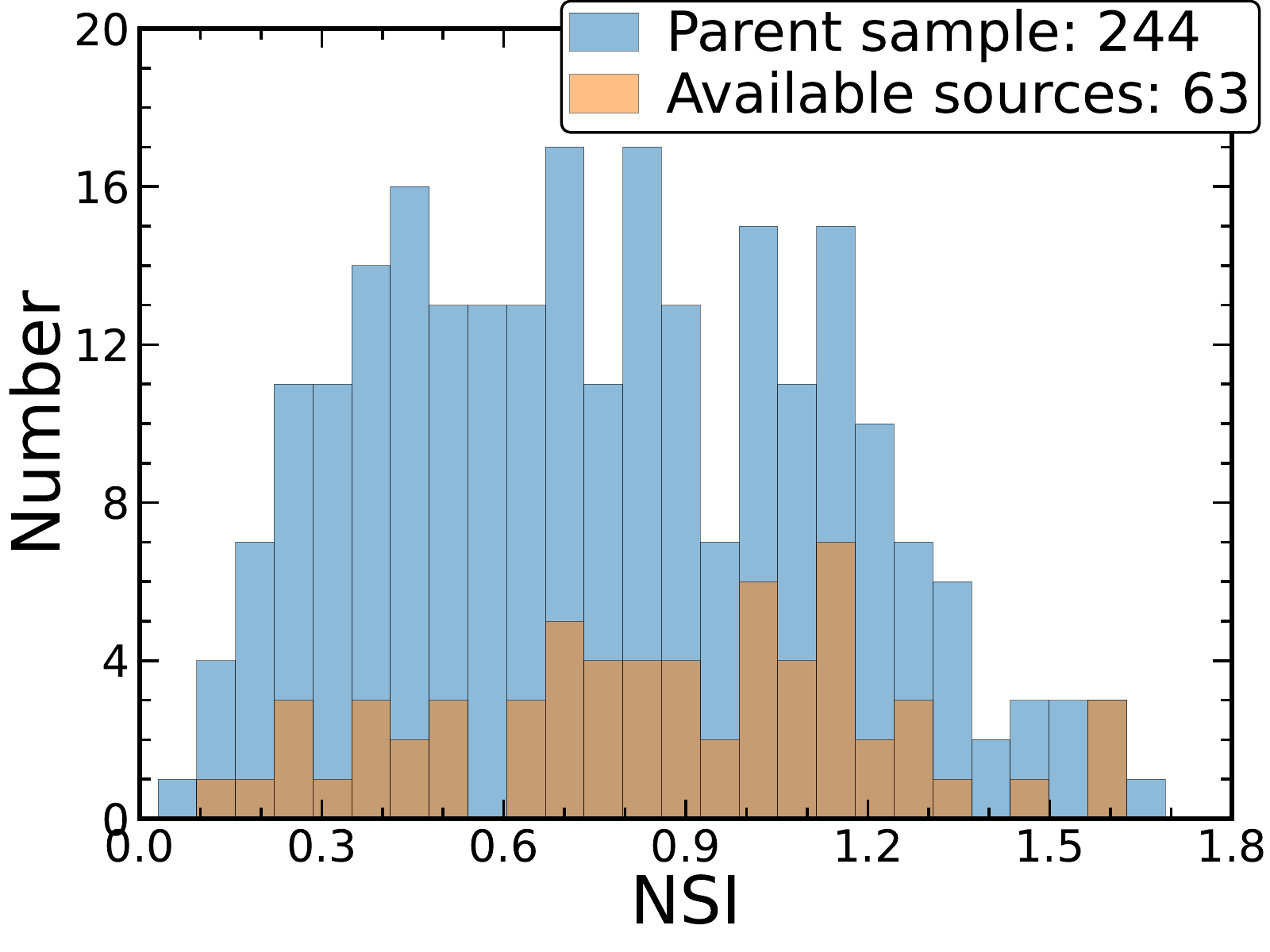}
\caption{The distributions of spectral index at MHz frequencies from the GLEAM (\textit{left}), spectral index at GHz frequencies from the NVSS and VLASS (\textit{middle}), and normalized scintillation index (NSI) of the interplanetary scintillation sample (\textit{right}).}
\label{fig:hist-2}
\end{figure*}

Figure~\ref{fig:hist-2} shows  $\alpha_{\rm GLEAM}$, $\alpha_{\rm GHz}$ and NSI values for the VLBI-detected sample and the parent sample. The estimated spectral indices in the GLEAM bands range from $-1.1$ to $1.0$. The histogram of $\alpha_{\rm GLEAM}$ of the parent sample shows a peak around $\alpha = -0.7$, a typical value of optically thin steep spectrum; while there is a secondary peak around $\alpha = -0.4$. The histogram of $\alpha_{\rm 
GHz}$ of the parent sample shows a narrow peak around $\alpha = -1.0$, steeper than the peak of the $\alpha_{\rm GLEAM}$ histogram. 
$\alpha_{\rm GLEAM}$ of both the VLBI-detected and parent samples shows a relatively flatter distribution, but the relative proportion of VLBI-detected flat-spectrum sources in the parent sample is higher than that of steep-spectrum sources, which is consistent with the bias of VLBI observations towards flat-spectrum radio-loud AGN. 
The NSI values of the VLBI-detected sample sources are evenly distributed over the NSI range of 0.1--1.7, and similarly, the parent sample shows a flat distribution of NSI values between 0.2 and 1.4. However, the available 63 VLBI sources show a weak tendency to cluster towards higher NSI values compared to the parent sample; this of course needs to be confirmed with more data. 
Among the 63 VLBI sources, 31 are strongly scintillating sources with NSI~$\ge 0.9$, 25 moderately scintillating sources with $0.9 >$~NSI~$\ge 0.4$ and 7 weakly scintillating sources with $0.4 >$~NSI~$\ge 0.1$.



\section{VLBI data analysis}\label{sec:data}
The calibrated VLBI archival data for the $S$ (2.3 GHz), $C$ (4.3 GHz), $X$ (7.6 and 8.7 GHz) and $U$ (15.3 GHz) bands for the 63 sources were obtained from the Astrogeo database. The calibration had been performed with the VLBI data processing software \textsc{Pima} \citep{2011AJ....142...35P}. The calibrated data were further analysed using the Caltech \textsc{Difmap} software package \citep{1994BAAS...26..987S} to deal with any residual phase errors, imaging, and parameter estimation. If the on-source time exceeded 1~minute, any data with significant scattering were averaged in time over 30~seconds. Any bad data points due to the presence of radio frequency interference and/or any antenna not working properly were identified and flagged. The edited visibility data were then imaged and self-calibrated following standard procedures. Final images were generated using self-calibrated data by applying natural weighting. Deconvolution of the images used the CLEAN algorithm \citep{1974A&A....33..289H}. In the final images, the CLEAN procedure was performed down to three times the residual noise. In many cases, the on-source time was less than 1 minute, so CLEAN boxes were carefully used during the processing of such data with poor (u,v)-coverage. It is important to note that the VLBI data with poor (u,v)-coverage caused strong side-lobes and prevent deep CLEANing. The observing and imaging parameters are given in Table~\ref{tab:image}, which lists the quality of the archival data in the different bands and epochs. Based on the availability of the data, we used single-epoch data at $S$, $C$ and $U$ bands, and multiple-epoch data at $X$ band for this study. As the archive data were observed with different on-source times and at different frequency bandwidths in different epochs, the final images have different sensitivities.

By using the {\sc modelfit} task in {\sc Difmap}, we fitted circular Gaussian models to the self-calibrated visibility data until no structure with brightness higher than 5$\sigma$ appears in the residual image and the amplitude radial profile of the model matches the visibility data. We assume the brightest component to be the core and other prominent components as jet components. The integrated flux densities and angular sizes of the source components are given in Table~\ref{tab:fitting}. The parameters derived from fitting using circular Gaussian models are consistent with those obtained by using elliptical Gaussian models. When the fitted Gaussian size $d$ of a component was found to be smaller than the minimum resolvable size $d_{\rm min}$ estimated using the \citet{2005astro.ph..3225L} relation, we used $d_{\rm min}$ as the upper limit for the component size.
The uncertainties in the model fitted parameters were estimated using the relations given in \citet{2012A&A...537A..70S}, which were based on the approximations introduced by \citet{1999ASPC..180..301F} and were modified for the strong sidelobes in VLBI measurements. In addition to the random statistical error, a systematic amplitude calibration  error of 5 per cent of the flux density value was also included in the final flux density error. 

\section{Results}\label{sec:results}

\subsection{Morphology}\label{sec:results:morpho}
Figure~\ref{fig:morph} shows the $X$- or $C$-band VLBI images of the IPS sources obtained with the most sensitive data available. These sources can be classified into four categories based on their VLBI morphologies: single compact core (SC), core and one-sided jet (CJ), core and two-sided jets (CdJ), and core with extended emission (CE). Note that the classification of sources into different morphological groups is dependent on the detection sensitivity and resolution of the VLBI images. If observed at higher resolutions, some SC sources would be resolved into CJ or CdJ structures. 
\begin{itemize}
    \item The images include 25 SC sources and 31 CJ sources detected with the sensitivity of the VLBI data.
    \item Four sources are identified in the category of CdJ (J002549$-$260211, J003829$-$211957, J011651$-$205202 and J014237$-$074232). The VLBI $X$-band image of the source J002549$-$260211 denotes one of the jet components of this compact symmetric object (CSO) whose other jet component, as identified in the $S$-band VLBI image, is located at $\sim0.01^\prime$ away to the southeast. 
    \item  Three sources (J000905$-$205630, J002125$-$005540 and J011857$-$071855) are classified as CE sources.  Some sources may have extended arcsec-scale structure which is resolved in VLBI images.
\end{itemize}

\subsection{Spectral index}\label{sec:results:spix}
The spectral indices of VLBI cores, $\alpha$  (defined as $S_\nu \propto \nu^{+\alpha}$), are estimated using contemporaneous dual-frequency VLBI observations in the $C$ and $X$ bands (see Table~\ref{tab:fitting}). If a source does not have contemporaneous $C$- and $X$-band data, its spectral index is measured using $S$- and $X$-band data instead. 19 sources have only single C-band data and their spectral indices are not calculated. Of the other 44 sources, 23 are found to have flat spectra ($-0.5 \ge \alpha \le +0.5$), 18 have steep spectra ($\alpha < -0.5$), and 3 are identified as inverted spectrum sources ($\alpha > +0.5$). However, the MWA spectral indices given in Table~\ref{tab:sample} show 29 flat-spectrum and 34 steep-spectrum sources in the IPS sample. The flat-spectrum VLBI sources are all flat-spectrum MWA sources, suggesting that the radio emission from these sources is dominated by compact, pc-scale radio structure. 

\subsection{Brightness temperature}\label{sec:results:temp}
The brightness temperatures, $T_{\rm B}$ (see Table~\ref{tab:fitting}), of VLBI cores are estimated for the 31 sources whose redshift information is known, using the \citet{1988gera.book..563K} relation:
\begin{equation}
T_{\rm B} = 1.22 \times 10^{12}~{\rm K} \left(\frac{S_{\rm total}}{\mathrm{Jy}}\right) \left(\frac{d}{\mathrm{mas}}\right)^{-2} \left(\frac{\nu}{\mathrm{GHz}}\right)^{-2} (1+z),
\end{equation}
where $S_{\rm total}$ is the total flux density of the source component, $d$ its size (full width at half maximum, FWHM), $\nu$ the observing frequency, and $z$ the redshift. The relationship between the Doppler factor $\delta$ and the apparent brightness temperatures $T_{\rm B}$ is $T_{\rm B} = \delta \times T_{\rm int}$, where $T_{\rm int}$ is the intrinsic brightness temperature of the source in the source's rest frame. The intrinsic brightness temperature is given as the equipartition brightness temperature $\sim 5 \times 10^{10}$~K \citep{1994ApJ...426...51R} or the inverse Compton limit $\sim 10^{11}$~K \citep{1969ApJ...155L..71K}. Based on the estimates of the core brightness temperatures,  13 sources have $T_{\rm B} > 10^{11}$ K which are attributed to Doppler boosting, consistent with their blazar identification (see Section~\ref{sec:discuss:SED}).

\subsection{Variablity}\label{sec:results:variability}
The variability of the IPS sources with well-separated multi-epoch VLBI observations in the $X$-band is investigated. The variability index (VI), defined as $(S_{\rm max} - S_{\rm min})/(S_{\rm max} + S_{\rm min})$, where $S_{\rm max}$ and $S_{\rm min}$ are the total flux densities at the epoch of maximum and minimum emission, respectively, indicates the variable nature of these sources on the pc scale.
The variability index measurements of 19 IPS sources using available multi-epoch VLBI data indicate that a large fraction of these sources are highly variable in nature and are consistent with their blazar identification. Except for J003829$-$211957 (with a CSO-type morphology and a GPS-type spectrum), the other 18 sources with available multi-epoch X-band VLBI data have a variability index $\ge 0.1$, with 12 of them having variability indices $\ge 0.2$. 
In addition to the variability of the parsec-scale radio emission, these sources may also have spectral variability. Recently, \citet{2021MNRAS.501.6139R} investigated the GaLactic and Extragalactic All-sky Murchison Widefield Array (GLEAM) year 1 and year 2 data and found 323 sources with significant spectral variability on timescales up to one year. They speculate that this variability could be due to interstellar scintillation or jet evolution. We need to note that the variability indices obtained from the VLBI measurements represent intrinsic variability on smaller spatial scales and longer timescales than the variability caused by the IPS. Therefore, these two variability mechanisms are independent of each other.

\section{Discussion}\label{sec:discuss}

\subsection{Effect of compactness on scintillation}\label{sec:discuss:compactness}
\begin{figure}
    \centering
    \includegraphics[scale=0.41]{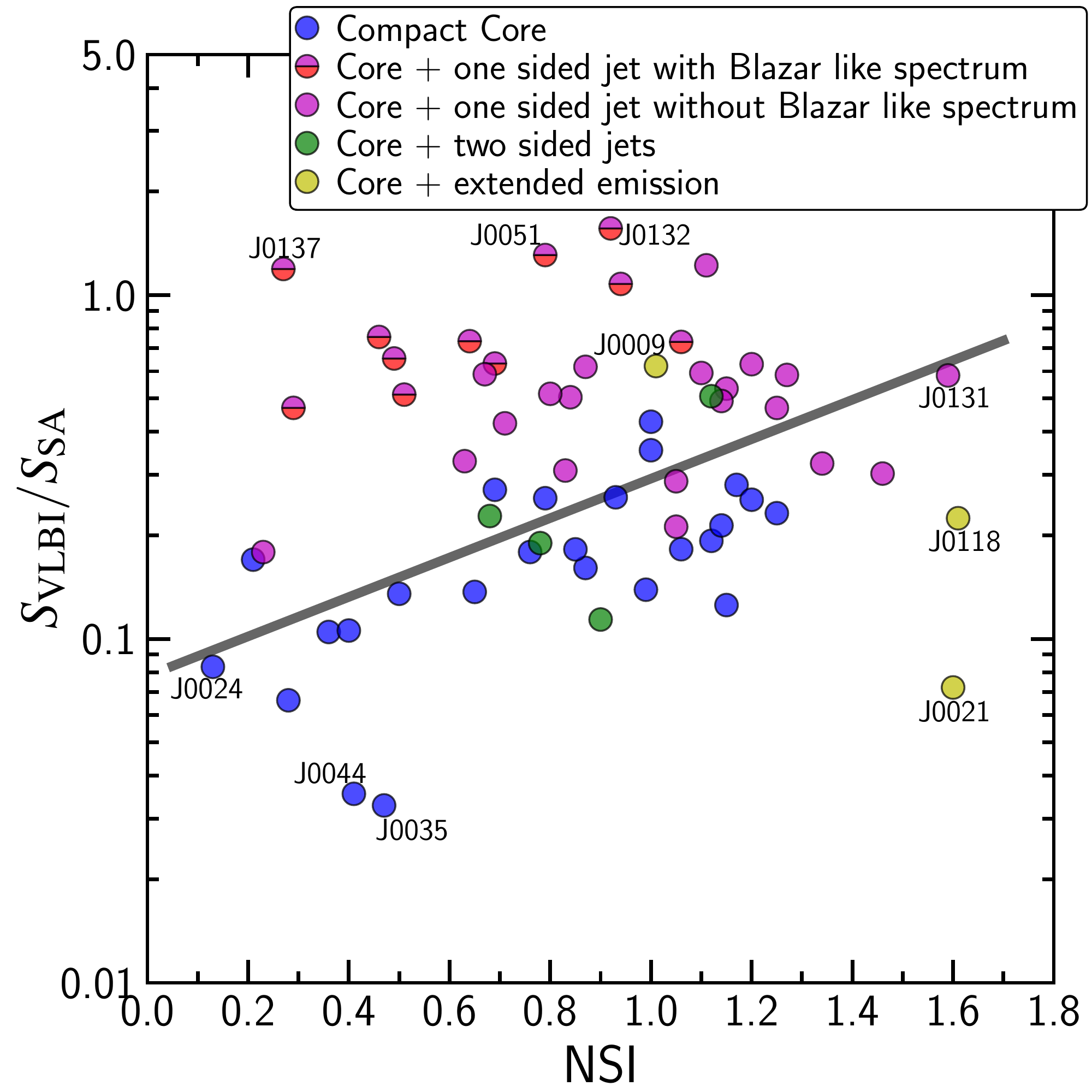}
    \caption{Variation of $S_{\rm VLBI} / S_{\rm SA}$ with normalized scintillation index  for different morphological classes. 25 `compact core', 31 `core + one sided jet', 4 `core + two sided jets' and 3 `core + extended emission'.  }
    \label{fig:fluxratio_nsi_1}
\end{figure}

\begin{figure}
    \centering
    \includegraphics[scale=0.41]{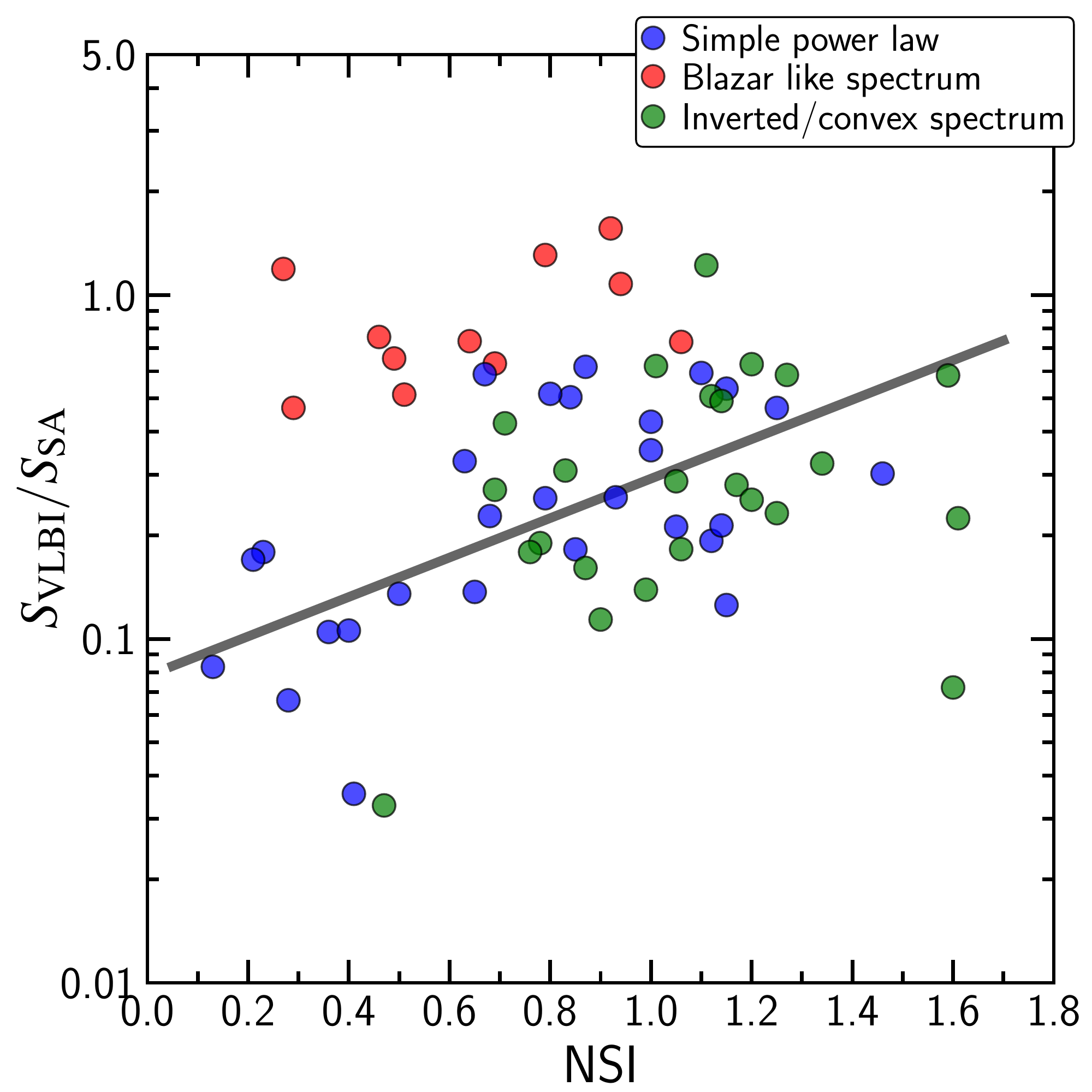}
    \caption{Variation of $S_{\rm VLBI} / S_{\rm SA}$ with normalized scintillation index for  different spectral classes. 28 `simple power law', 24 `inverted/convex spectrum', 11 `blazar like spectrum'.}
    \label{fig:fluxratio_nsi_2}
\end{figure}

Morphological analysis of the IPS sources (see Section~\ref{sec:results:morpho}) shows that most of these sources contain compact components. The compactness of the sources can be further analysed by measuring the ratio of the pc-scale flux density to the total flux density. The former can be expressed by the 4.3-GHz VLBI flux density, and the latter as the short baseline synthesis array (SA) measurement. The 4.3-GHz synthesis array flux density is estimated using 1.4-GHz flux density from NVSS and 3-GHz flux density from VLASS.
If 4.3-GHz VLBI data are not available, we use the 8.4 GHz flux density and the spectral index (Figure~\ref{fig:sed}) to estimate the 4.3 GHz value. This ratio $R = S_{\rm VLBI}/S_{\rm SA}$ denotes the compactness of the source if the source variability can be neglected. The ratio $R>1$ can be explained in terms of the variability. For example, two sources, J005108$-$065001 (J0051, in short) and  J013243$-$165444 (J0132, in short), have $R$ above 1 and are classified as blazars and they show a compact core and one-sided jet (Figures \ref{fig:fluxratio_nsi_1} and \ref{fig:fluxratio_nsi_2}). If $R$ is close to unity, that means that the source is compact and if $R$ is much less than unity, it indicates a significant contribution of extended diffuse emission. We found one source in the bottom-left corner, J003508$-$200354 (J0035, $R = 0.033$), showing a resolved Fanaroff-Riley (FR) type II morphology in the Very Large Array image \citep{1998ApJS..118..275K}. Although two radio lobes dominate the emission on arcsecond scales, they are resolved in VLBI images, leaving only a weak compact core. Imaging observations of another source J004441$-$353029 (J0044, $R = 0.035$) near J0035 in Figure \ref{fig:fluxratio_nsi_1} at sub-arcsec resolution should also reveal a resolved extended jet structure.

\begin{figure}
    \centering
    \includegraphics[scale=0.41]{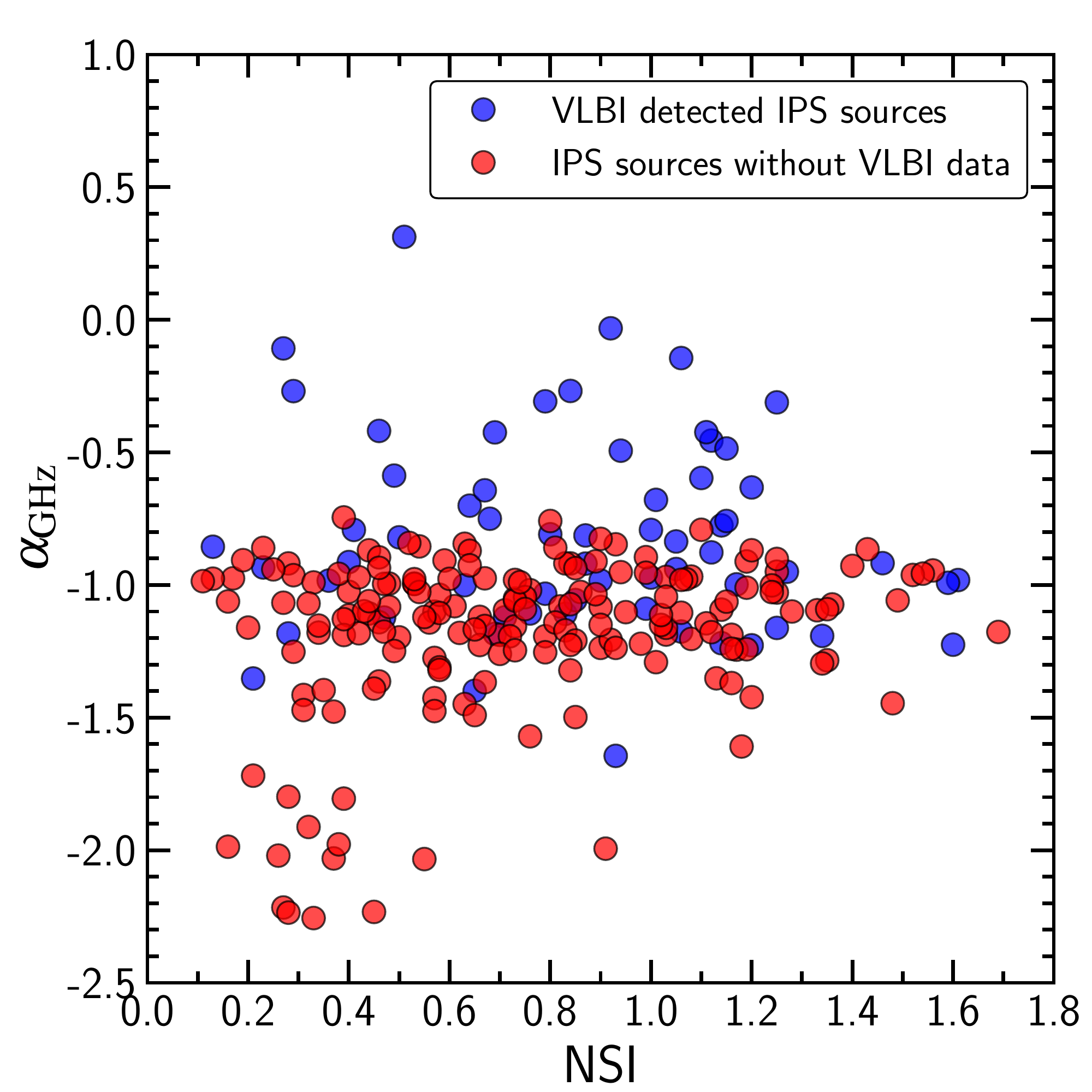}
    \caption{Variation of $\alpha_{\rm VLBI}$ with normalized scintillation index.}
    \label{fig:spix_nsi_1}
\end{figure}

The variation of the source compactness parameter $R$ with the normalized scintillation index is shown in Figure~\ref{fig:fluxratio_nsi_1} (note the log scale vertically). The data points are shown in different colors corresponding to four VLBI morphological types (see Section~\ref{sec:results:morpho}).
Figure~\ref{fig:fluxratio_nsi_2}, on the other hand, shows the variation of $R$ with the normalized scintillation index for different spectrum classes (see Section~\ref{sec:discuss:SED} for detailed discussion). The following inferences can be made from these plots:
\begin{itemize}
    \item 
Three sources, J000905$-$205630 (J0009), J002125$-$005540 (J0021), and J011857$-$071855 (J0118) show a compact core and diffuse emission in VLBI images. Both J0009 and J0118 show inverted spectra, while J0021 show a broken power law spectrum. J0009 has a high $R$ ($=0.62$) and a higher turnover frequency ($\sim$900 MHz), suggesting that its overall radio structure is compact ($<0.3\arcsec$) and that the VLBI core dominates the total emission. In contrast, J0118 has a high $\mathrm{NSI} = 1.6$, but a lower $R = 0.22$ and a lower turnover frequency (around 300 MHz), indicating that the source structure is compact but the emission is dominated by sub-arcsec structures. These implications are consistent with their VLBI morphologies shown in Figure~\ref{fig:morph}.
    \item
The spectra of four sources (J002549$-$260211, J003829$-$211957, J011651$-$205202 and  J014237$-$074232), morphologically classified as CdJ, are all of inverted shape. From their triple VLBI morphology and overall peaked-spectrum type, we consider them as CSOs, belonging to a class of young radio sources in the earliest transitional stage toward large-size symmetric objects (\textit{i.e.}, Fanaroff-Riley type 2 galaxies) \citep[e.g.,][]{1996ApJ...460..634R,2012ApJ...760...77A}.  
    \item 
As can be seen in the Figure~\ref{fig:fluxratio_nsi_1},  there is a positive correlation between the compactness parameter $R$ and the normalized scintillation index of the sources with a correlation coefficient $r=0.47$, with the exception of the CJ sources with blazar-like spectrum and two core+extended emission sources (J0118 and J0021) as the scatters, indicating that weakly scintillating (smaller NSIs) sources generally have larger sizes and strongly scintillating sources are more compact. The scatter in this correlation could be caused by the intrinsic variability of these sources (see Section~\ref{sec:results:variability}). The typical uncertainties in $R$ and NSI values are $\sim$15 per cent and 10 per cent, respectively. The interplanetary scintillation index shows a power-law dependence on the radial distance of scattering screen and the observing frequency \citep[e.g.,][]{2019PASA...36....2M}. It is therefore reasonable to fit a power-law model on the compactness-NSI graph.
The fitted power law has the form: ${\rm log}(S_{\rm VLBI}/S_{\rm SA}) = 0.57~{\rm NSI} - 1.11$. The rms of the residual values of the fit is 0.26, characterising the dispersion of the correlation due to the intrinsic variability of the sources. The spectrum analysis is given in the Section~\ref{sec:discuss:SED}.
After examining the VLASS and the Faint Images of the Radio Sky at Twenty-Centimeters (FIRST)\footnote{\url{http://sundog.stsci.edu/}} image archives, we found that the SC sources with $NSI > 0.6$  have compact and unresolved structure on arcsecond scales, and the VLBI core accounts for 10--30 per cent of the total flux density. 
However, sources with $NSI < 0.6$ have large extended structures with the VLBI core contributing to  $\lesssim$10 per cent of the total flux density. The minimal NSI is associated with J002430$-$292847 (J0024) which shows a compact core in the VLBI images but an FR-II morphology with a size of $\sim$50~arcsec in the VLASS image. Its overall size might be the largest among this IPS sample.
These results are consistent with the findings of \citet{2019PASA...36....2M} who have shown that sources smaller than 0.3~arcsec in size exhibit scintillation like a point source, while larger sources have lower scintillation indices than  point sources. 
    \item 
    The NSI distribution of the CJ sources is broad, with no clear indication of correlation with $R$. The lack of correlation is due to their complex composition, consisting of blazars, peaked-spectrum sources and steep-spectrum sources.
    The flatter the spectral index, the more compact the source structure. The variation of the GHz spectral index with the normalised scintillation index shows a random distribution, with VLBI-detected sources having flatter spectra (see Figure~\ref{fig:spix_nsi_1}).
\end{itemize}

\subsection{Radio spectrum analysis}\label{sec:discuss:SED}
We have plotted the spectrum ($S_\nu$ vs $\nu$) of each IPS source using VLBI based flux densities, and GLEAM- and NED-based flux densities, and fitted them separately (see Figure~\ref{fig:sed}). We used the closest (if possible) epoch multi-band VLBI measurements in the spectrum plots to minimise the source variability. However, in many cases the closest epoch multi-band observations are not available and the variability effect may have an important influence in the spectrum. The fitted lines are found to cross each other in most cases, indicating that the spectral index from the VLBI data is different from that derived from the GLEAM and NED flux densities, which in turn suggests the pc-scale and kpc-scale characteristics of these sources are different.

We note that the spectra of 14 sources are highly inverted, showing turnovers in the range of 0.05--1~GHz. The remaining sources show steep spectral indices below 1 GHz. For sources of steep spectra without any spectral turnover, the spectrum is fitted with a simple power-law model ($S_\nu = S_0 \nu^\alpha$). If the sources have more than two frequency bands of VLBI data, the VLBI flux densities are also fitted with a power law. Free–free absorption (FFA) and synchrotron self-absorption (SSA) are the most commonly proposed mechanisms thought to be responsible for the spectral turnovers \citep{2009AN....330..120F,2012ApJS..198....5A,2017ApJ...836..174C,2021A&ARv..29....3O}. We thus attempted to fit those radio spectra with peaked spectrum models as described in \citealt{1998A&AS..131..435S} for estimating the turnover frequency.
\begin{equation}
  S_\nu = \frac{S_{\rm p}}{(1-e^{-1})} \left(1 - e^{-(\nu/\nu_{\rm p})^{\alpha_{\rm thin}-\alpha_{\rm thick}}}\right)\left(\frac{\nu}{\nu_{\rm p}}\right)^{\alpha_{\rm thick}}
\end{equation}
where $S_p$ is the peak flux density at the turnover frequency $\nu_p$, and $\alpha_{\rm thick}$ and $\alpha_{\rm thin}$ are the spectral indices in the optically thick and optically thin regimes of the spectrum, respectively. This model fits the spectrum with power laws in the optically thick and thin regimes separately, without considering whether the underlying absorption mechanism is SSA or FFA. The spectra, along with the fitted models, are shown in Figure~\ref{fig:sed} and relevant fitting parameters are given in Table~\ref{tab:sedfit}. We used a similar spectral fitting for the 9 `broken power law' sources having spectral break in place of the spectral turnover. 

The spectra of 11 sources show prominent variability at GHz frequencies, and are identified as blazars.  The blazar identifications are verified through the literature \citep[e.g.][]{2009A&A...495..691M,2011ApJS..194...29R,2014ApJS..215...14D,2014ApJS..213....3M,2018ApJS..234...12L,2019ApJS..242....4D}. All these blazars have a steep spectrum below 1 GHz, indicating that these sources much have optically-thin extended components (e.g., lobes or extended jets) which become prominent at low frequencies. Five blazars with $\mathrm{NSI}<0.6$ are detected with extended jet and lobe structures. In the VLBI images, all these blazars show either compact core or core-jet structures. Their flux densities are dominated by the Doppler-boosted cores. 

Based on the variation of $R = S_{\rm VLBI}/S_{\rm SA}$ with the normalized scintillation index for different spectral classes shown in Figure~\ref{fig:fluxratio_nsi_2}, the `simple power law' sources (blue coloured symbols) and `inverted/convex spectrum' sources (green colored symbols) show a correlation (with a coefficient of $r=0.47$) between $R$ and NSI. They do not differ much in the distribution and the correlation trend, indicating that the scintillation properties of two classes are essentially similar. The only difference is that the `broken power law' sources are generally more compact with higher $R$s and higher NSIs. The `simple power law' and `convex spectrum' sources  fall in the conventional classification of radio galaxies and steep-spectrum quasars. They show a correlation between their sizes in the FIRST and VLASS images with the NSIs. All `inverted spectrum' sources have spectrum turnovers in the frequency range of 0.05--1 GHz, but their turnover frequencies are much lower than GPS sources. They can be referred to as MHz-peaked spectrum (MPS) objects. These MPS sources generally have high NSIs, suggesting that their radio structures are compact on sub-arcsecond scales, although they show diverse morphologies in VLBI images: compact core, core and one-sided jet, core and two-sided jets (Figure~\ref{fig:fluxratio_nsi_1} and Figure~\ref{fig:morph}). MPS sources with sub-arcsecond structure represent an intermediate evolutionary phase between the youngest CSOs and the evolved FR type galaxies. Combining scintillation and spectral properties provides an approach to search for radio galaxies or quasars with sub-arcsecond size and turnover frequency at tens to hundreds MHz.

\section{Summary}\label{sec:summary}
We have performed a high-resolution imaging study of 63 interplanetary scintillation sources with available VLBI data selected from the total 244 IPS sources detected in the $30\degr \times 30\degr$ MWA field-of-view at 162~MHz \citep{2018MNRAS.474.4937C,2018MNRAS.479.2318C}. The main conclusion of this study is that for the `compact core', `core + one-sided jets' and `core + two-sided jets' sources (excluding the blazar-like sources), there is a positive correlation between the source compactness and the scintillation index, indicating stronger scintillation for more compact sources. A similar positive correlation is found for the sources with simple power law or convex spectrum shapes. The scatter of these correlations might be due to the intrinsic variability and complex structure of the sources.

Observations of IPS sources at meter wavelengths with connected synthesis arrays have provided information on the presence of compact components with sizes smaller than $\sim$0.3~arcsec. Moreover, details of these compact components and how interplanetary scintillation properties vary with compactness can be obtained from VLBI studies of IPS sources. 
This comprehensive study of the radio properties of IPS sources on mas scales is the first of its kind to use the most extensive archival VLBI database. 
New VLBI observations of other IPS sources in the sample for which VLBI data are not available would be a necessary complement to improve the statistical confidence of the results presented here. Future IPS survey with the extended MWA Phase II will help to extend this study.

\section*{Acknowledgements}

We thank the referee Hayley Bignall for her valuable comments and suggestions which help to improve the quality of the manuscript. 
SJ and TA thank Rajan Chhetri for providing the catalog of 247 IPS sources used in this study.
This work is supported by the SKA grant from the National Key R\&D Programme of China (2018YFA0404603), the Chinese Academy of Sciences (CAS, 114231KYSB20170003).  SJ is supported by the CAS-PIFI (grant No. 2020PM0057) postdoctoral fellowship. TA is supported by the Youth Innovation Promotion Association of CAS. We are grateful to the Astrogeo team for making their fully calibrated VLBI FITS data publicly available. The NASA/IPAC Extragalactic Database (NED) is operated by the Jet Propulsion Laboratory, California Institute of Technology, under contract with the National Aeronautics and Space Administration. The National Radio Astronomy Observatory is a facility of the National Science Foundation operated under cooperative agreement by Associated Universities, Inc.

\section*{Data availability}

The VLBI datasets underlying this article were derived from the public domain in Astrogeo archive (\url{http://astrogeo.org/}). The IPS data can be found in the published literature. 




\bibliographystyle{mnras}
\bibliography{references} 



\appendix

\section{Tables and Figures}

\begin{table*}
\caption{The sample properties}
\label{tab:sample}
\scalebox{0.98}{
\begin{tabular}{ccccccccc}
\hline
GLEAM Name & RA & DEC & Redshift & $S_{\rm MWA}^{\rm 162MHz}$ & $\alpha_{\rm GLEAM}$ & $S_{\rm NVSS}^{\rm 1.4GHz}$ & $S_{\rm VLASS}^{\rm 3GHz}$ & NSI\\
 & ($^\mathrm{h}$:$^\mathrm{m}$:$^\mathrm{s}$) & (\degr:$'$:$''$) &  & (Jy) &  & (Jy) & (Jy) & \\
\hline
J000106-174125 & 00:01:06.49 & $-$17:41:25.20 &                & $1.214  \pm 0.021$ & $-0.53 \pm 0.13$ & $0.447 \pm 0.013$ & 0.244 & $1.00 \pm 0.13$ \\
J000125-132630 & 00:01:25.85 & $-$13:26:30.41 &                & $1.778  \pm 0.021$ & $\sim -0.19$*    & $0.446 \pm 0.013$ & 0.227 & $0.87 \pm 0.11$ \\
J000322-172708 & 00:03:22.07 & $-$17:27:08.75 & $1.465^a$      & $10.307 \pm 0.021$ & $-0.68 \pm 0.07$ & $2.415 \pm 0.072$ & 1.185 & $0.23 \pm 0.02$ \\
J000905-205630 & 00:09:05.98 & $-$20:56:30.76 & $0.910^b$      & $1.322  \pm 0.017$ & $-0.10 \pm 0.13$ & $0.545 \pm 0.016$ & 0.325 & $1.01 \pm 0.10$ \\
J001913-120958 & 00:19:13.60 & $-$12:09:58.77 &                & $2.590  \pm 0.019$ & $-0.66 \pm 0.08$ & $0.524 \pm 0.016$ & 0.280 & $0.50 \pm 0.05$ \\
J002125-005540 & 00:21:25.37 & $-$00:55:40.45 &                & $4.776  \pm 0.021$ & $-0.58 \pm 0.08$ & $1.164 \pm 0.035$ & 0.458 & $1.60 \pm 0.26$ \\
J002208-104133 & 00:22:08.87 & $-$10:41:33.23 &                & $1.774  \pm 0.019$ & $\sim 0.1$*      & $0.417 \pm 0.013$ & 0.168 & $1.34 \pm 0.12$ \\
J002223-070230 & 00:22:23.11 & $-$07:02:30.36 &                & $3.821  \pm 0.016$ & $\sim -0.55$*    & $0.530 \pm 0.016$ & 0.219 & $1.25 \pm 0.10$ \\
J002430-292847 & 00:24:30.15 & $-$29:28:47.81 & $0.406^c$      & $15.271 \pm 0.018$ & $-0.86 \pm 0.07$ & $2.923 \pm 0.095$ & 1.523 & $0.13 \pm 0.01$ \\
J002546-124724 & 00:25:46.71 & $-$12:47:24.23 &                & $1.388  \pm 0.021$ & $-0.37 \pm 0.12$ & $0.394 \pm 0.012$ & 0.218 & $1.14 \pm 0.10$ \\
J002549-260211 & 00:25:49.15 & $-$26:02:11.40 & $0.322^d$      & $18.786 \pm 0.020$ & $\sim 0.2$*      & $8.753 \pm 0.263$ &       & $0.78 \pm 0.06$ \\
J003201-085133 & 00:32:01.02 & $-$08:51:33.18 &                & $2.125  \pm 0.020$ & $\sim -0.01$*    & $0.581 \pm 0.017$ & 0.282 & $1.27 \pm 0.11$ \\
J003246-293107 & 00:32:46.19 & $-$29:31:07.25 &                & $2.632  \pm 0.016$ & $\sim -0.31$*    & $0.572 \pm 0.017$ & 0.243 & $0.71 \pm 0.06$ \\
J003332-165009 & 00:33:32.30 & $-$16:50:09.62 &                & $1.276  \pm 0.017$ & $-0.28 \pm 0.12$ & $0.341 \pm 0.010$ & 0.146 & $0.83 \pm 0.08$ \\
J003454-230328 & 00:34:54.74 & $-$23:03:28.85 &                & $0.662  \pm 0.016$ & $-0.23 \pm 0.18$ & $0.395 \pm 0.012$ & 0.322 & $0.84 \pm 0.10$ \\
J003502-083437 & 00:35:02.72 & $-$08:34:37.38 &                & $2.360  \pm 0.019$ & $-0.87 \pm 0.08$ & $0.378 \pm 0.013$ & 0.179 & $0.36 \pm 0.05$ \\
J003508-200354 & 00:35:08.64 & $-$20:03:54.11 & $\sim 0.518^e$ & $12.487 \pm 0.017$ & $-0.71 \pm 0.07$ & $1.957 \pm 0.059$ & 0.832 & $0.47 \pm 0.03$ \\
J003829-211957 & 00:38:29.77 & $-$21:19:57.41 & $0.338^b$      & $1.081  \pm 0.017$ & $0.15  \pm 0.15$ & $0.828 \pm 0.025$ & 0.585 & $1.12 \pm 0.10$ \\
J003931-111057 & 00:39:31.43 & $-$11:10:57.87 & $0.553^f$      & $1.006  \pm 0.019$ & $-0.66 \pm 0.13$ & $0.308 \pm 0.009$ & 0.195 & $1.10 \pm 0.12$ \\
J004441-353029 & 00:44:41.38 & $-$35:30:29.28 & $\sim 0.980^g$ & $8.499  \pm 0.013$ & $-0.54 \pm 0.07$ & $2.560 \pm 0.077$ & 1.400 & $0.41 \pm 0.04$ \\
J004644-052157 & 00:46:44.31 & $-$05:21:57.00 & $1.869^h$      & $2.933  \pm 0.016$ & $-0.79 \pm 0.08$ & $0.452 \pm 0.014$ & 0.206 & $0.79 \pm 0.07$ \\
J004744-121026 & 00:47:44.53 & $-$12:10:26.01 &                & $2.188  \pm 0.018$ & $-0.60 \pm 0.08$ & $0.470 \pm 0.014$ & 0.202 & $0.76 \pm 0.06$ \\
J004807-183838 & 00:48:07.57 & $-$18:38:38.28 &                & $1.340  \pm 0.015$ & $-0.56 \pm 0.10$ & $0.308 \pm 0.009$ & 0.158 & $1.12 \pm 0.09$ \\
J004858-062832 & 00:48:58.30 & $-$06:28:32.04 & $\sim 2.063^i$ & $3.059  \pm 0.015$ & $-0.79 \pm 0.08$ & $0.532 \pm 0.016$ & 0.265 & $0.40 \pm 0.04$ \\
J004954-100613 & 00:49:54.09 & $-$10:06:13.43 & $2.237^f$      & $2.257  \pm 0.018$ & $-0.31 \pm 0.09$ & $0.825 \pm 0.025$ & 0.436 & $1.05 \pm 0.08$ \\
J005026-120115 & 00:50:26.04 & $-$12:01:15.84 &                & $1.950  \pm 0.020$ & $-0.51 \pm 0.09$ & $0.588 \pm 0.018$ & 0.278 & $1.05 \pm 0.08$ \\
J005108-065001 & 00:51:08.08 & $-$06:50:01.00 & $\sim 1.975^j$ & $1.773  \pm 0.016$ & $-0.50 \pm 0.10$ & $0.904 \pm 0.027$ & 0.715 & $0.79 \pm 0.08$ \\
J005214-161712 & 00:52:14.95 & $-$16:17:12.63 &                & $0.461  \pm 0.018$ & $-0.40 \pm 0.24$ & $0.227 \pm 0.007$ & 0.157 & $1.15 \pm 0.15$ \\
J005242-215540 & 00:52:42.84 & $-$21:55:40.97 & $0.654^b$      & $2.454  \pm 0.016$ & $-0.73 \pm 0.08$ & $0.436 \pm 0.013$ & 0.195 & $0.85 \pm 0.06$ \\
J005433-195255 & 00:54:33.08 & $-$19:52:55.67 &                & $0.726  \pm 0.016$ & $-0.53 \pm 0.15$ & $0.324 \pm 0.010$ & 0.290 & $1.06 \pm 0.11$ \\
J005533-214817 & 00:55:33.90 & $-$21:48:17.60 &                & $1.237  \pm 0.015$ & $-0.94 \pm 0.10$ & $0.241 \pm 0.007$ & 0.141 & $0.64 \pm 0.06$ \\
J005551-124424 & 00:55:51.86 & $-$12:44:24.52 &                & $2.068  \pm 0.019$ & $-0.54 \pm 0.08$ & $0.324 \pm 0.010$ & 0.131 & $0.69 \pm 0.06$ \\
J005805-053952 & 00:58:05.04 & $-$05:39:52.70 & $1.246^k$      & $1.994  \pm 0.016$ & $-0.63 \pm 0.09$ & $0.742 \pm 0.022$ & 0.537 & $0.69 \pm 0.09$ \\
J010152-283118 & 01:01:52.42 & $-$28:31:18.02 & $\sim 1.600^l$ & $0.983  \pm 0.014$ & $0.29  \pm 0.13$ & $0.661 \pm 0.020$ & 0.408 & $1.20 \pm 0.12$ \\
J010335-271506 & 01:03:35.58 & $-$27:15:06.73 &                & $3.077  \pm 0.015$ & $-0.42 \pm 0.08$ & $0.568 \pm 0.017$ & 0.248 & $0.99 \pm 0.07$ \\
J010452-252052 & 01:04:52.33 & $-$25:20:52.31 &                & $3.391  \pm 0.016$ & $-0.57 \pm 0.07$ & $0.602 \pm 0.018$ & 0.246 & $1.06 \pm 0.08$ \\
J010527-182843 & 01:05:27.10 & $-$18:28:43.00 &                & $1.028  \pm 0.016$ & $-0.50 \pm 0.12$ & $0.289 \pm 0.009$ & 0.138 & $1.00 \pm 0.09$ \\
J010837-285124 & 01:08:37.97 & $-$28:51:24.88 &                & $5.040  \pm 0.014$ & $\sim 0.03$*     & $1.302 \pm 0.039$ & 0.609 & $1.17 \pm 0.09$ \\
J010938-144228 & 01:09:38.68 & $-$14:42:28.88 &                & $2.864  \pm 0.020$ & $-0.73 \pm 0.08$ & $0.424 \pm 0.013$ & 0.121 & $0.93 \pm 0.07$ \\
J011049-074142 & 01:10:49.87 & $-$07:41:42.96 & $\sim 1.776^j$ & $0.723  \pm 0.015$ & $-0.27 \pm 0.17$ & $0.846 \pm 0.025$ & 0.581 & $0.94 \pm 0.16$ \\
J011312-101419 & 01:13:12.28 & $-$10:14:19.84 & $1.323^f$      & $1.565  \pm 0.019$ & $-0.39 \pm 0.10$ & $0.542 \pm 0.016$ & 0.269 & $1.46 \pm 0.13$ \\
J011612-113610 & 01:16:12.45 & $-$11:36:10.13 & $\sim 0.670^j$ & $4.291  \pm 0.019$ & $-0.72 \pm 0.07$ & $1.785 \pm 0.054$ & 1.141 & $0.49 \pm 0.04$ \\
J011651-205202 & 01:16:51.25 & $-$20:52:02.10 & $1.410^b$      & $14.387 \pm 0.020$ & $\sim -0.02$*    & $4.091 \pm 0.123$ & 1.934 & $0.90 \pm 0.07$ \\
J011738-150750 & 01:17:38.91 & $-$15:07:50.15 &                & $0.921  \pm 0.021$ & $-0.40 \pm 0.16$ & $0.320 \pm 0.010$ & 0.179 & $1.15 \pm 0.12$ \\
J011815-012037 & 01:18:15.32 & $-$01:20:37.61 & $\sim 1.162^m$ & $5.357  \pm 0.020$ & $\sim -0.18$*    & $1.249 \pm 0.044$ & 0.490 & $1.20 \pm 0.15$ \\
J011834-184910 & 01:18:34.19 & $-$18:49:10.15 & $\sim 0.280^n$ & $5.897  \pm 0.016$ & $-0.79 \pm 0.07$ & $1.170 \pm 0.045$ & 0.417 & $0.21 \pm 0.02$ \\
J011857-071855 & 01:18:57.29 & $-$07:18:55.38 &                & $1.109  \pm 0.017$ & $-0.24 \pm 0.14$ & $0.325 \pm 0.010$ & 0.154 & $1.61 \pm 0.21$ \\
J012031-270125 & 01:20:31.49 & $-$27:01:25.11 & $\sim 0.559^o$ & $1.866  \pm 0.016$ & $-0.56 \pm 0.09$ & $0.934 \pm 0.028$ & 1.185 & $0.51 \pm 0.05$ \\
J012227-042123 & 01:22:27.82 & $-$04:21:23.33 & $1.925^p$      & $4.770  \pm 0.018$ & $-0.42 \pm 0.07$ & $1.472 \pm 0.052$ & 0.795 & $0.80 \pm 0.07$ \\
J012340-104900 & 01:23:40.73 & $-$10:49:00.00 &                & $0.631  \pm 0.018$ & $-0.30 \pm 0.18$ & $0.300 \pm 0.009$ & 0.237 & $1.25 \pm 0.21$ \\
J012614-222227 & 01:26:14.81 & $-$22:22:27.47 & $0.720^a$      & $2.567  \pm 0.016$ & $-0.76 \pm 0.08$ & $0.612 \pm 0.018$ & 0.499 & $0.29 \pm 0.03$ \\
J012730-195607 & 01:27:30.32 & $-$19:56:07.72 &                & $1.878  \pm 0.016$ & $-0.76 \pm 0.08$ & $0.359 \pm 0.011$ & 0.167 & $0.63 \pm 0.06$ \\
J013112-121058 & 01:31:12.45 & $-$12:10:58.85 &                & $0.571  \pm 0.020$ & $0.43  \pm 0.23$ & $0.374 \pm 0.011$ & 0.176 & $1.59 \pm 0.30$ \\
J013243-165444 & 01:32:43.42 & $-$16:54:44.55 & $\sim 1.020^j$ & $1.453  \pm 0.018$ & $\sim 0.25$*     & $0.830 \pm 0.025$ & 0.810 & $0.92 \pm 0.09$ \\
J013537-200844 & 01:35:37.40 & $-$20:08:44.48 & $\sim 1.144^j$ & $0.644  \pm 0.016$ & $0.16  \pm 0.2$  & $0.559 \pm 0.017$ & 0.405 & $1.11 \pm 0.18$ \\
J013737-243048 & 01:37:37.82 & $-$24:30:48.60 & $0.838^q$      & $3.760  \pm 0.016$ & $-0.77 \pm 0.07$ & $1.181 \pm 0.041$ & 1.088 & $0.27 \pm 0.03$ \\
J014127-270606 & 01:41:27.05 & $-$27:06:06.73 & $1.440^b$      & $8.202  \pm 0.017$ & $-0.67 \pm 0.07$ & $1.566 \pm 0.047$ & 0.539 & $0.65 \pm 0.05$ \\
J014237-074232 & 01:42:37.52 & $-$07:42:32.53 &                & $2.426  \pm 0.018$ & $-0.64 \pm 0.08$ & $0.606 \pm 0.018$ & 0.342 & $0.68 \pm 0.08$ \\
J015050-090126 & 01:50:50.27 & $-$09:01:26.99 &                & $8.408  \pm 0.023$ & $-0.81 \pm 0.07$ & $1.036 \pm 0.031$ & 0.421 & $0.28 \pm 0.03$ \\
J015231-141237 & 01:52:31.60 & $-$14:12:37.57 & $\sim 1.350^r$ & $3.792  \pm 0.020$ & $-0.90 \pm 0.07$ & $0.745 \pm 0.022$ & 0.541 & $0.46 \pm 0.05$ \\
J015455-204023 & 01:54:55.71 & $-$20:40:23.34 & $1.921^s$      & $2.471  \pm 0.014$ & $-0.55 \pm 0.08$ & $0.453 \pm 0.014$ & 0.179 & $1.14 \pm 0.11$ \\
J015843-141308 & 01:58:43.43 & $-$14:13:08.10 &                & $4.456  \pm 0.018$ & $-0.37 \pm 0.07$ & $1.641 \pm 0.049$ & 0.541 & $0.87 \pm 0.09$ \\
J020157-113234 & 02:01:57.14 & $-$11:32:34.17 & $0.671^q$      & $7.579  \pm 0.021$ & $-0.44 \pm 0.07$ & $2.753 \pm 0.083$ & 1.686 & $0.67 \pm 0.07$ \\
\hline
\end{tabular}}
\begin{flushleft}
$^*$ Approximate spectral index across GLEAM band estimated through a least squares fit.\\ 
References for redshifts: $^a$~\citet{1978MNRAS.185..149H}; $^b$~\citet{1996ApJS..107...19M}; $^c$~\citet{2009ApJS..184..398H}; $^d$~\citet{2008MNRAS.387..639H}; $^e$~\citet{1995MNRAS.277..553B}; $^f$~\citet{2010MNRAS.405.2302H}; $^g$~\citet{2006AJ....131..114B}; $^h$~\citet{1976ApJS...31..143W}; $^i$~\citet{1989QSO...M...0000H}; $^j$~\citet{1983MNRAS.205..793W}; $^k$~\citet{2011AJ....142..165T}; $^l$~\citet{1998ApJS..118..275K}; $^m$~\citet{2002AA...386...97J}; $^n$~\citet{1994ApJ...428...65H}; $^o$~\citet{1992NED11.R......1N}; $^p$~\citet{1994ApJ...436..678O}; $^q$~\citet{20096dF...C...0000J}; $^r$~\citet{2000AAS..144..247C}; $^s$~\citet{2011ApJ...734L..25E}. Here, $^g$ is photometric redshift.
\end{flushleft}
\end{table*}

\begin{table*}
\caption{Observing and imaging parameters}
\label{tab:image}
\scalebox{0.9}{
\begin{tabular}{ccccccccc}
\hline
Name & Observing date & Project code & Frequency (Band) & Bandwidth & On-source time & Beam$^a$ & \rm{rms} noise & Peak intensity\\
 & (yyyy-mm-dd) &  & (GHz) & (MHz) & (min) & (mas$\times$mas, \degr) & (mJy~beam$^{-1}$) & (mJy~beam$^{-1}$)\\
\hline
J000106-174125 & 2016-05-15 & bp192f1 & 4.3 (C)  & 256 & 0.9  & $20.4\times 12.8, 3.9$   & 0.19 & 32 \\
               & 2016-05-15 & bp192f1 & 7.6 (X)  & 256 & 0.9  & $11.4\times 6.7, -7.5$   & 0.18 & 20 \\
\hline
J000125-132630 & 2016-07-26 & bp192h6 & 4.3 (C)  & 256 & 0.9  & $9.0\times 2.5, -11.0$   & 0.15 & 25 \\
\hline
J000322-172708 & 2015-08-24 & bp192a3 & 4.3 (C)  & 256 & 0.9  & $4.2\times 1.7, -0.4$    & 0.30 & 93 \\
               & 2015-08-24 & bp192a3 & 7.6 (X)  & 256 & 0.9  & $2.4\times 0.9, -1.8$    & 0.30 & 99 \\
               & 2016-06-28 & s7104a1 & 7.6 (X)  & 256 & 3.0  & $3.0\times 1.2, -2.3$    & 0.20 & 88 \\
\hline
\end{tabular}}\\
$^a$ Restoring beam major and minor axes (FWHM) and the major axis position angle measured from north through east.
\end{table*}
\begin{table*}
\contcaption{Observing and imaging parameters}
\scalebox{0.9}{
\begin{tabular}{ccccccccc}
\hline
Name & Observing date & Project code & Frequency (Band) & Bandwidth & On-source time & Beam$^a$ & \rm{rms} noise & Peak intensity\\
 & (yyyy-mm-dd) &  & (GHz) & (MHz) & (min) & (mas$\times$mas, \degr) & (mJy~beam$^{-1}$) & (mJy~beam$^{-1}$)\\
\hline
J005108-065001 & 1997-05-07 & bb023   & 8.3 (X)  & 32  & 5.0  & $2.6\times 1.0, -4.9$    & 0.59 & 463 \\
               & 2011-02-20 & bl149db & 15.4 (U) & 64  & 31   & $1.3\times 0.5, -6.7$    & 0.24 & 916 \\
               & 2014-08-09 & bg219e  & 8.7 (X)  & 384 & 2.0  & $3.1\times 1.2, 19.0$    & 0.33 & 667 \\
               & 2017-06-10 & uf001k  & 2.3 (S)  & 128 & 2.6  & $9.3\times 3.5, -4.8$    & 0.64 & 408 \\
               & 2017-06-10 & uf001k  & 8.7 (X)  & 384 & 2.6  & $2.2\times 0.9, -1.0$    & 0.32 & 632 \\
               & 2018-05-01 & bs262a2 & 7.6 (X)  & 256 & 5.1  & $2.5\times 1.1, 5.3$     & 0.30 & 562 \\
\hline
\end{tabular}}
\end{table*}
\begin{table*}
\contcaption{Observing and imaging parameters}
\scalebox{0.9}{
\begin{tabular}{ccccccccc}
\hline
Name & Observing date & Project code & Frequency (Band) & Bandwidth & On-source time & Beam$^a$ & \rm{rms} noise & Peak intensity\\
 & (yyyy-mm-dd) &  & (GHz) & (MHz) & (min) & (mas$\times$mas, \degr) & (mJy~beam$^{-1}$) & (mJy~beam$^{-1}$)\\
\hline
J011612-113610 & 1997-01-11 & bf025   & 8.3 (X)  & 16  & 6.9 & $2.7\times 1.1, -1.5$    & 0.80 & 378 \\
               & 1997-05-07 & bb023   & 8.3 (X)  & 32  & 4.6 & $2.6\times 1.0, -0.5$    & 0.81 & 399 \\
               & 1999-08-02 & rdv17   & 8.6 (X)  & 32  & 3.8 & $8.6\times 5.4, 34.0$    & 0.63 & 580 \\
               & 2002-01-16 & rdv31   & 8.6 (X)  & 32  & 29  & $2.7\times 0.6, -6.7$    & 0.58 & 614 \\
               & 2005-05-12 & bp118   & 8.6 (X)  & 32  & 1.5 & $2.2\times 0.8, 0.1$     & 1.52 & 751 \\
               & 2011-06-07 & bc196t  & 8.4 (X)  & 128 & 4.8 & $2.6\times 1.2, 11.5$    & 0.47 & 569 \\
\hline
\end{tabular}}
\end{table*}
\begin{table*}
\contcaption{Observing and imaging parameters}
\scalebox{0.9}{
\begin{tabular}{ccccccccc}
\hline
Name & Observing date & Project code & Frequency (Band) & Bandwidth & On-source time & Beam$^a$ & \rm{rms} noise & Peak intensity\\
 & (yyyy-mm-dd) &  & (GHz) & (MHz) & (min) & (mas$\times$mas, \degr) & (mJy~beam$^{-1}$) & (mJy~beam$^{-1}$)\\
\hline
               & 2018-07-21 & bp222a  & 8.7 (X)  & 384 & 5.5 & $2.6\times 1.0, -7.8$    & 0.53 & 821 \\
               & 2018-08-25 & sb072a0 & 7.6 (X)  & 256 & 1.1 & $2.7\times 1.1, 6.4$     & 0.87 & 694 \\
               & 2019-07-24 & bp242a  & 4.3 (C)  & 256 & 3.1 & $5.1\times 1.9, -1.0$    & 0.36 & 973 \\
               & 2019-07-24 & bp242a  & 7.6 (X)  & 256 & 3.1 & $3.0\times 1.2, -1.4$    & 0.49 & 969 \\
\hline
J013537-200844 & 1997-07-02 & bb023   & 8.3 (X)  & 32  & 4.1 & $2.6\times 1.0, -1.5$    & 0.60 & 436 \\
               & 1997-08-27 & bb023   & 8.3 (X)  & 32  & 4.7 & $2.4\times 1.0, -4.8$    & 0.56 & 413 \\
\hline
\end{tabular}}
\end{table*}

\clearpage
\begin{table*}
\caption{Fitting parameters}
\label{tab:fitting}
\begin{tabular}{cccccccc}
\hline
Name & Frequency band & Observing date & Component & Flux density & Size & $T_{\rm B}$ & $\alpha_{\rm VLBI}$\\
     &                & (yyyy-mm-dd)   &           & (mJy)        & (mas) & ($10^{10}$ K) &     \\
\hline
J000106-174125 & C & 2016-05-15 & Core & $78 \pm 9$   & $12.97 \pm 1.95$ &     & $-1.28 \pm 0.15$\\
               & X & 2016-05-15 & Core & $38 \pm 5$   & $5.61 \pm 0.84$  &     &                 \\
\hline
J000125-132630 & C & 2016-07-26 & Core & $26 \pm 4$   & $1.01 \pm 0.15$  &     &                 \\
\hline
J000322-172708 & C & 2015-08-24 & Core & $100 \pm 8$  & $0.64 \pm 0.04$  & $3.83 \pm 0.61$ & $0.19 \pm 0.25$\\
               &   &            & Jet  & $50 \pm 7$   &                  &                 & \\ 
               & X & 2015-08-24 & Core & $111 \pm 11$ & $0.46 \pm 0.04$  & $2.78 \pm 0.51$ & \\
\hline
\end{tabular}\\
Here, one-sided single component jet is denoted as `Jet', one-sided multiple component jet is noted as `J1,J2,J3' for different components going away from the core and two-sided single component jets are `Jet1,Jet2' for east and north jets from the core. There is no source with two-sided multiple component jet.
\end{table*}
\begin{table*}
\contcaption{Fitting parameters}
\begin{tabular}{cccccccc}
\hline
Name & Frequency band & Observing date & Component & Flux density & Size & $T_{\rm B}$ & $\alpha_{\rm VLBI}$\\
     &                & (yyyy-mm-dd)   &           & (mJy)        & (mas) & ($10^{10}$ K) & \\
\hline
               & X & 1997-08-27 & Core & $112 \pm 17$ & $1.32 \pm 0.20$  & $0.15 \pm 0.05$ & \\
               &   &            & Jet  & $13 \pm 2$   &                  &                 & \\
               & X & 2014-08-09 & Core & $114 \pm 15$ & $1.06 \pm 0.12$  & $0.22 \pm 0.06$ & \\
               &   &            & Jet  & $17 \pm 2$   &                  &                 & \\
               & X & 2017-06-15 & Core & $115 \pm 17$ & $1.32 \pm 0.16$  & $0.14 \pm 0.04$ & \\
               &   &            & Jet  & $14 \pm 2$   &                  &                 & \\
\hline
\end{tabular}
\end{table*}
\begin{table*}
\contcaption{Fitting parameters}
\begin{tabular}{cccccccc}
\hline
Name & Frequency band & Observing date & Component & Flux density & Size & $T_{\rm B}$ & $\alpha_{\rm VLBI}$\\
     &                & (yyyy-mm-dd)   &           & (mJy)        & (mas) & ($10^{10}$ K) & \\
\hline
               & X & 2018-08-25 & Core & $617 \pm 36$ & $<0.05$          & $>1411$         & \\
               &   &            & J1   & $125 \pm 9$  &                  &                 & \\
               &   &            & J2   & $66 \pm 9$   &                  &                 & \\
               &   &            & J3   & $33 \pm 5$   &                  &                 & \\
               & C & 2019-07-24 & Core & $448 \pm 28$ & $0.122 \pm 0.008$& $578 \pm 63$    & \\
               &   &            & J1   & $206 \pm 14$ &                  &                 & \\
\hline
\end{tabular}
\end{table*}
\begin{table*}
\contcaption{Fitting parameters}
\begin{tabular}{cccccccc}
\hline
Name & Frequency band & Observing date & Component & Flux density & Size & $T_{\rm B}$ & $\alpha_{\rm VLBI}$\\
     &                & (yyyy-mm-dd)   &           & (mJy)        & (mas) & ($10^{10}$ K) & \\
\hline
               & X & 2013-06-08 & Core & $541 \pm 51$ & $0.57 \pm 0.04$  & $7.77 \pm 1.40$ & \\
               &   &            & Jet  & $31 \pm 5$   &                  &                 & \\
               & X & 2013-12-11 & Core & $401 \pm 45$ & $0.49 \pm 0.05$  & $6.17 \pm 1.37$ & \\
               &   &            & Jet  & $30 \pm 5$   &                  &                 & \\
               & X & 2014-02-23 & Core & $474 \pm 47$ & $0.47 \pm 0.04$  & $10.00 \pm 1.96$& \\
               &   &            & Jet  & $14 \pm 2$   &                  &                 & \\
\hline
\end{tabular}
\end{table*}
\begin{table*}
\contcaption{Fitting parameters}
\begin{tabular}{cccccccc}
\hline
Name & Frequency band & Observing date & Component & Flux density & Size & $T_{\rm B}$ & $\alpha_{\rm VLBI}$\\
     &                & (yyyy-mm-dd)   &           & (mJy)        & (mas) & ($10^{10}$ K) & \\
\hline
               & X & 1997-05-07 & Core & $439 \pm 49$ & $0.54 \pm 0.05$  & $4.39 \pm 0.98$ & \\
               &   &            & J1   & $147 \pm 20$ &                  &                 & \\
               &   &            & J2   & $36 \pm 5$   &                  &                 & \\
               & X & 1999-08-02 & Core & $595 \pm 49$ & $0.81 \pm 0.05$  & $2.46 \pm 0.38$ & \\
               &   &            & J2   & $94 \pm 14$  &                  &                 & \\
               & X & 2002-01-16 & Core & $571 \pm 37$ & $0.15 \pm 0.01$  & $72.64 \pm 7.65$& \\
\hline
\end{tabular}
\end{table*}
\begin{table*}
\contcaption{Fitting parameters}
\begin{tabular}{cccccccc}
\hline
Name & Frequency band & Observing date & Component & Flux density & Size & $T_{\rm B}$ & $\alpha_{\rm VLBI}$\\
     &                & (yyyy-mm-dd)   &           & (mJy)        & (mas) & ($10^{10}$ K) & \\
\hline
               & X & 2000-10-23 & Core & $375 \pm 42$ & $0.70 \pm 0.07$  & $1.93 \pm 0.43$ & \\
               &   &            & Jet  & $35 \pm 5$   &                  &                 & \\
               & X & 2002-01-31 & Core & $225 \pm 26$ & $0.33 \pm 0.03$  & $5.14 \pm 1.21$ & \\
               &   &            & Jet  & $303 \pm 45$ &                  &                 & \\
               & X & 2003-03-12 & Core & $270 \pm 41$ & $0.50 \pm 0.08$  & $2.71 \pm 0.91$ & \\
               &   &            & Jet  & $284 \pm 43$ &                  &                 & \\
\hline
\end{tabular}
\end{table*}
\begin{table*}
\contcaption{Fitting parameters}
\begin{tabular}{cccccccc}
\hline
Name & Frequency band & Observing date & Component & Flux density & Size & $T_{\rm B}$ & $\alpha_{\rm VLBI}$\\
     &                & (yyyy-mm-dd)   &           & (mJy)        & (mas) & ($10^{10}$ K) & \\
\hline
               & X & 2014-01-29 & Core &$1640 \pm 115$& $0.56 \pm 0.03$  & $21.85 \pm 2.62$& \\
               &   &            & J1   & $39 \pm 6$   &                  &                 & \\
               &   &            & J2   & $78 \pm 12$  &                  &                 & \\
               & X & 2015-08-07 & Core & $721 \pm 61$ & $1.02 \pm 0.07$  & $2.95 \pm 0.45$ & \\
               &   &            & J1   & $11 \pm 2$   &                  &                 & \\
               &   &            & J2   & $77 \pm 12$  &                  &                 & \\
\hline
\end{tabular}
\end{table*}
\begin{table*}
\contcaption{Fitting parameters}
\begin{tabular}{cccccccc}
\hline
Name & Frequency band & Observing date & Component & Flux density & Size & $T_{\rm B}$ & $\alpha_{\rm VLBI}$\\
     &                & (yyyy-mm-dd)   &           & (mJy)        & (mas) & ($10^{10}$ K) & \\
\hline
               & S & 1999-06-21 & Core & $394 \pm 29$ & $0.95 \pm 0.08$  & $18.72 \pm 2.64$& \\
               &   &            & Jet  & $109 \pm 13$ &                  &                 & \\
               & X & 1999-06-21 & Core & $289 \pm 22$ & $0.33 \pm 0.03$  & $7.74 \pm 1.12$ & \\
               &   &            & Jet  & $171 \pm 16$ &                  &                 & \\
\hline
J014127-270606 & C & 2015-08-31 & Core & $44 \pm 7$   & $15.65 \pm 2.35$ &$0.003 \pm 0.001$& \\
\hline
J014237-074232 & C & 2015-08-18 & Core & $59 \pm 6$   & $5.86 \pm 0.64$  &                 & $-1.53 \pm 0.14$\\
\hline
\end{tabular}
\end{table*}

\clearpage
\begin{figure*}
\centering
\begin{tabular}{cccc}
\includegraphics[scale=0.24]{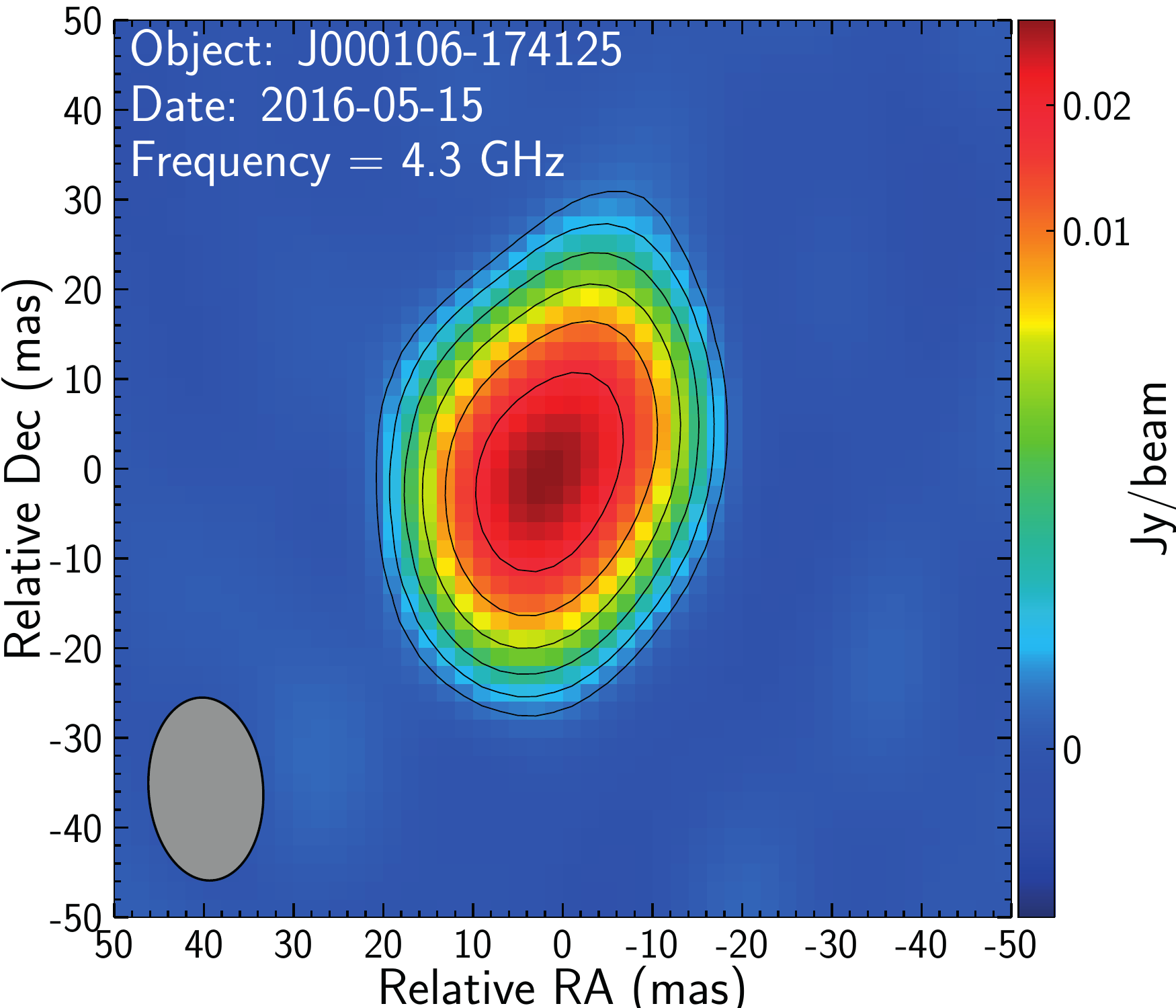}&
\includegraphics[scale=0.24]{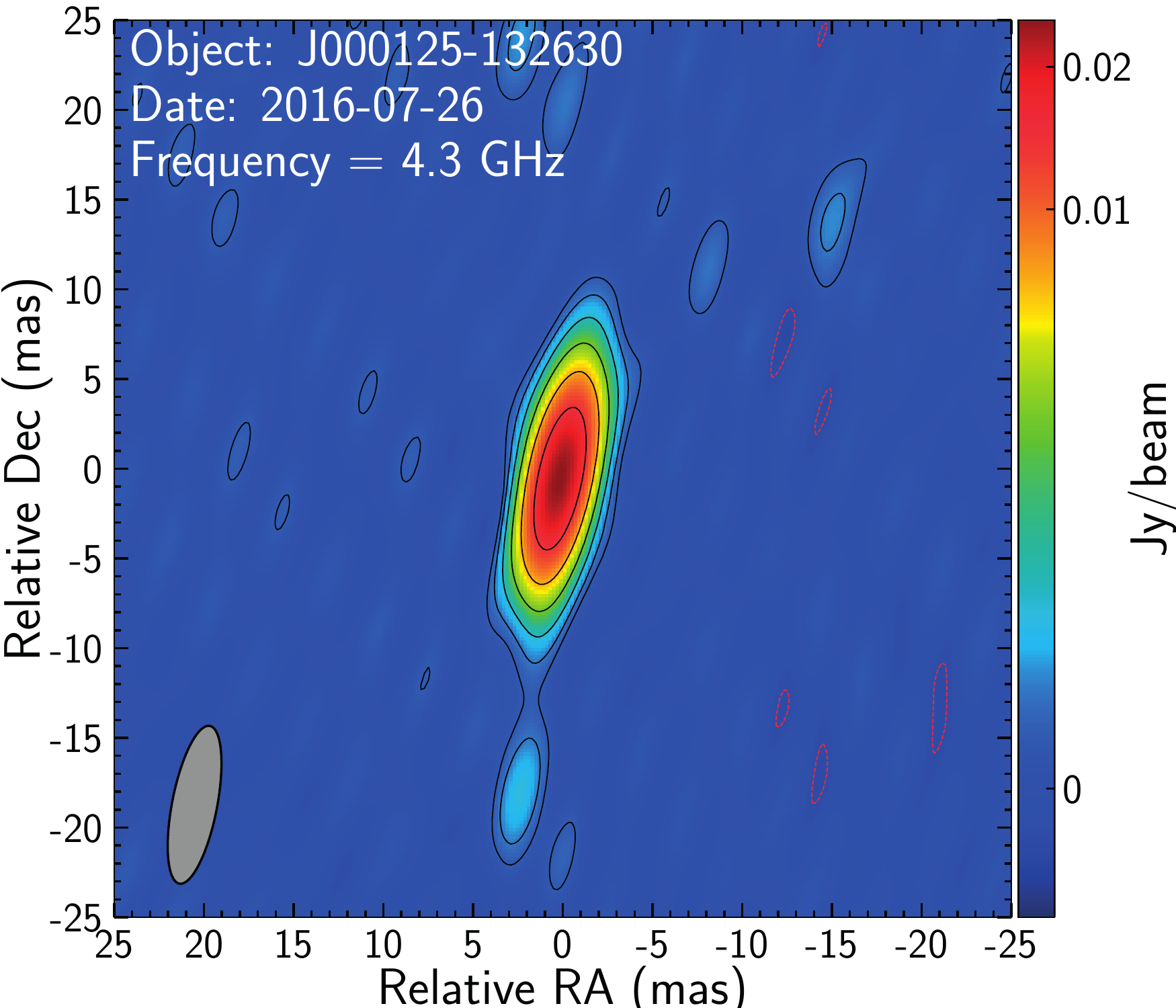}&
\includegraphics[scale=0.24]{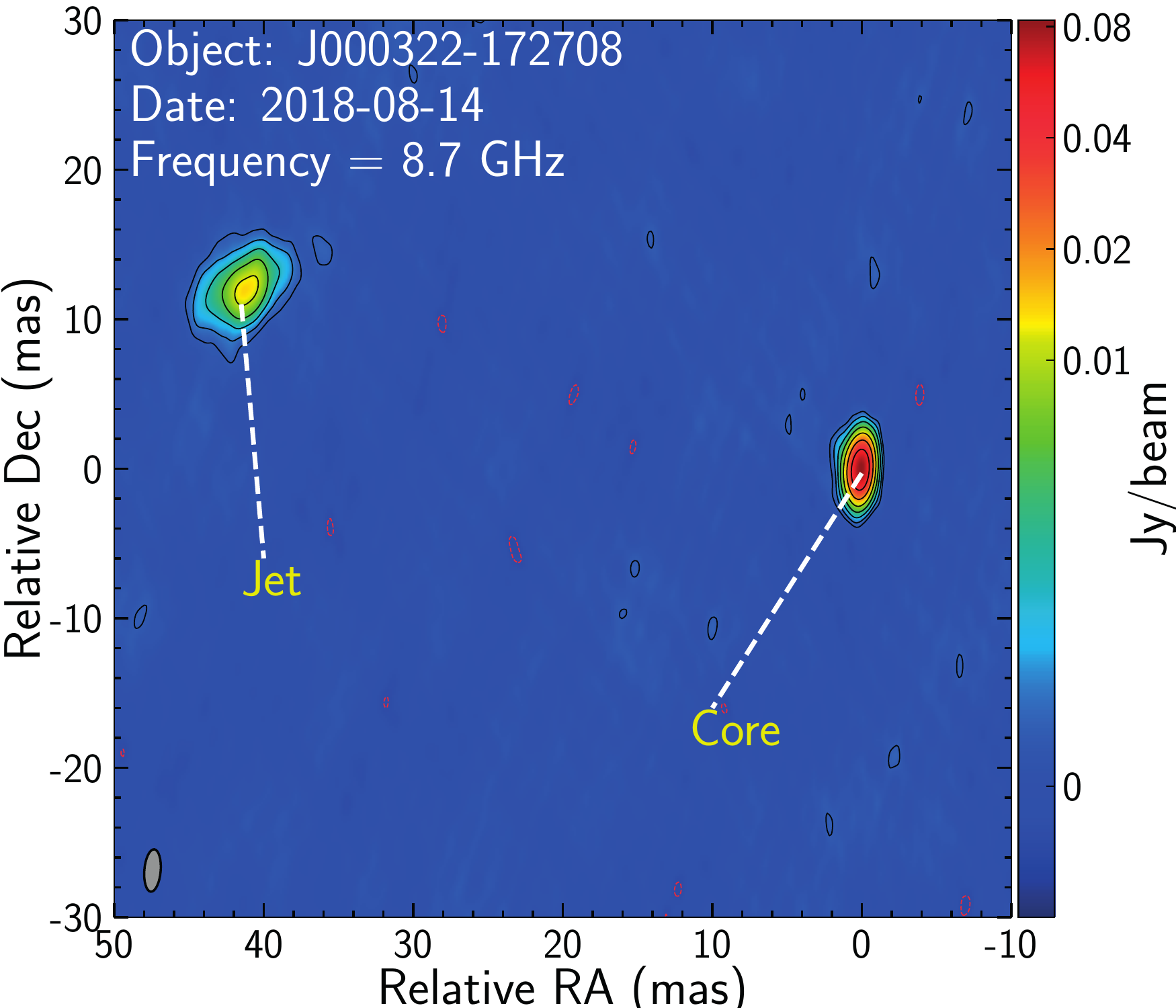}&
\includegraphics[scale=0.24]{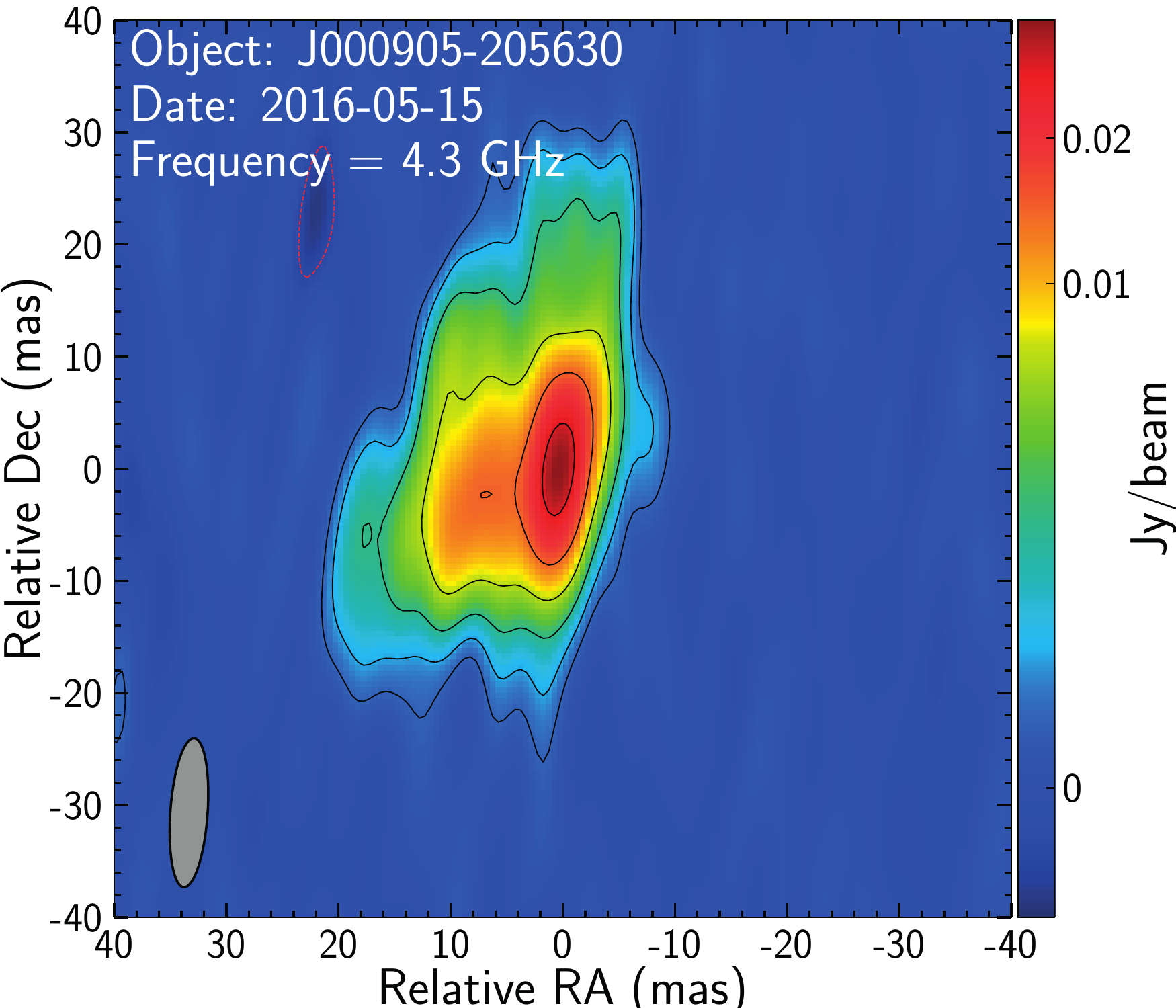}\\
\includegraphics[scale=0.24]{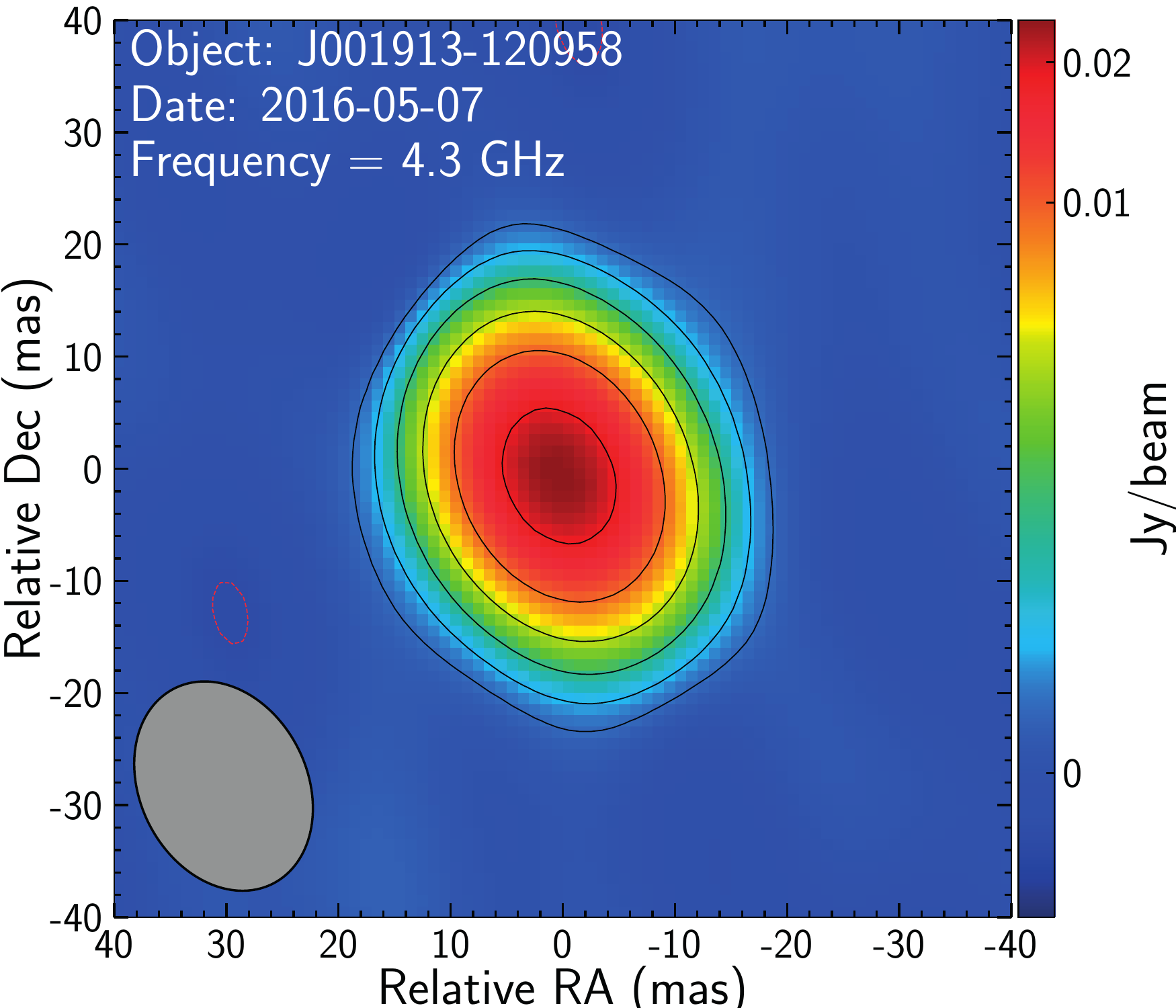}&
\includegraphics[scale=0.24]{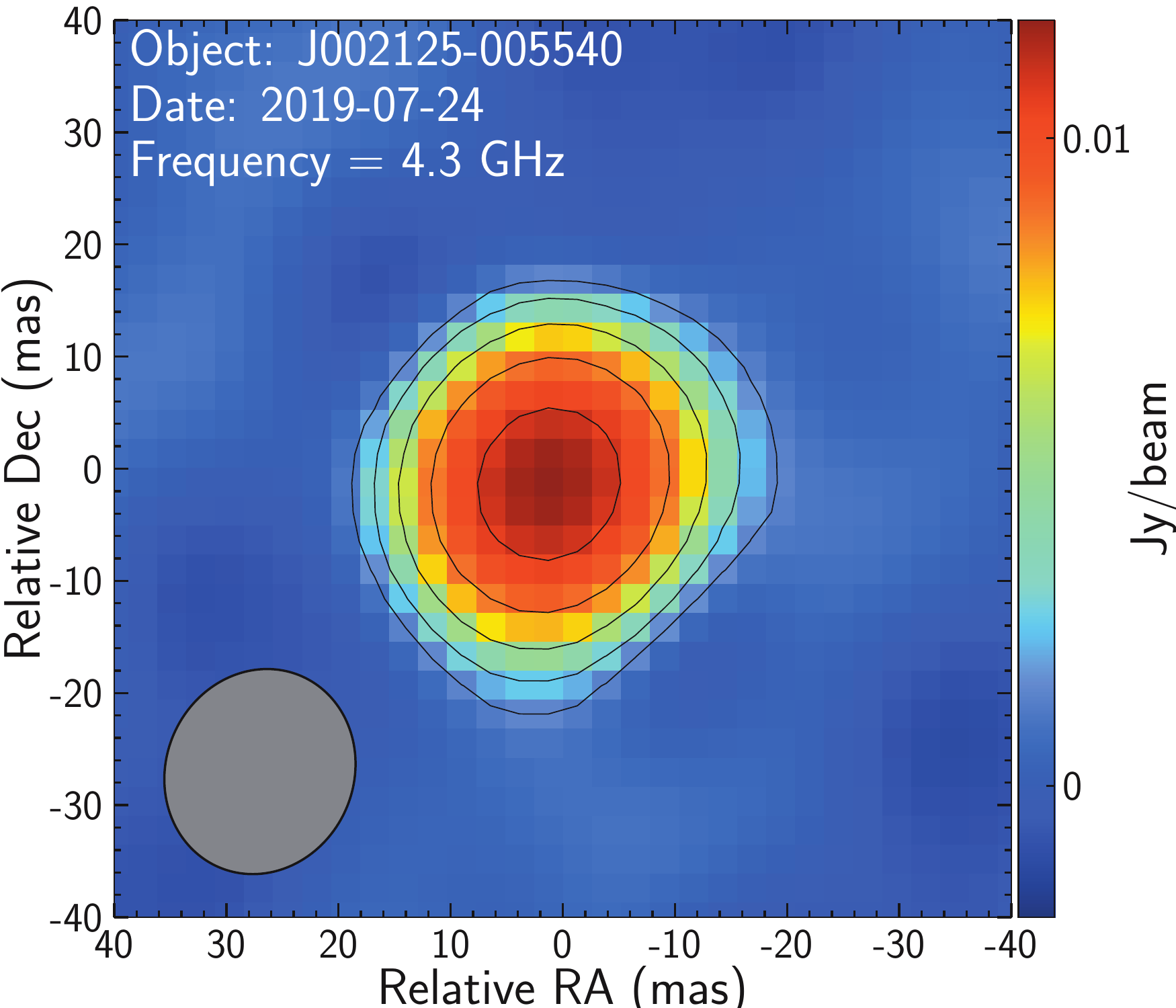}&
\includegraphics[scale=0.24]{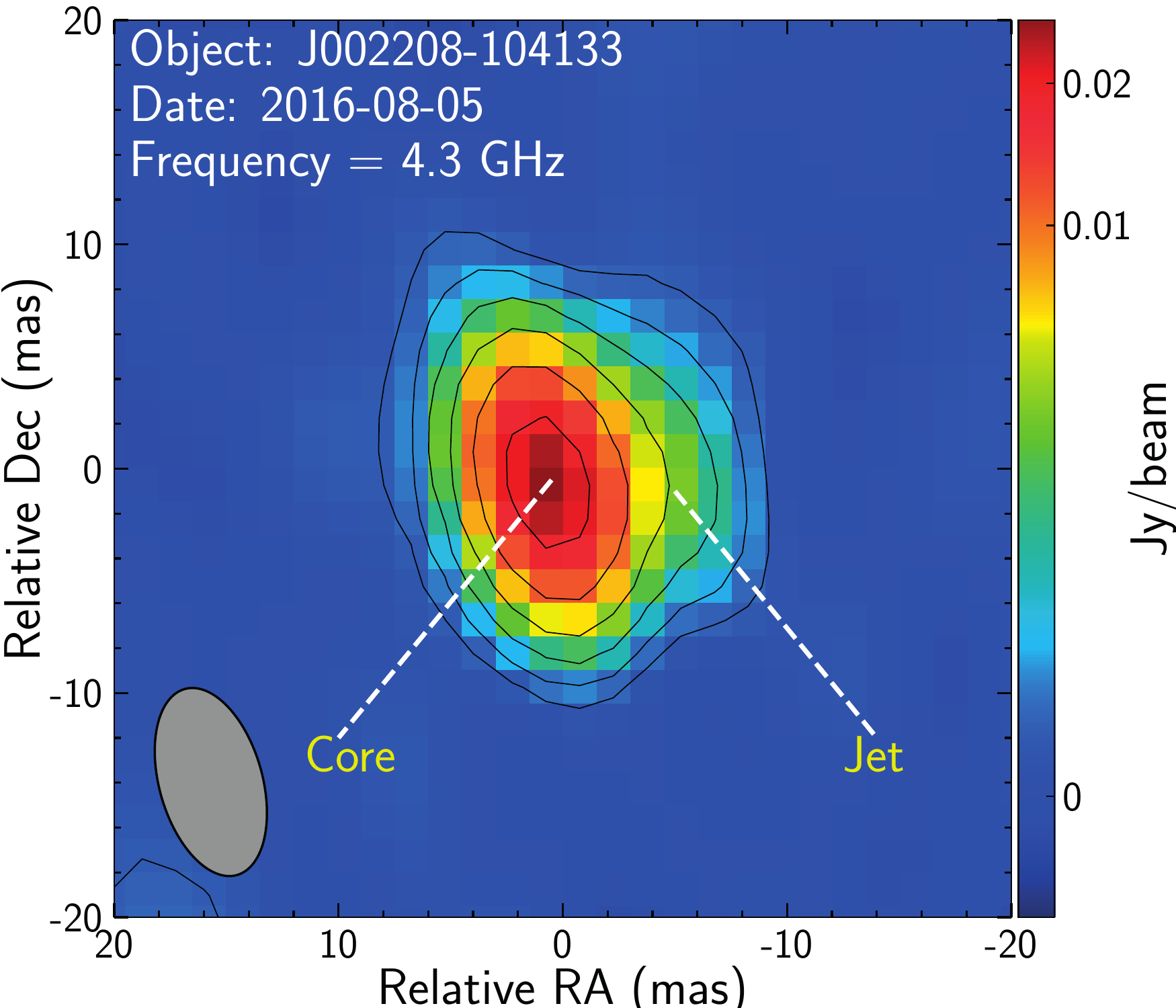}&
\includegraphics[scale=0.24]{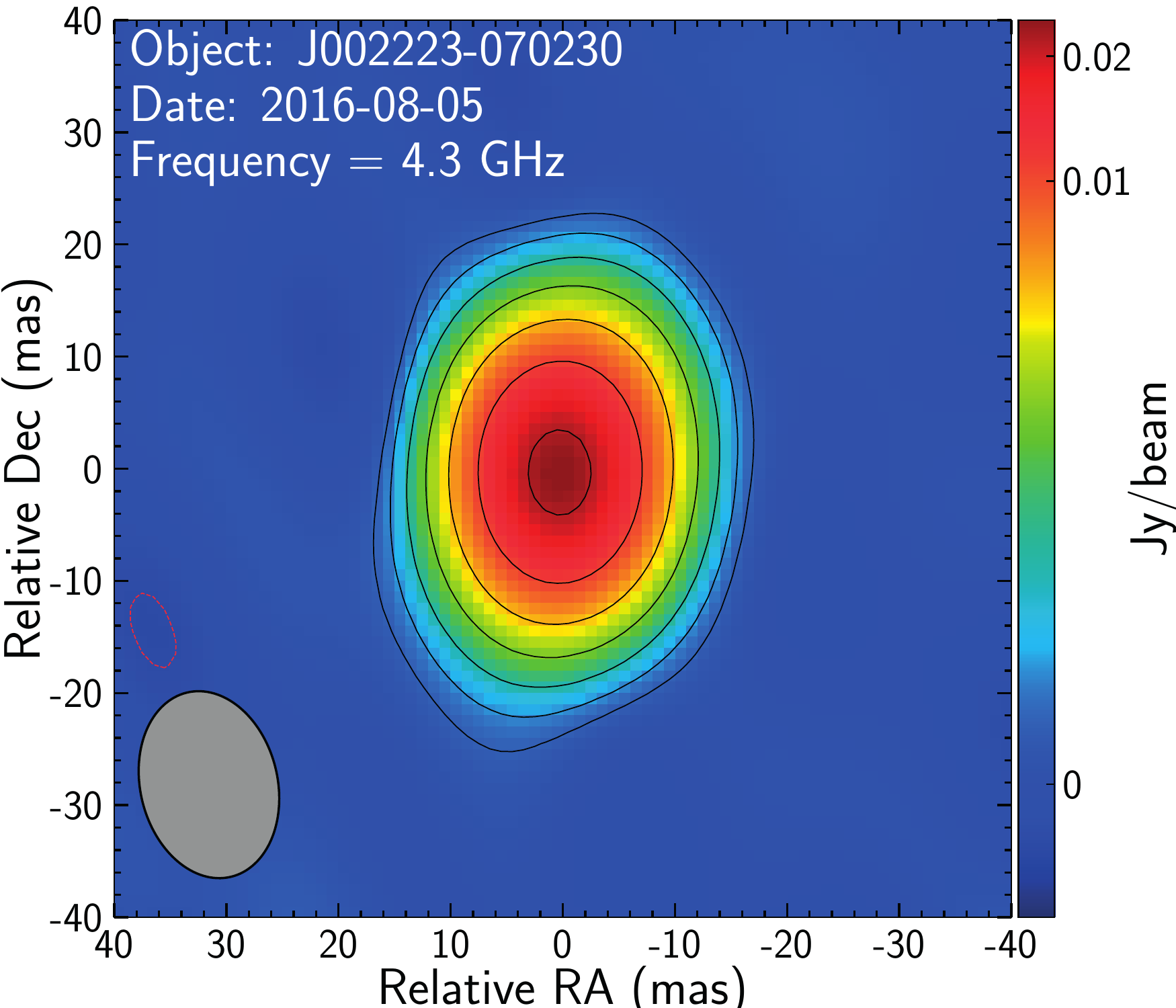}\\
\includegraphics[scale=0.24]{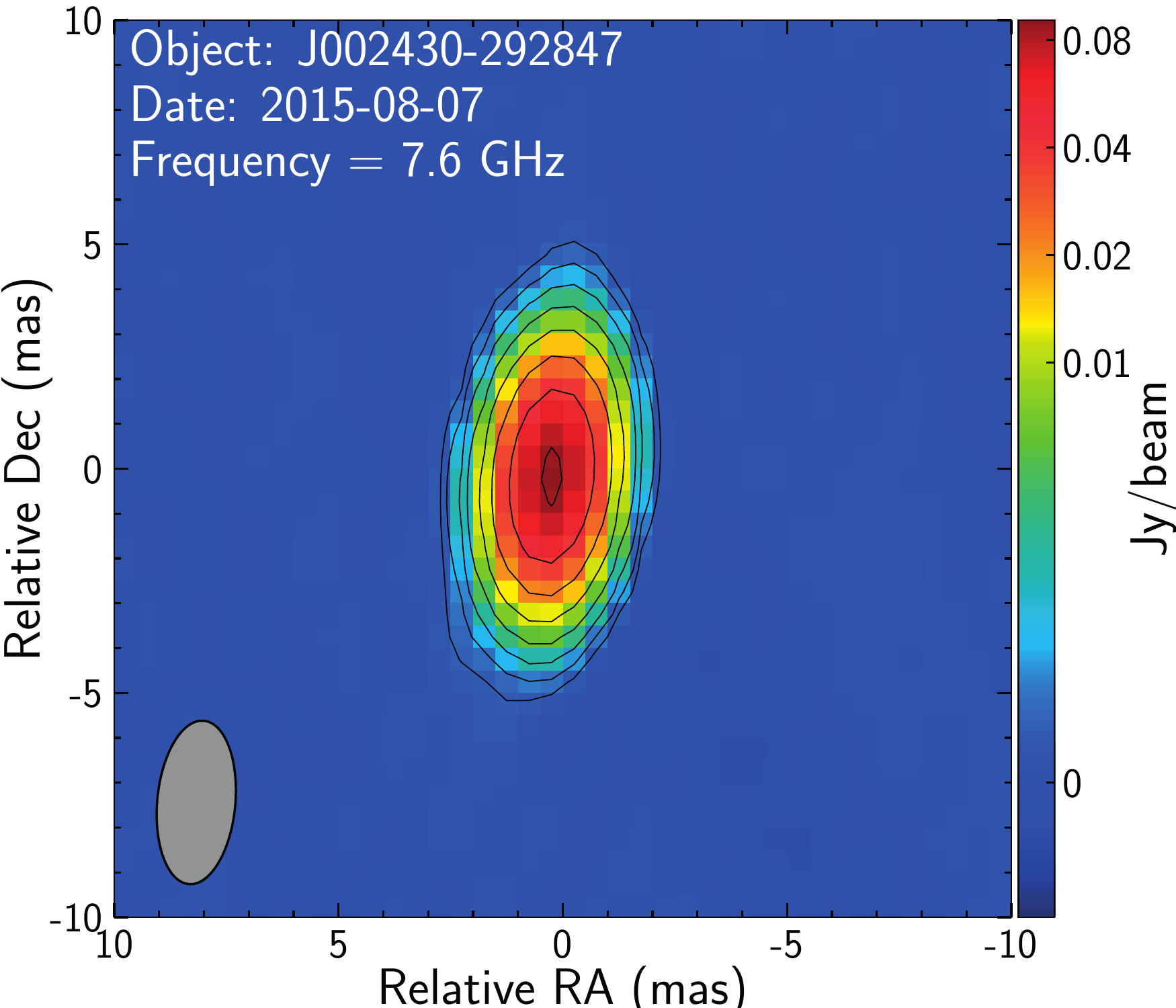}&
\includegraphics[scale=0.24]{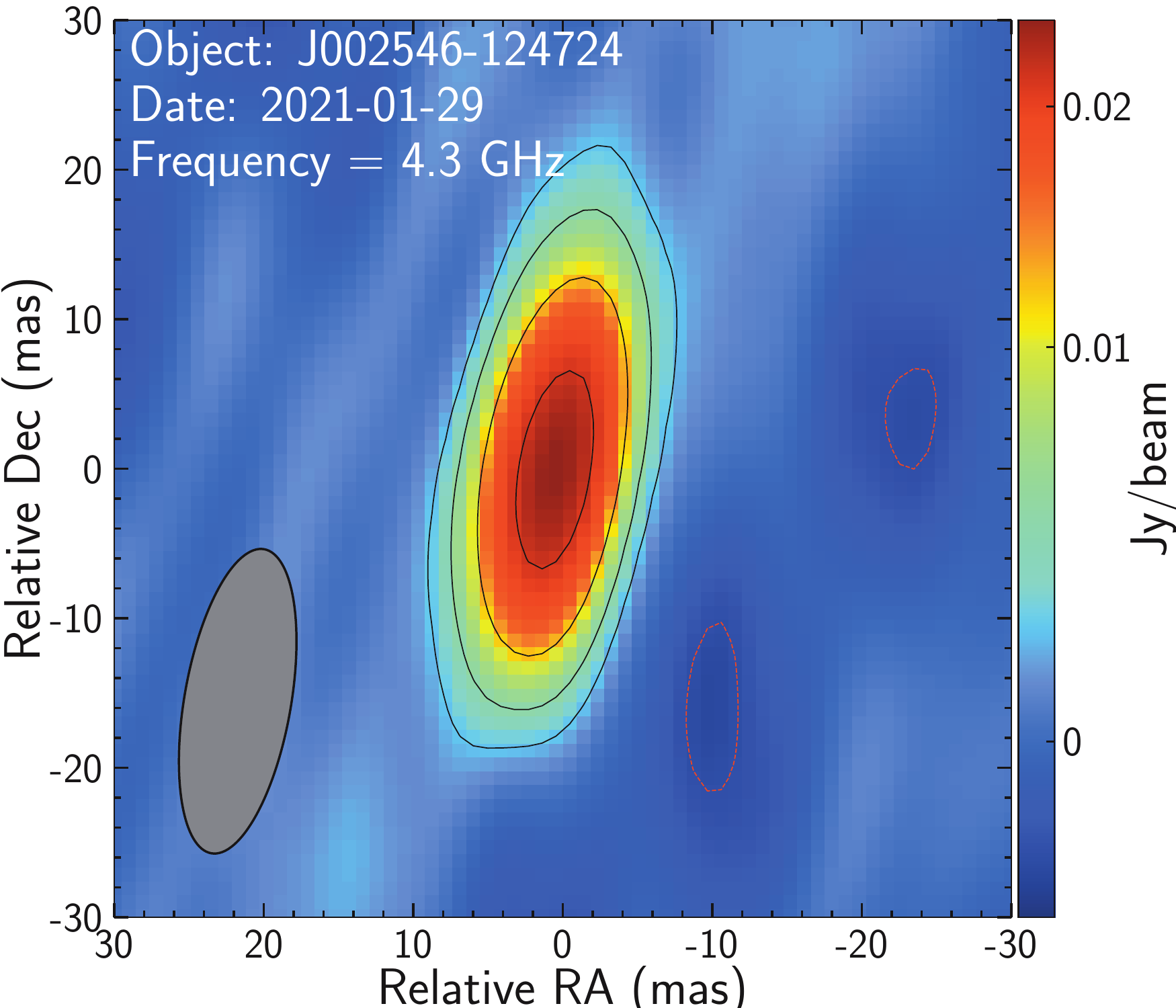}&
\includegraphics[scale=0.24]{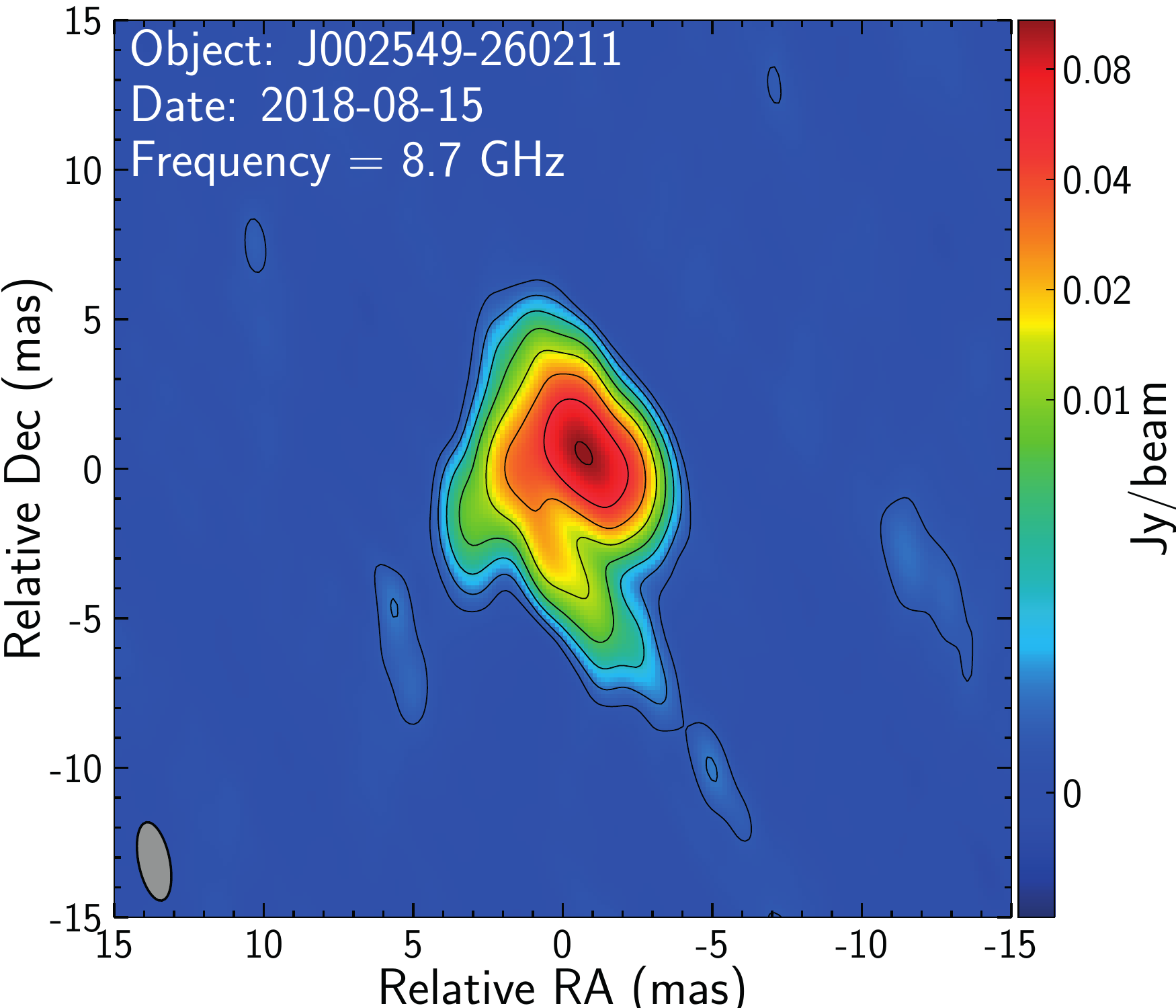}&
\includegraphics[scale=0.24]{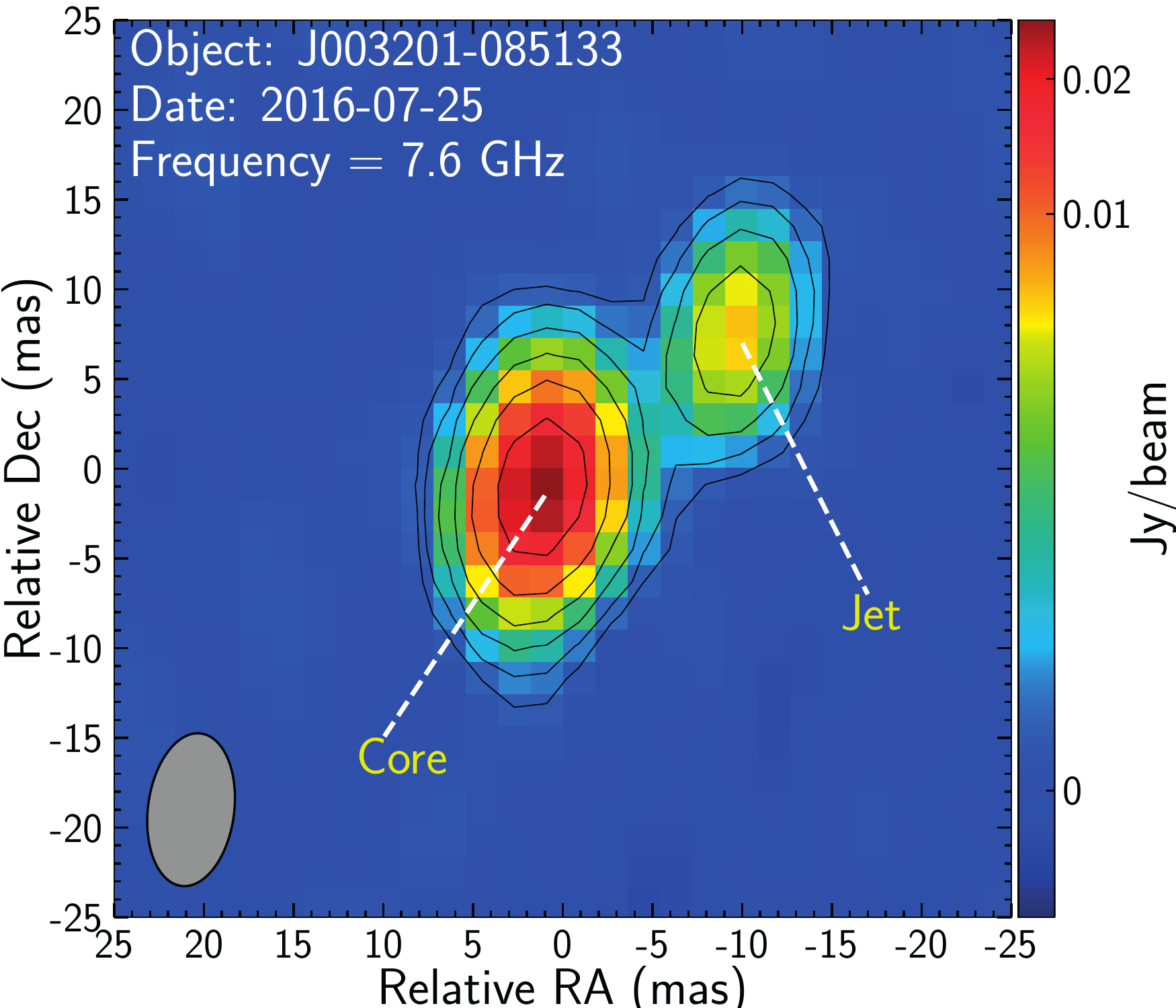}\\
\includegraphics[scale=0.24]{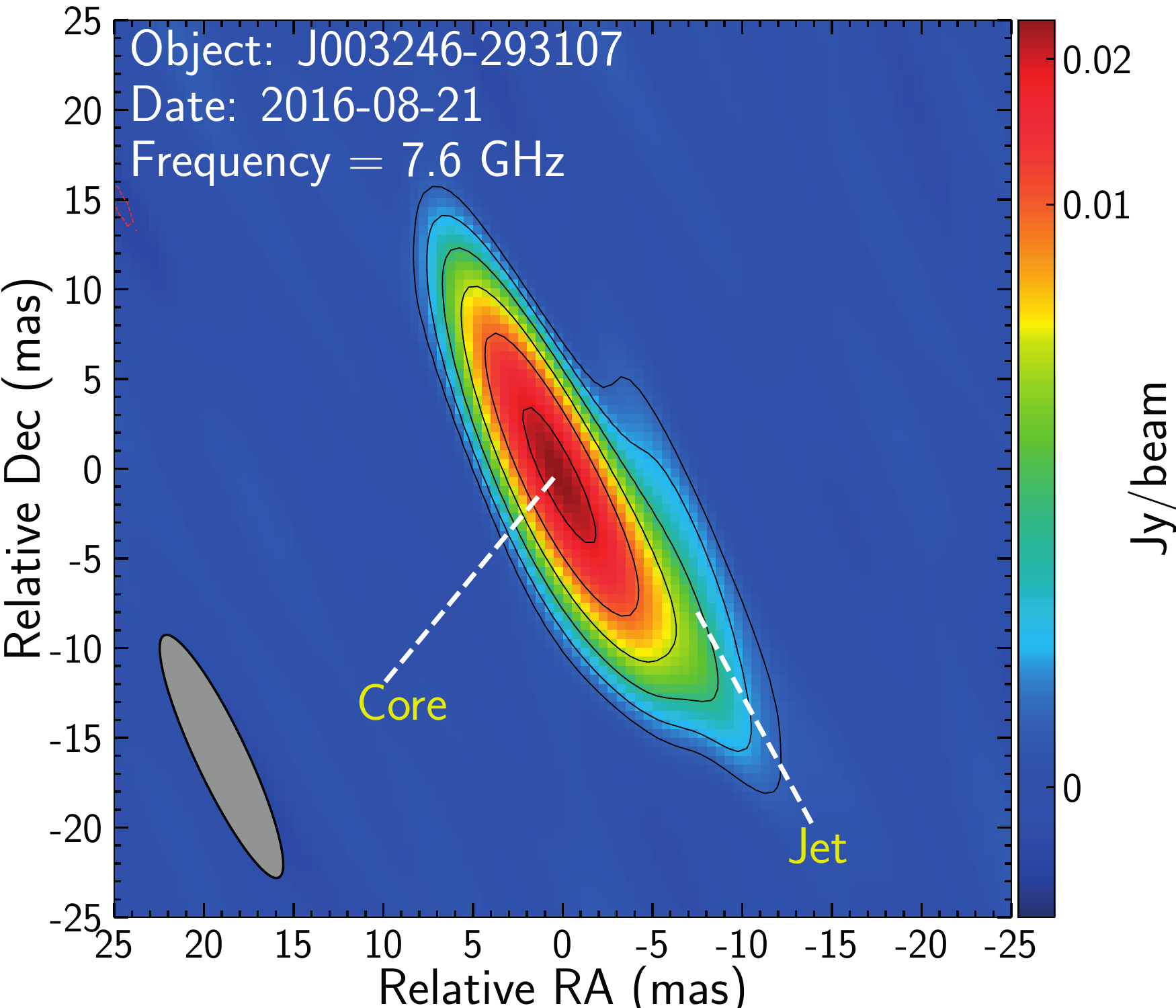}&
\includegraphics[scale=0.24]{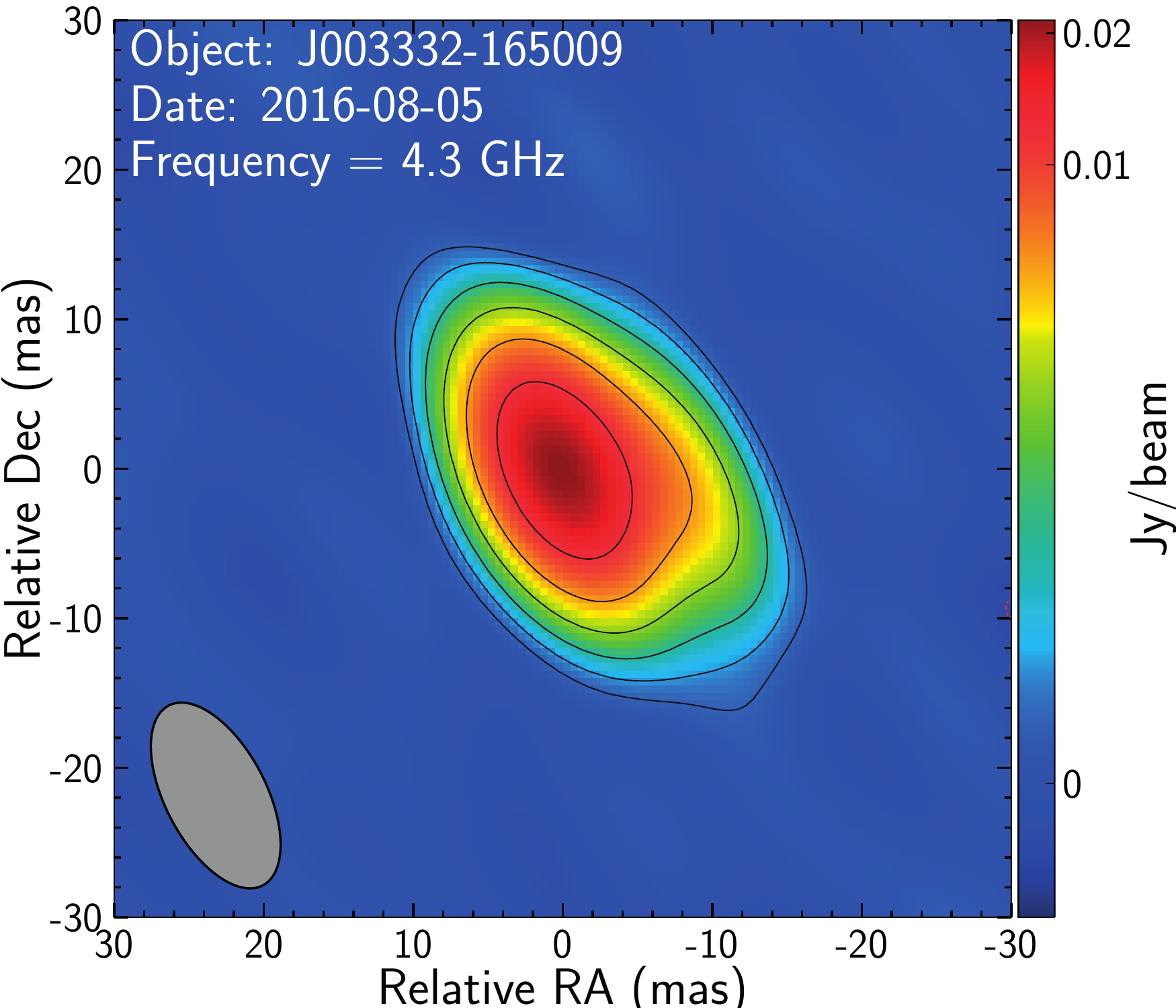}&
\includegraphics[scale=0.24]{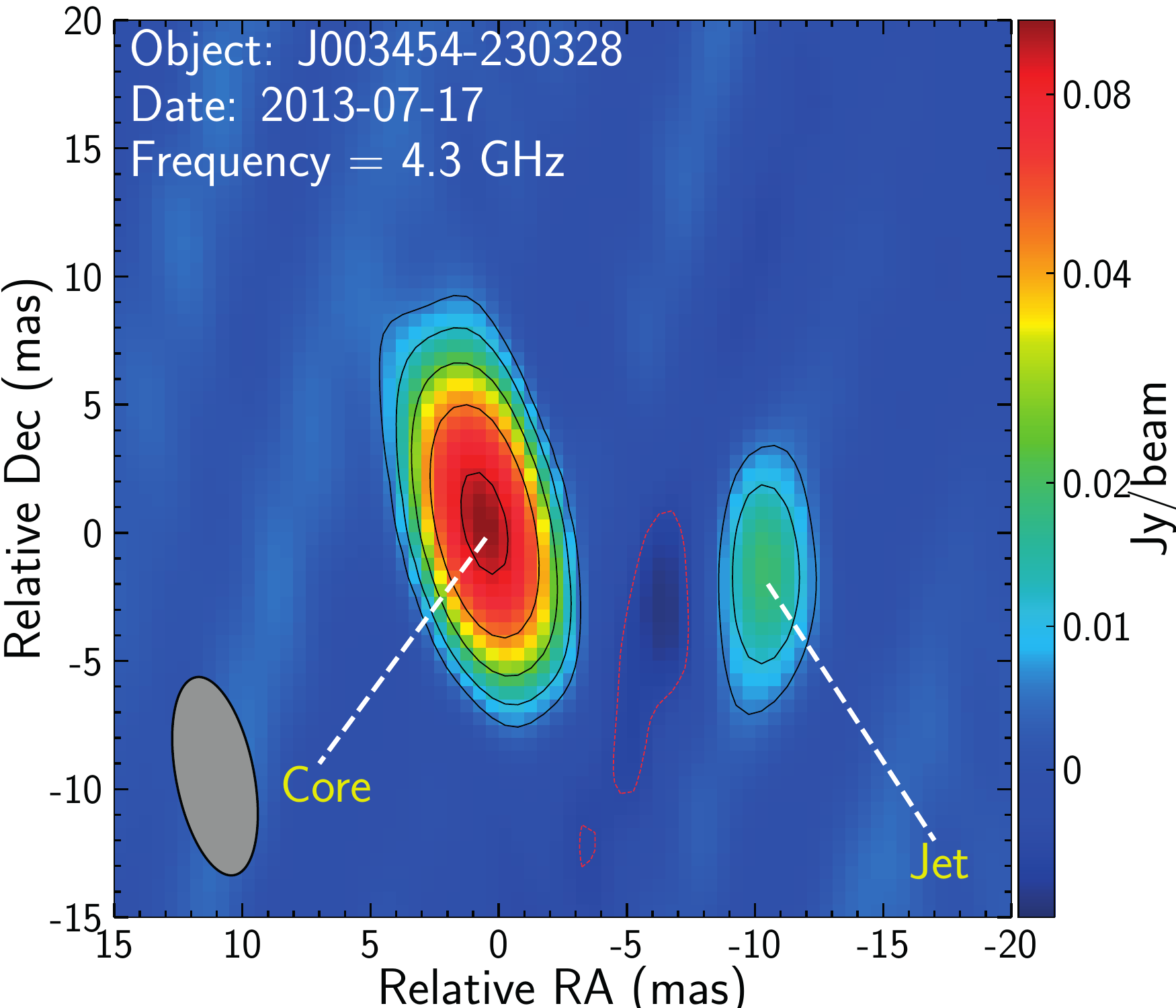}&
\includegraphics[scale=0.24]{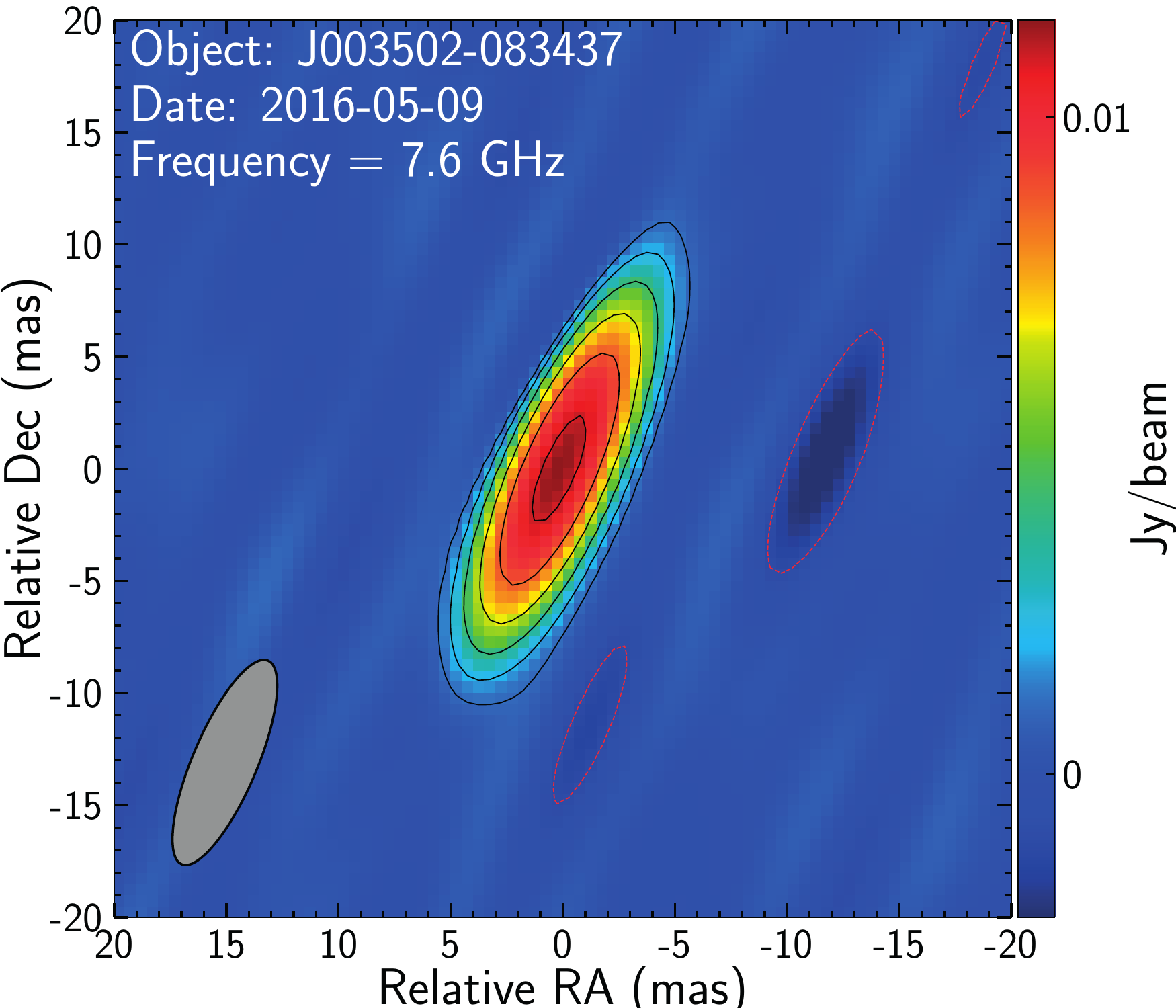}\\
\includegraphics[scale=0.24]{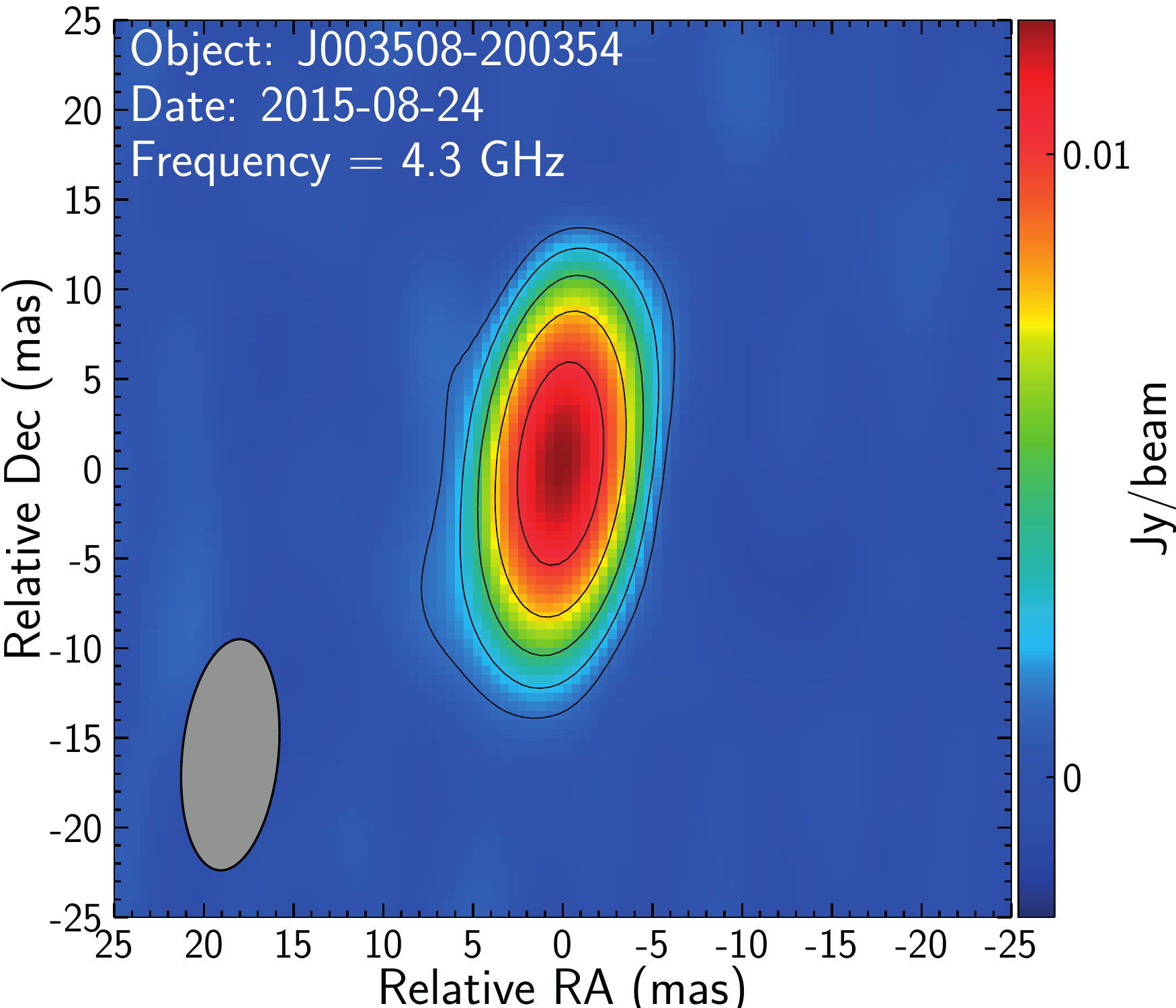}&
\includegraphics[scale=0.24]{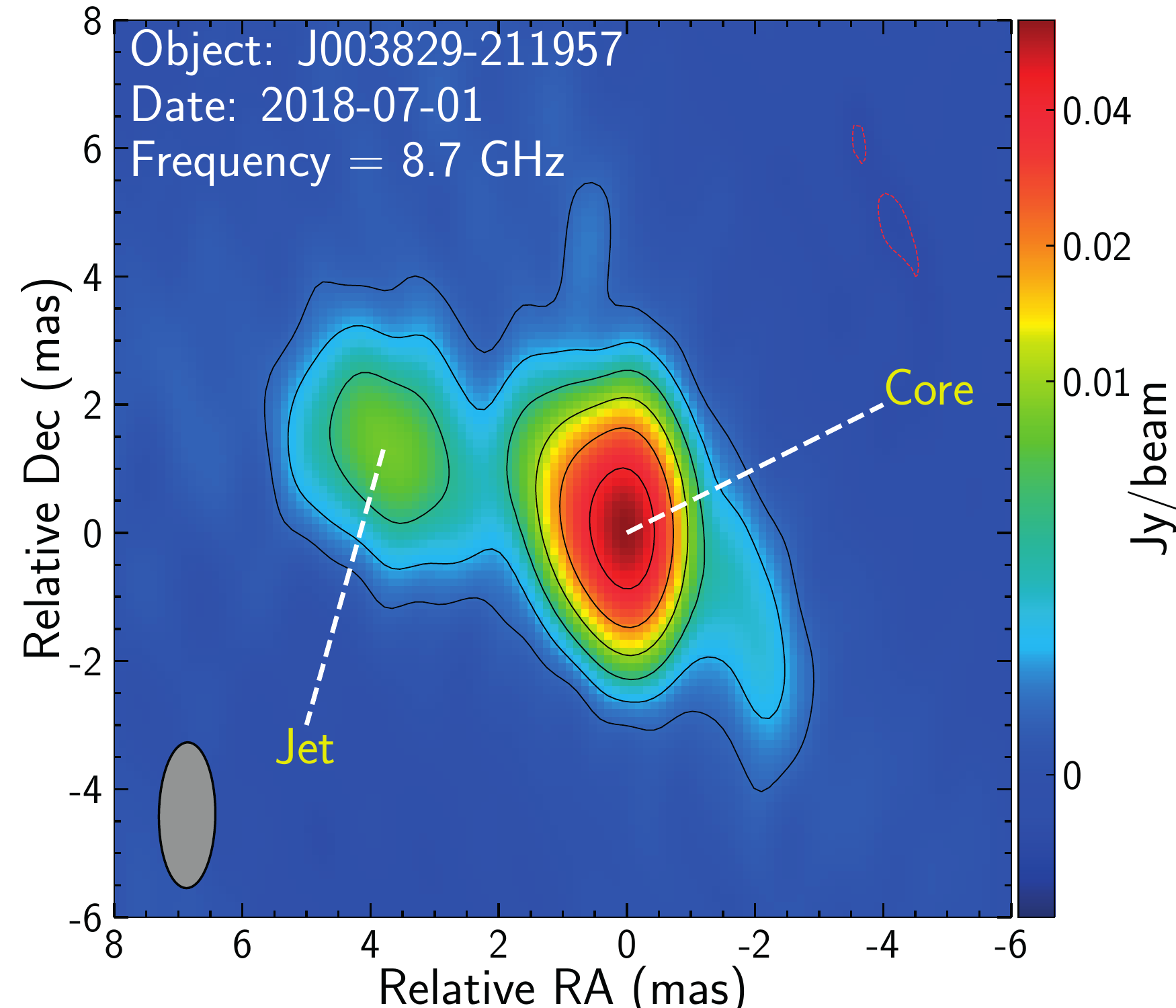}&
\includegraphics[scale=0.24]{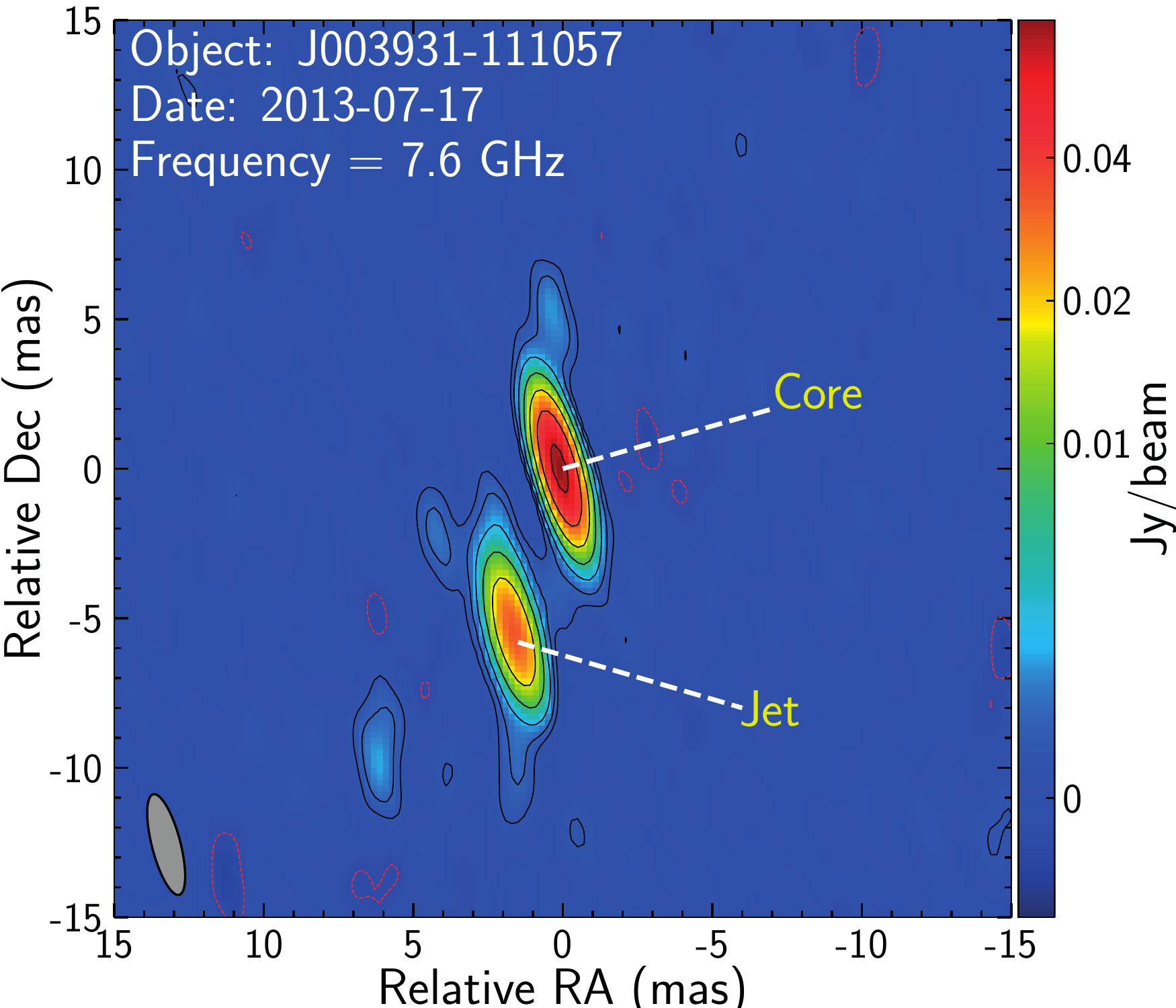}&
\includegraphics[scale=0.24]{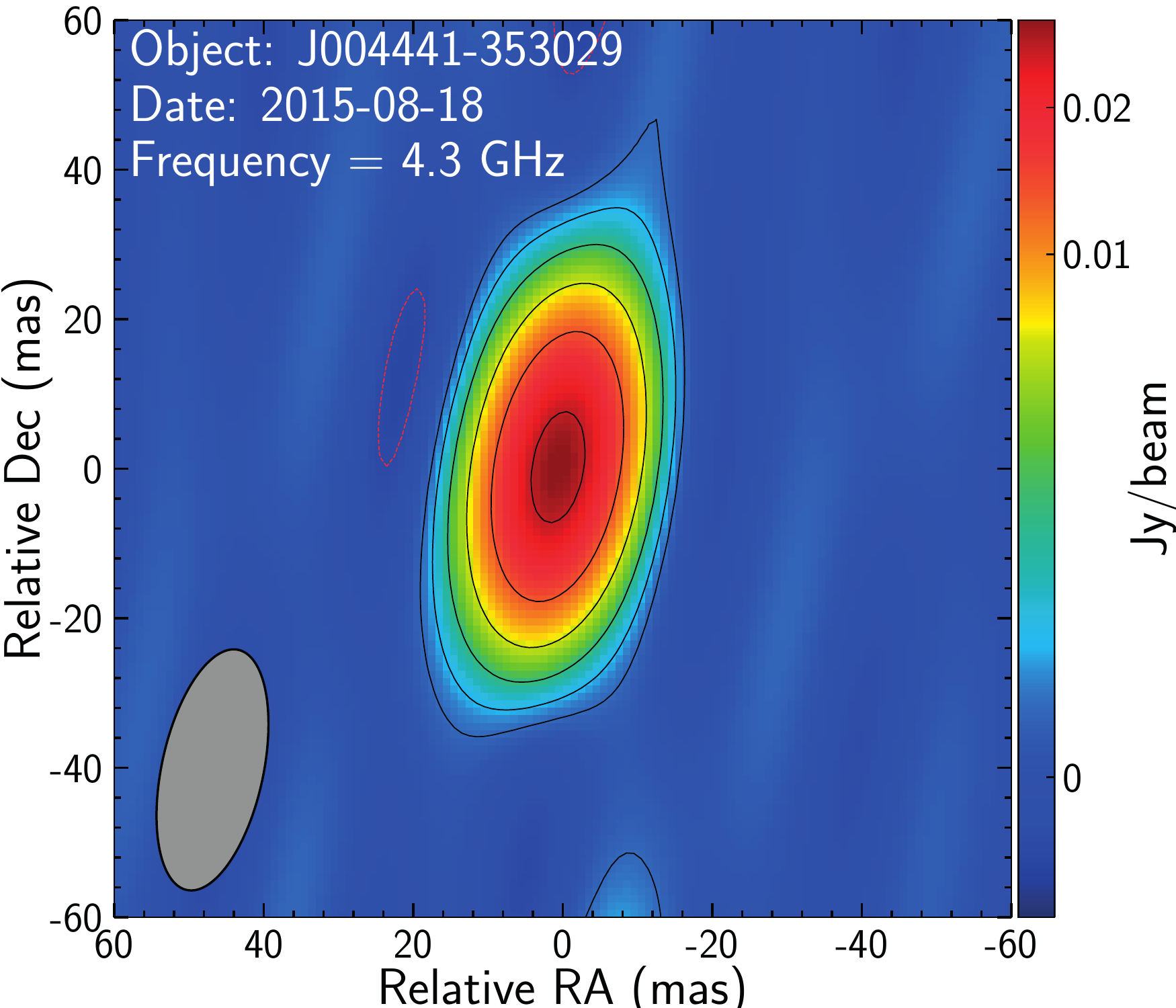}\\
\includegraphics[scale=0.24]{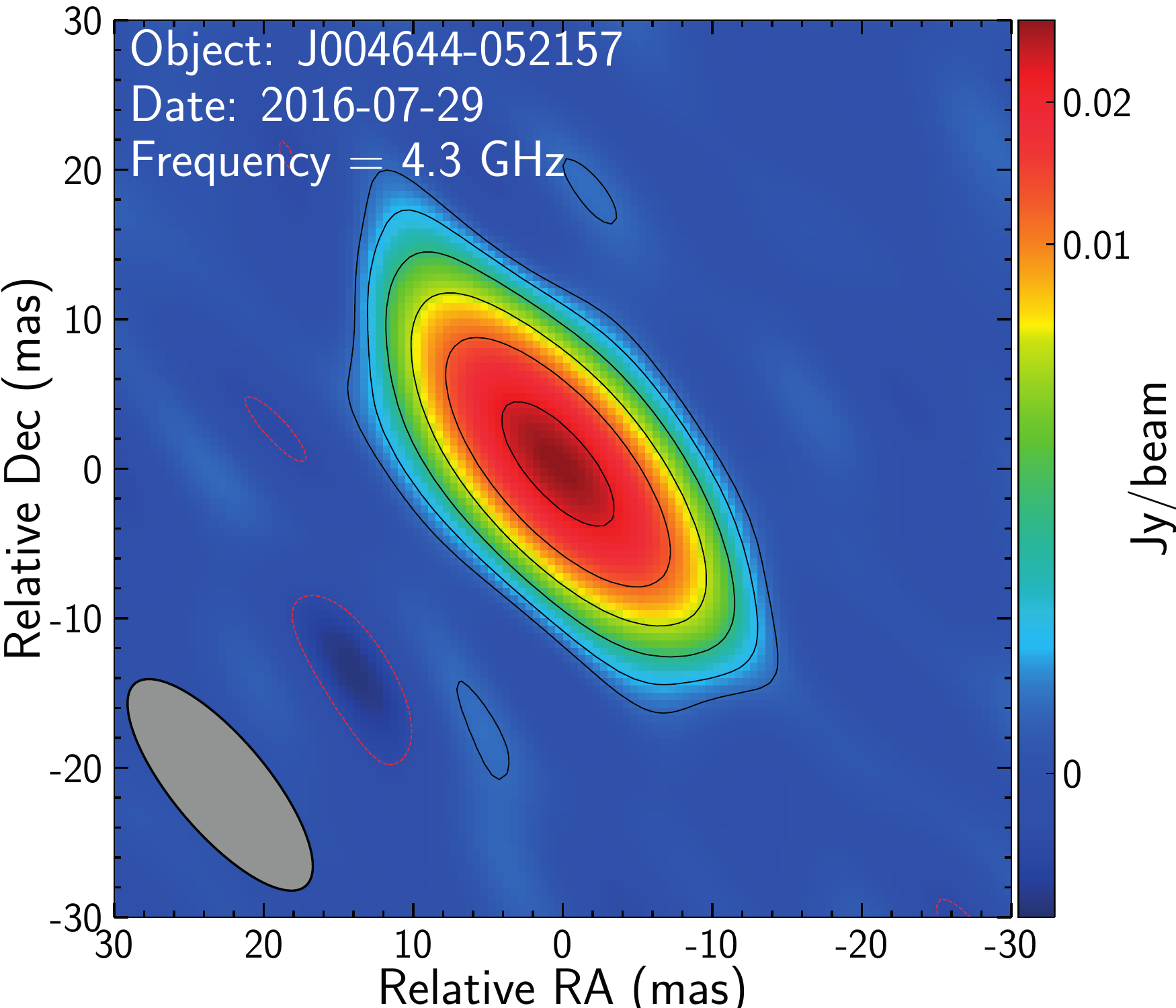}&
\includegraphics[scale=0.24]{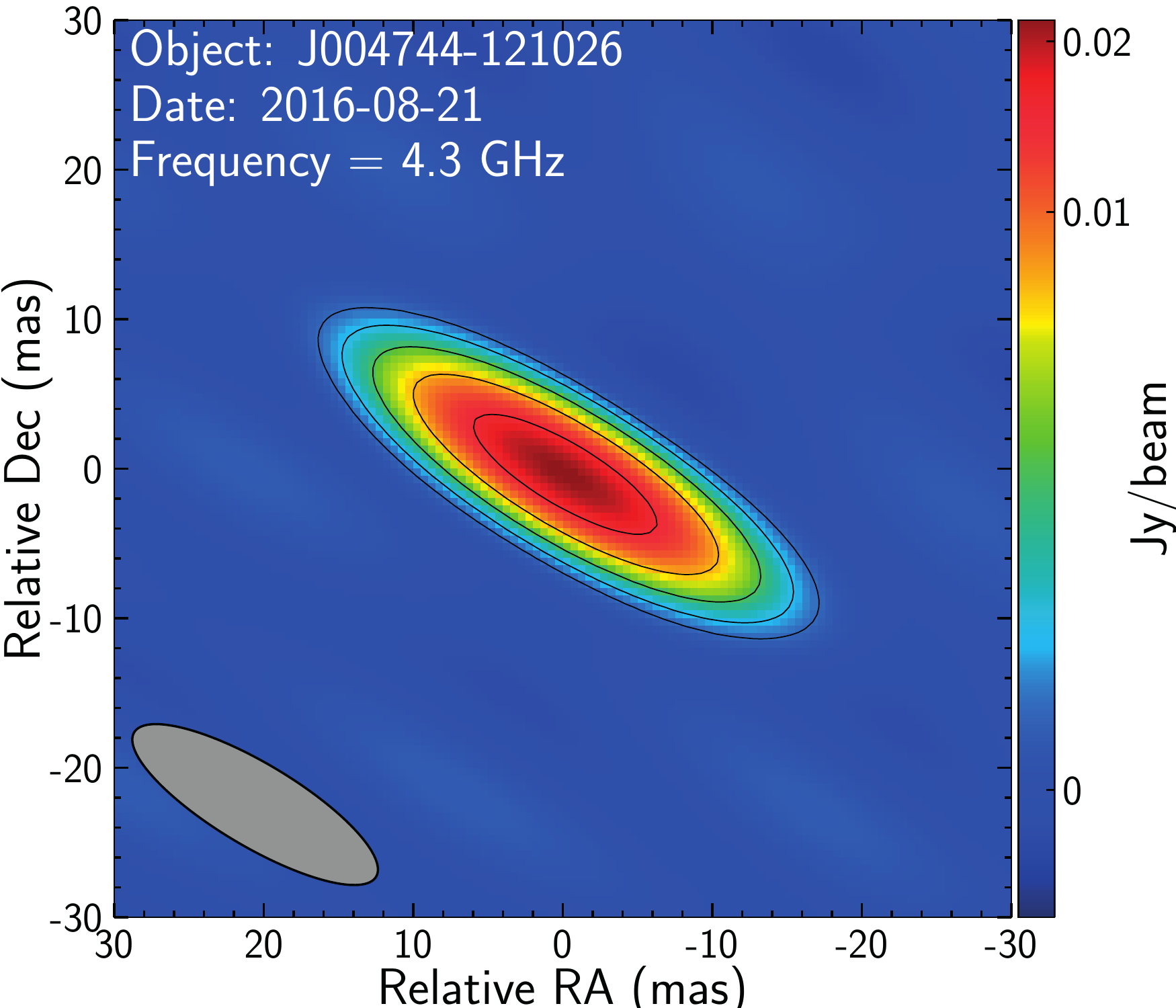}&
\includegraphics[scale=0.24]{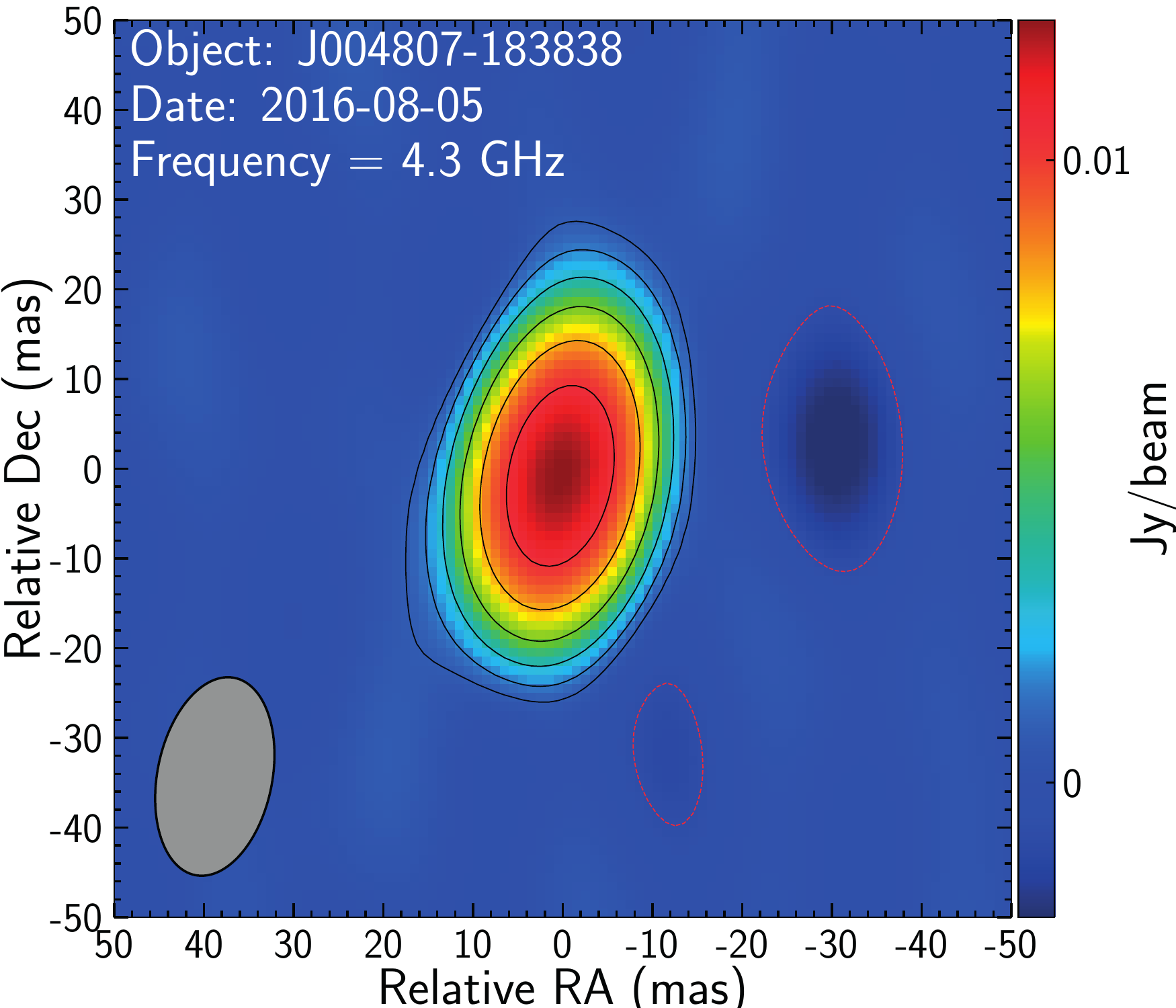}&
\includegraphics[scale=0.24]{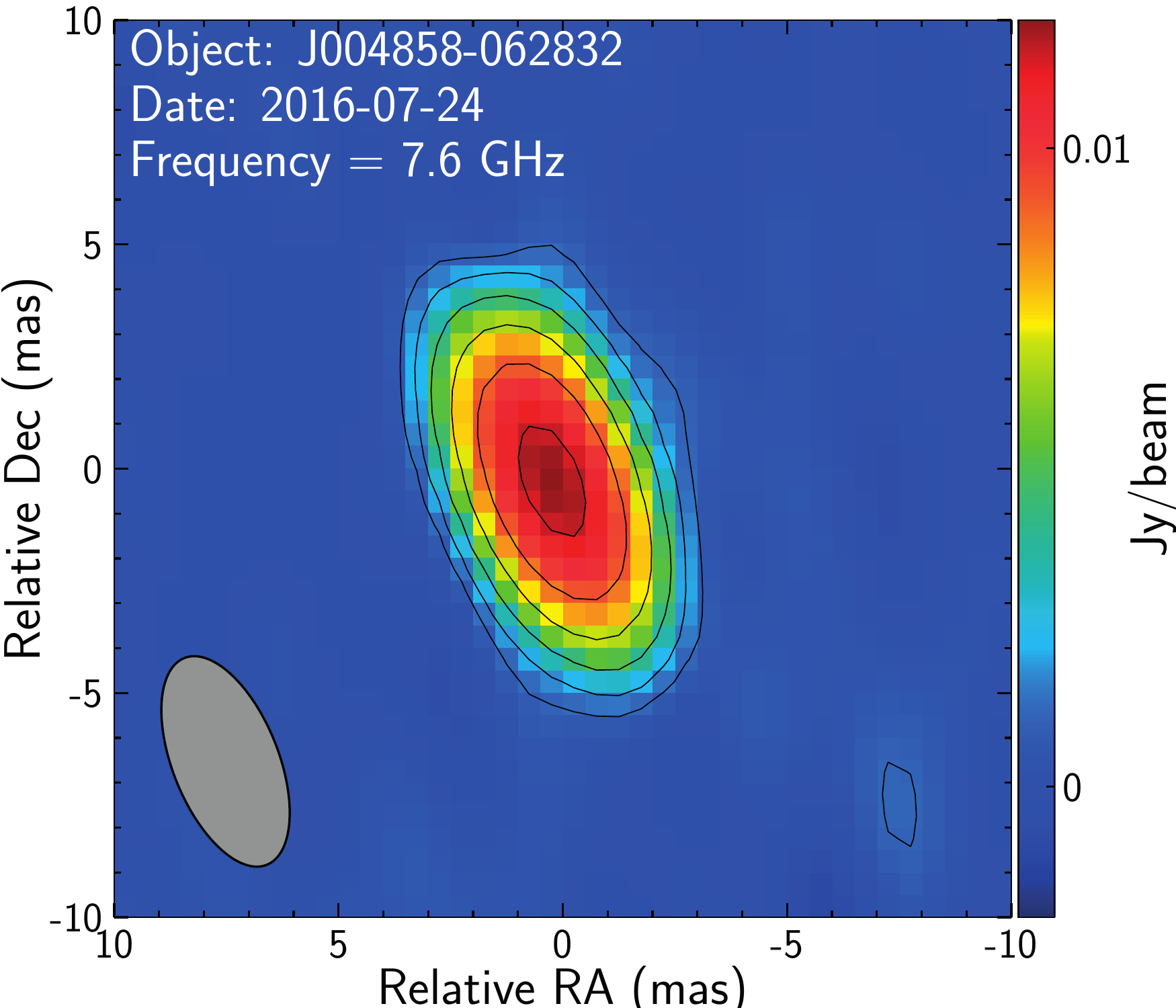}
\end{tabular}
\caption{\small VLBI morphology of the IPS sample sources in $X$ or $C$ band. The contours of each image are overlaid on the corresponding color-scaled image. The contour levels are given as $3\times \mathrm{rms} \times (-1,1,2,4,8,16,32)$. The $\mathrm{rms}$ in each image is referred in Table~\ref{tab:image}. The grey-colored ellipse in the bottom-left corner of each panel represents the restoring beam of the color-scaled image. The restoring beams for all the images are quantitatively given in Table~\ref{tab:image}.}
\label{fig:morph}
\end{figure*}
\addtocounter{figure}{-1}
\begin{figure*}
\centering
\begin{tabular}{cccc}
\includegraphics[scale=0.24]{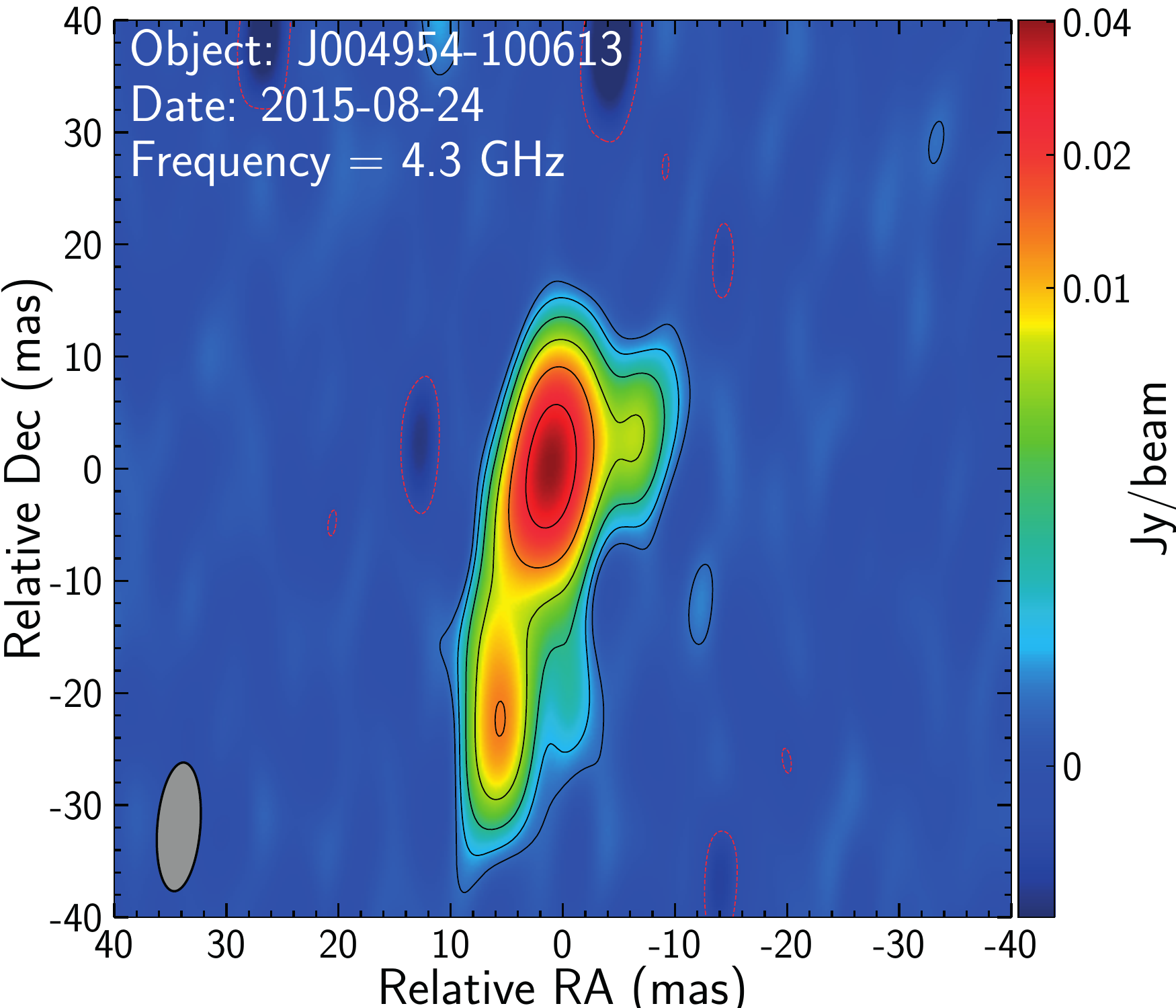}&
\includegraphics[scale=0.24]{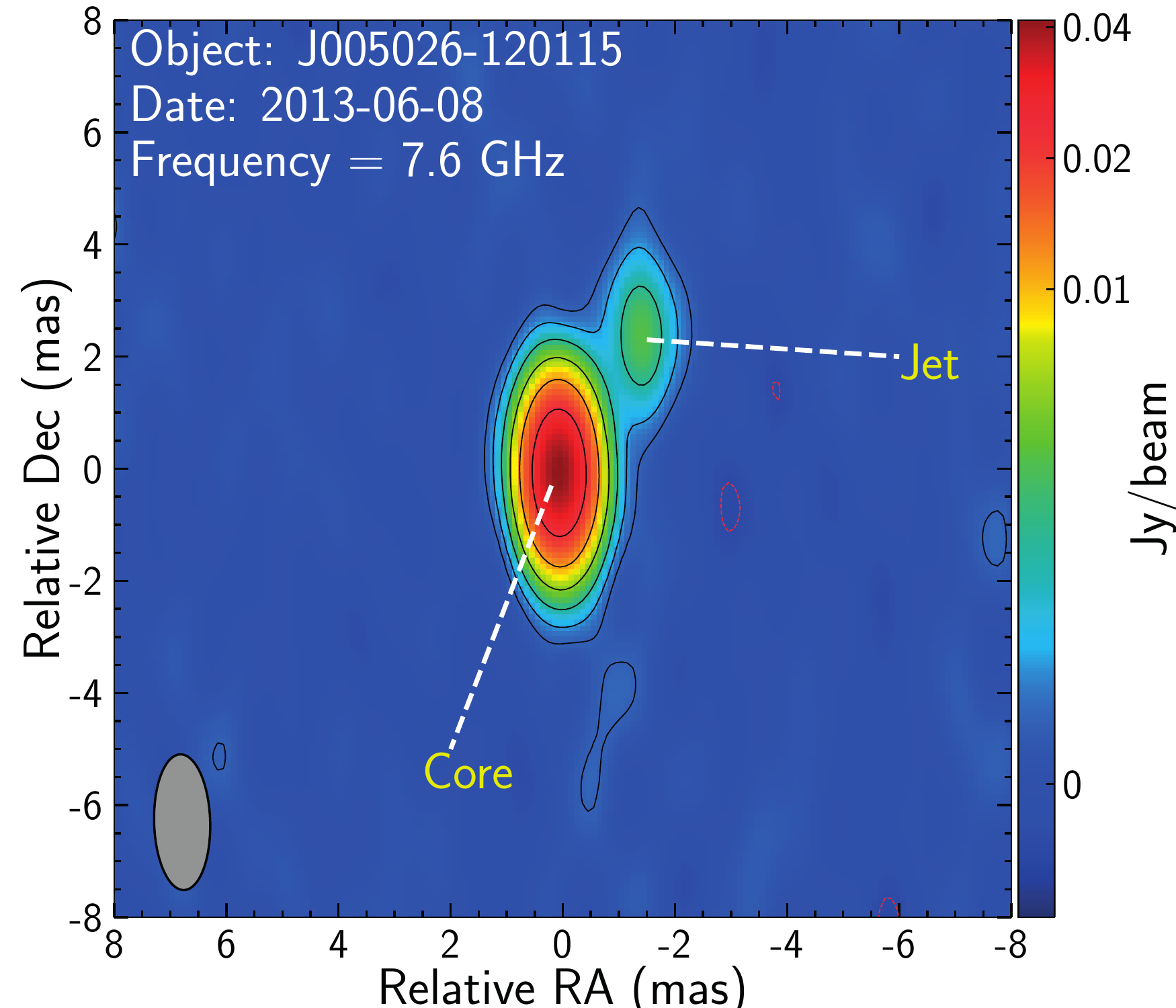}&
\includegraphics[scale=0.24]{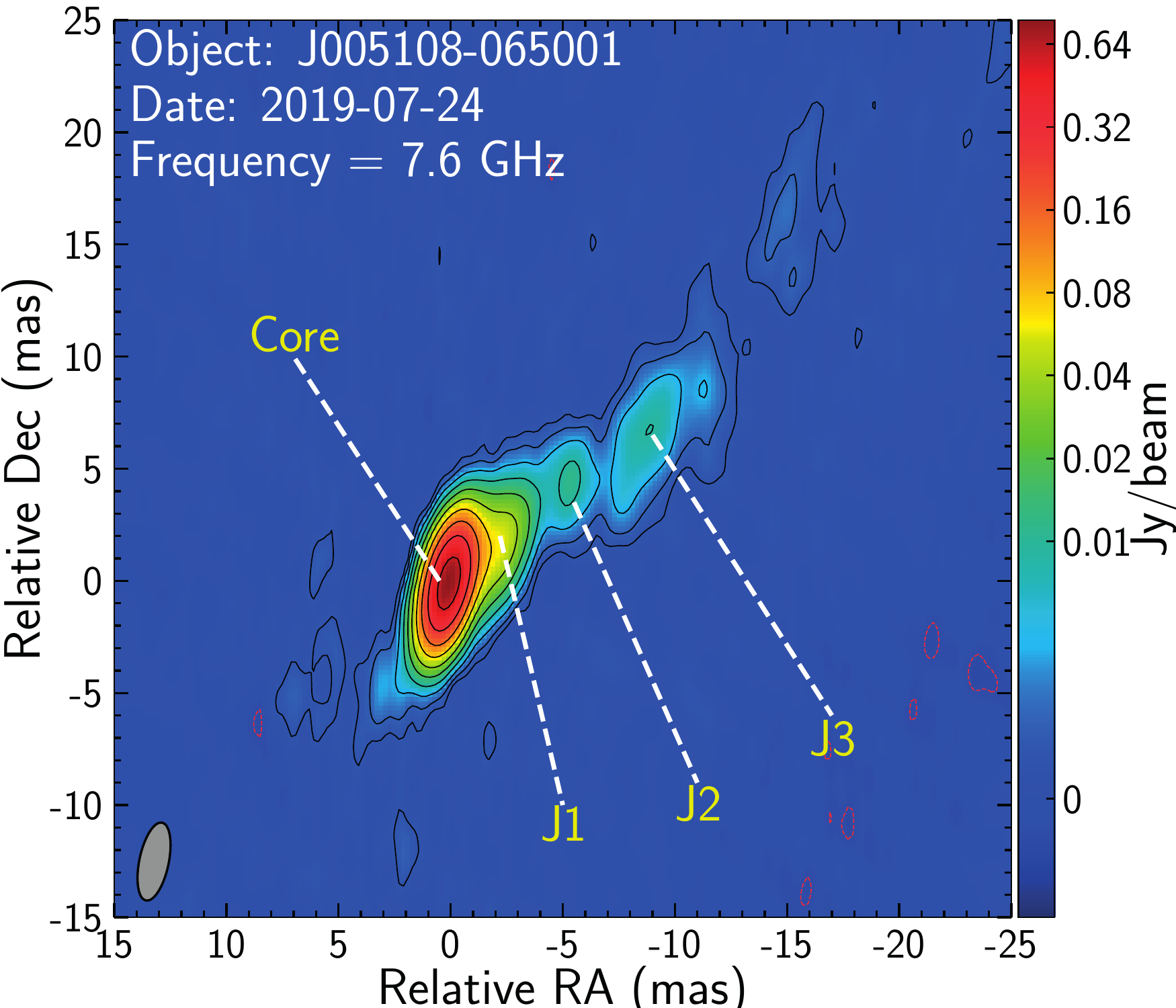}&
\includegraphics[scale=0.24]{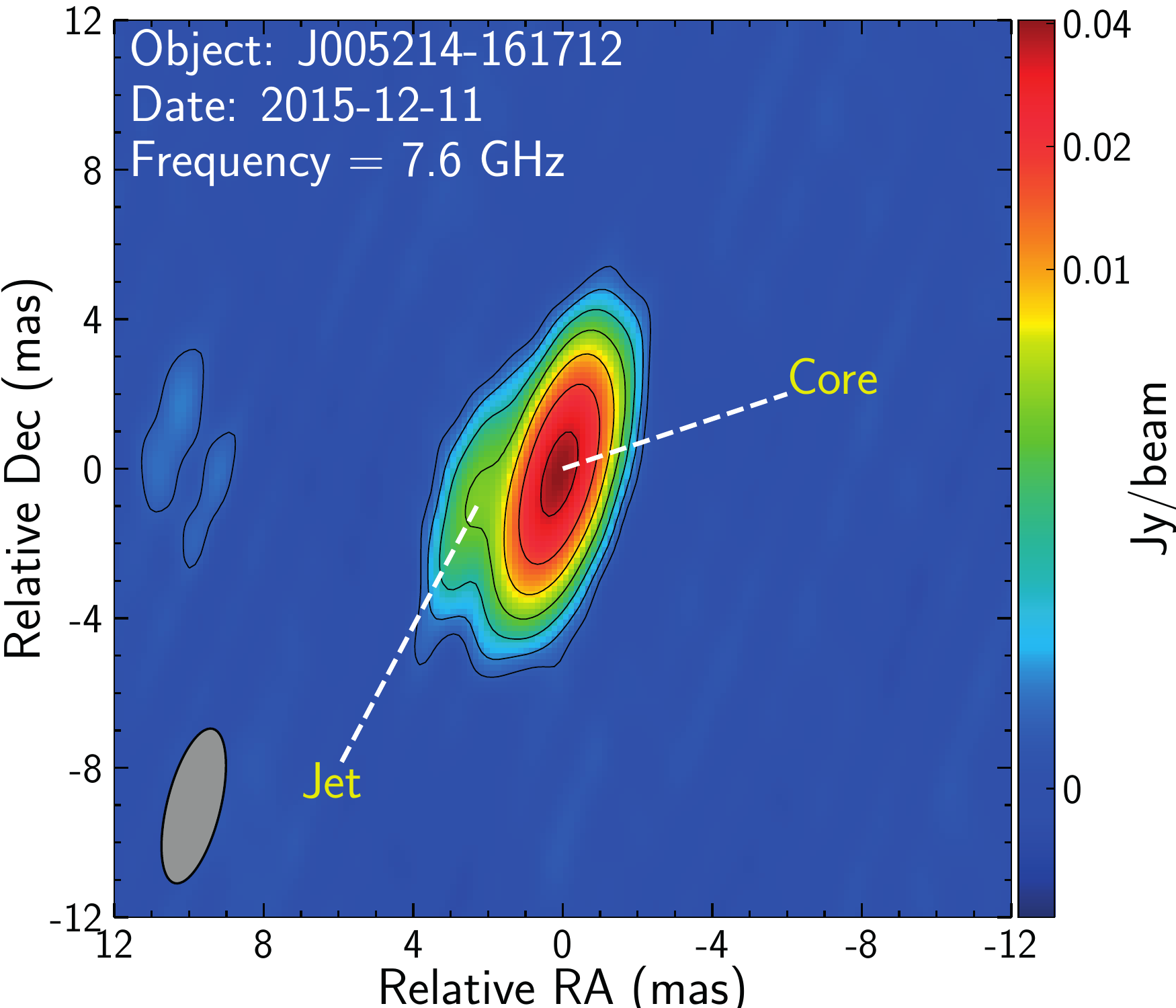}\\
\includegraphics[scale=0.24]{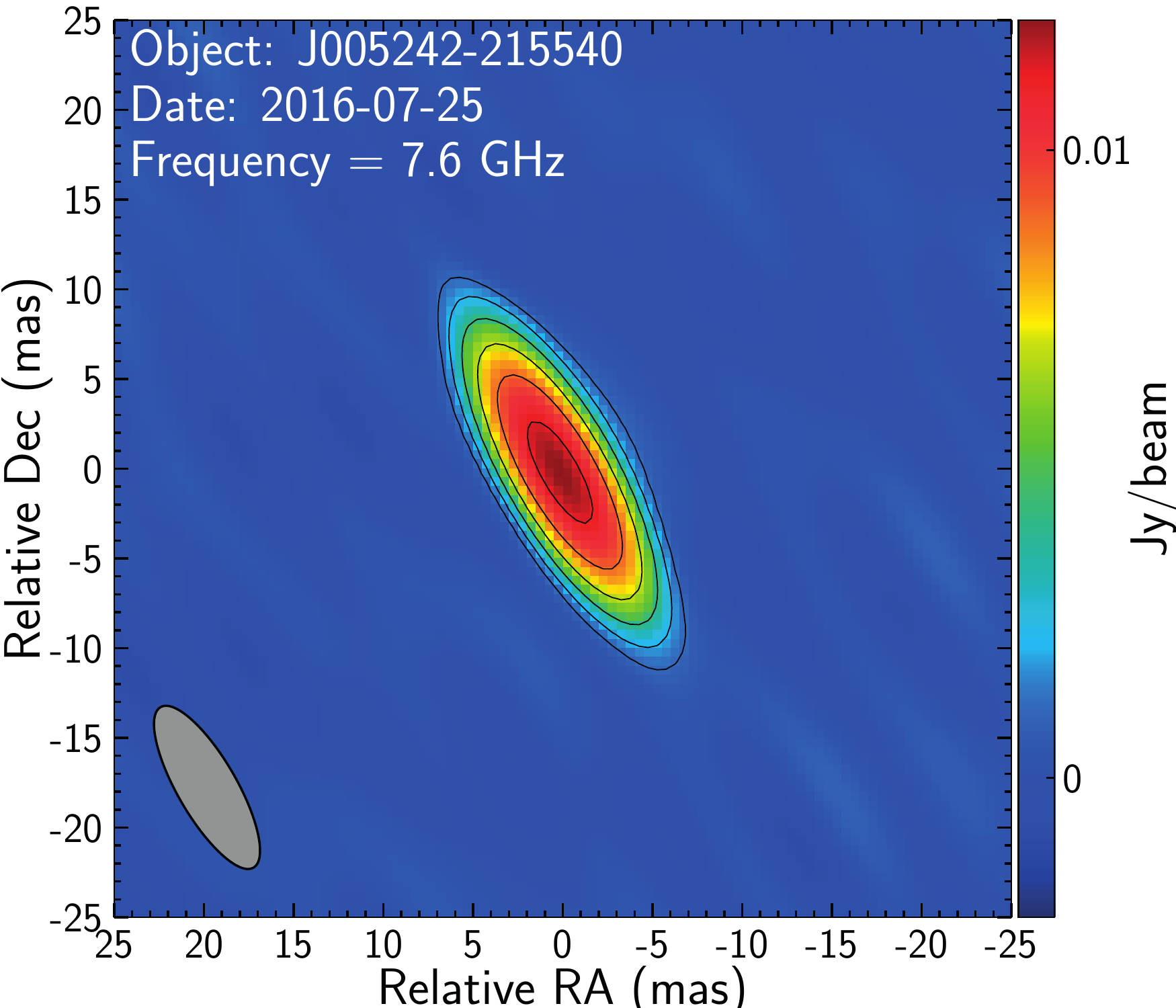}&
\includegraphics[scale=0.24]{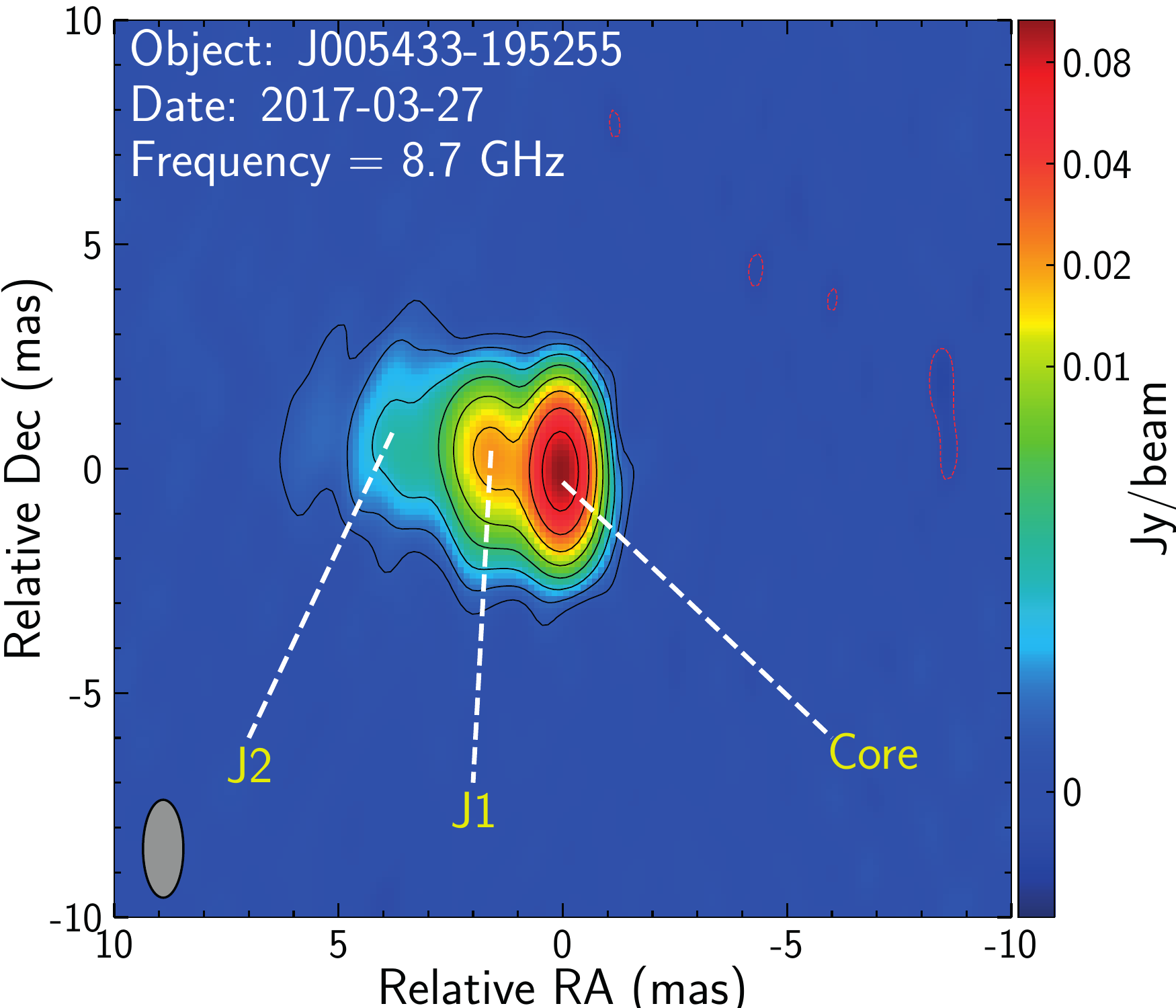}&
\includegraphics[scale=0.24]{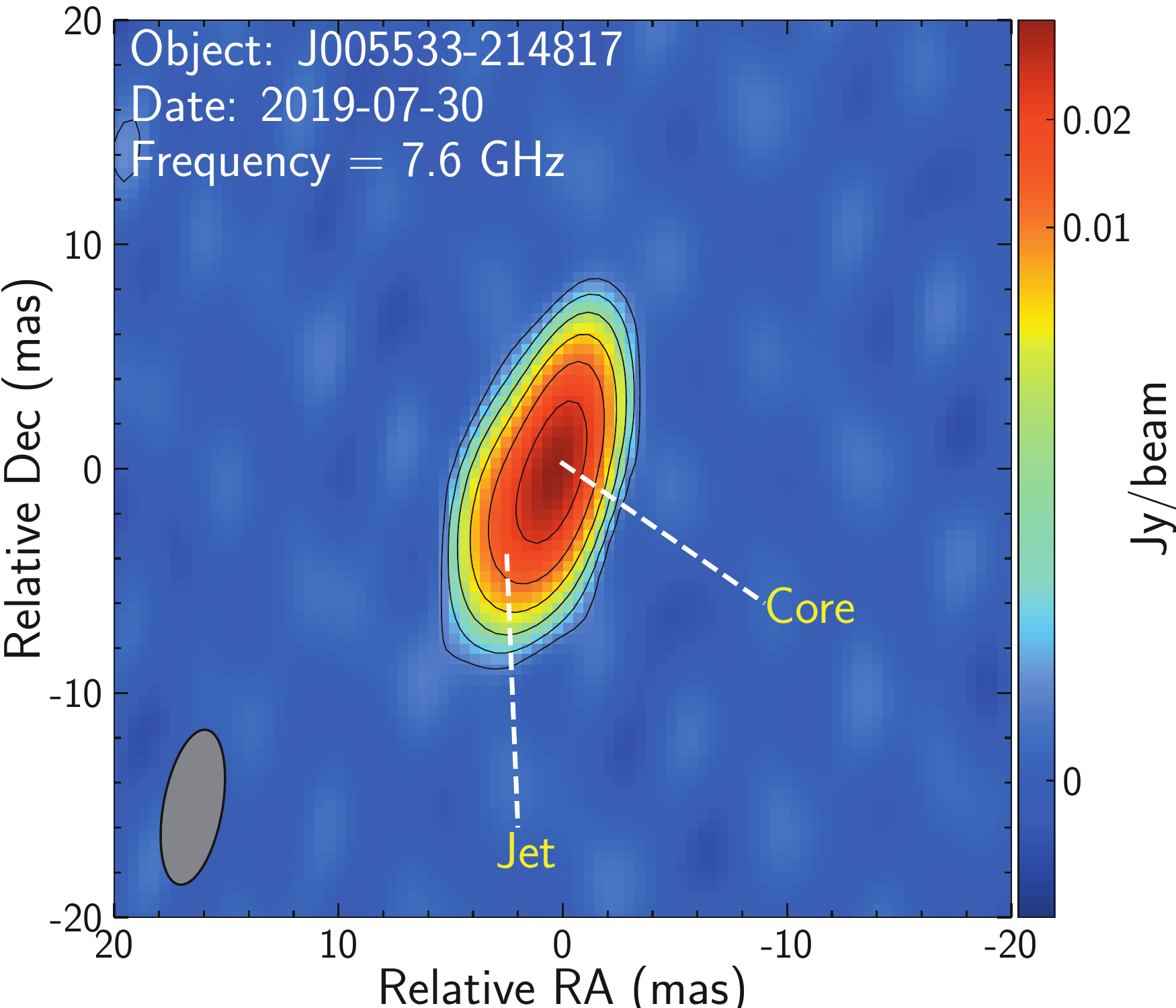}&
\includegraphics[scale=0.24]{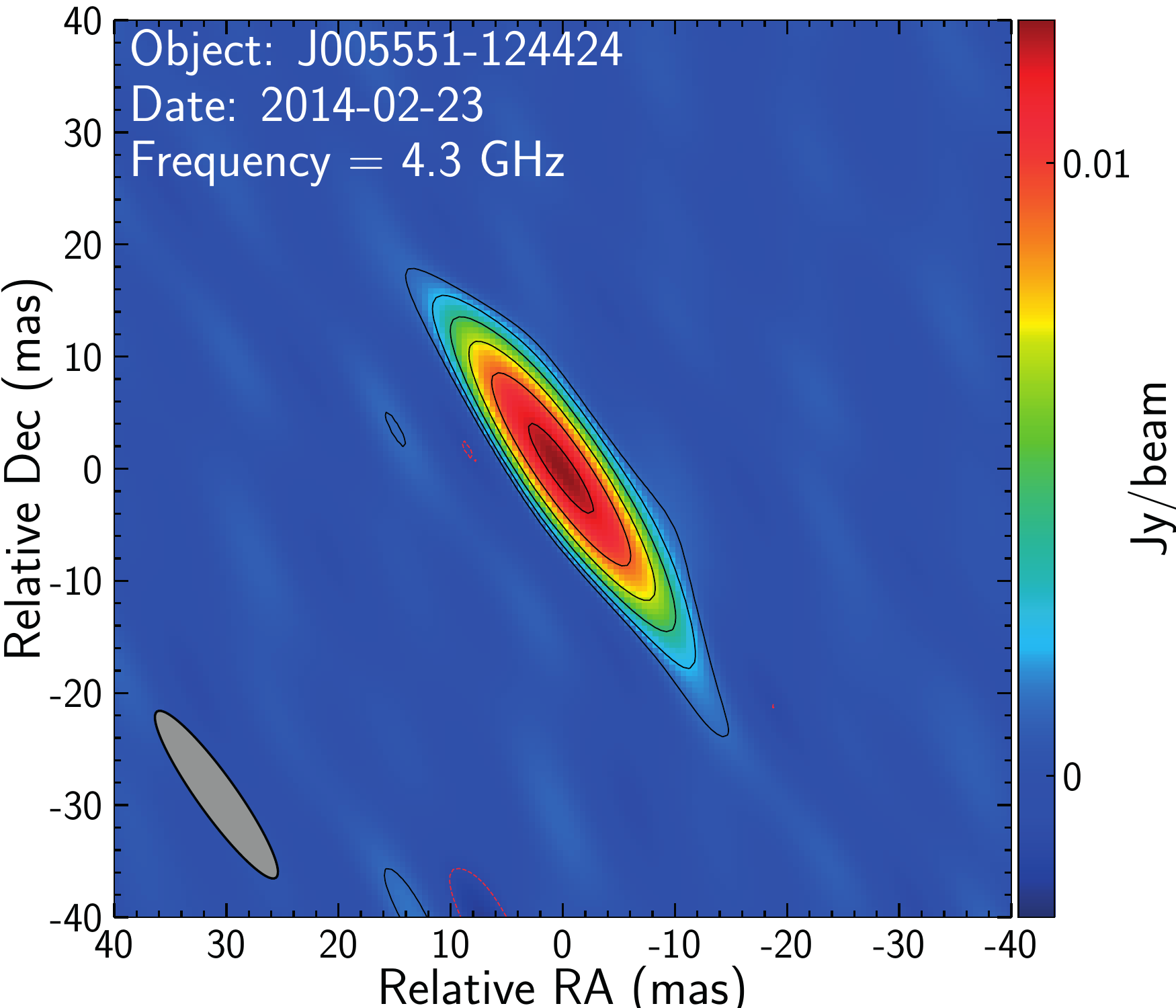}\\
\includegraphics[scale=0.24]{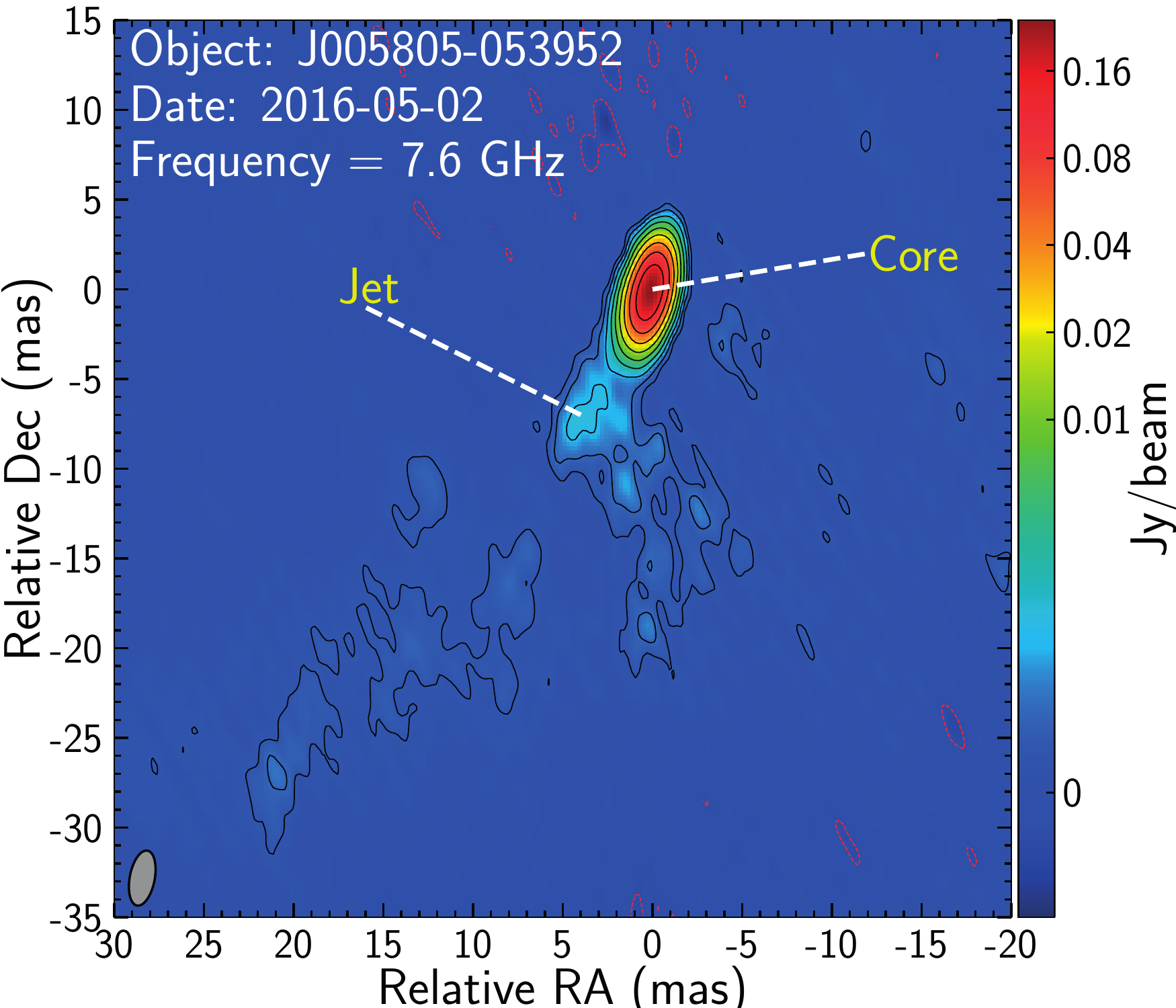}&
\includegraphics[scale=0.24]{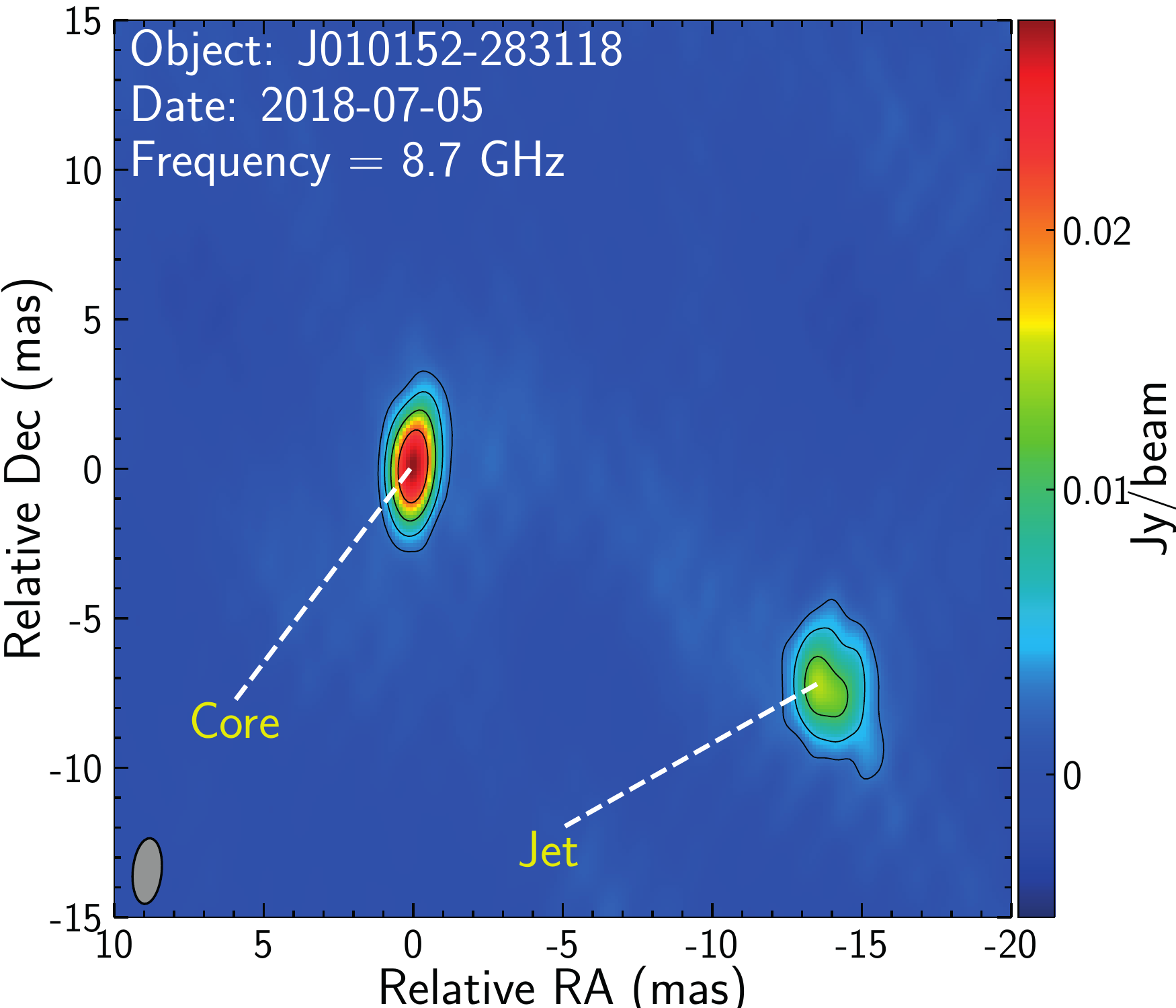}&
\includegraphics[scale=0.24]{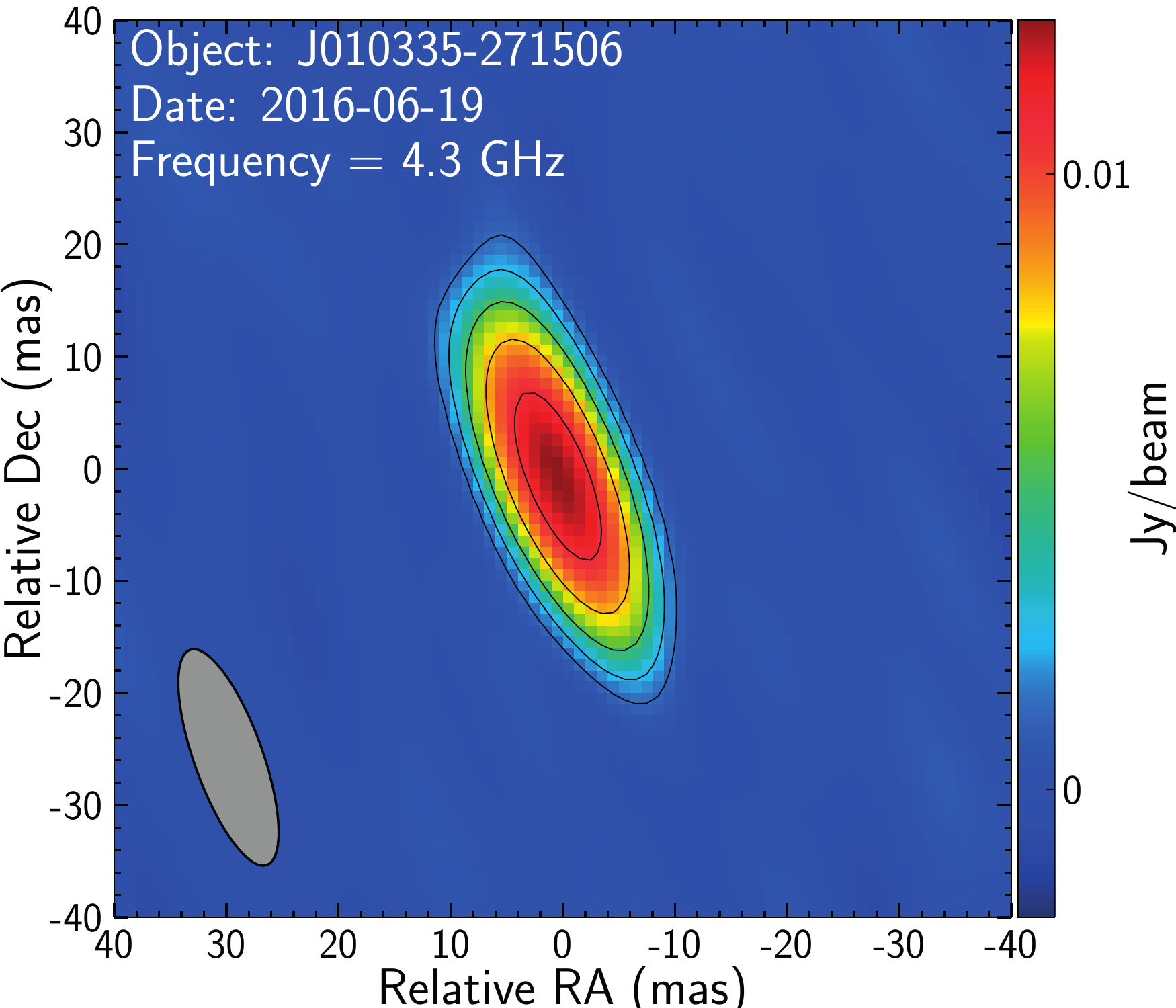}&
\includegraphics[scale=0.24]{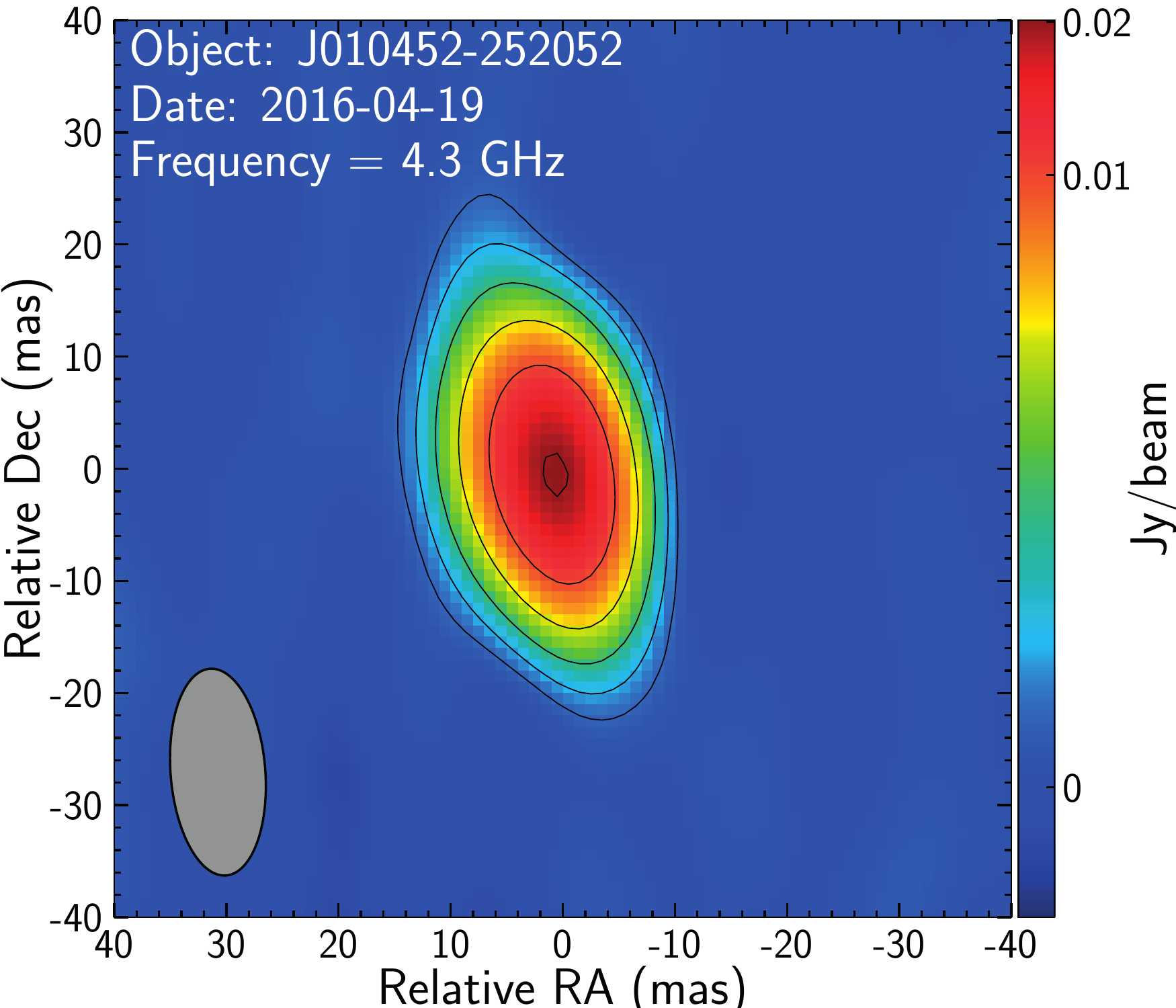}\\
\includegraphics[scale=0.24]{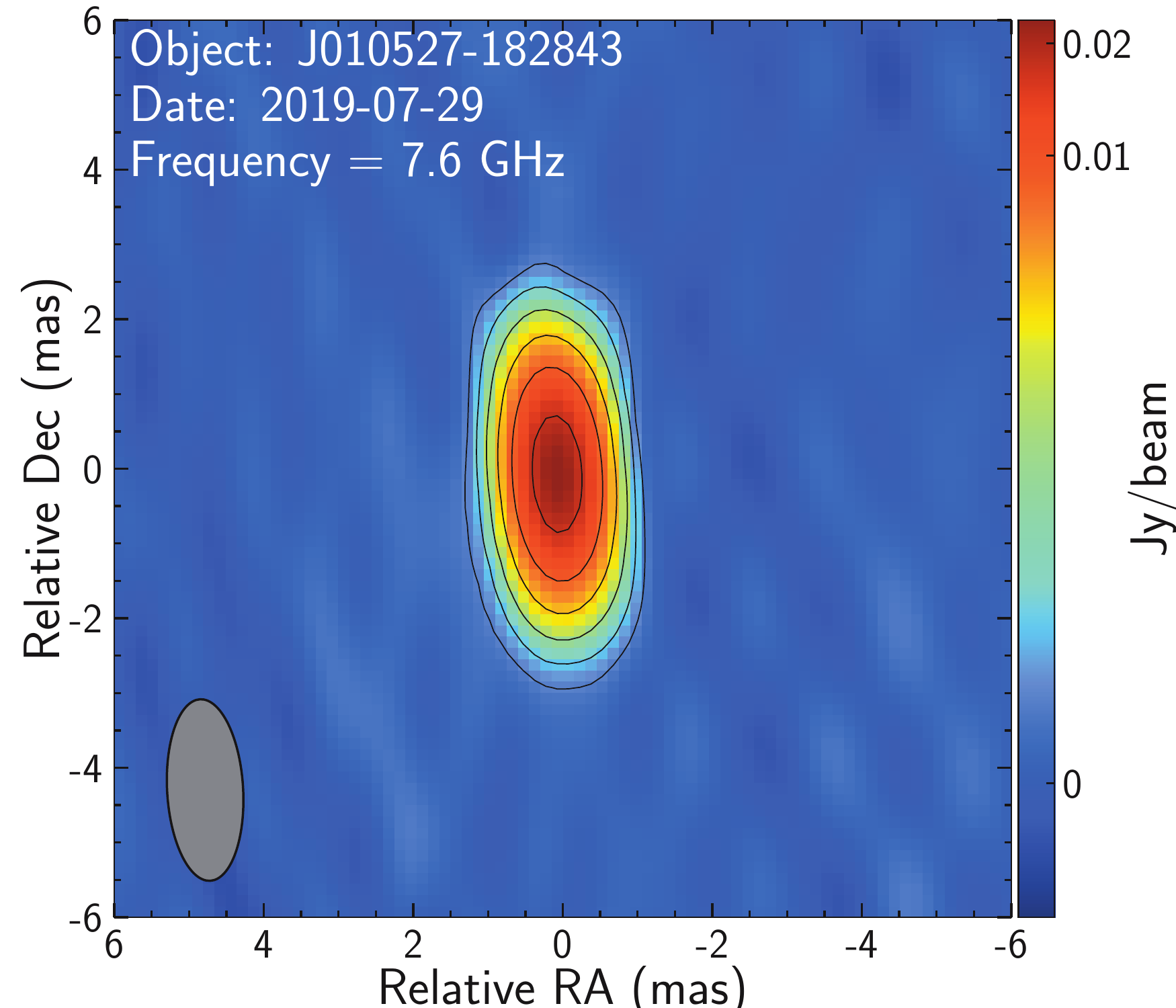}&
\includegraphics[scale=0.24]{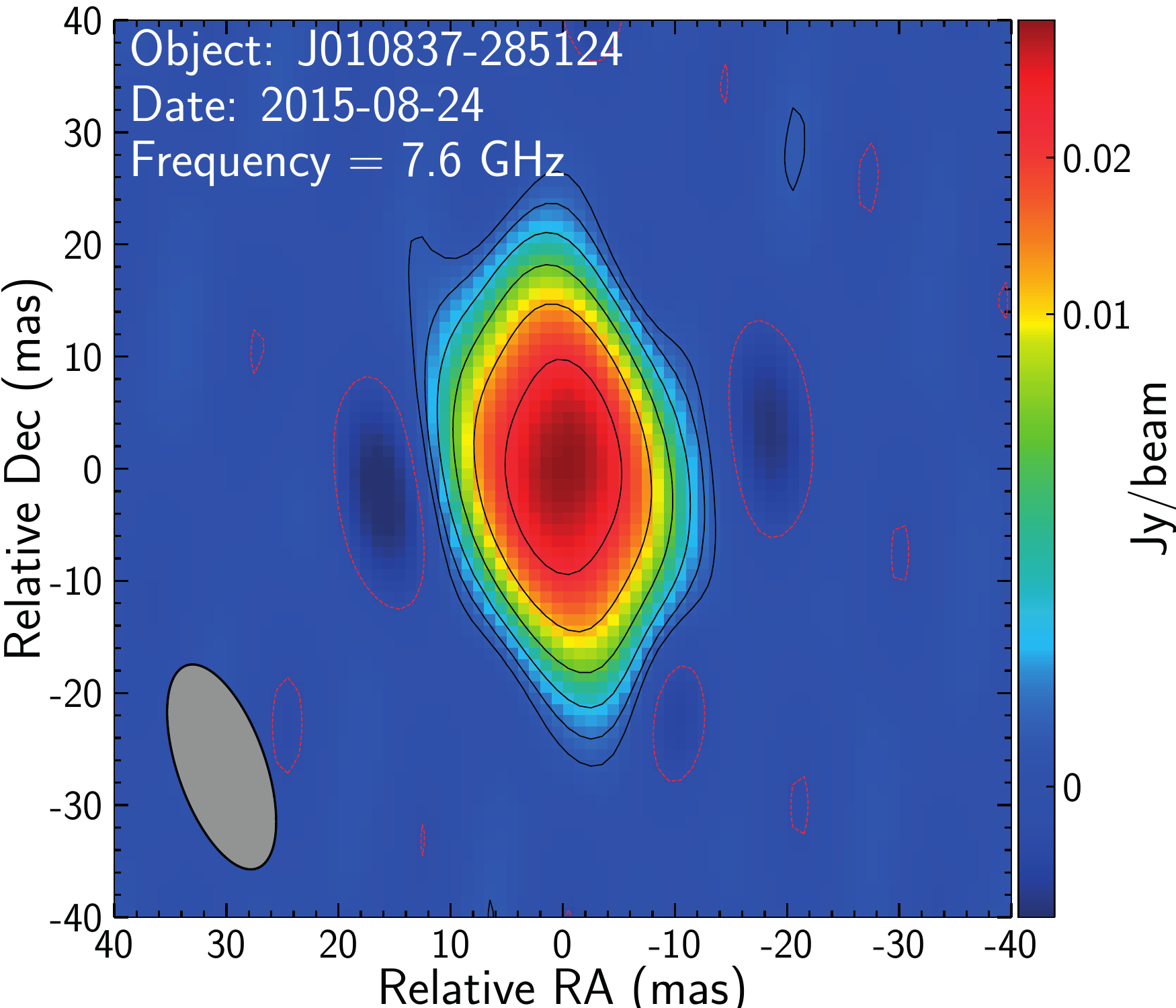}&
\includegraphics[scale=0.24]{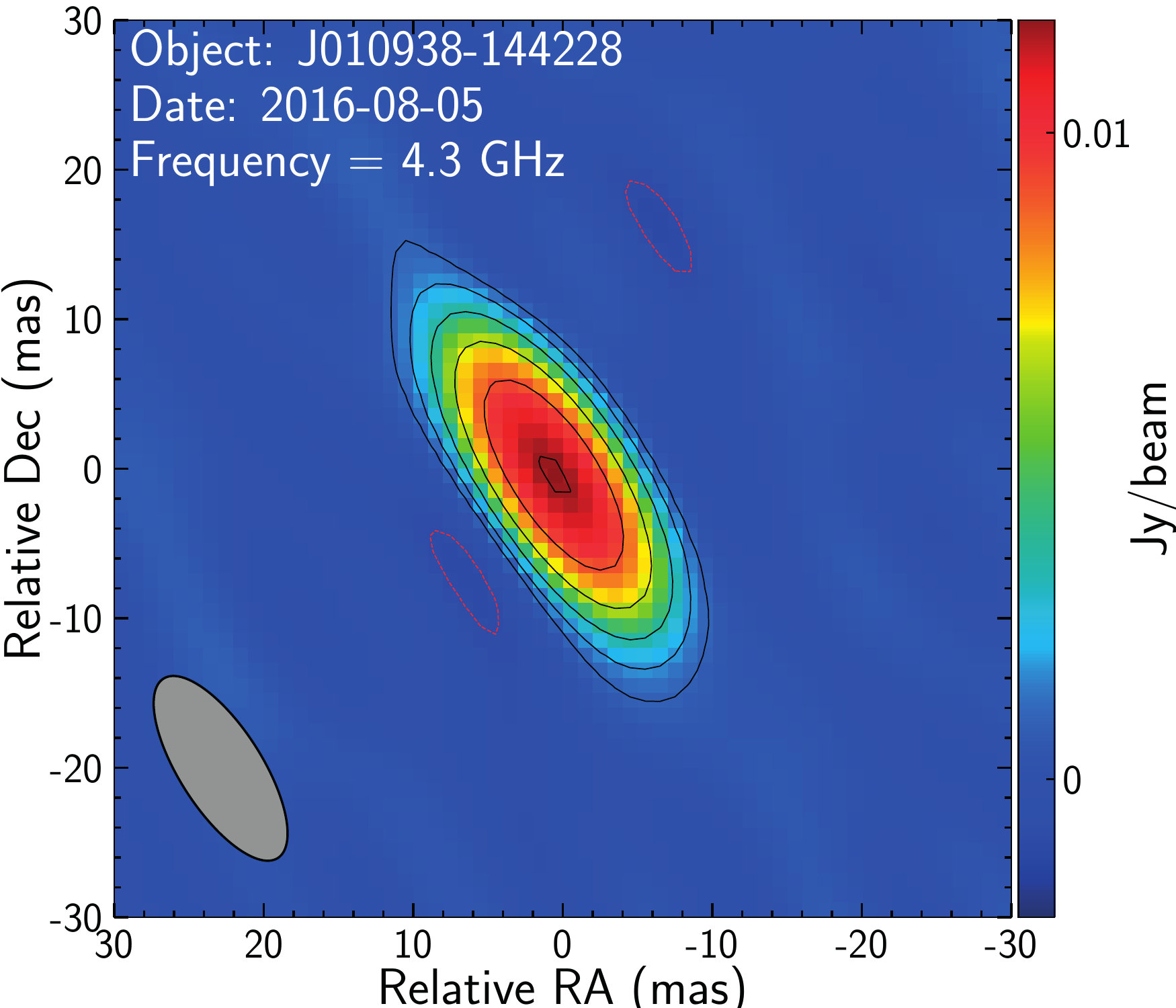}&
\includegraphics[scale=0.24]{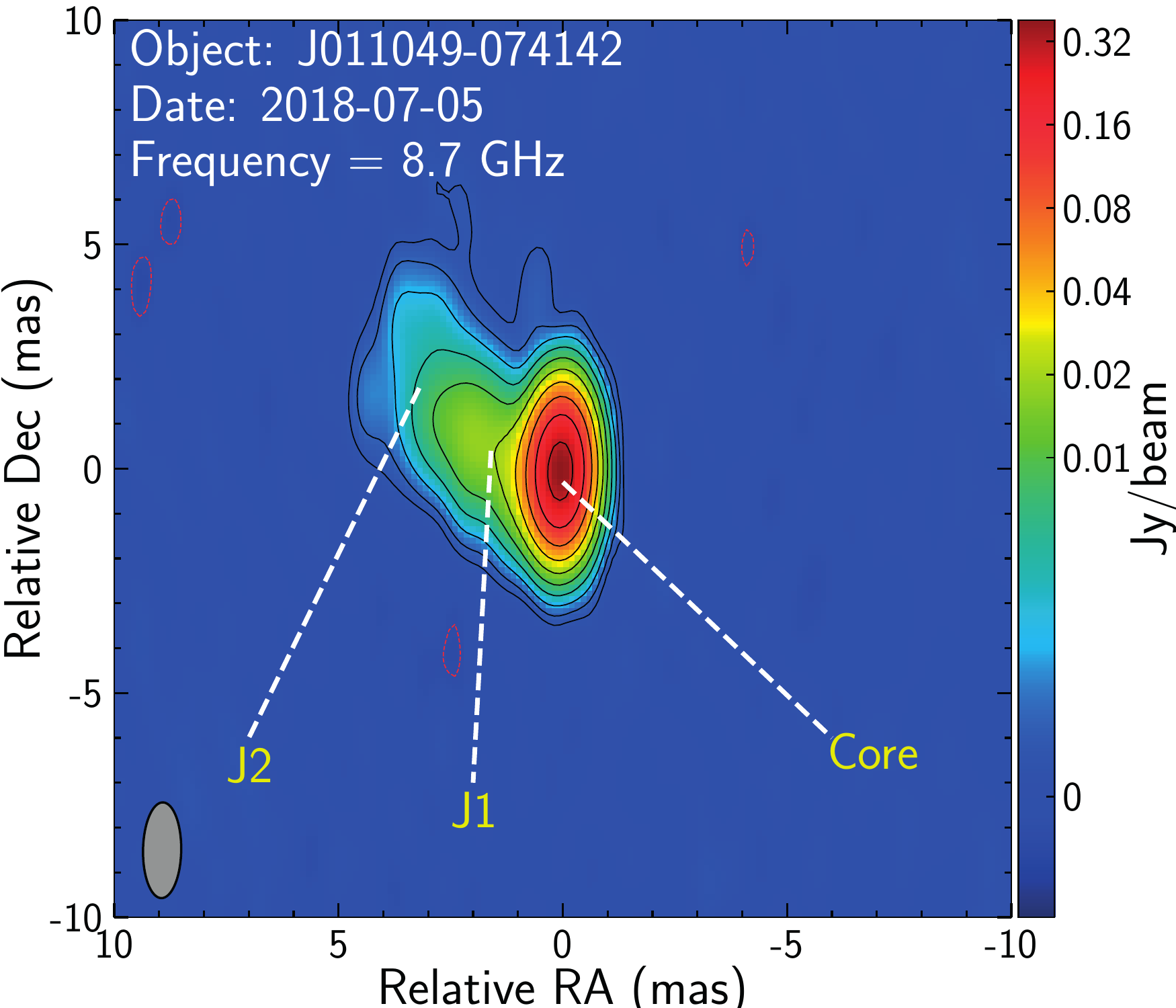}\\
\includegraphics[scale=0.24]{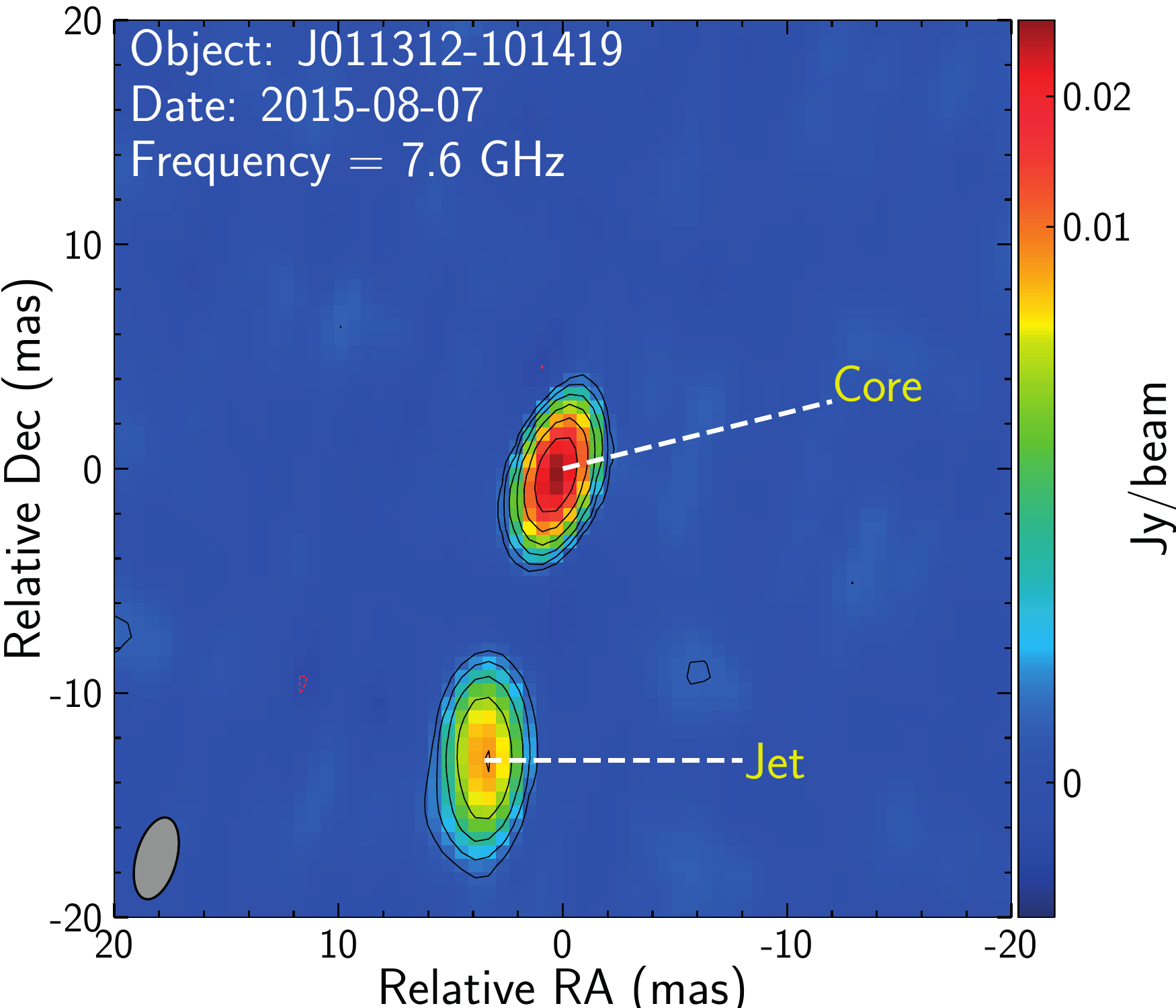}&
\includegraphics[scale=0.24]{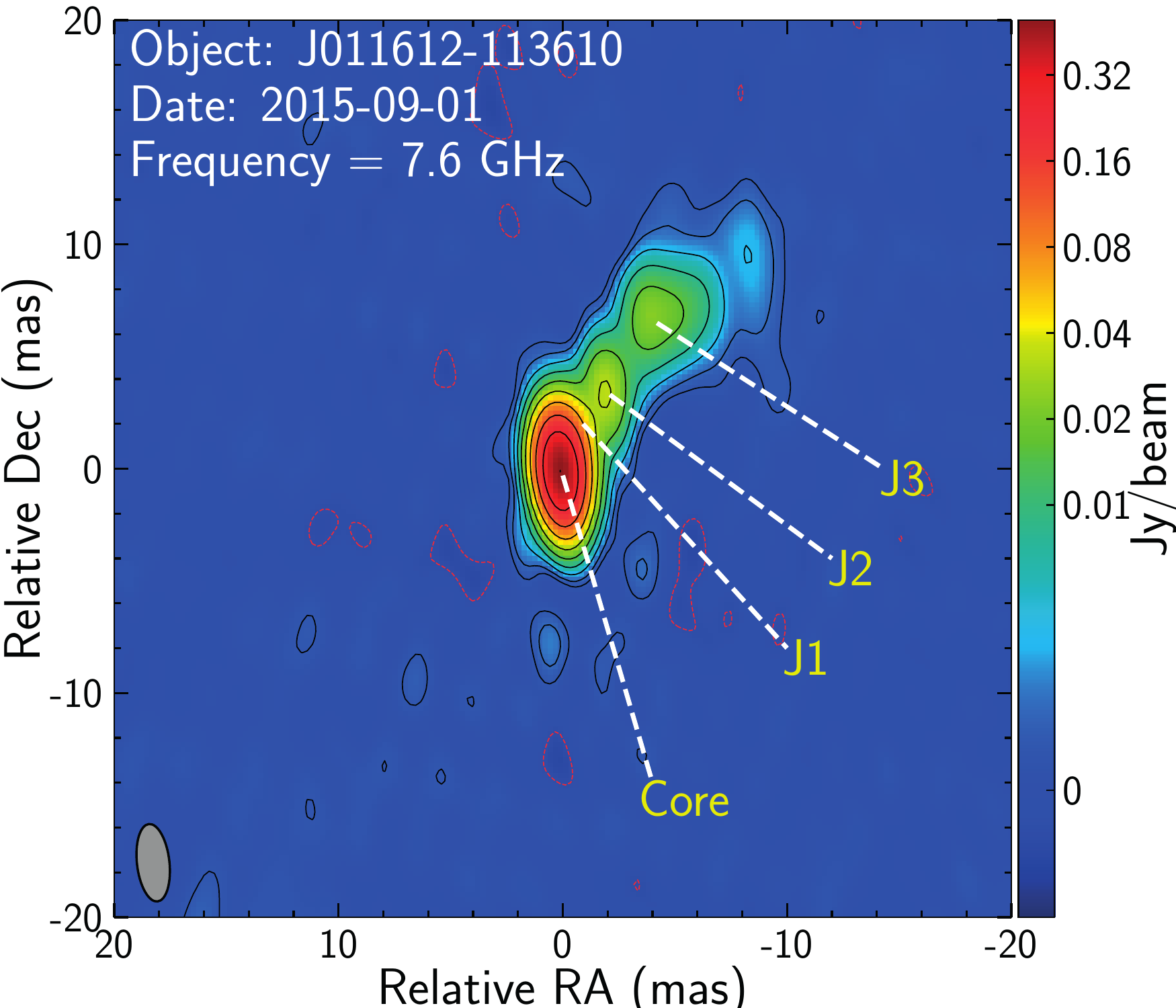}&
\includegraphics[scale=0.24]{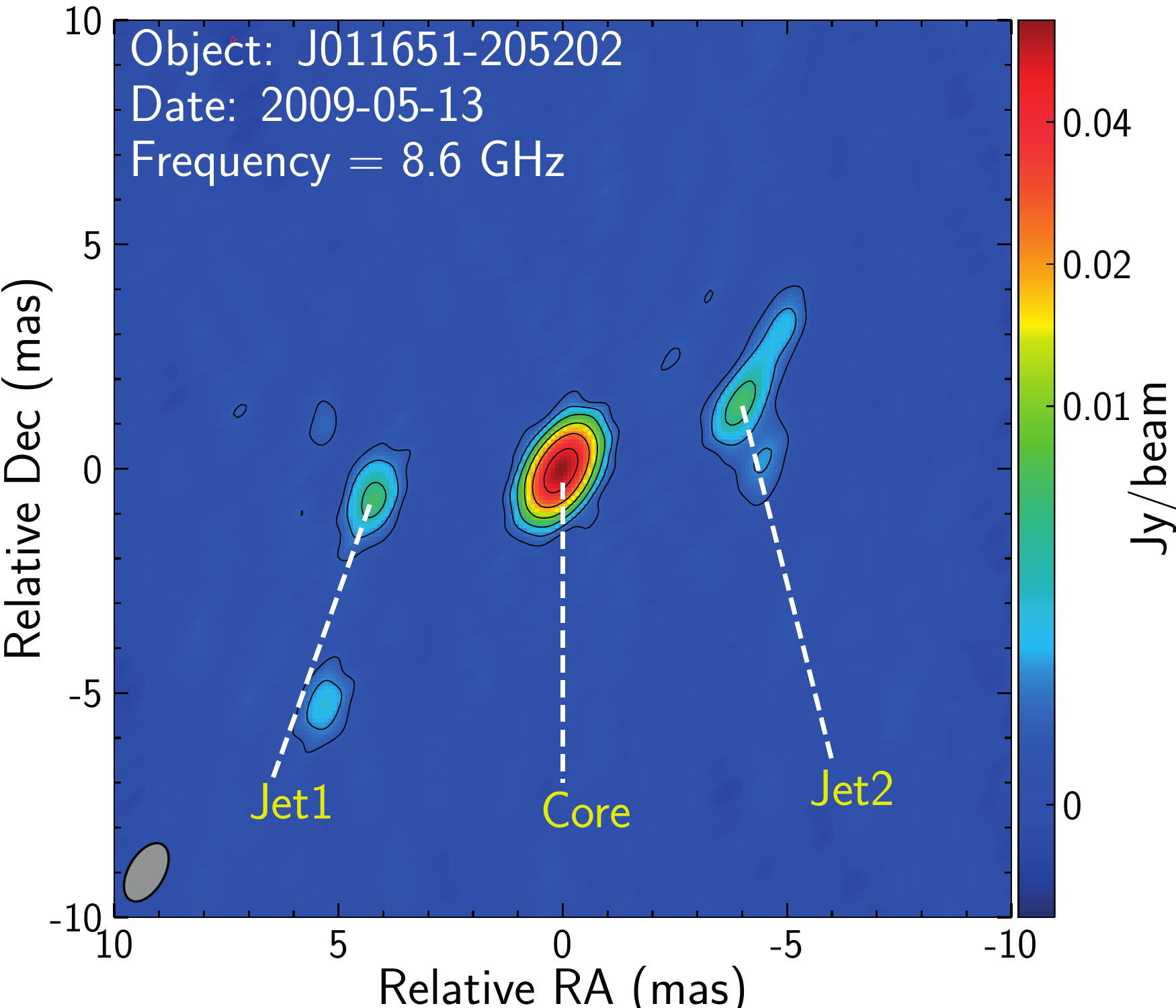}&
\includegraphics[scale=0.24]{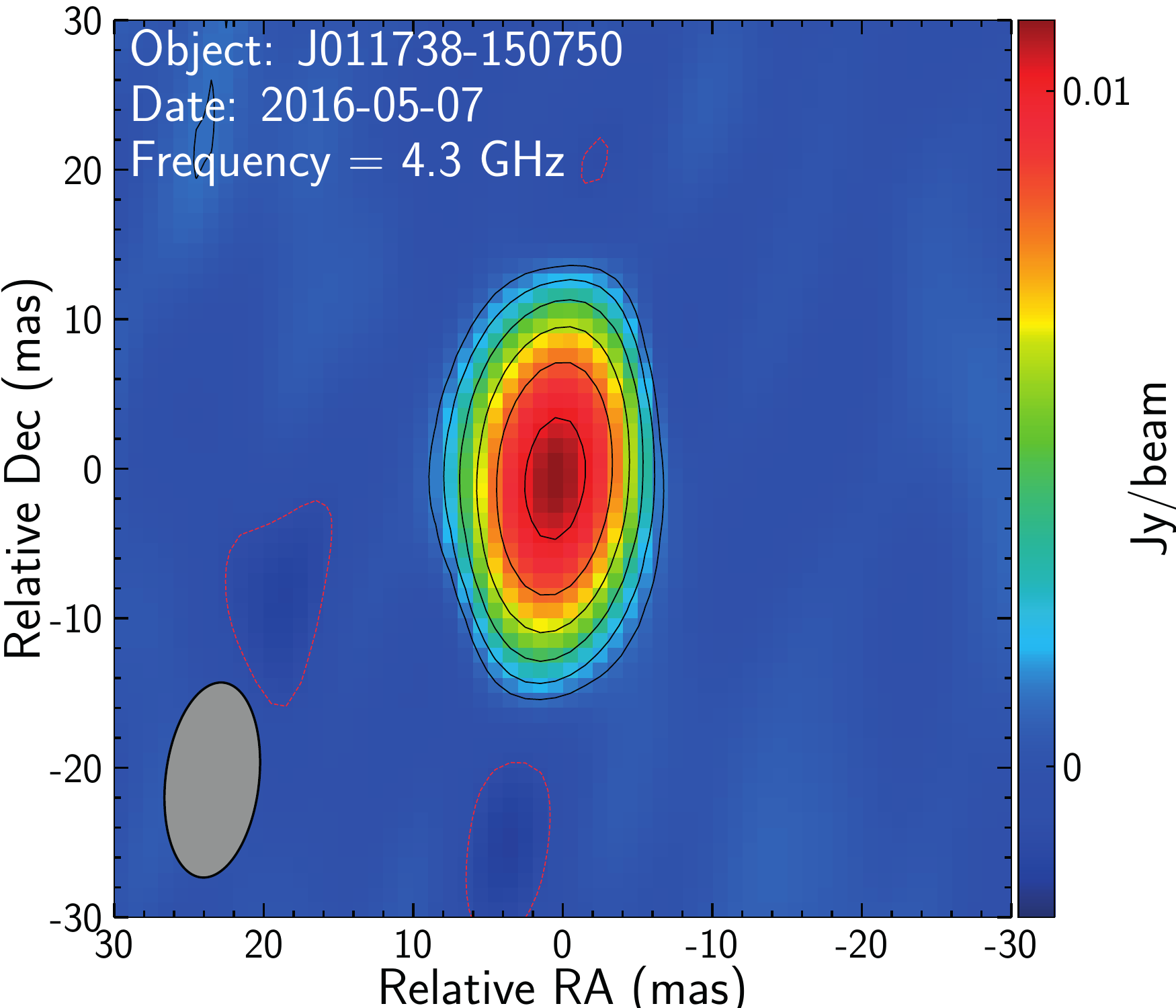}\\
\includegraphics[scale=0.24]{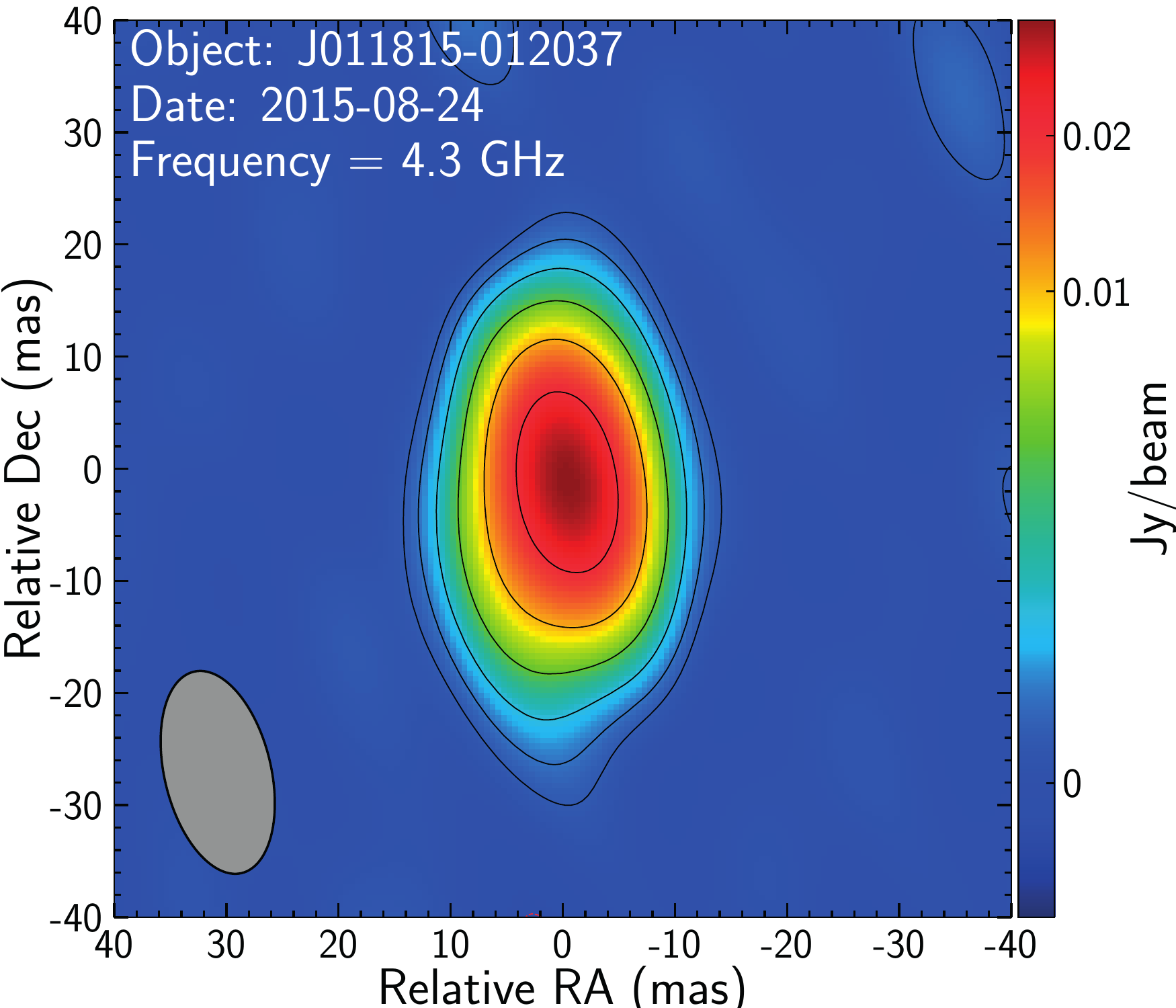}&
\includegraphics[scale=0.24]{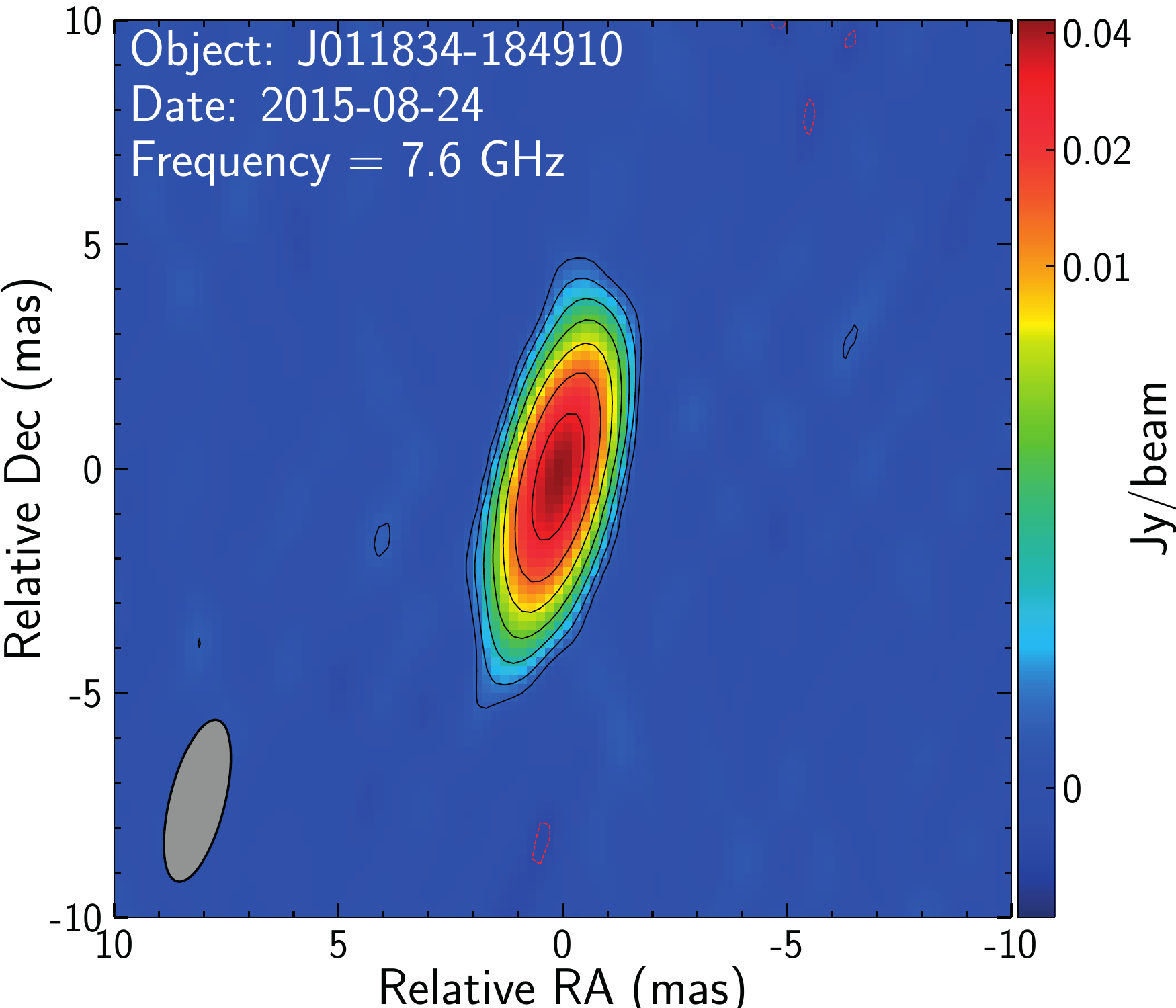}&
\includegraphics[scale=0.24]{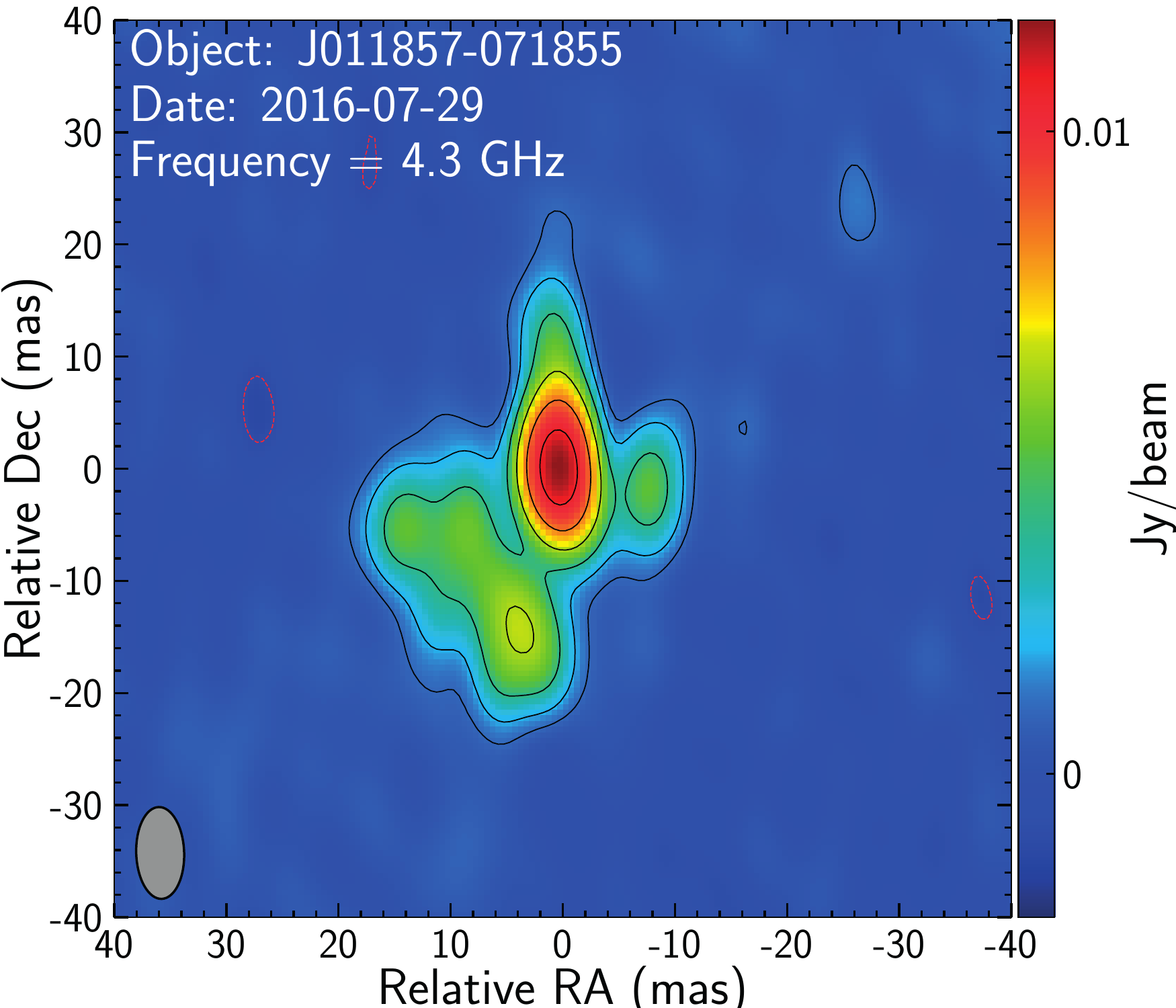}&
\includegraphics[scale=0.24]{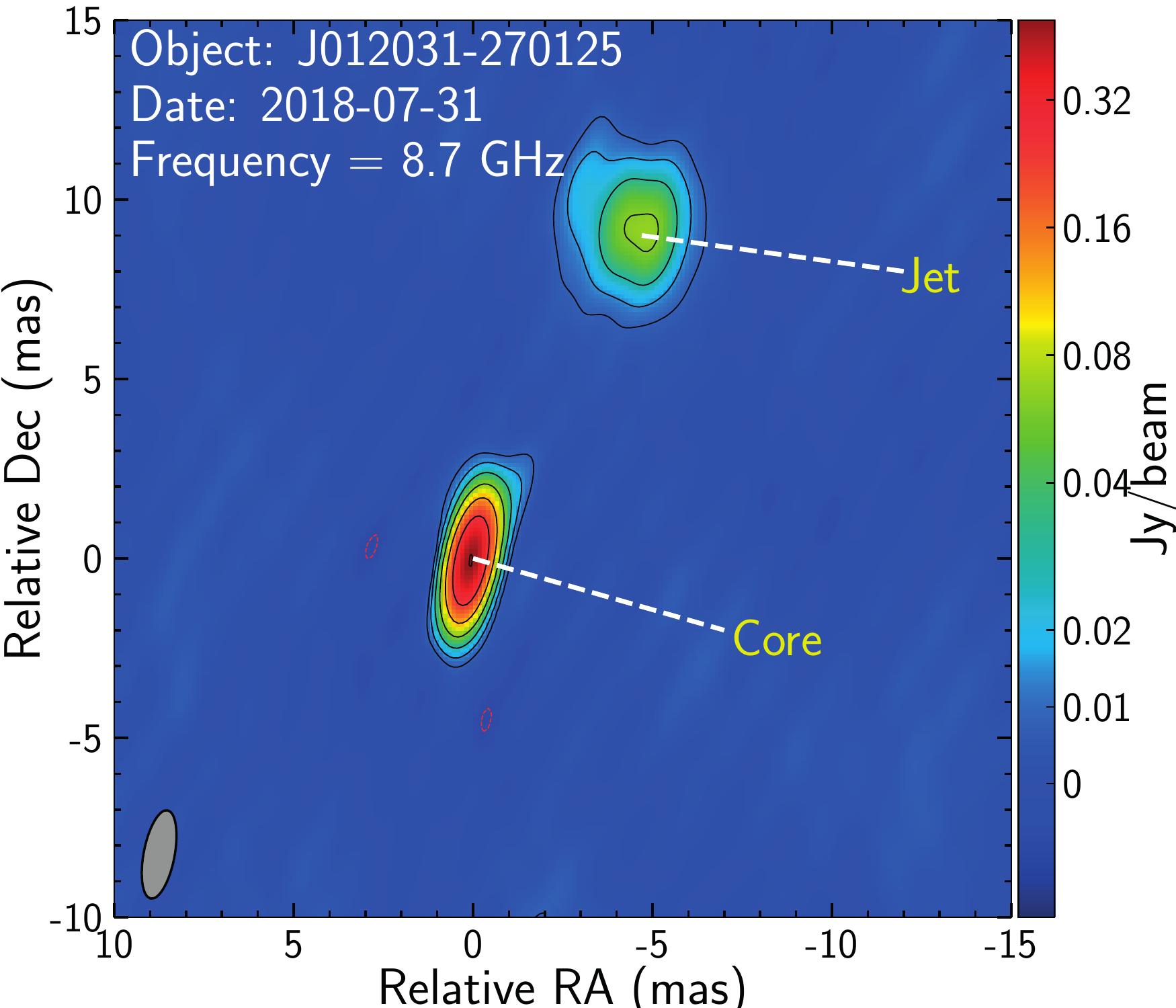}
\end{tabular}
\caption{\small Continued.}
\end{figure*}
\addtocounter{figure}{-1}
\begin{figure*}
\centering
\begin{tabular}{cccc}
\includegraphics[scale=0.24]{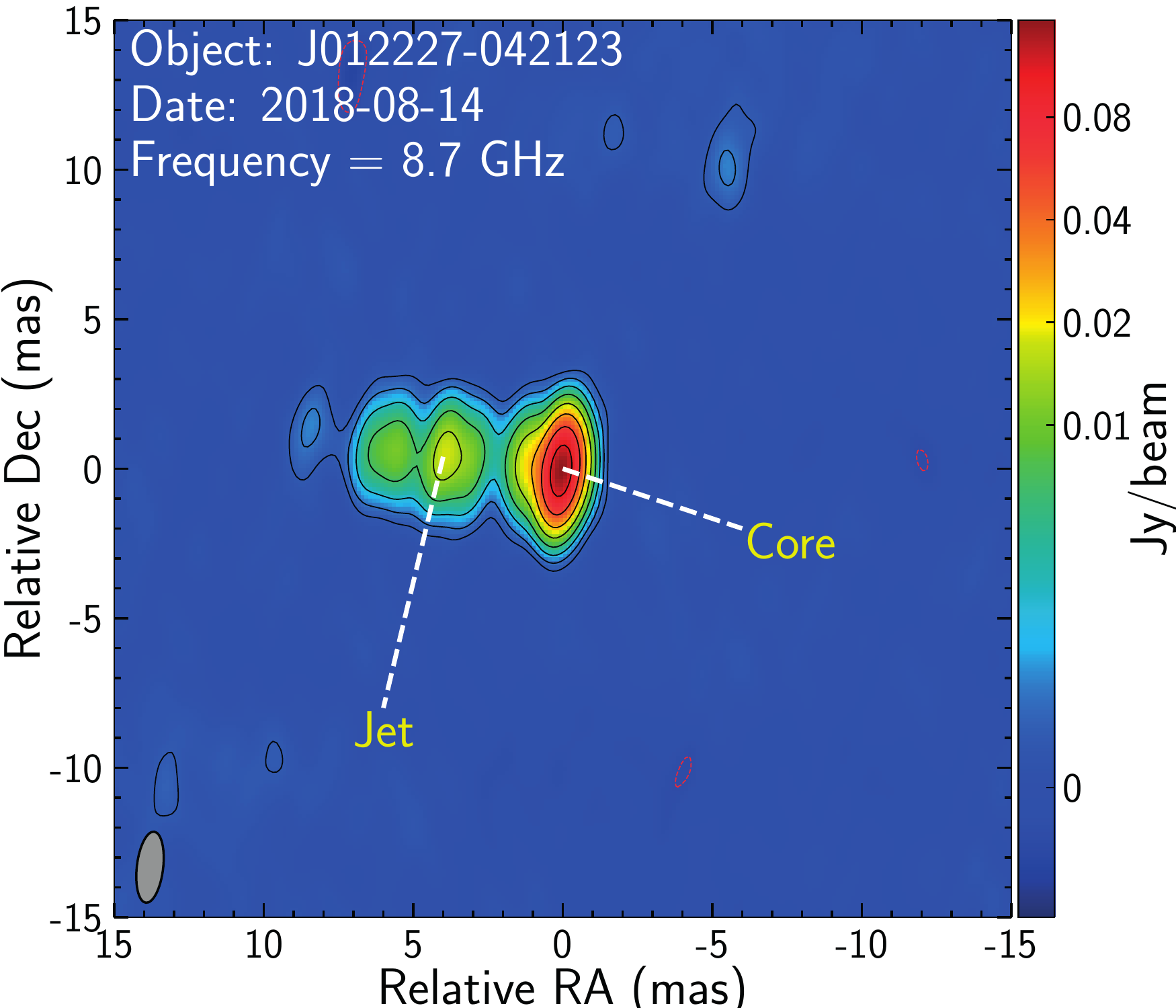}&
\includegraphics[scale=0.24]{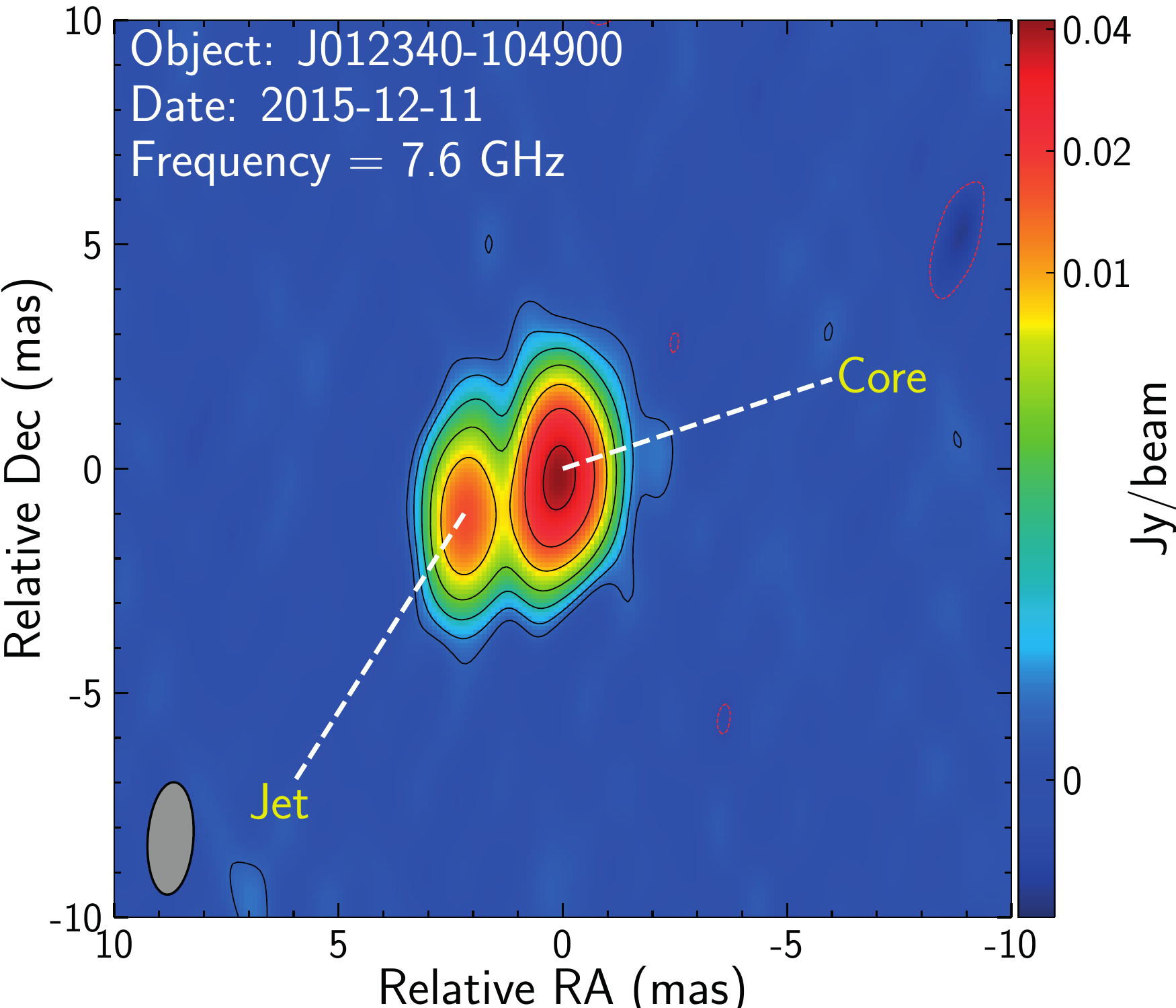}&
\includegraphics[scale=0.24]{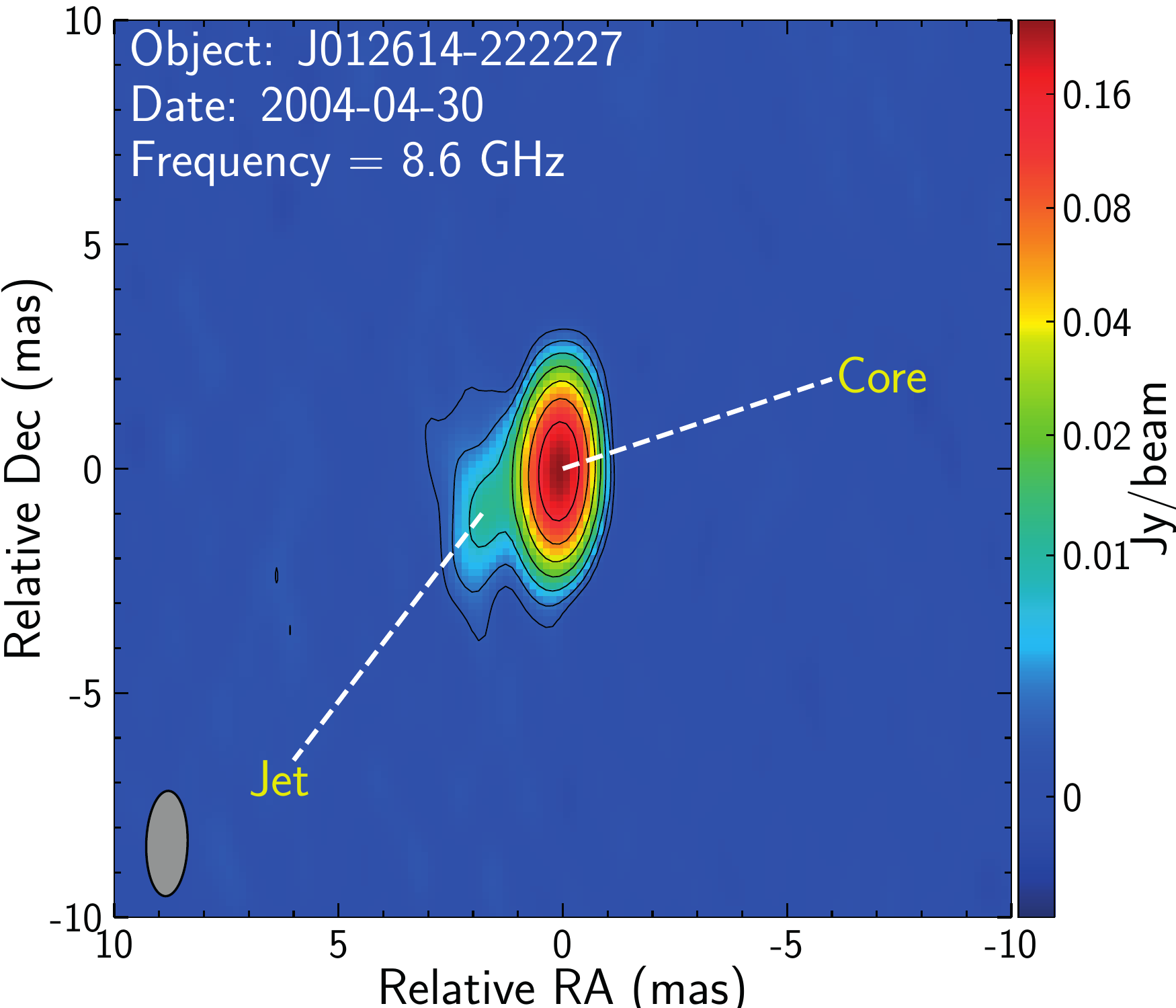}&
\includegraphics[scale=0.24]{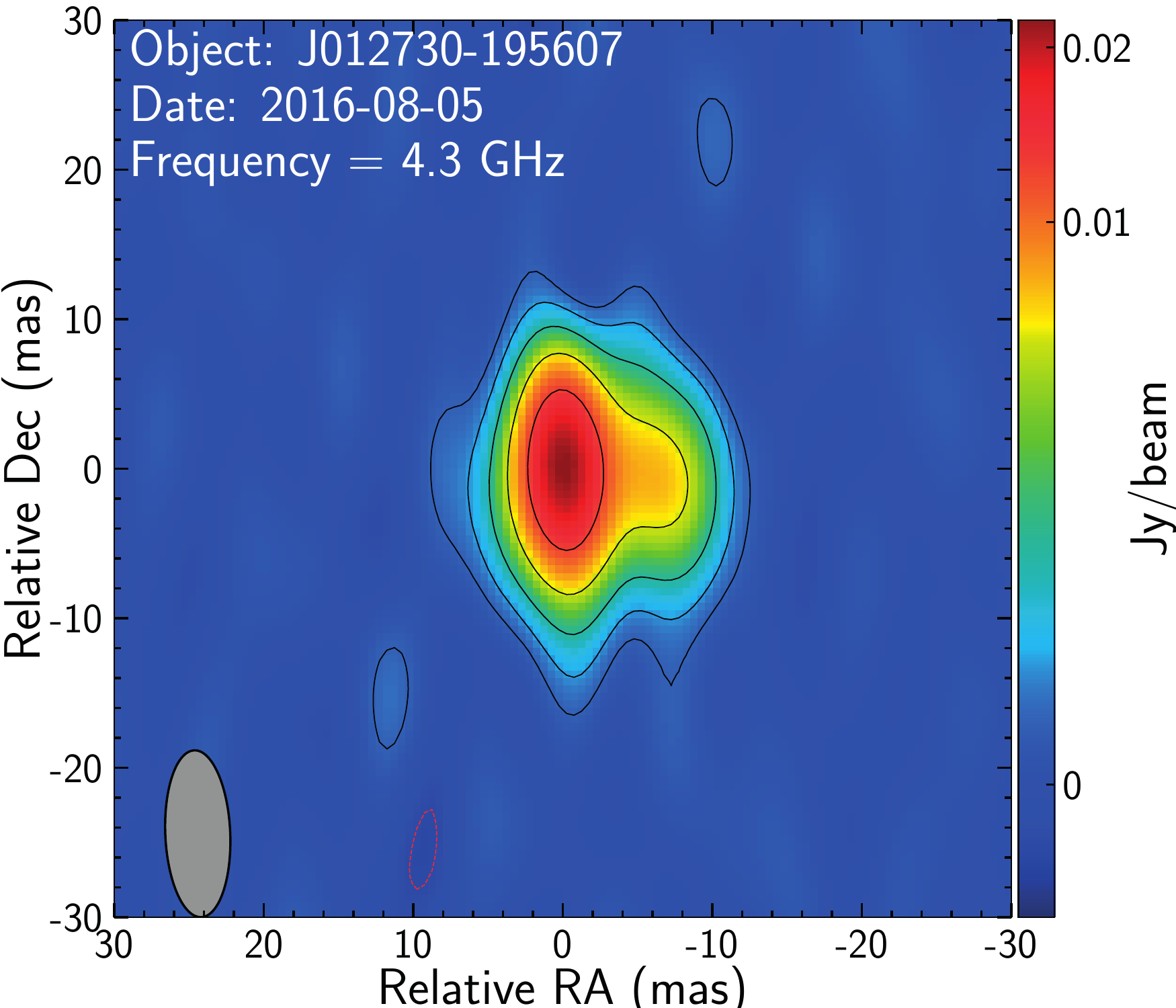}\\
\includegraphics[scale=0.24]{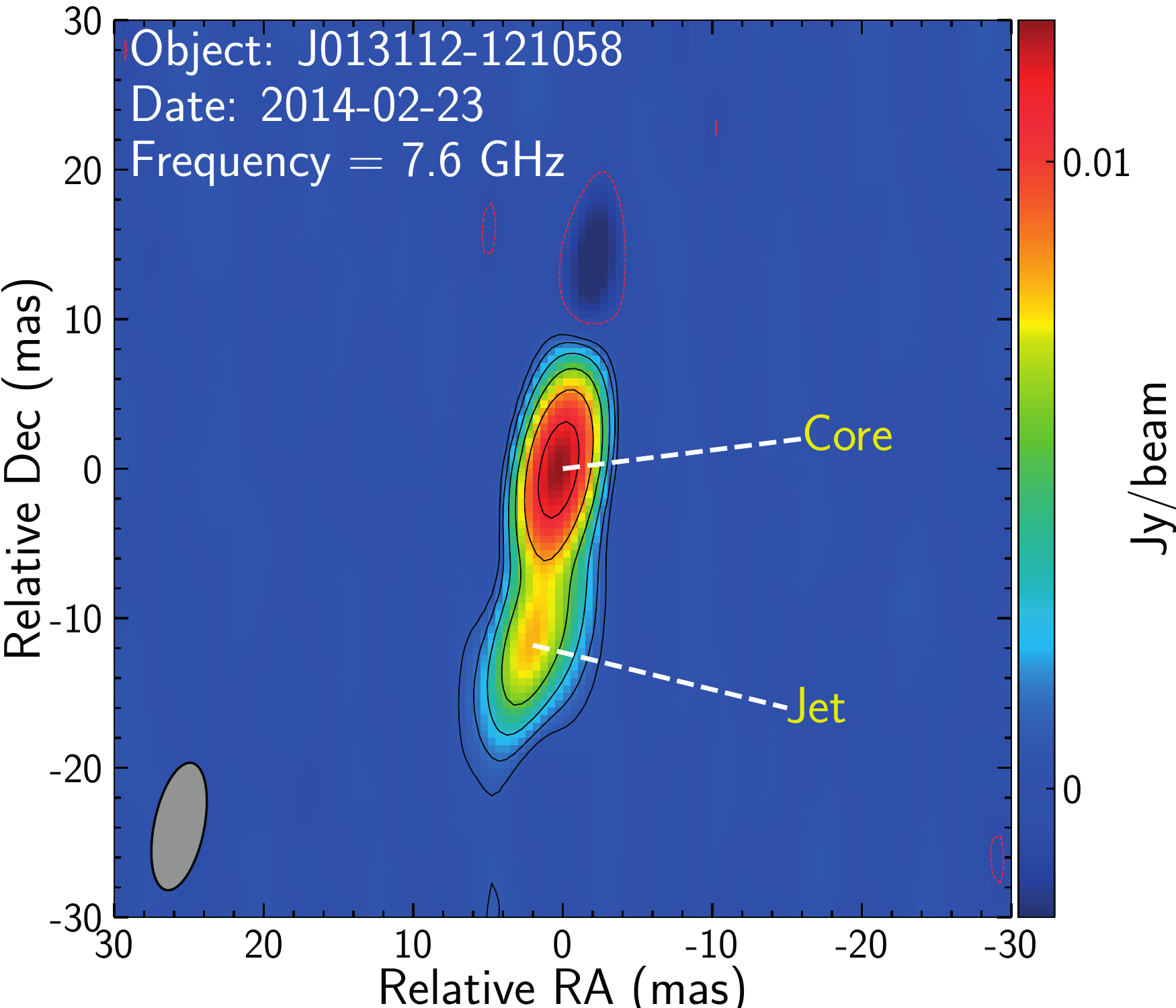}&
\includegraphics[scale=0.24]{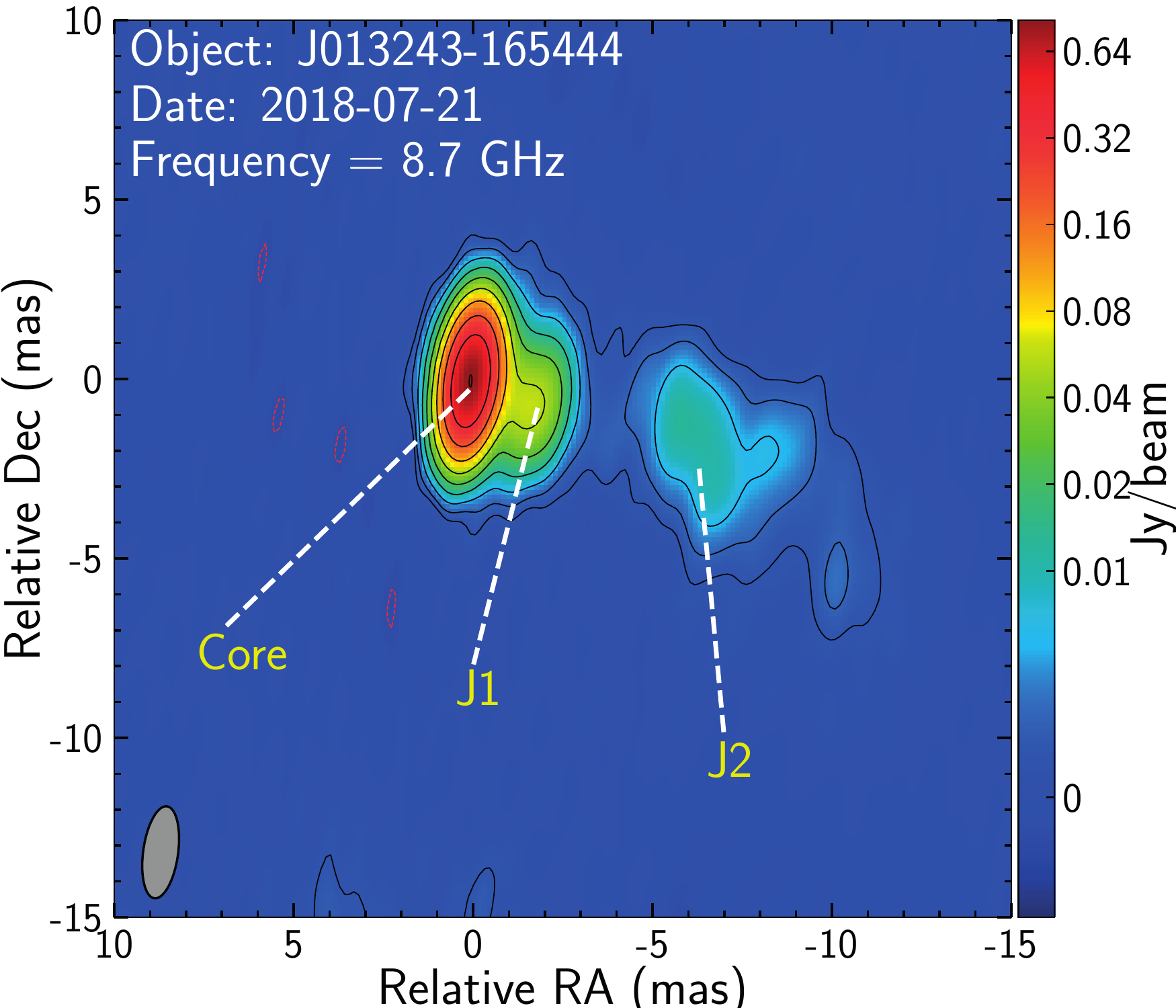}&
\includegraphics[scale=0.24]{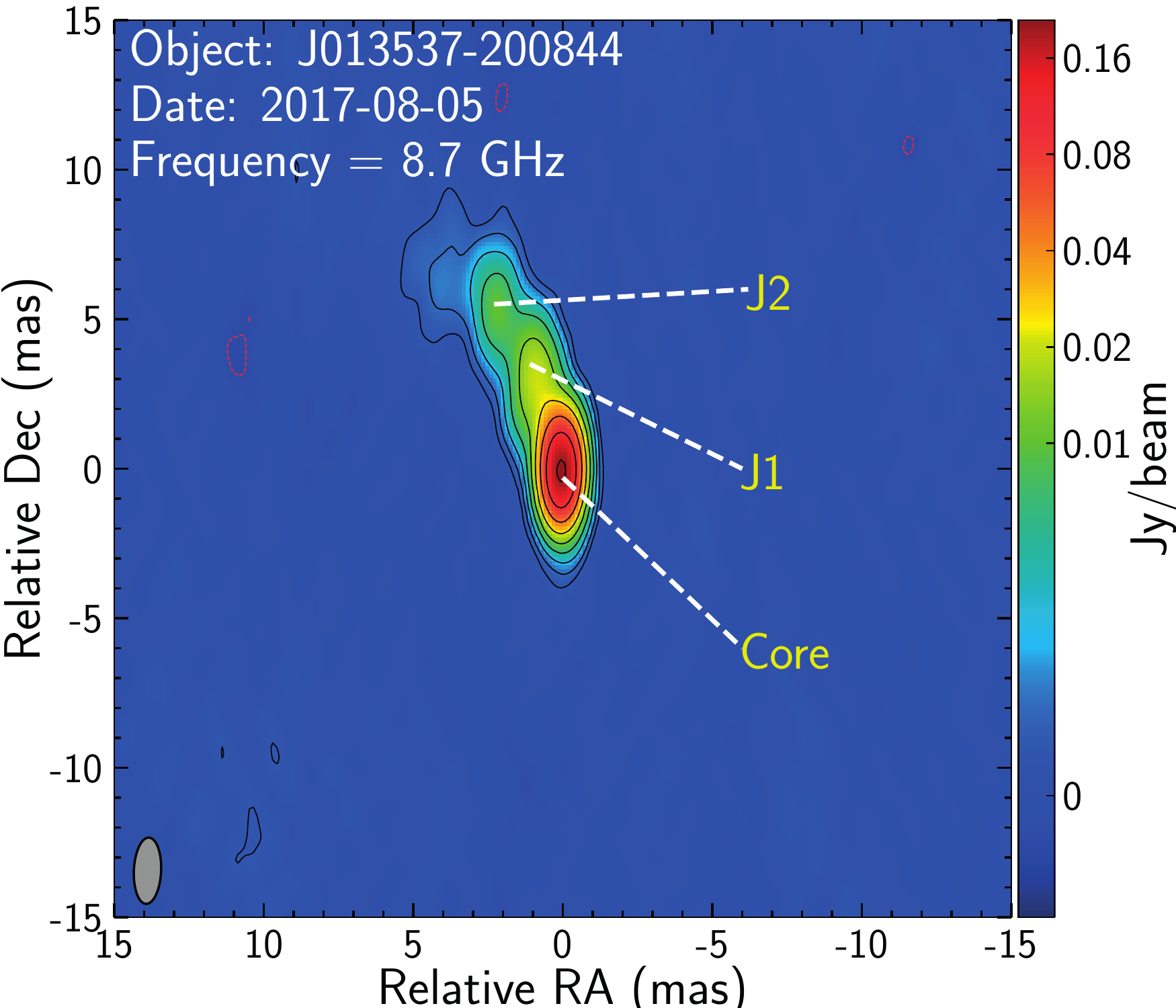}&
\includegraphics[scale=0.24]{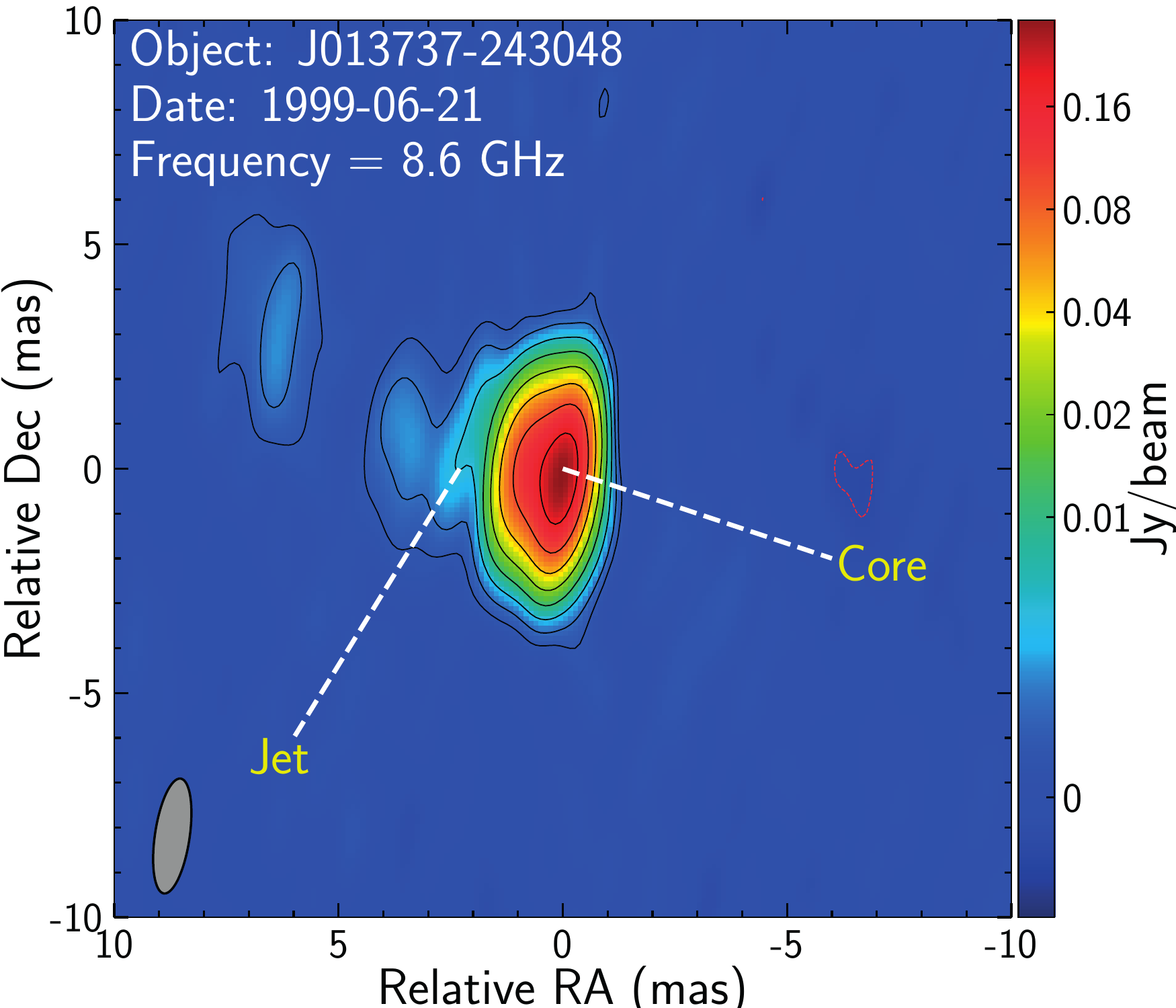}\\
\includegraphics[scale=0.24]{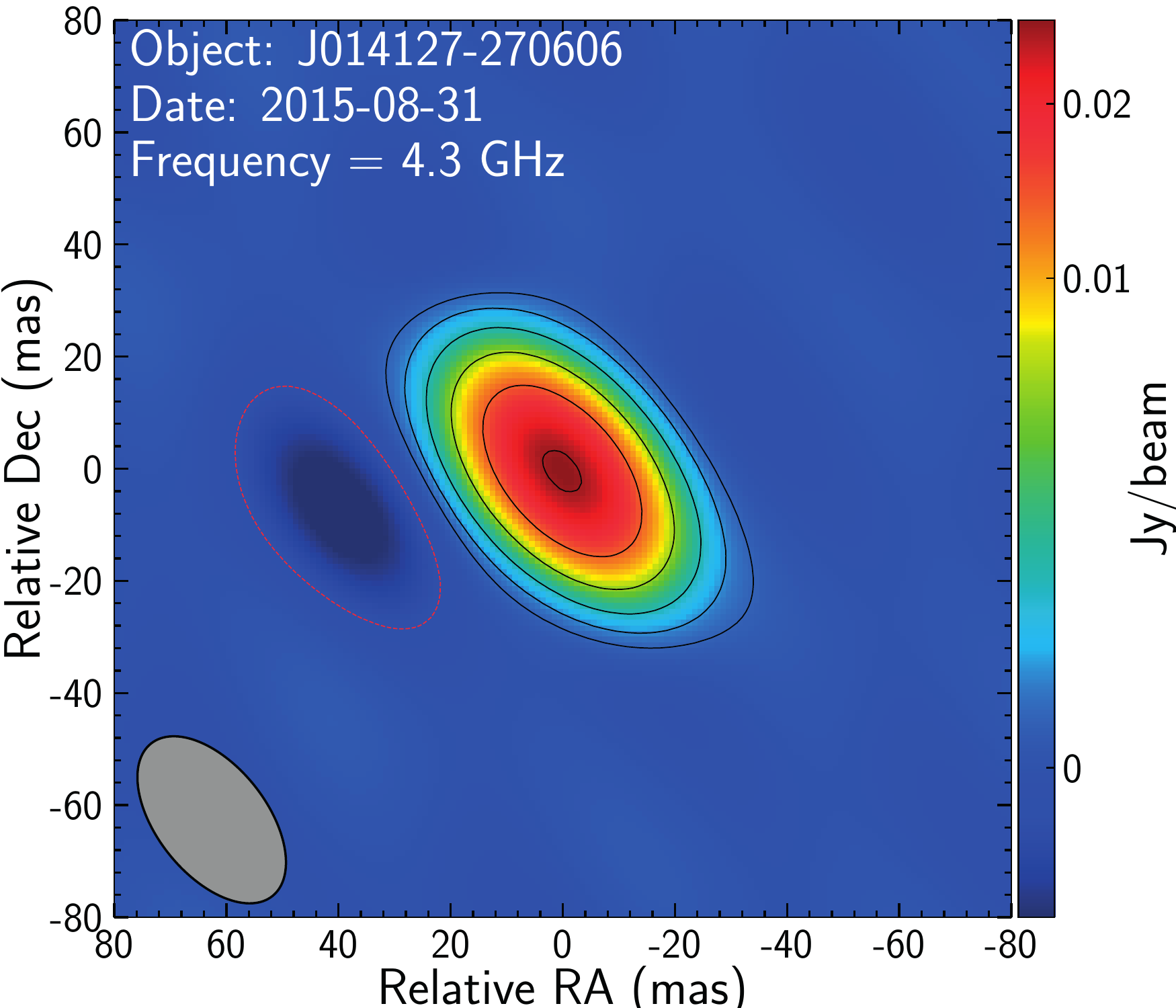}&
\includegraphics[scale=0.24]{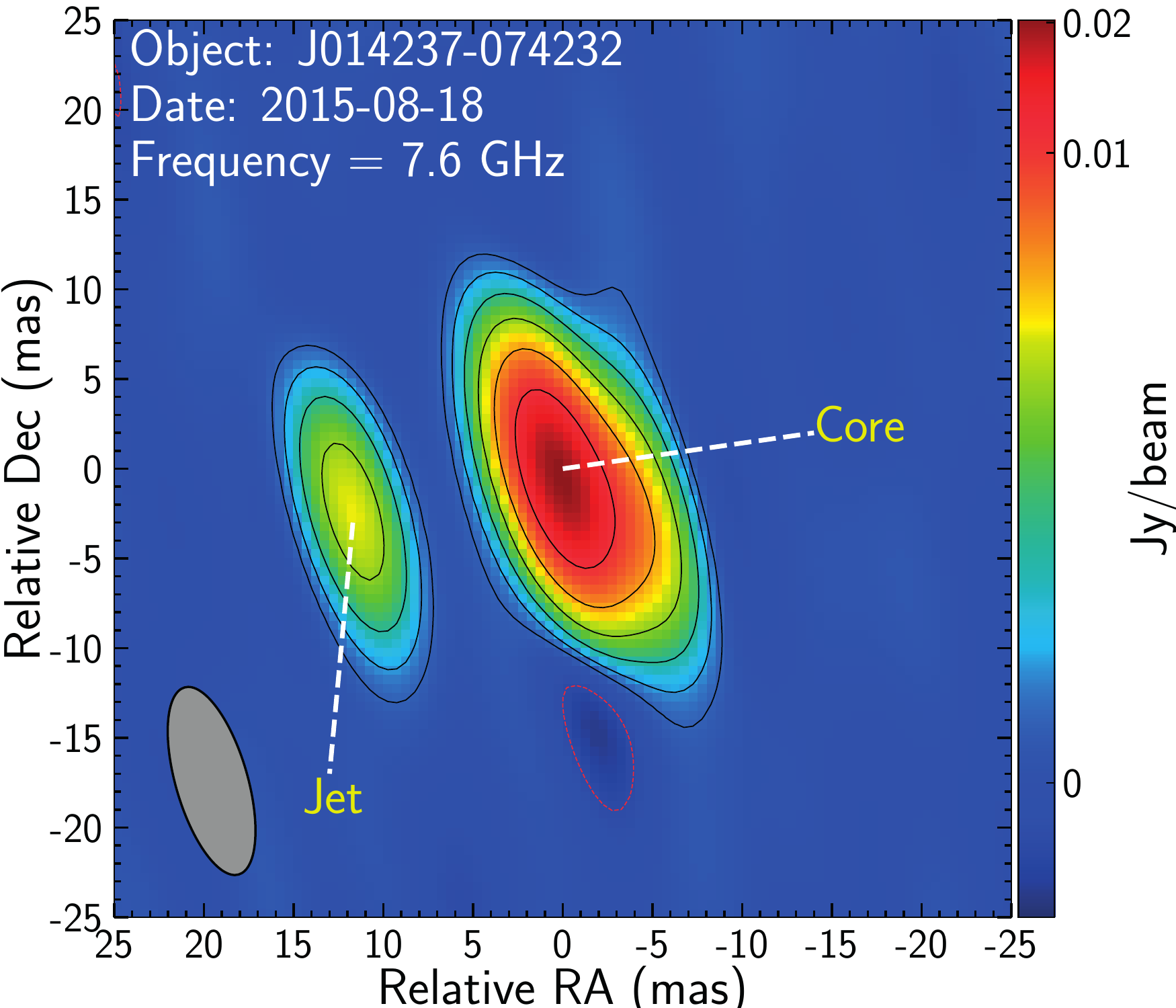}&
\includegraphics[scale=0.24]{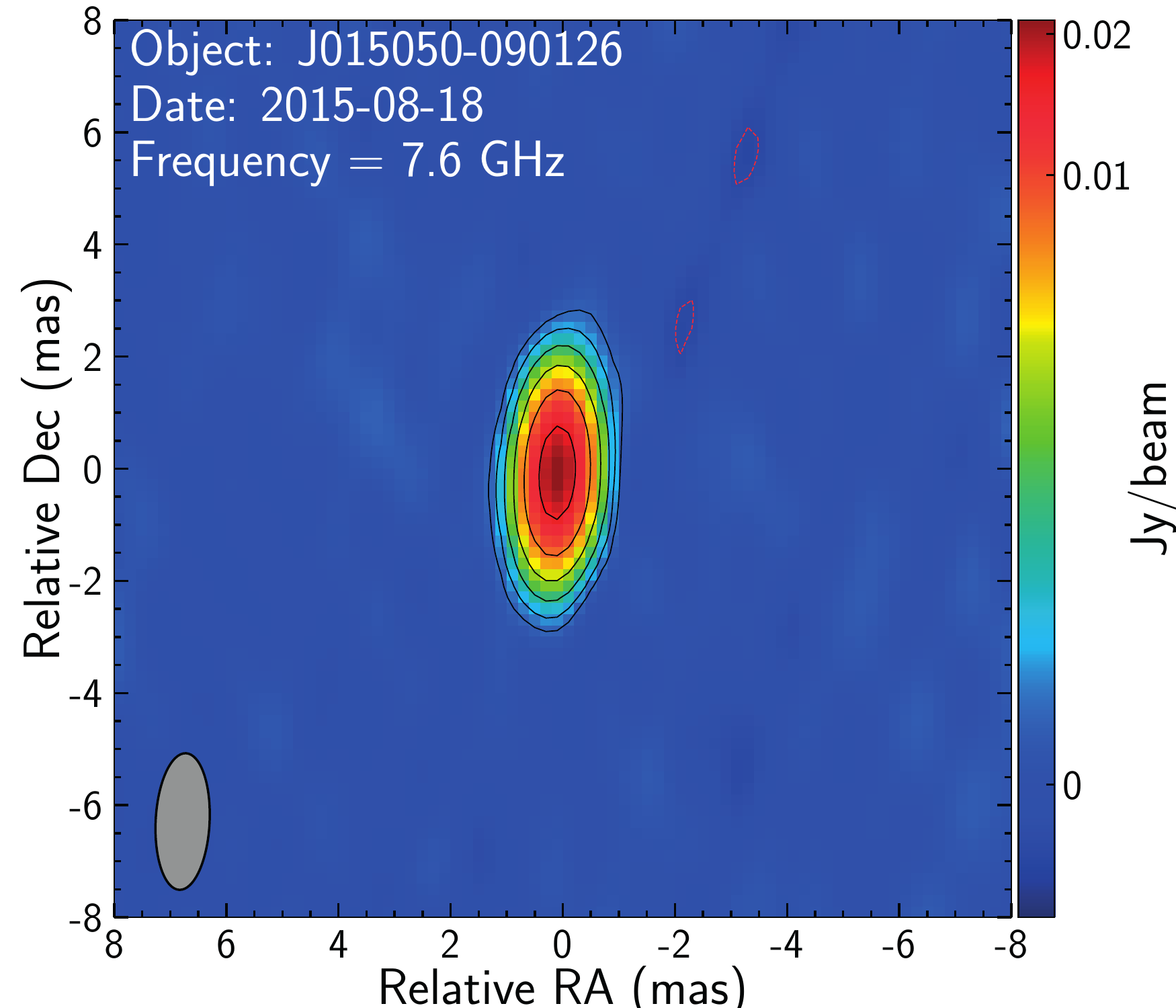}&
\includegraphics[scale=0.24]{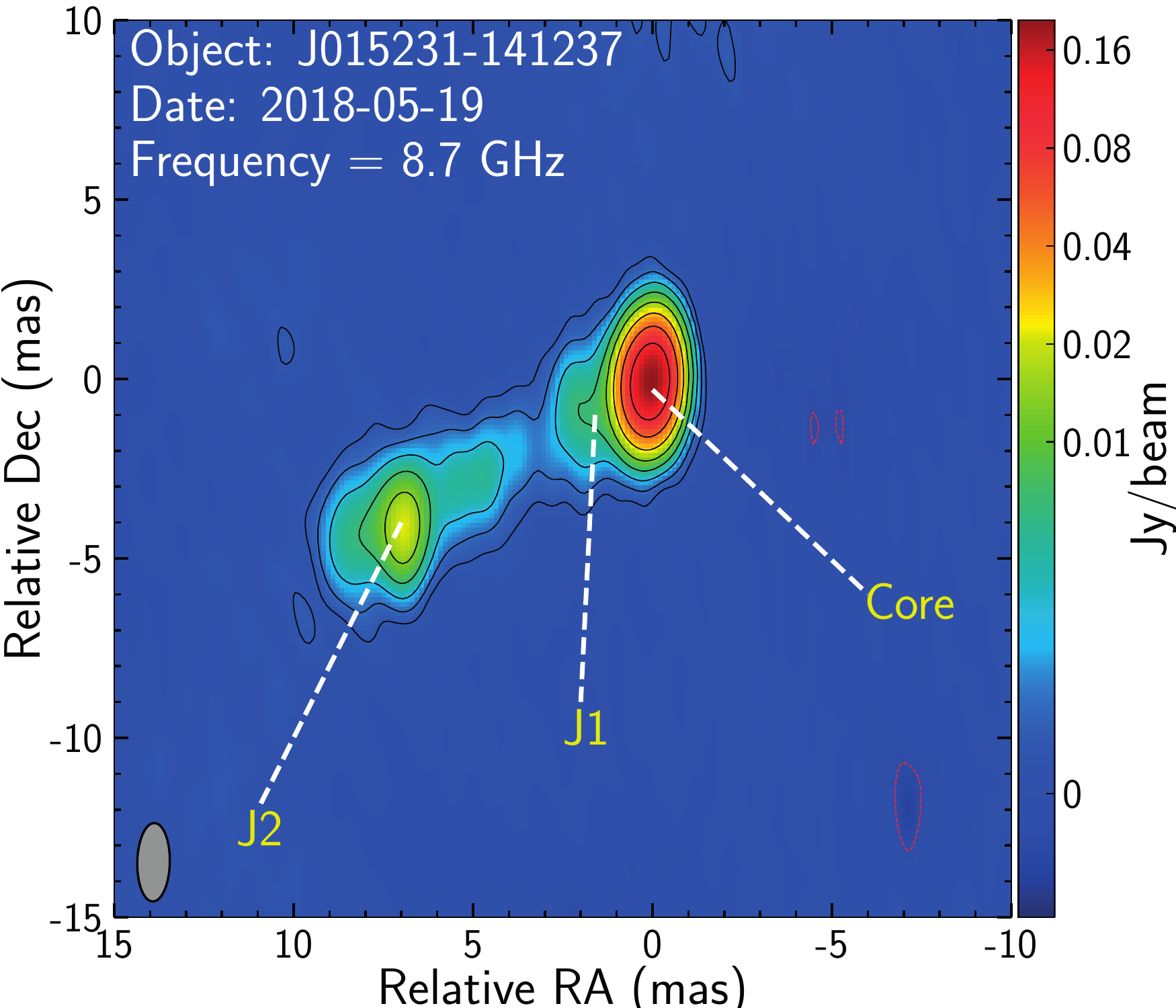}\\
\includegraphics[scale=0.24]{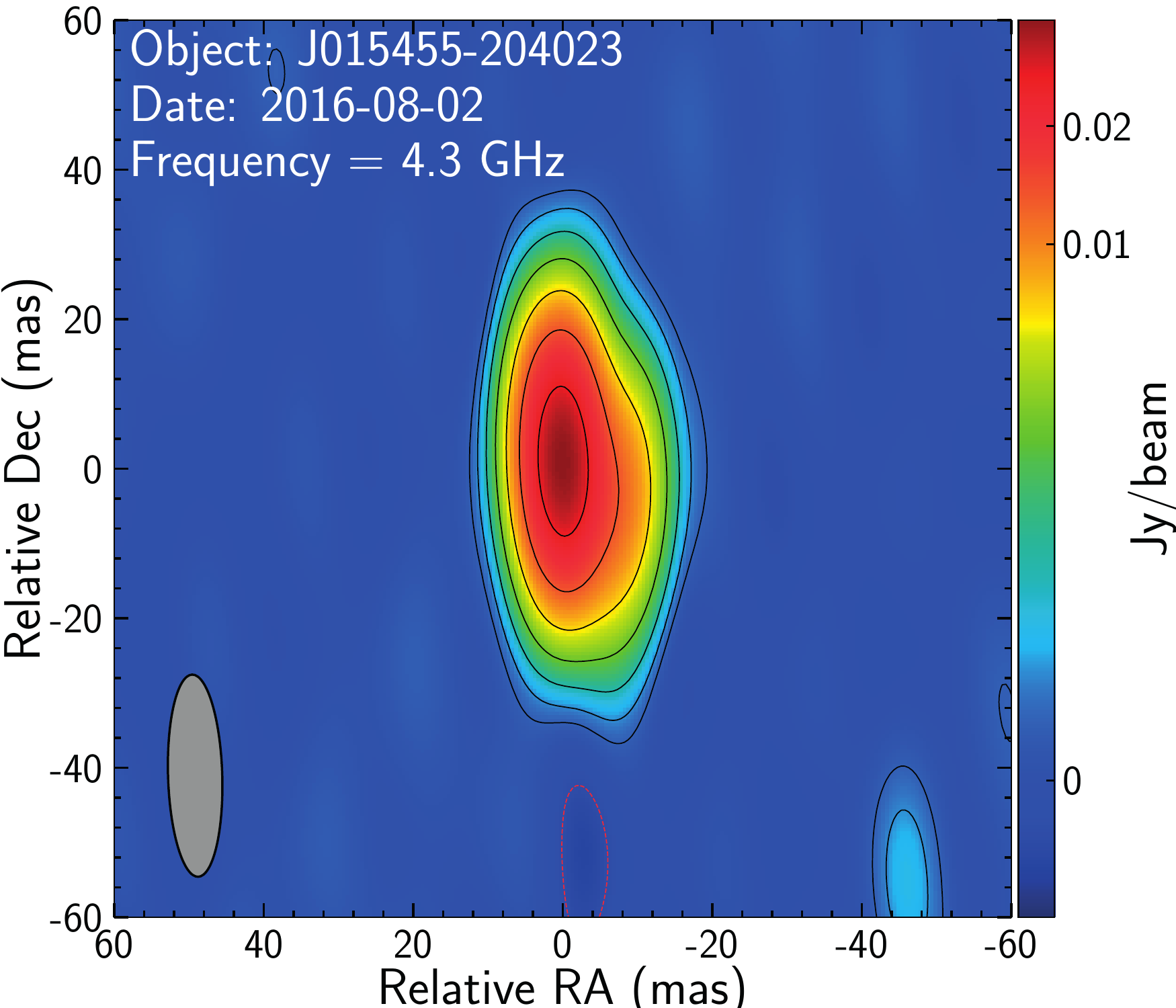}&
\includegraphics[scale=0.24]{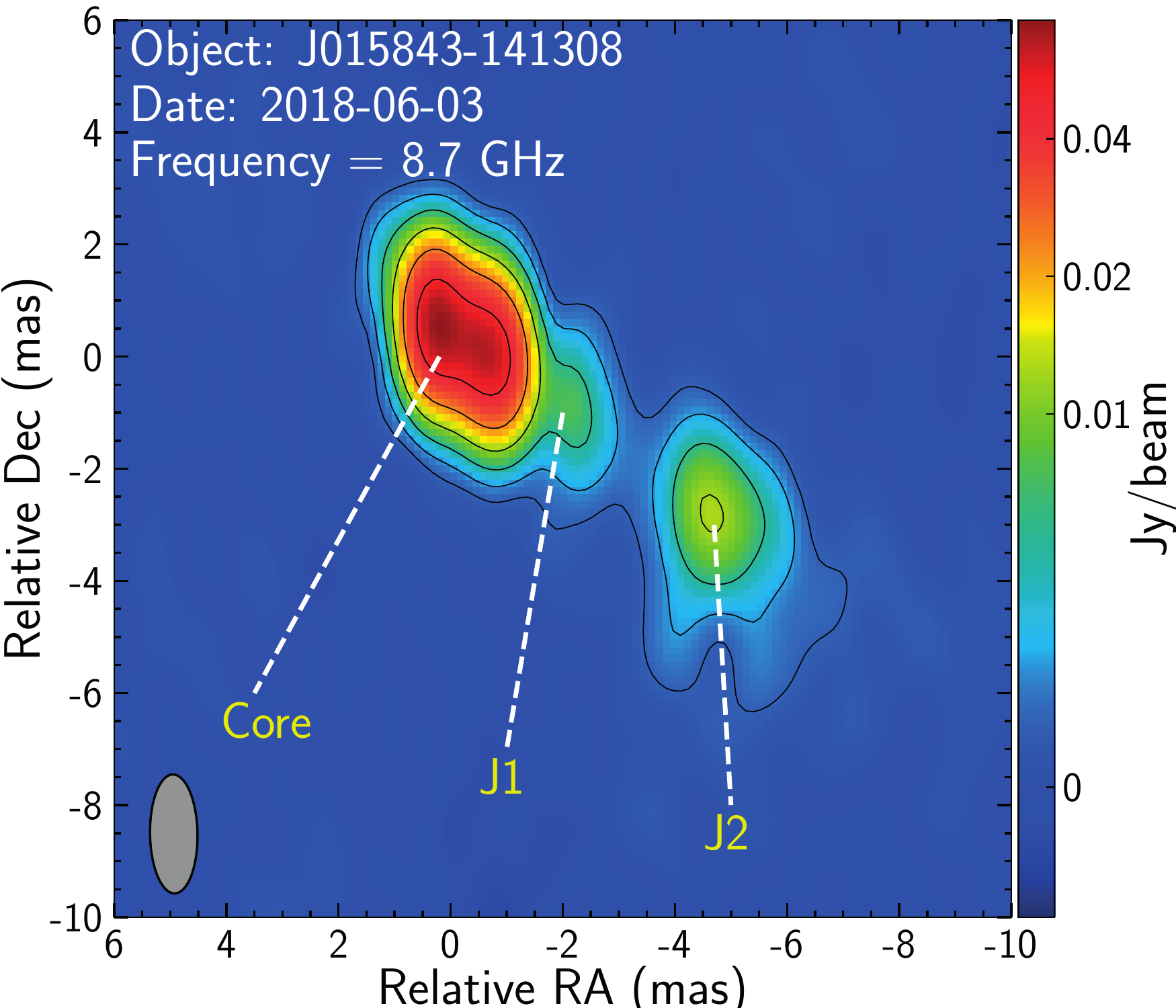}&
\includegraphics[scale=0.24]{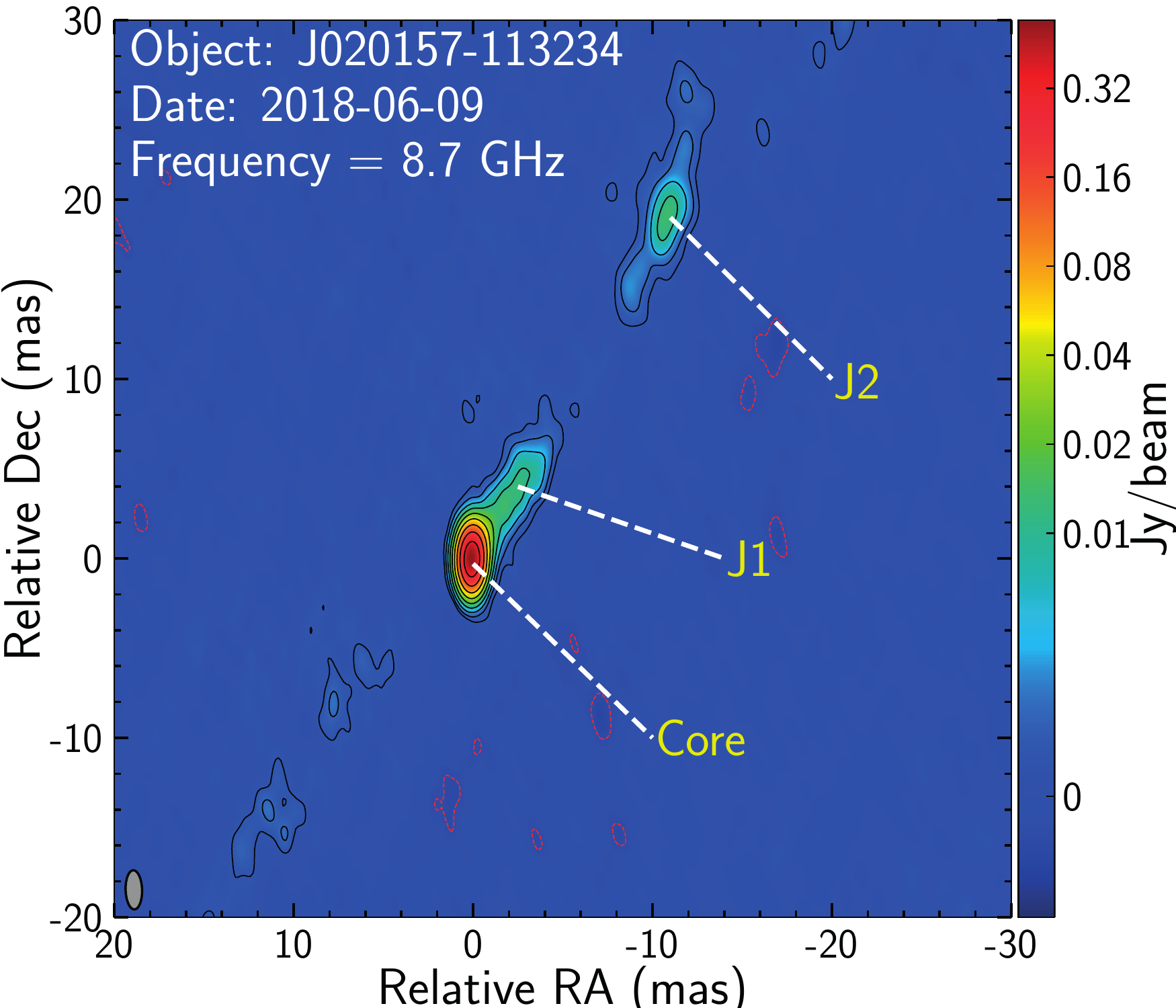}
\end{tabular}
\caption{\small Continued.}
\end{figure*}

\begin{table*}
\caption{SED fitting parameters}
\label{tab:sedfit}
\begin{tabular}{cccc}
\hline
Name & $\nu_\mathrm{p}$ & $\alpha_\mathrm{SED~fit}$ & $\alpha_\mathrm{VLBI~fit}$\\
     & (GHz)   &                    &                    \\
\hline
J000106-174125 &                 & $-0.51 \pm 0.01$ & \\
J000125-132630 & $0.12 \pm 0.02$ & $-0.76 \pm 0.03$ & \\
J000322-172708 &                 & $-0.68 \pm 0.01$ & $-0.18 \pm 0.07$ \\
J000905-205630 & $0.95 \pm 0.11$ & $-1.04 \pm 0.05$ & \\
J001913-120958 &                 & $-0.71 \pm 0.01$ & \\
J002125-005540 & $3.88 \pm 0.54$ & $-0.56 \pm 0.01$ & \\
J002208-104133 & $0.16 \pm 0.01$ & $-0.88 \pm 0.02$ & \\
J002223-070230 & $0.08 \pm 0.01$ & $-0.98 \pm 0.01$ & \\
J002430-292847 &                 & $-0.86 \pm 0.01$ & \\
J002546-124724 &                 & $-0.49 \pm 0.01$ & \\
J002549-260211 & $1.16 \pm 0.01$ & $-1.03 \pm 0.01$ & $-0.67 \pm 0.13$ \\
J003201-085133 & $0.14 \pm 0.01$ & $-0.75 \pm 0.02$ & \\
J003246-293107 & $0.11 \pm 0.01$ & $-0.86 \pm 0.02$ & \\
J003332-165009 & $0.96 \pm 0.14$ & $-1.23 \pm 0.10$ & \\
J003454-230328 &                 & $-0.34 \pm 0.01$ & \\
J003502-083437 &                 & $-0.85 \pm 0.01$ & \\
J003508-200354 & $0.24 \pm 0.04$ & $-1.10 \pm 0.02$ & \\
J003829-211957 & $1.33 \pm 0.10$ & $-0.89 \pm 0.03$ & \\
J003931-111057 &                 & $-0.50 \pm 0.01$ & \\
J004441-353029 &                 & $-0.60 \pm 0.01$ & \\
J004644-052157 &                 & $-0.84 \pm 0.01$ & \\
J004744-121026 & $2.16 \pm 0.14$ & $-1.52 \pm 0.08$ & \\
J004807-183838 &                 & $-0.62 \pm 0.01$ & \\
J004858-062832 &                 & $-0.79 \pm 0.01$ & \\
J004954-100613 & $1.96 \pm 0.20$ & $-1.01 \pm 0.03$ & \\
J005026-120115 &                 & $-0.60 \pm 0.01$ & \\
J005108-065001 &                 & $-0.50 \pm 0.01$ & $0.28 \pm 0.03$ \\
J005214-161712 &                 & $-0.38 \pm 0.01$ & \\
J005242-215540 &                 & $-0.77 \pm 0.01$ & \\
J005433-195255 &                 & $-0.53 \pm 0.03$ & \\
J005533-214817 &                 & $-0.82 \pm 0.01$ & \\
J005551-124424 & $0.34 \pm 0.13$ & $-0.36 \pm 0.05$ & \\
J005805-053952 &                 & $-0.42 \pm 0.01$ & $-0.44 \pm 0.09$ \\
J010152-283118 & $0.62 \pm 0.05$ & $-0.74 \pm 0.02$ & \\
J010335-271506 & $0.66 \pm 0.08$ & $-1.21 \pm 0.03$ & \\
J010452-252052 & $1.36 \pm 0.13$ & $-1.35 \pm 0.05$ & \\
J010527-182843 &                 & $-0.58 \pm 0.01$ & \\
J010837-285124 & $0.26 \pm 0.01$ & $-1.01 \pm 0.01$ & \\
J010938-144228 &                 & $-0.79 \pm 0.01$ & \\
J011049-074142 &                 & $-0.27 \pm 0.03$ & \\
J011312-101419 &                 & $-0.51 \pm 0.01$ & \\
J011612-113610 &                 & $-0.71 \pm 0.01$ & $0.29 \pm 0.06$ \\
J011651-205202 & $0.16 \pm 0.01$ & $-0.85 \pm 0.01$ & \\
J011738-150750 &                 & $-0.49 \pm 0.01$ & \\
J011815-012037 & $1.07 \pm 0.03$ & $-1.40 \pm 0.01$ & \\
J011834-184910 &                 & $-0.79 \pm 0.01$ & \\
J011857-071855 & $0.31 \pm 0.17$ & $-0.82 \pm 0.09$ & \\
J012031-270125 &                 & $-0.58 \pm 0.01$ & $-0.22 \pm 0.06$ \\
J012227-042123 &                 & $-0.62 \pm 0.01$ & \\
J012340-104900 &                 & $-0.33 \pm 0.01$ & \\
J012614-222227 &                 & $-0.76 \pm 0.01$ & $-0.05 \pm 0.06$ \\
J012730-195607 &                 & $-0.75 \pm 0.01$ & \\
J013112-121058 & $0.48 \pm 0.06$ & $-0.90 \pm 0.06$ & \\
J013243-165444 & $0.23 \pm 0.01$ & $-1.91 \pm 0.25$ & $0.23 \pm 0.05$ \\
J013537-200844 & $0.59 \pm 0.43$ & $-0.27 \pm 0.05$ & $-0.20 \pm 0.05$ \\
J013737-243048 &                 & $-0.77 \pm 0.01$ & $0.47 \pm 0.06$ \\
J014127-270606 &                 & $-0.76 \pm 0.01$ & \\
J014237-074232 &                 & $-0.68 \pm 0.01$ & \\
J015050-090126 &                 & $-0.85 \pm 0.01$ & \\
J015231-141237 &                 & $-0.91 \pm 0.01$ & $-0.36 \pm 0.07$ \\
J015455-204023 & $1.34 \pm 0.11$ & $-1.41 \pm 0.06$ & \\
J015843-141308 &                 & $-0.46 \pm 0.01$ & \\
J020157-113234 &                 & $-0.47 \pm 0.01$ & \\
\hline
\end{tabular}
\end{table*}

\begin{figure*}
\centering
\includegraphics[scale=0.26]{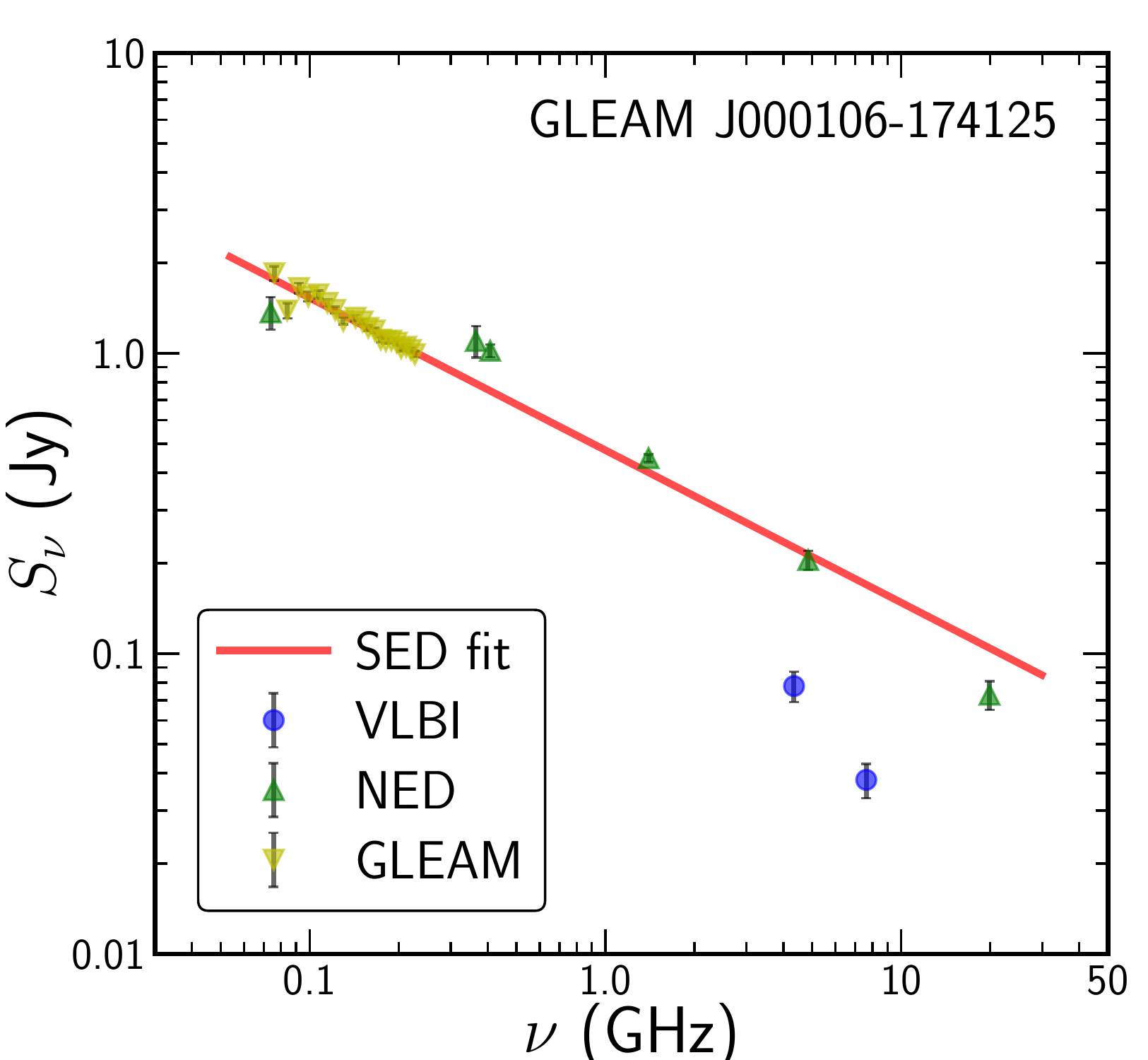}
\includegraphics[scale=0.26]{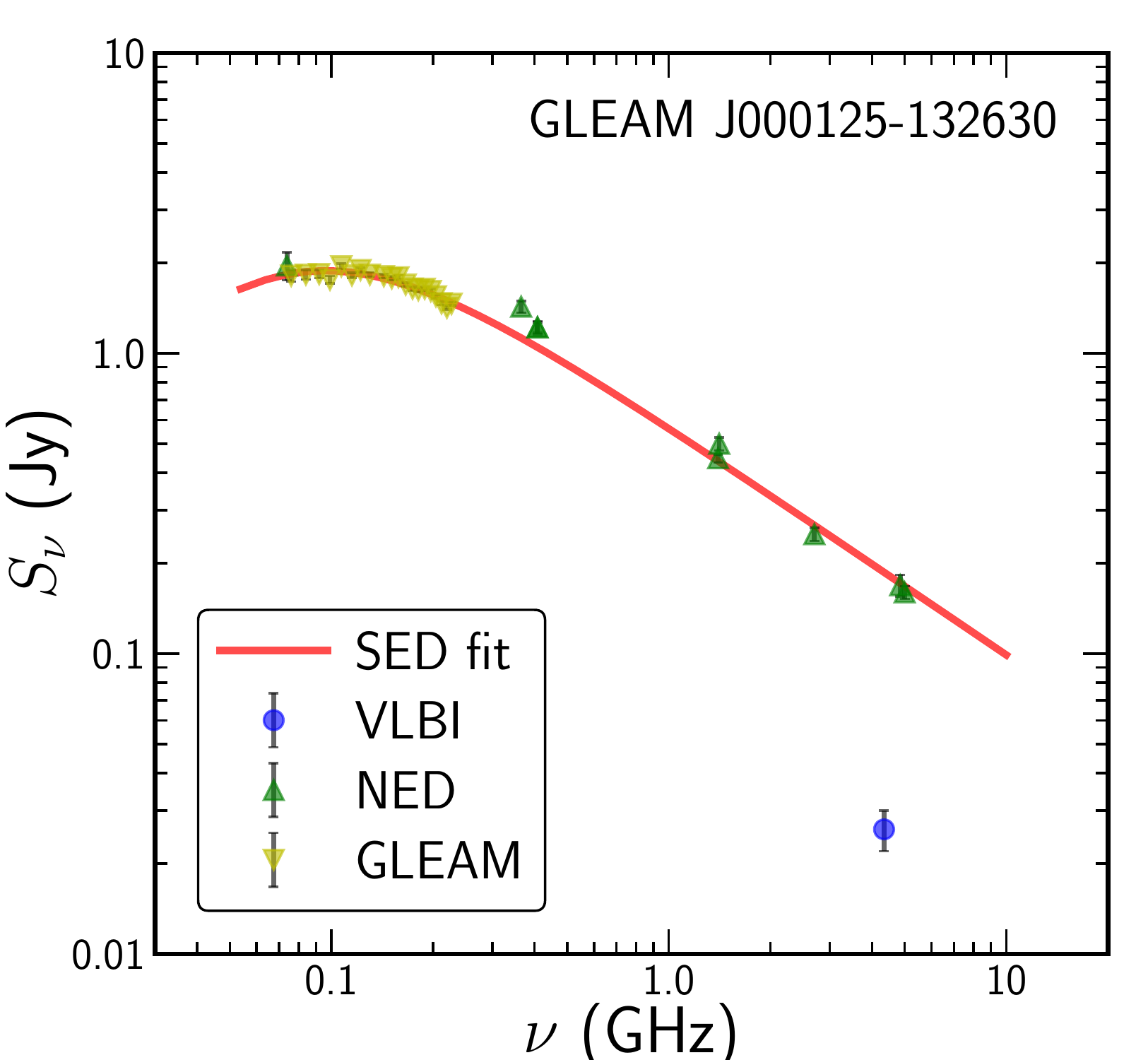}
\includegraphics[scale=0.26]{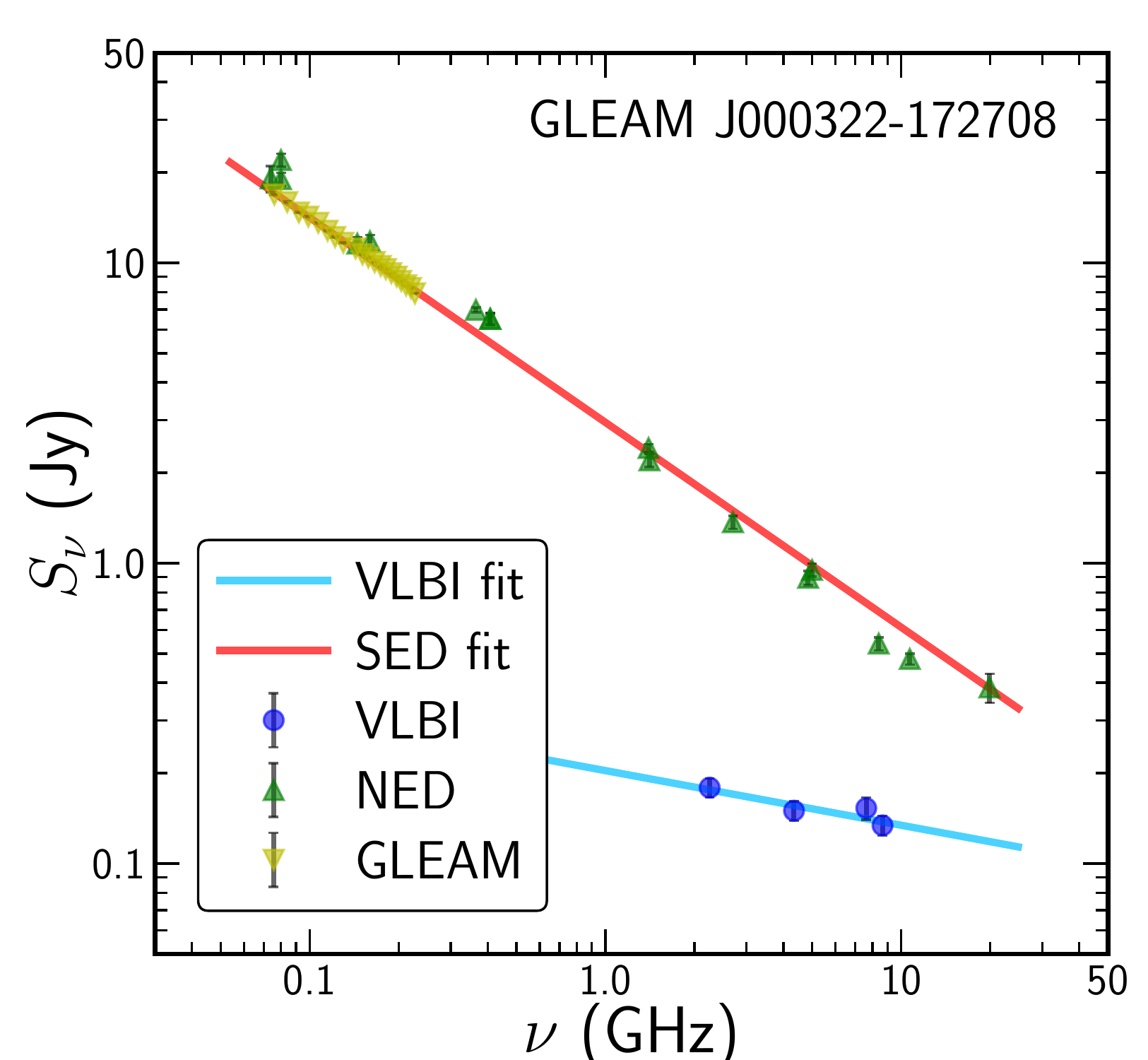}
\includegraphics[scale=0.26]{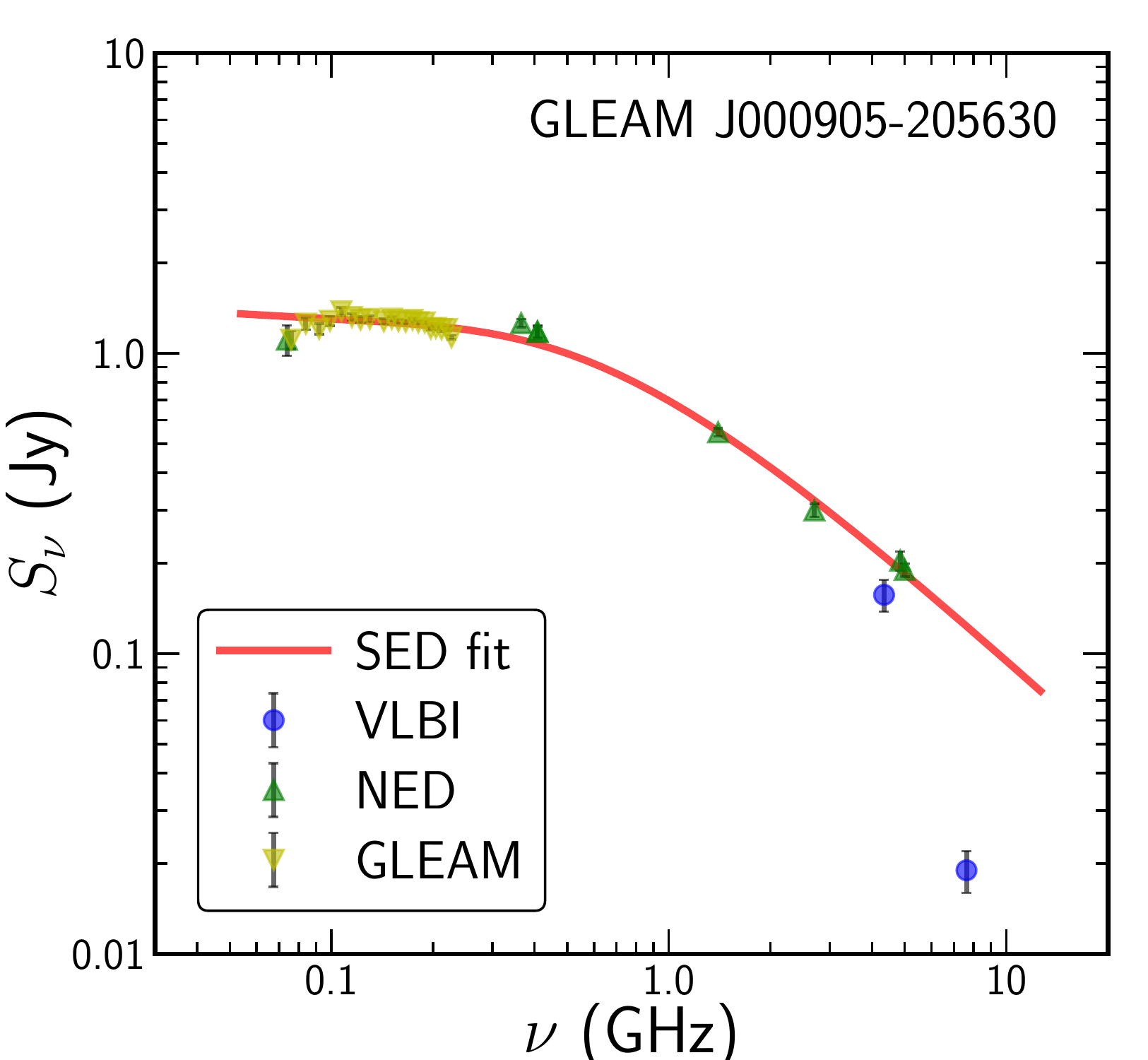}\\
\includegraphics[scale=0.26]{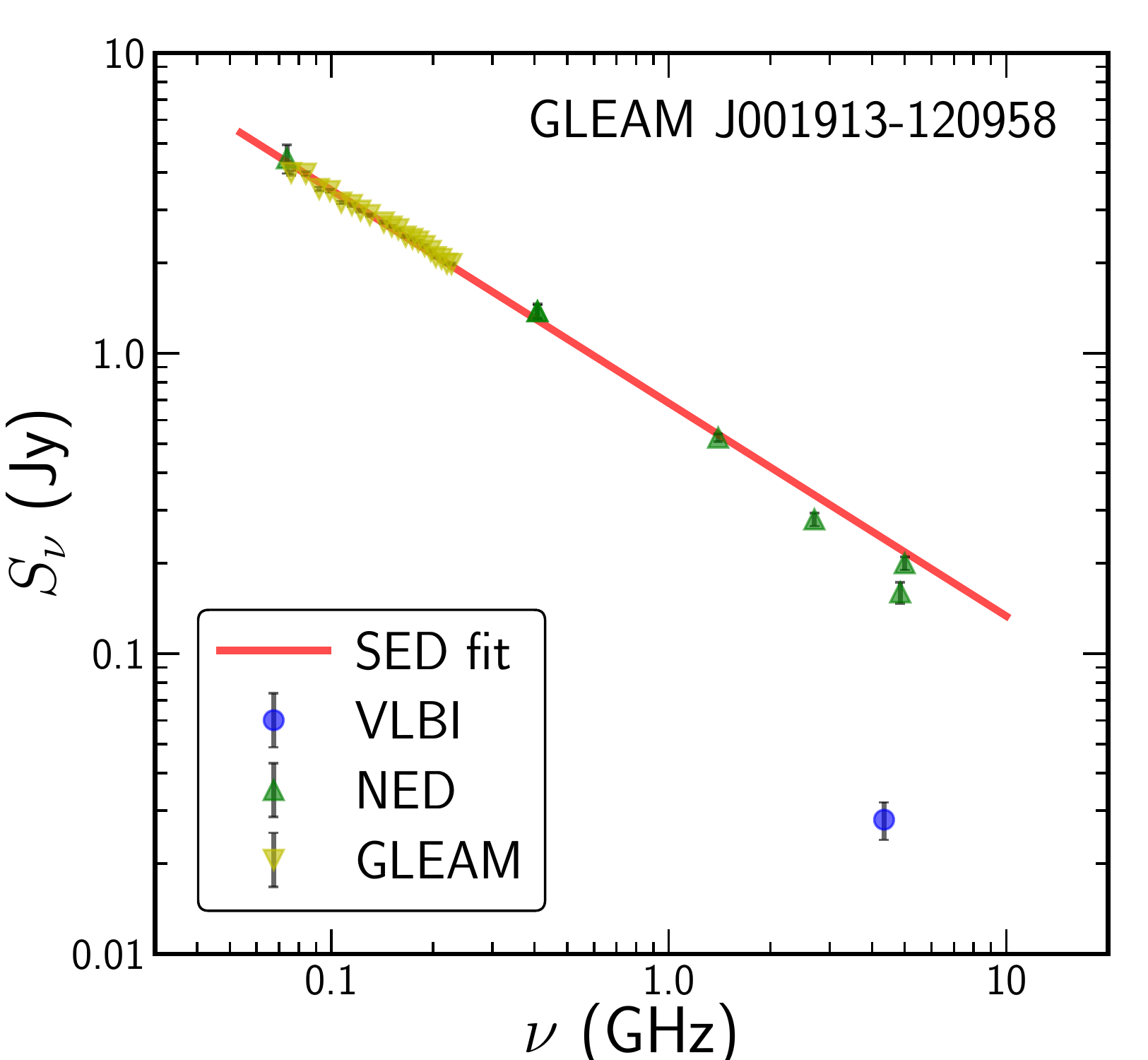}
\includegraphics[scale=0.26]{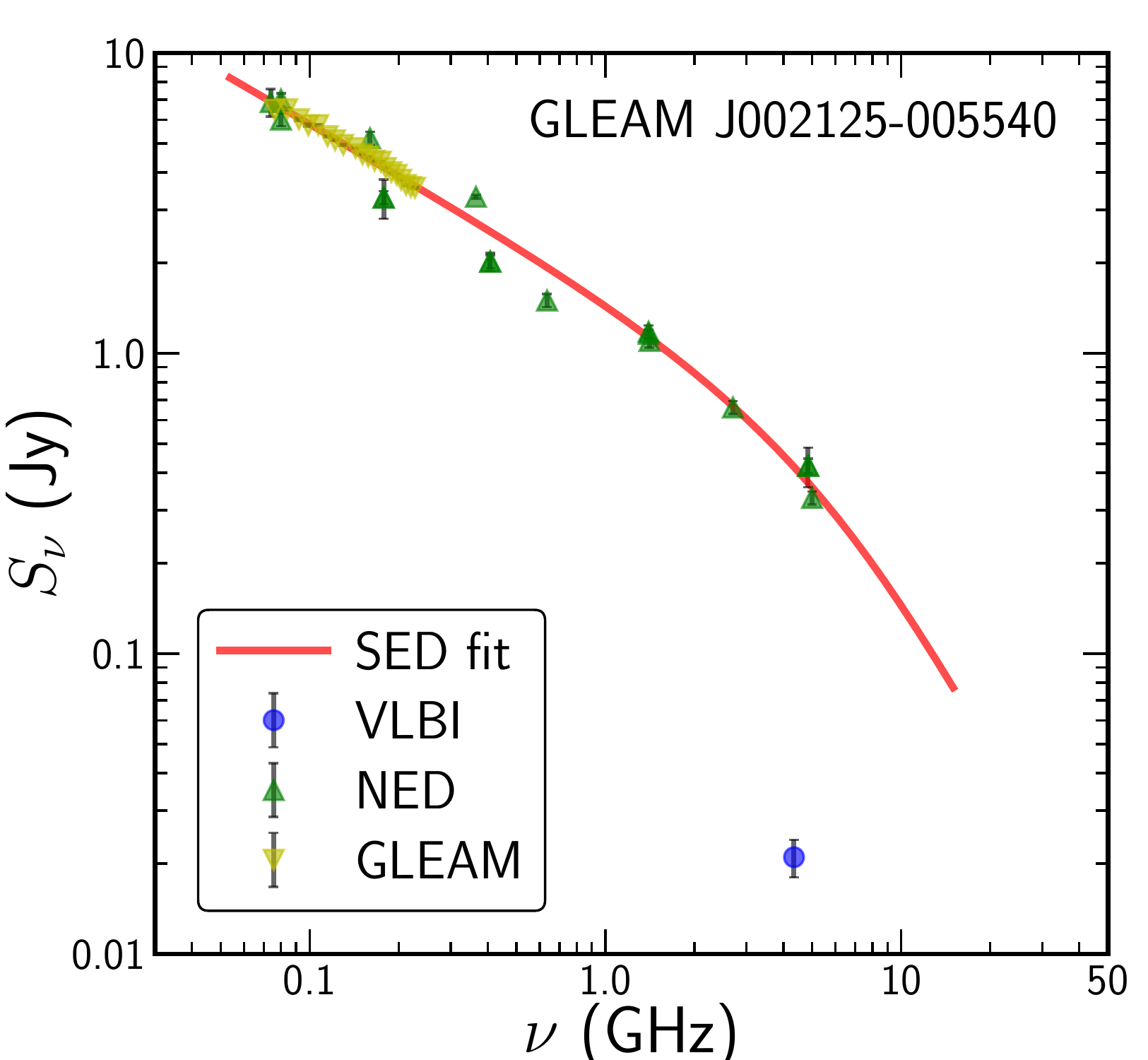}
\includegraphics[scale=0.26]{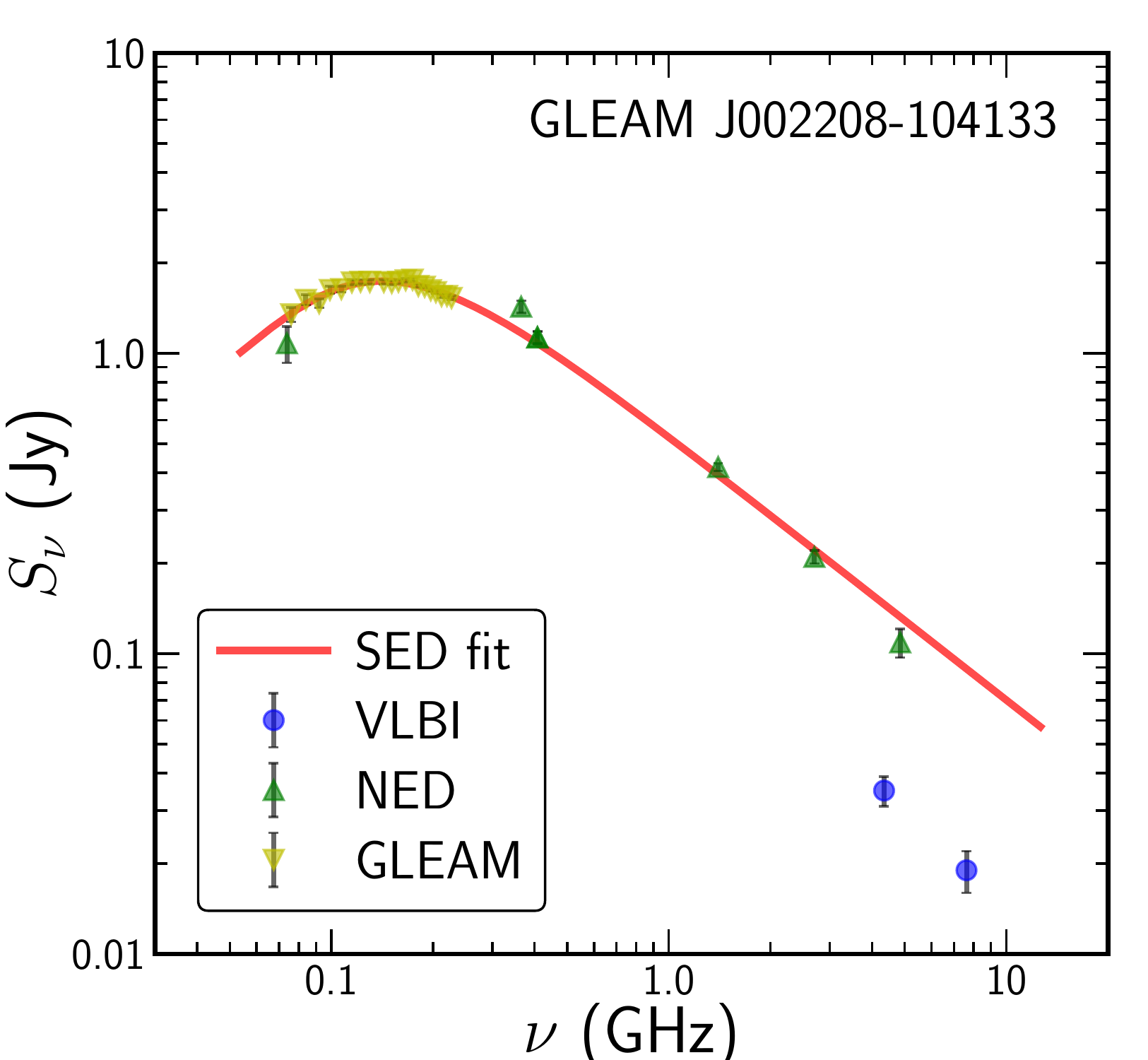}
\includegraphics[scale=0.26]{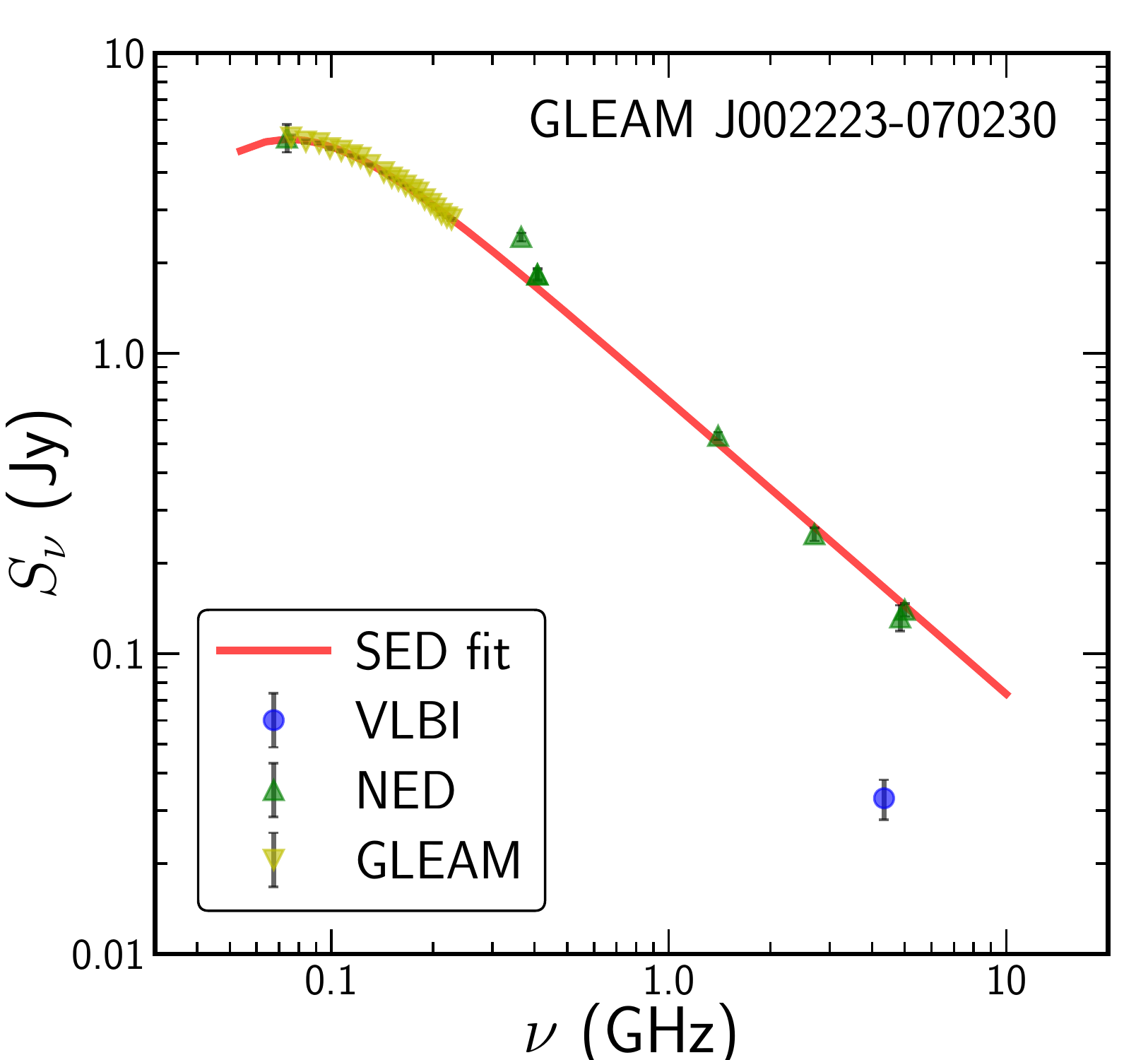}\\
\includegraphics[scale=0.26]{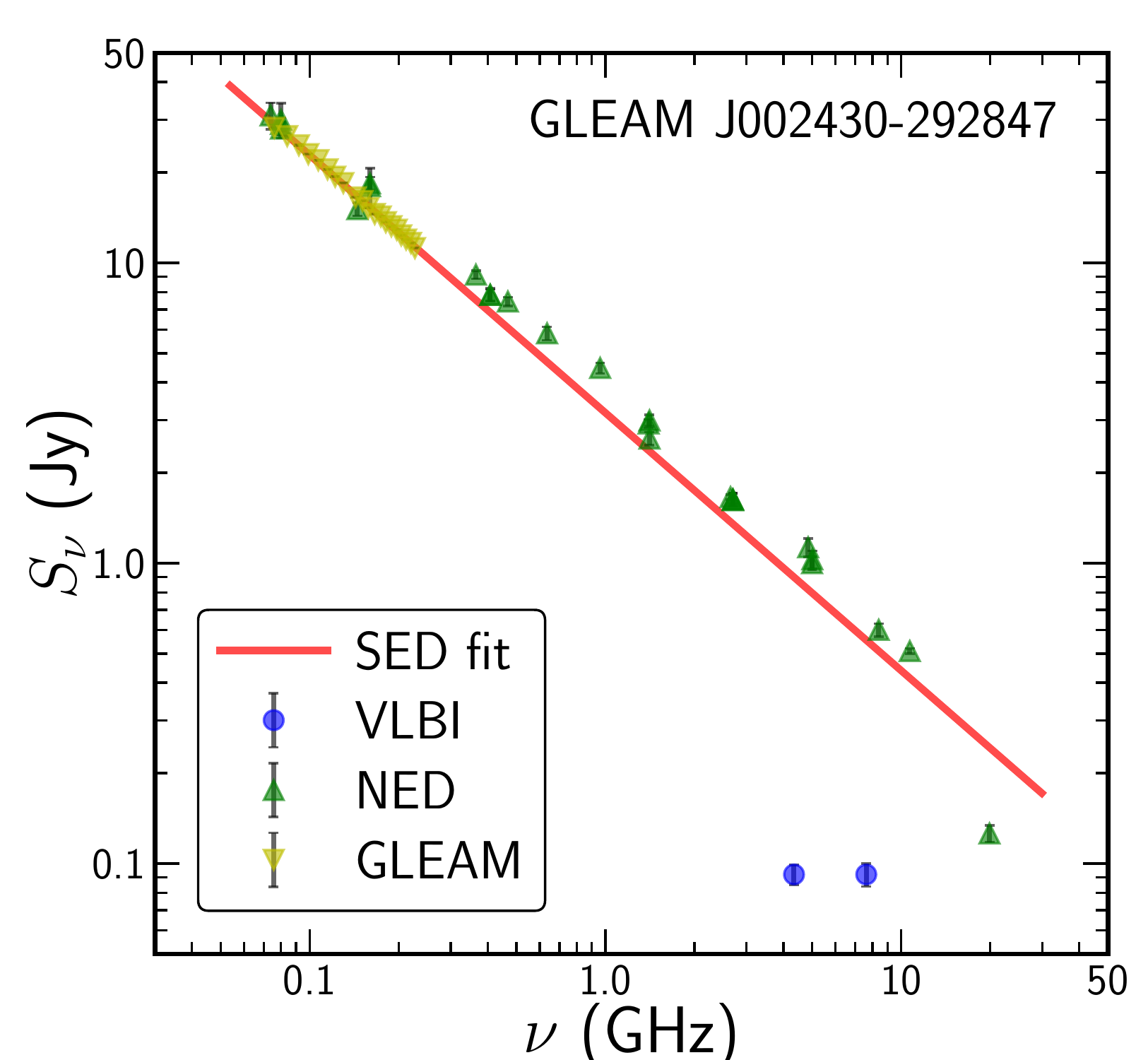}
\includegraphics[scale=0.26]{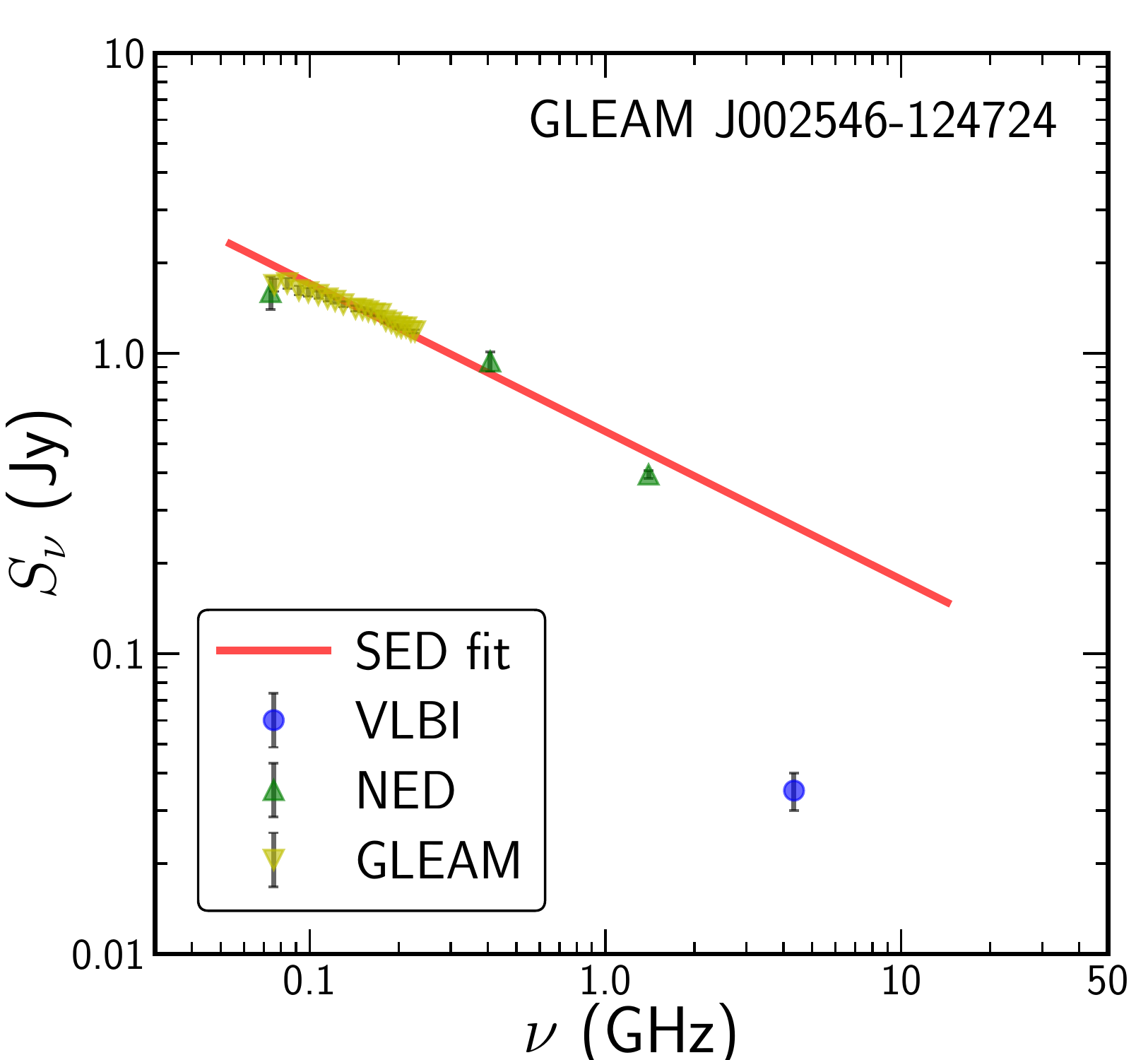}
\includegraphics[scale=0.26]{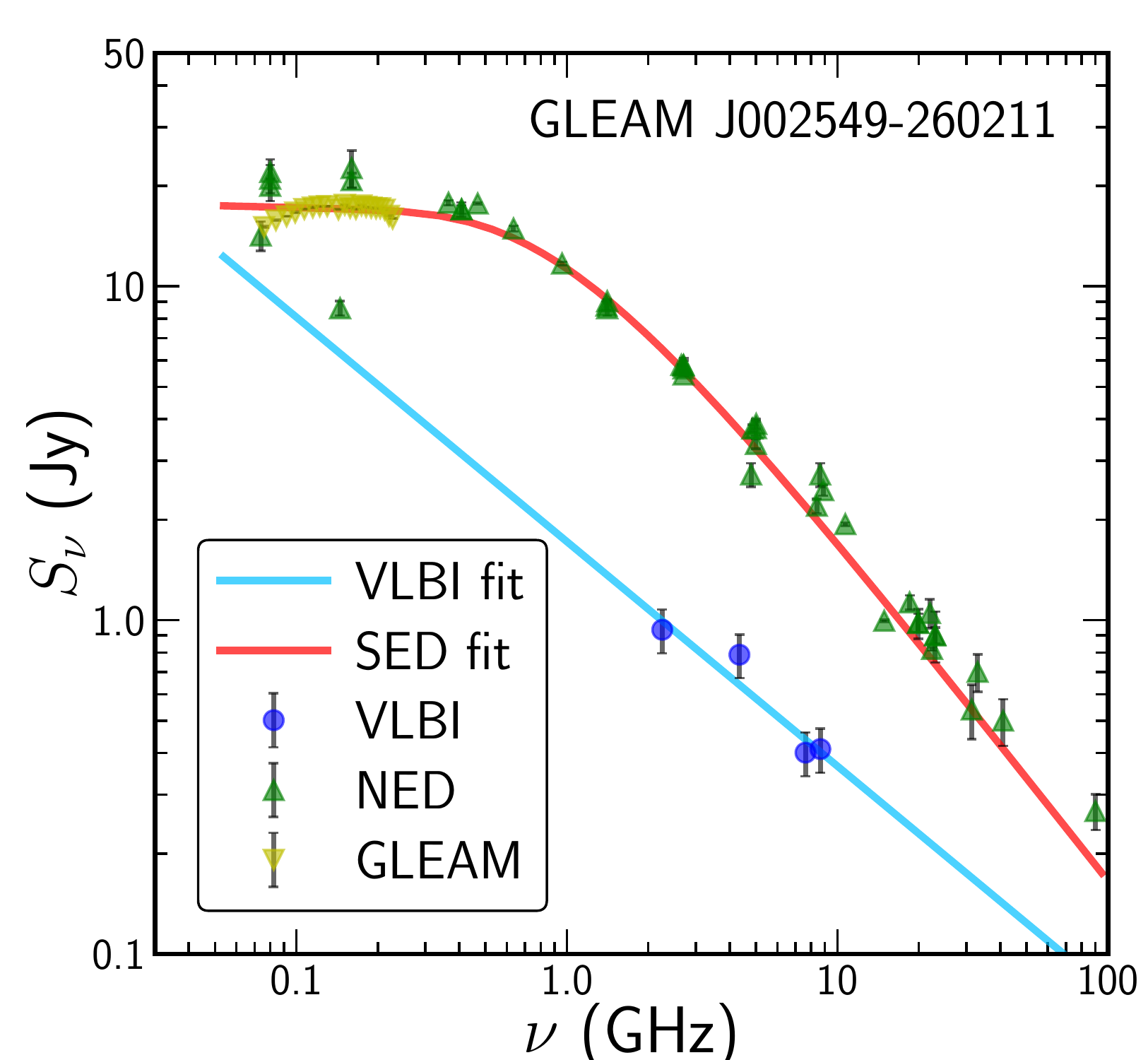}
\includegraphics[scale=0.26]{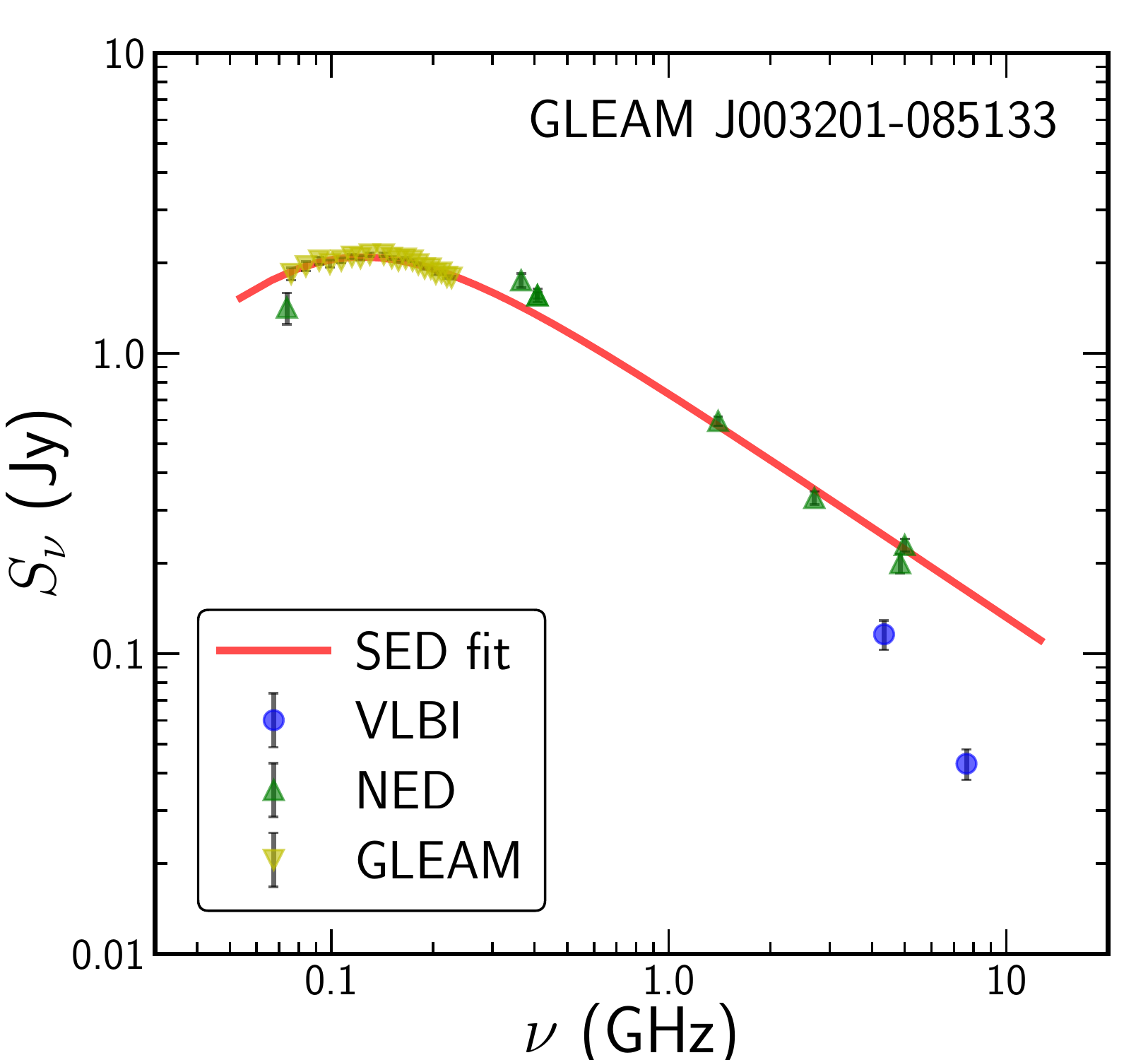}\\
\includegraphics[scale=0.26]{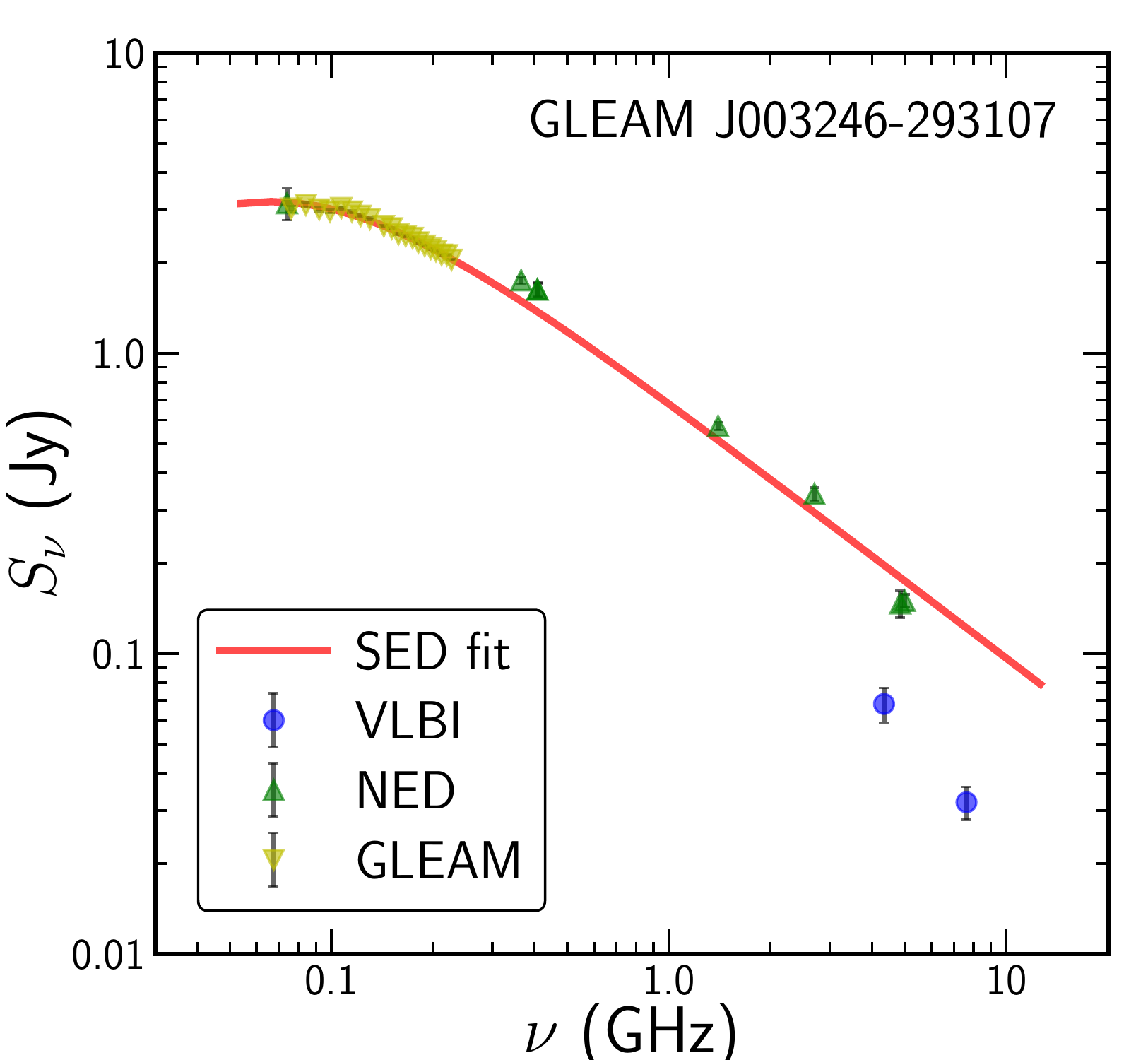}
\includegraphics[scale=0.26]{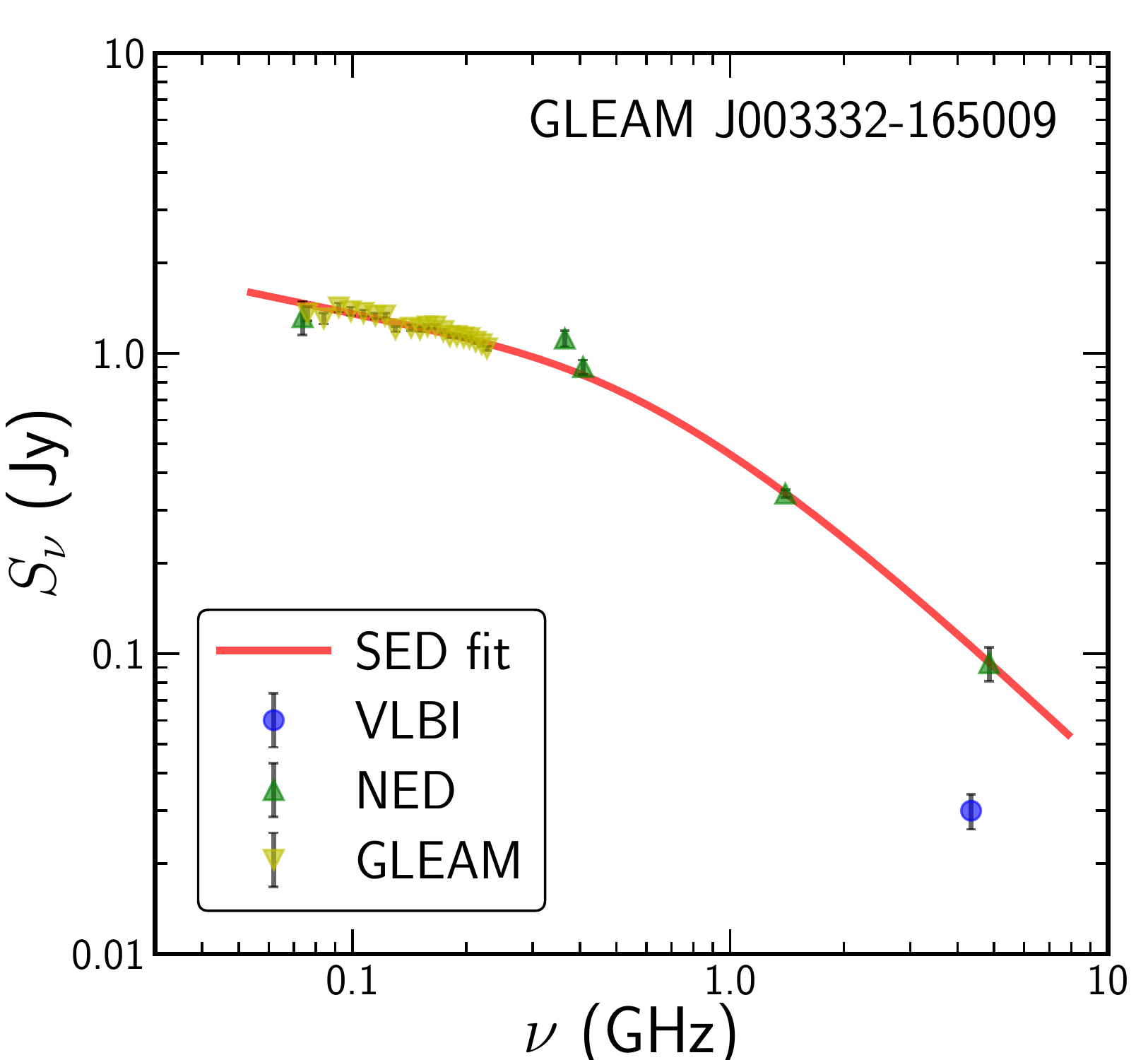}
\includegraphics[scale=0.26]{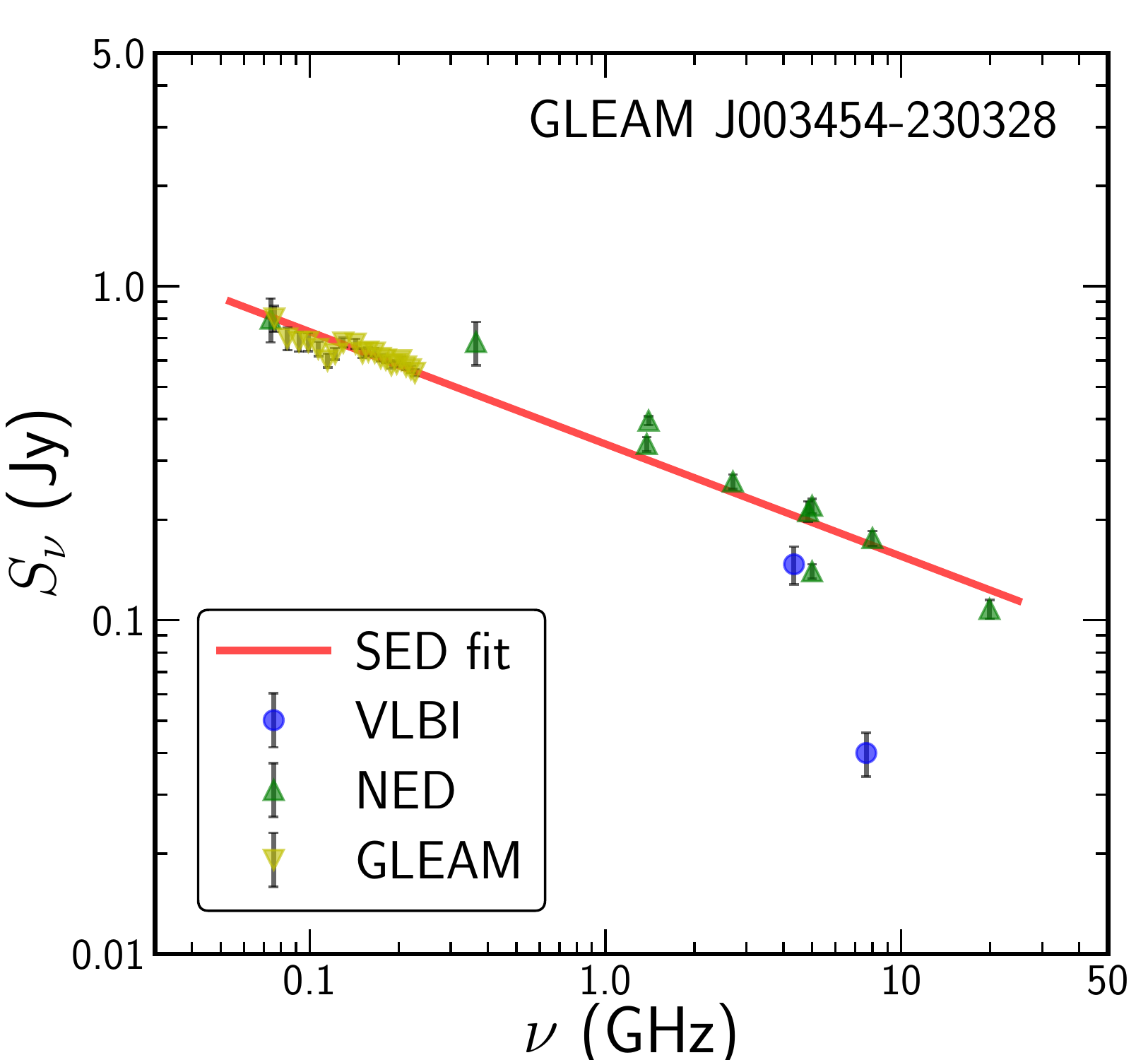}
\includegraphics[scale=0.26]{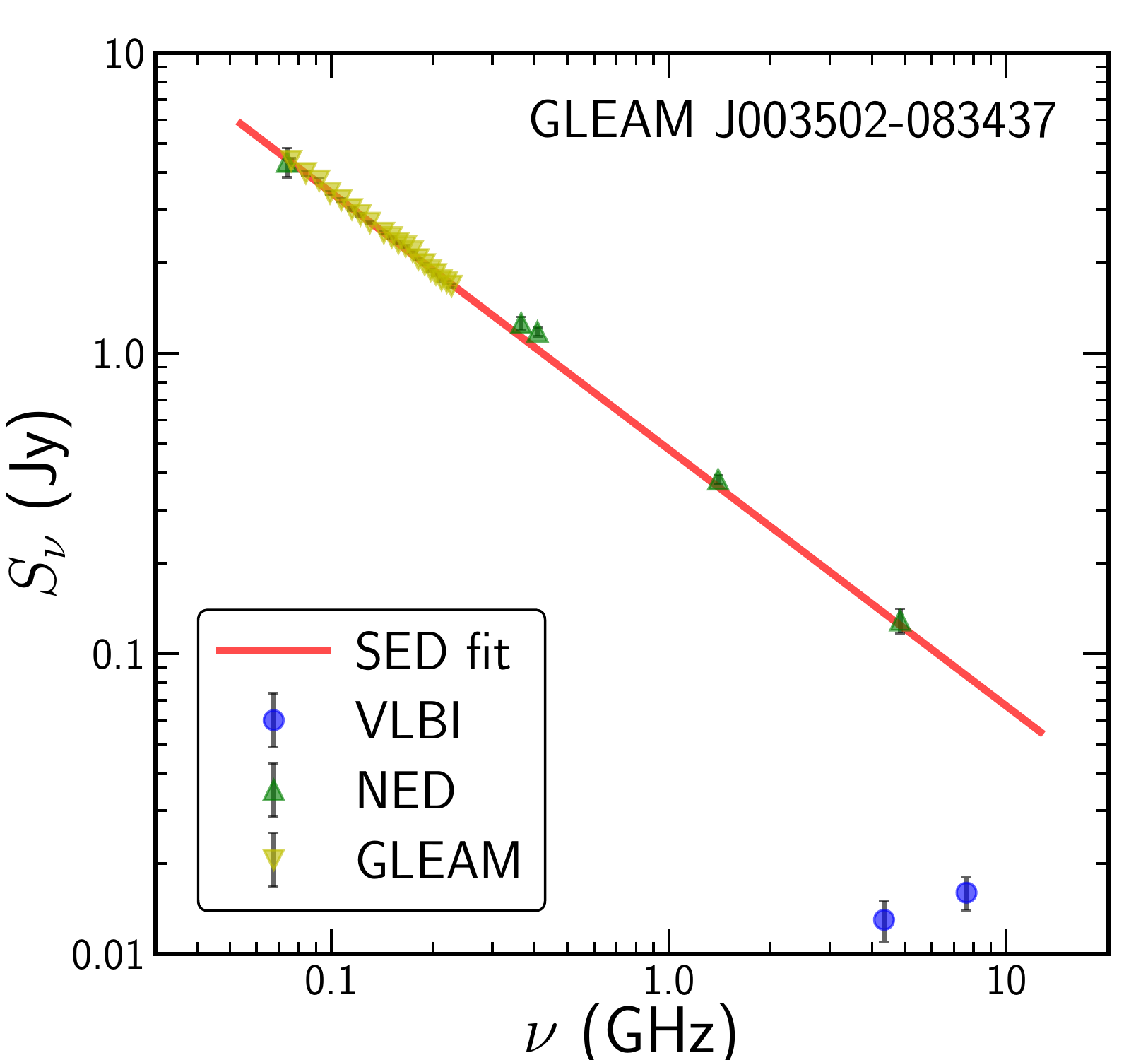}\\
\includegraphics[scale=0.26]{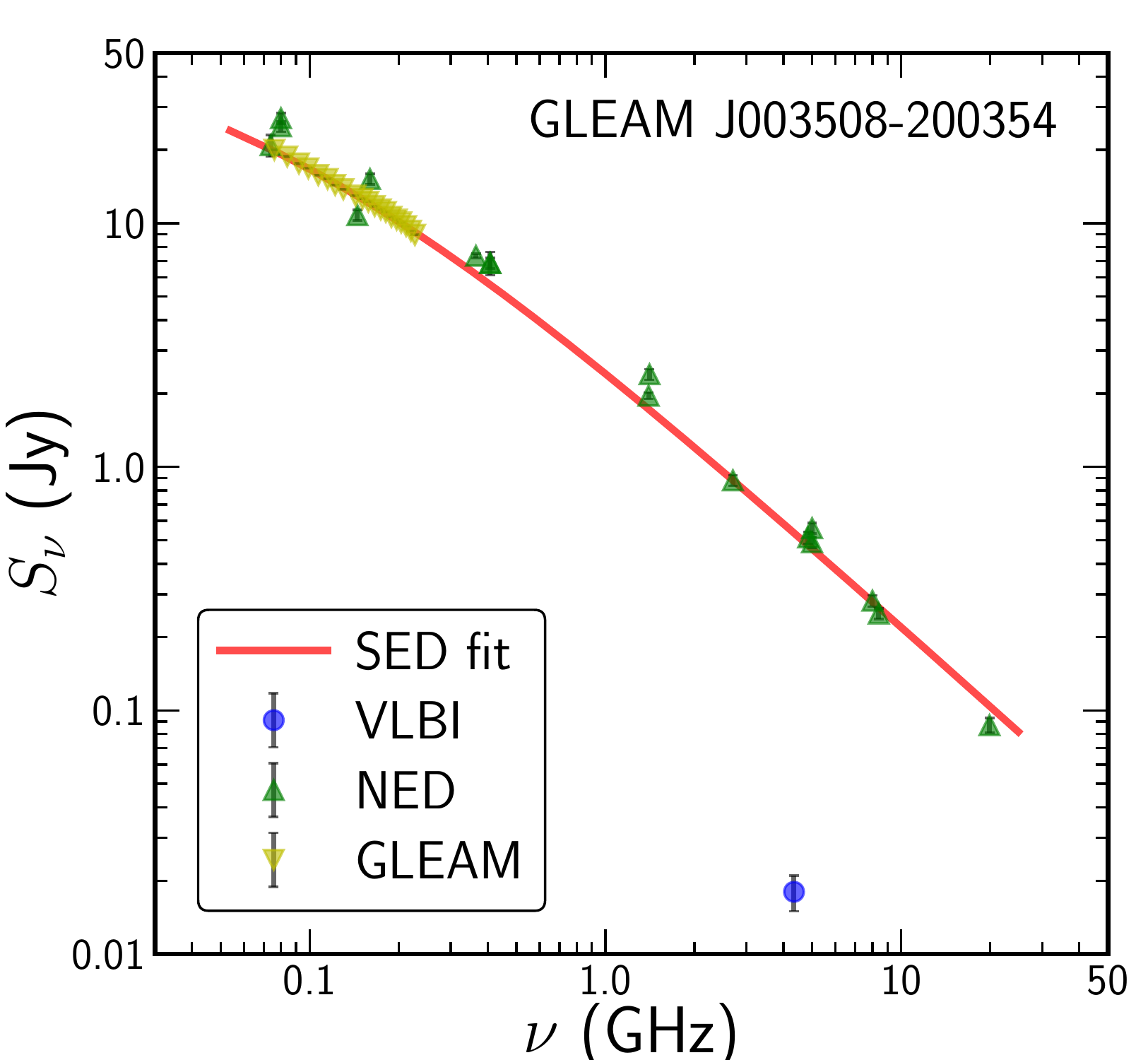}
\includegraphics[scale=0.26]{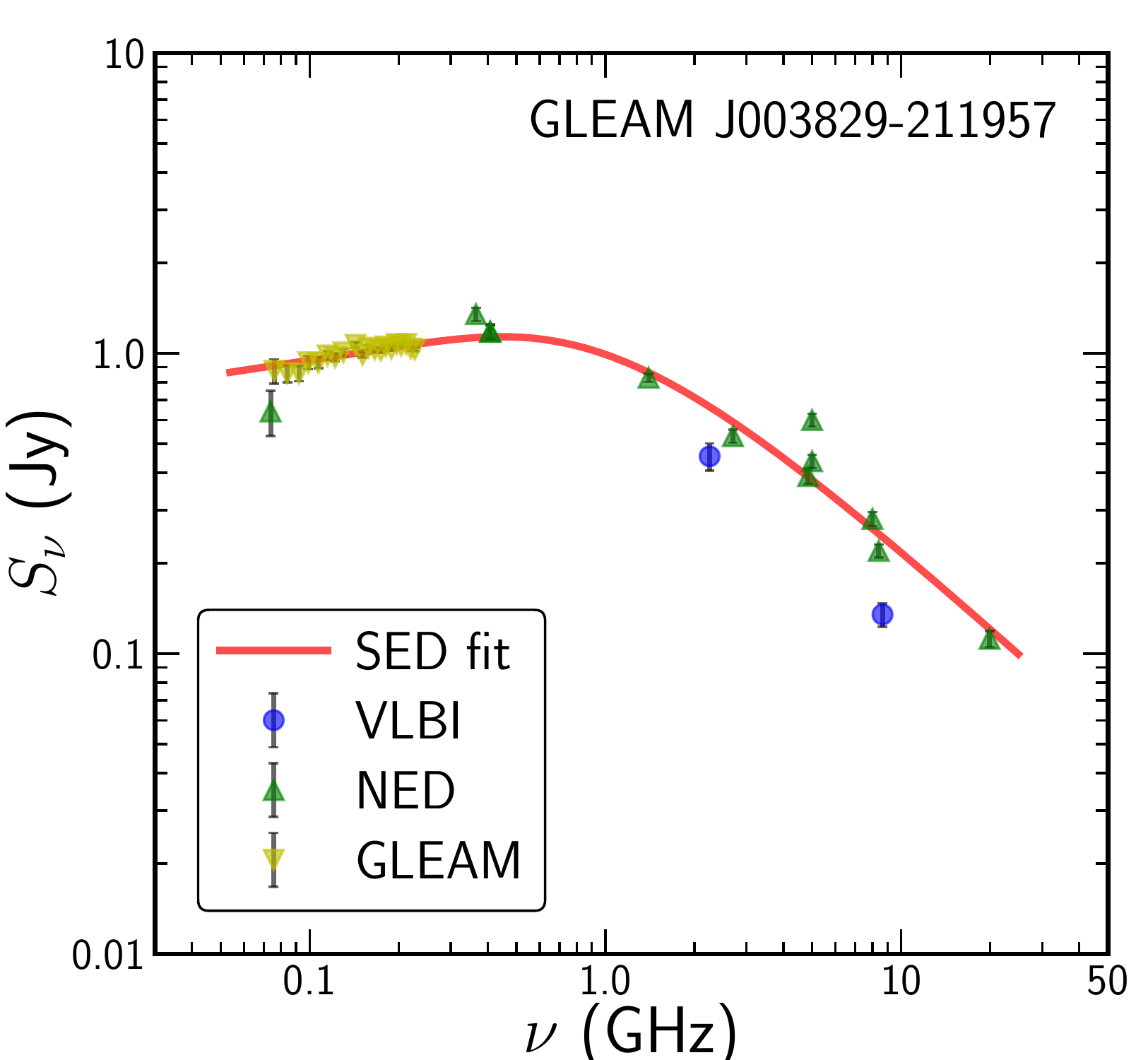}
\includegraphics[scale=0.26]{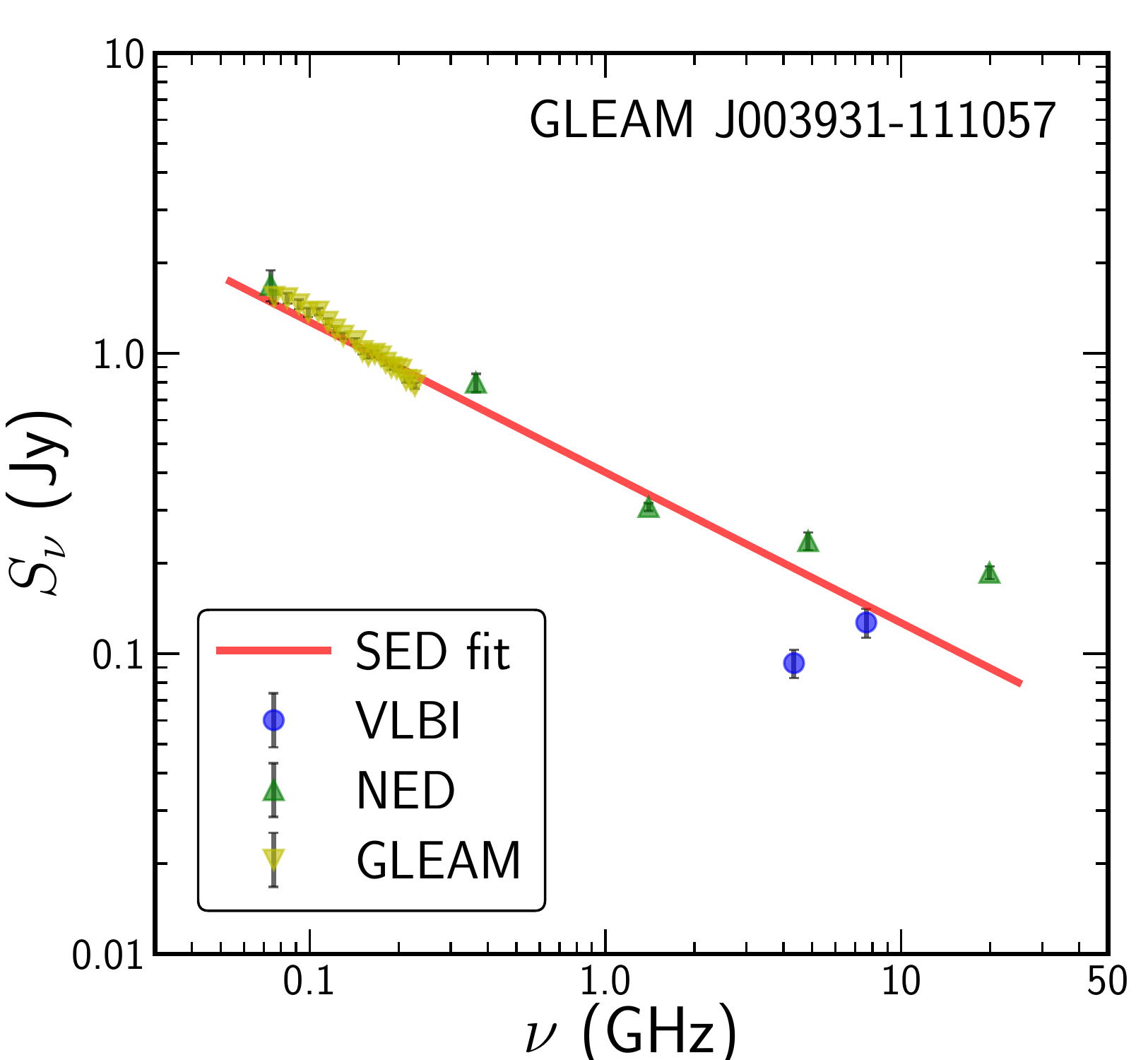}
\includegraphics[scale=0.26]{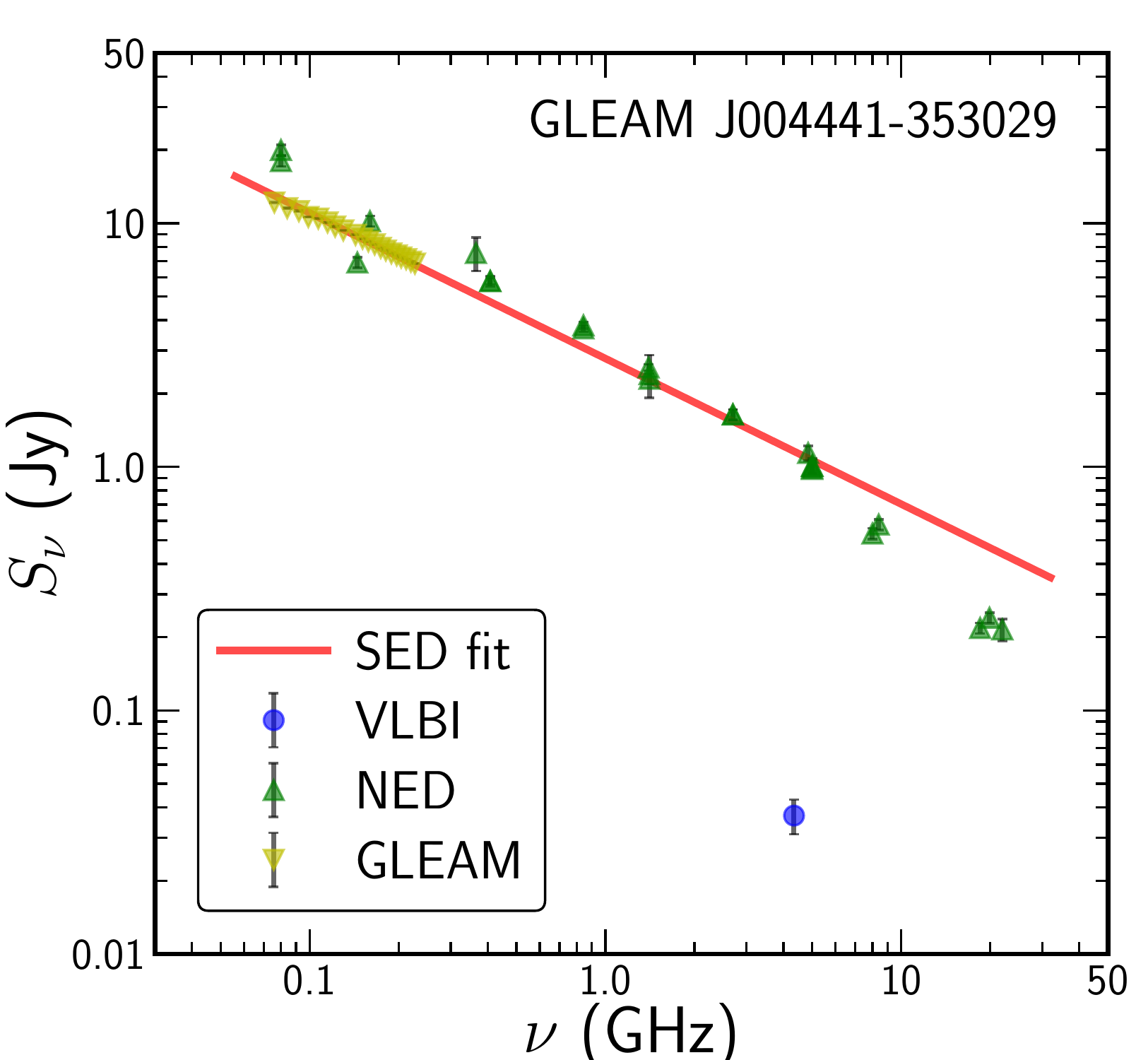}\\
\includegraphics[scale=0.26]{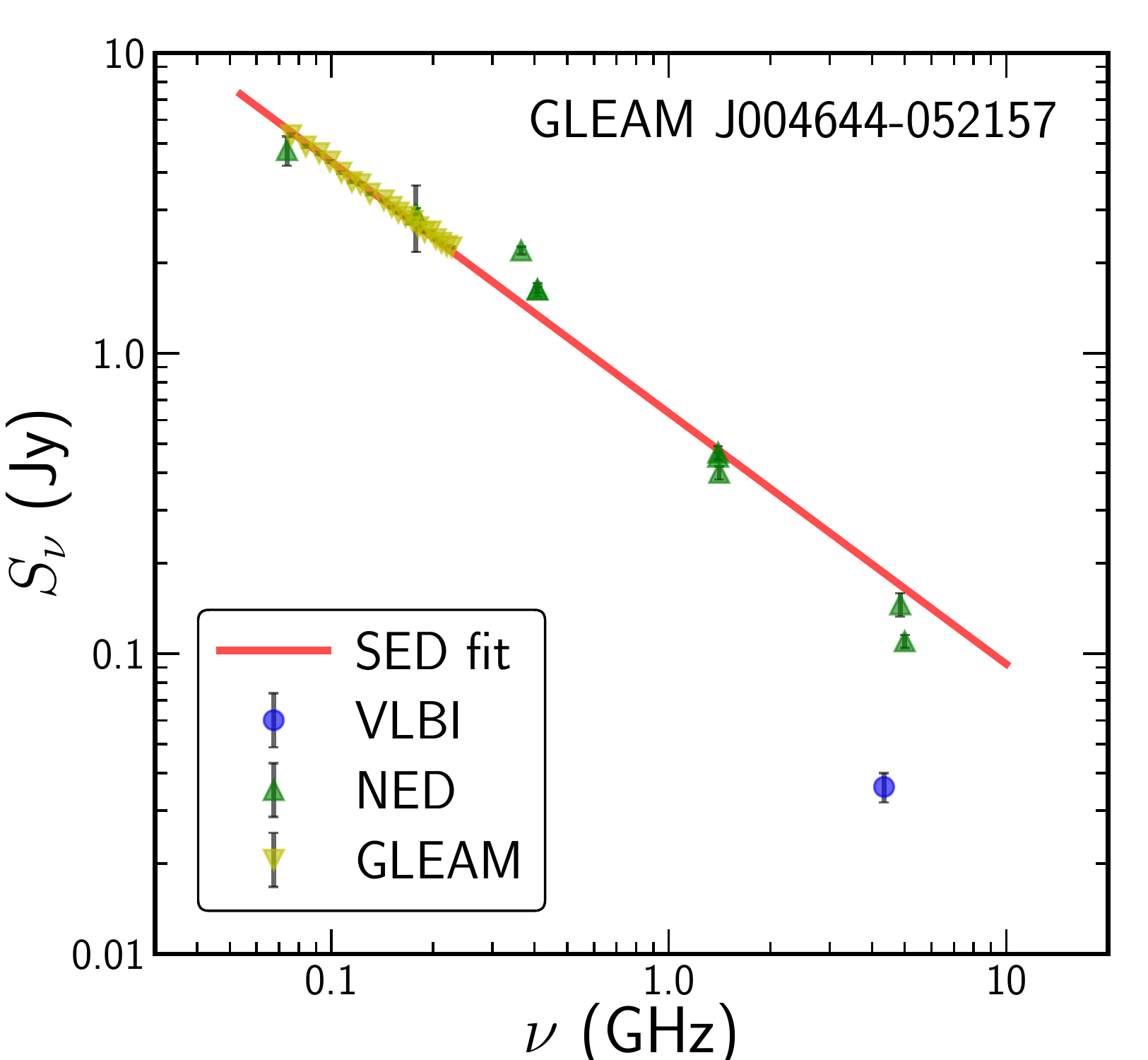}
\includegraphics[scale=0.26]{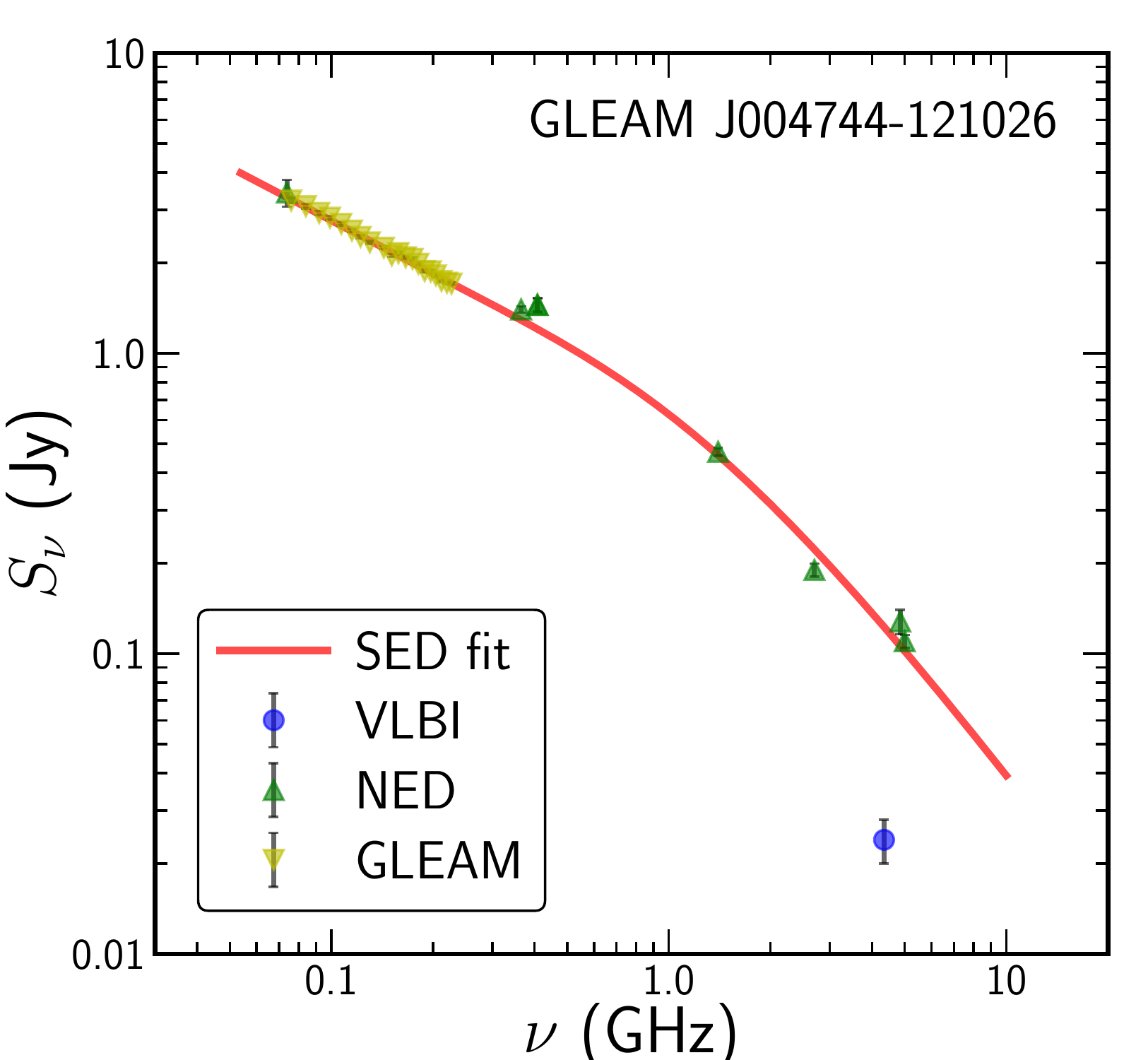}
\includegraphics[scale=0.26]{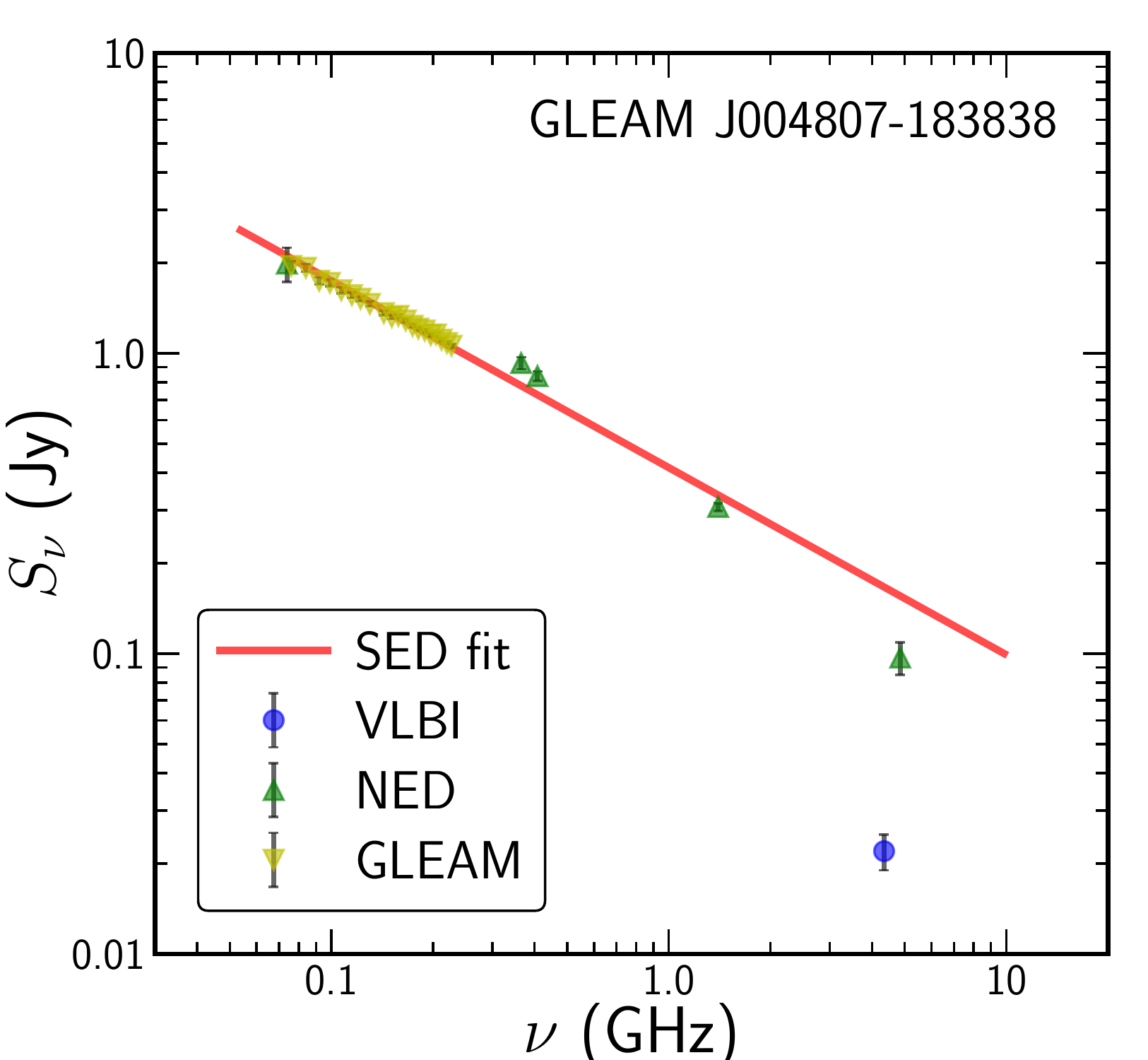}
\includegraphics[scale=0.26]{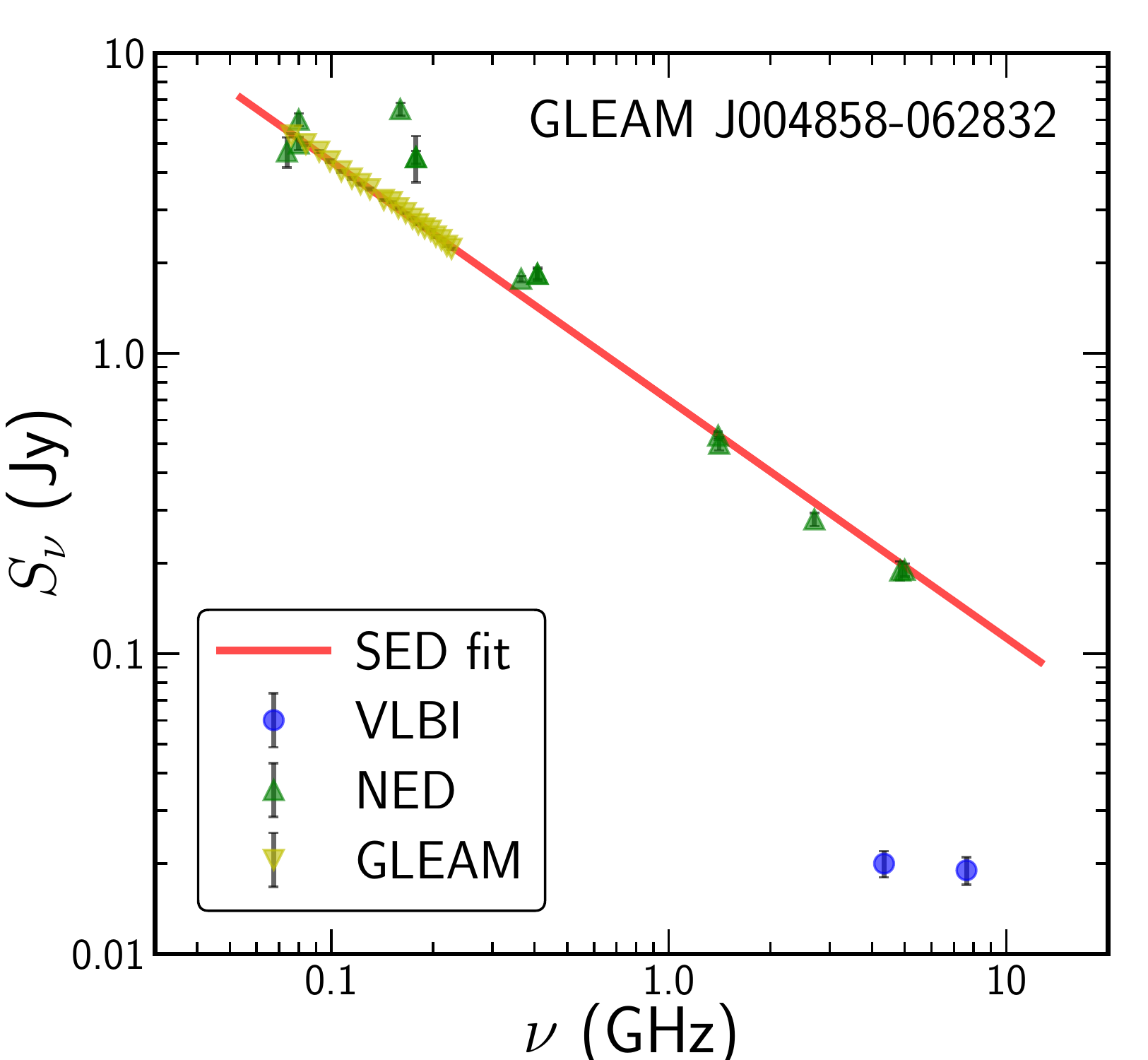}
\caption{Comparing pc-scale radio spectrum with kpc-scale radio spectrum for IPS sources.}
\label{fig:sed}
\end{figure*}
\addtocounter{figure}{-1}
\begin{figure*}
\centering
\includegraphics[scale=0.26]{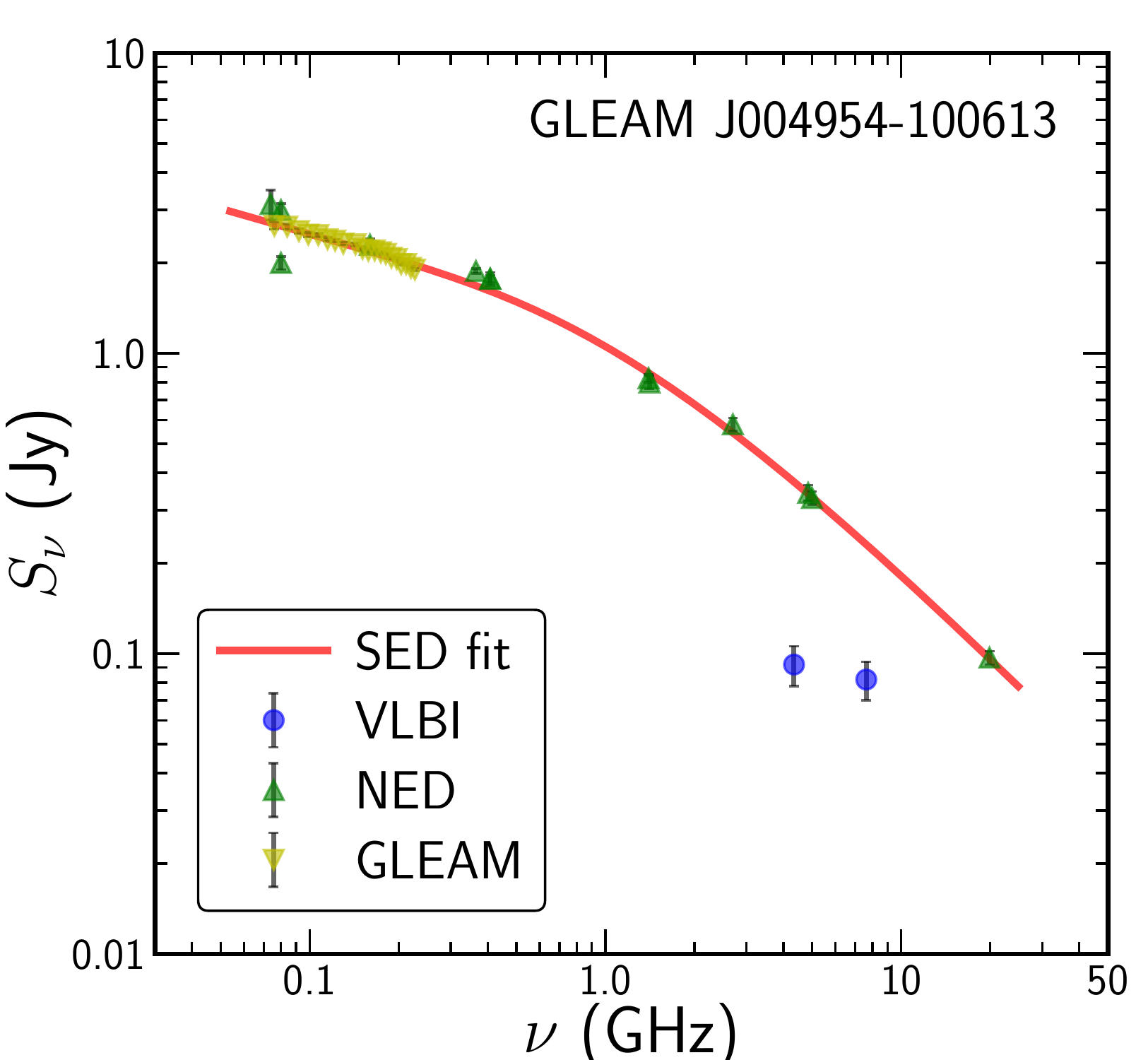}
\includegraphics[scale=0.26]{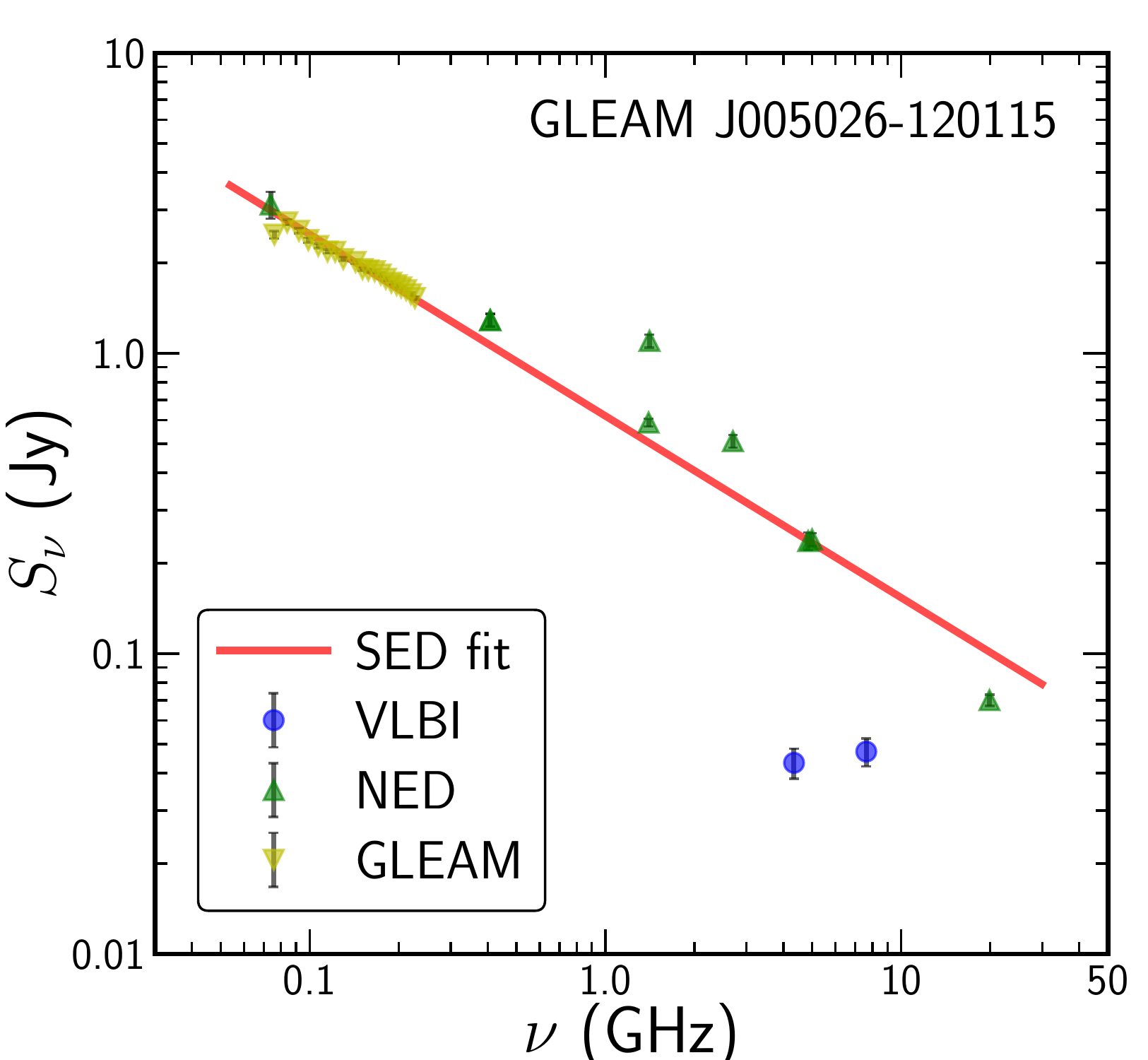}
\includegraphics[scale=0.26]{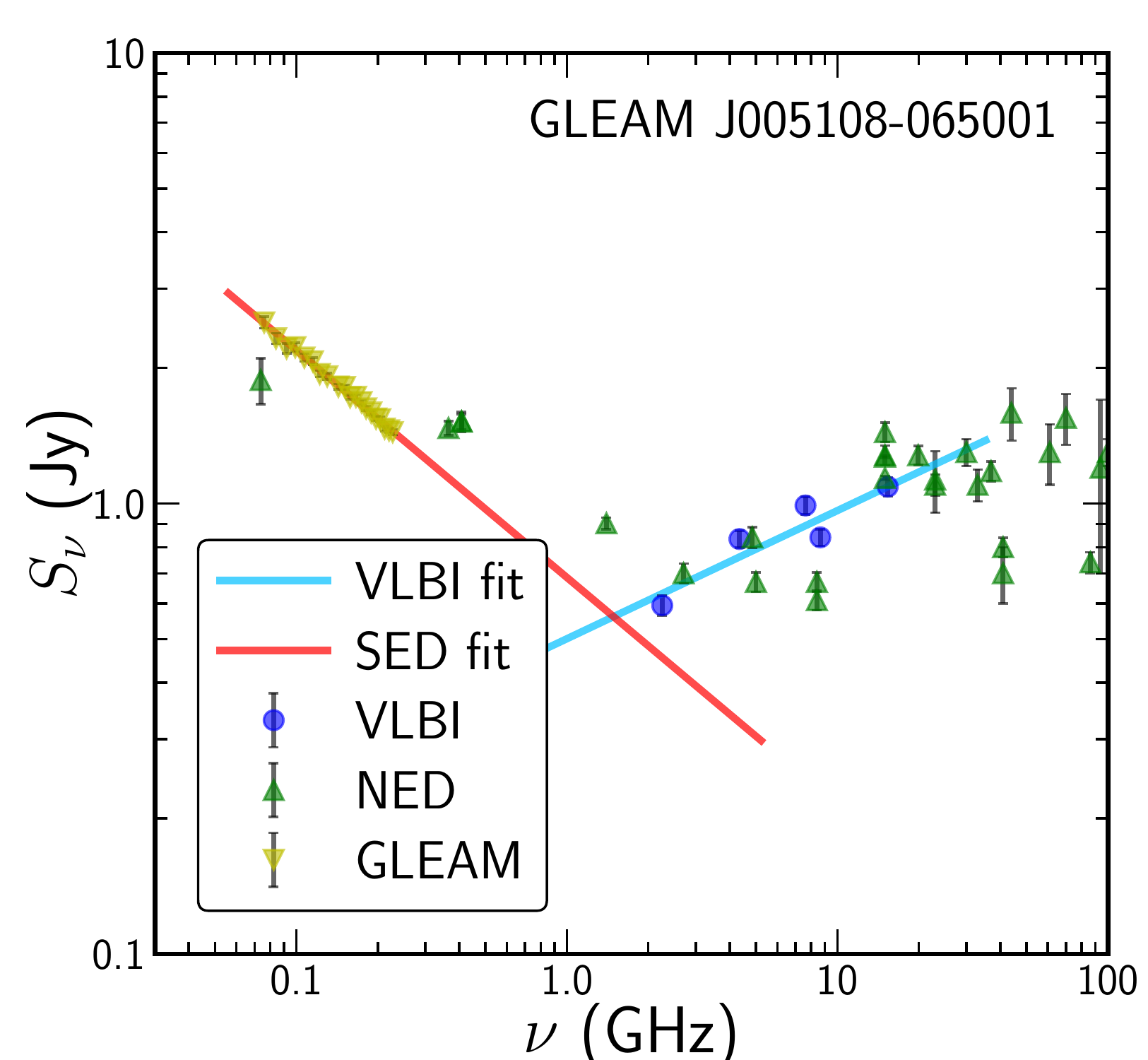}
\includegraphics[scale=0.26]{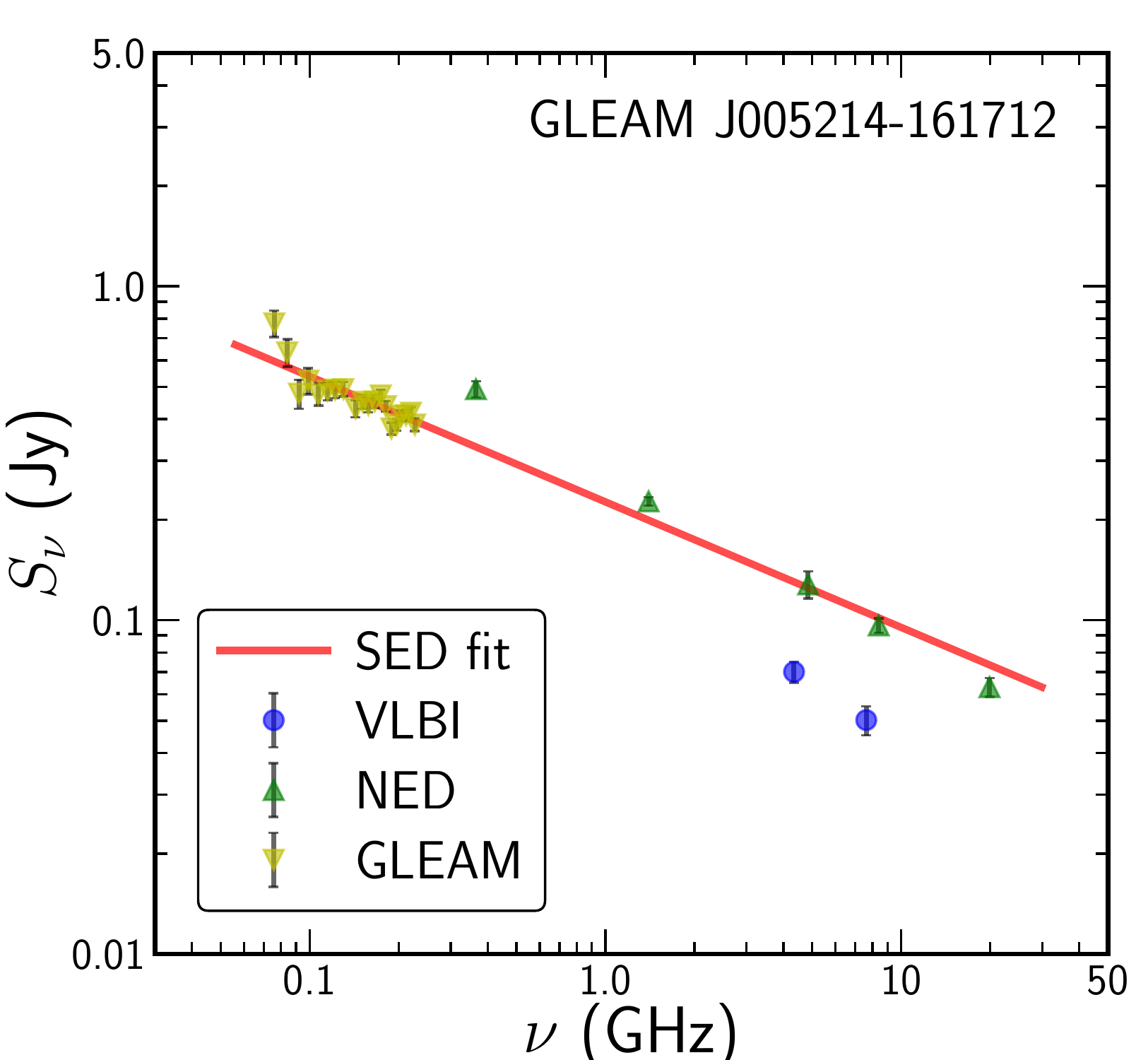}\\
\includegraphics[scale=0.26]{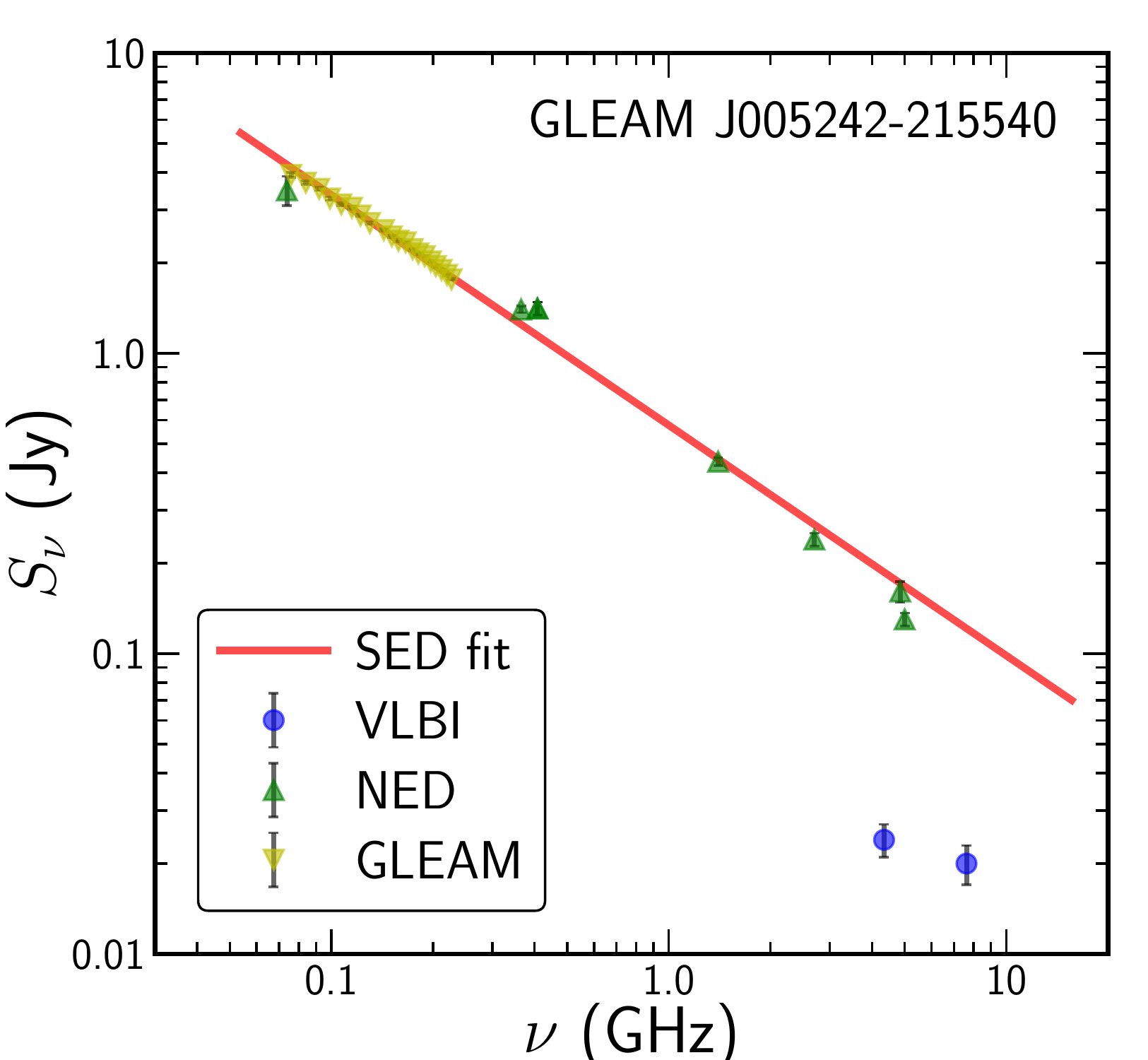}
\includegraphics[scale=0.26]{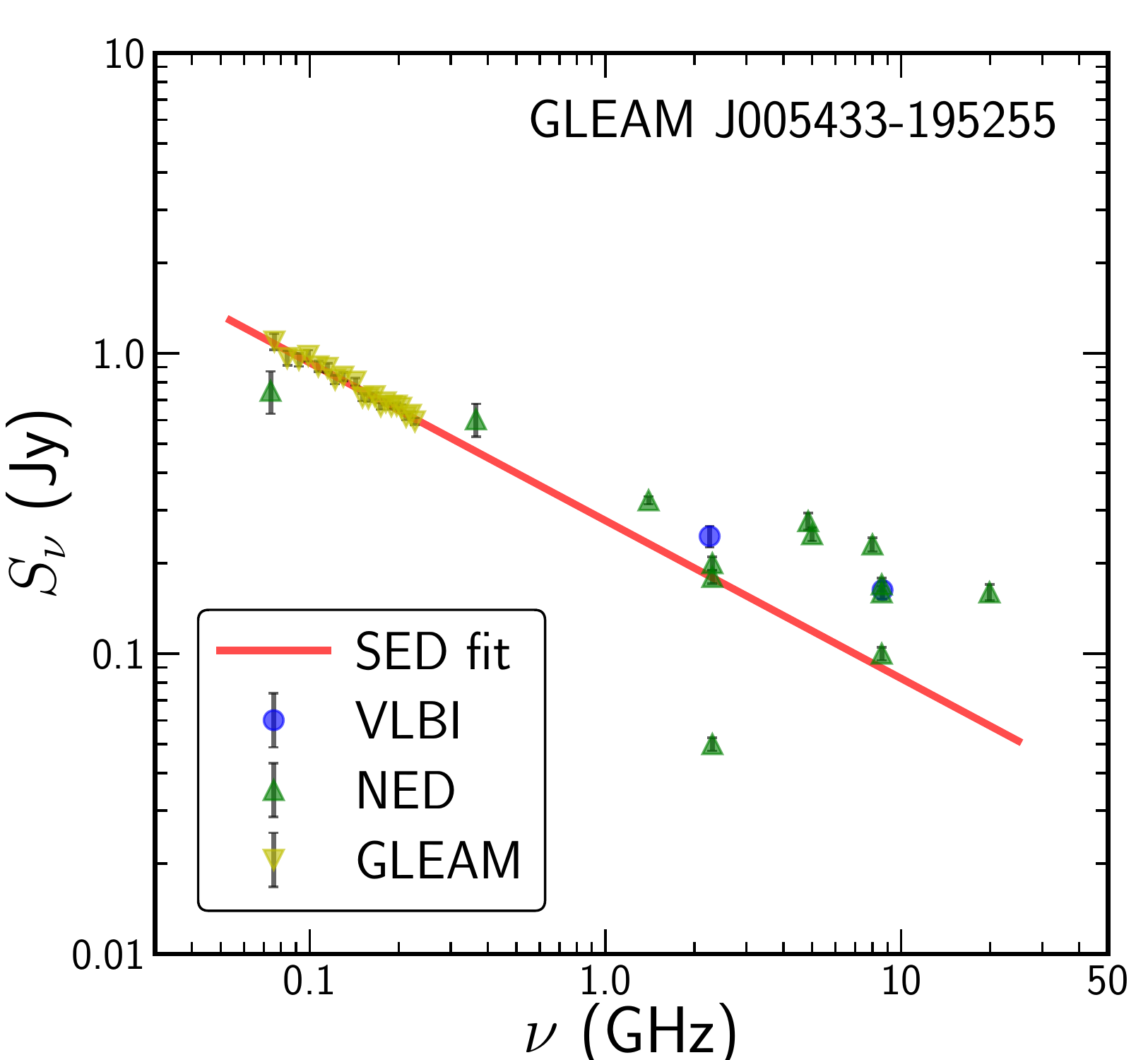}
\includegraphics[scale=0.26]{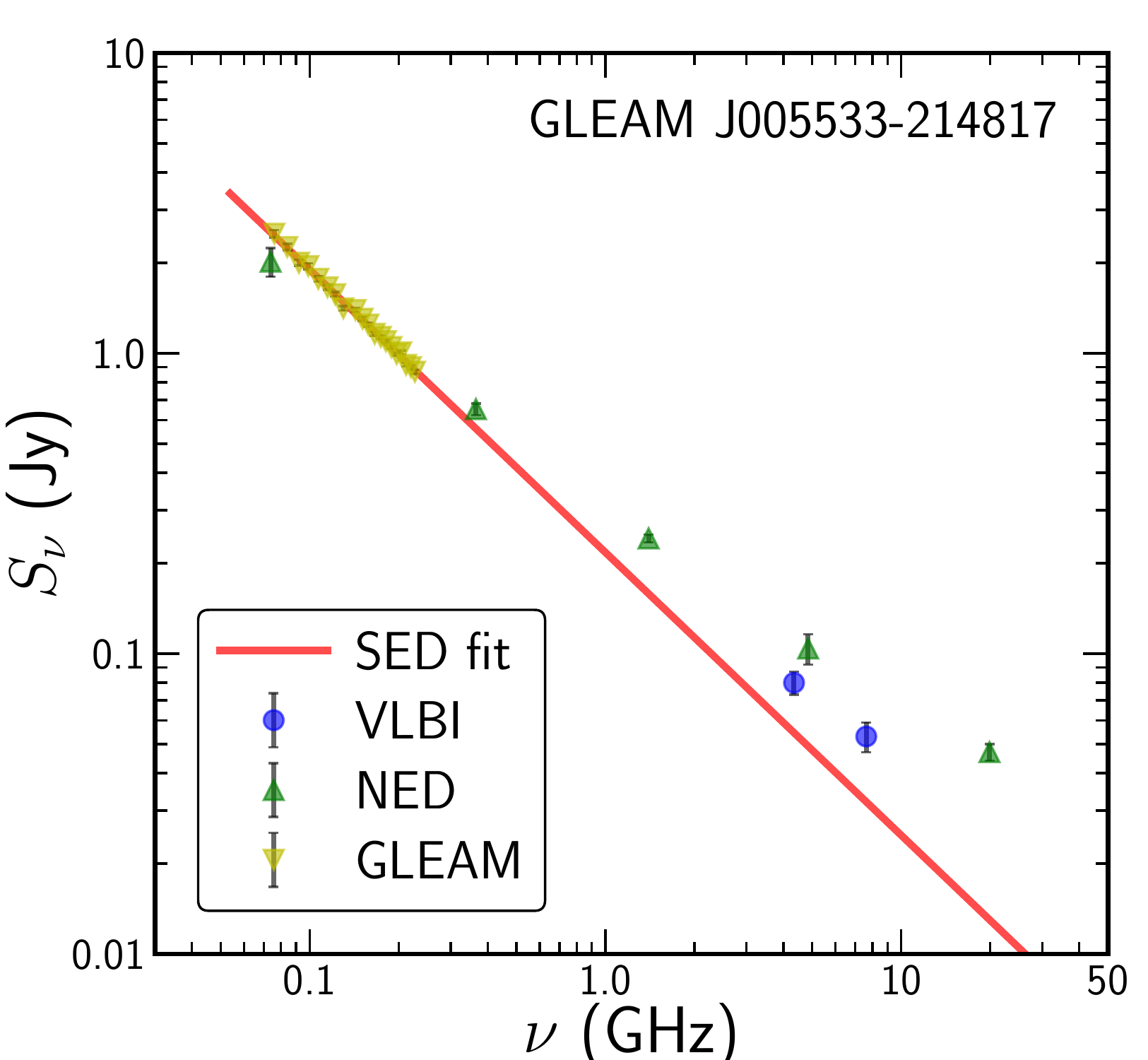}
\includegraphics[scale=0.26]{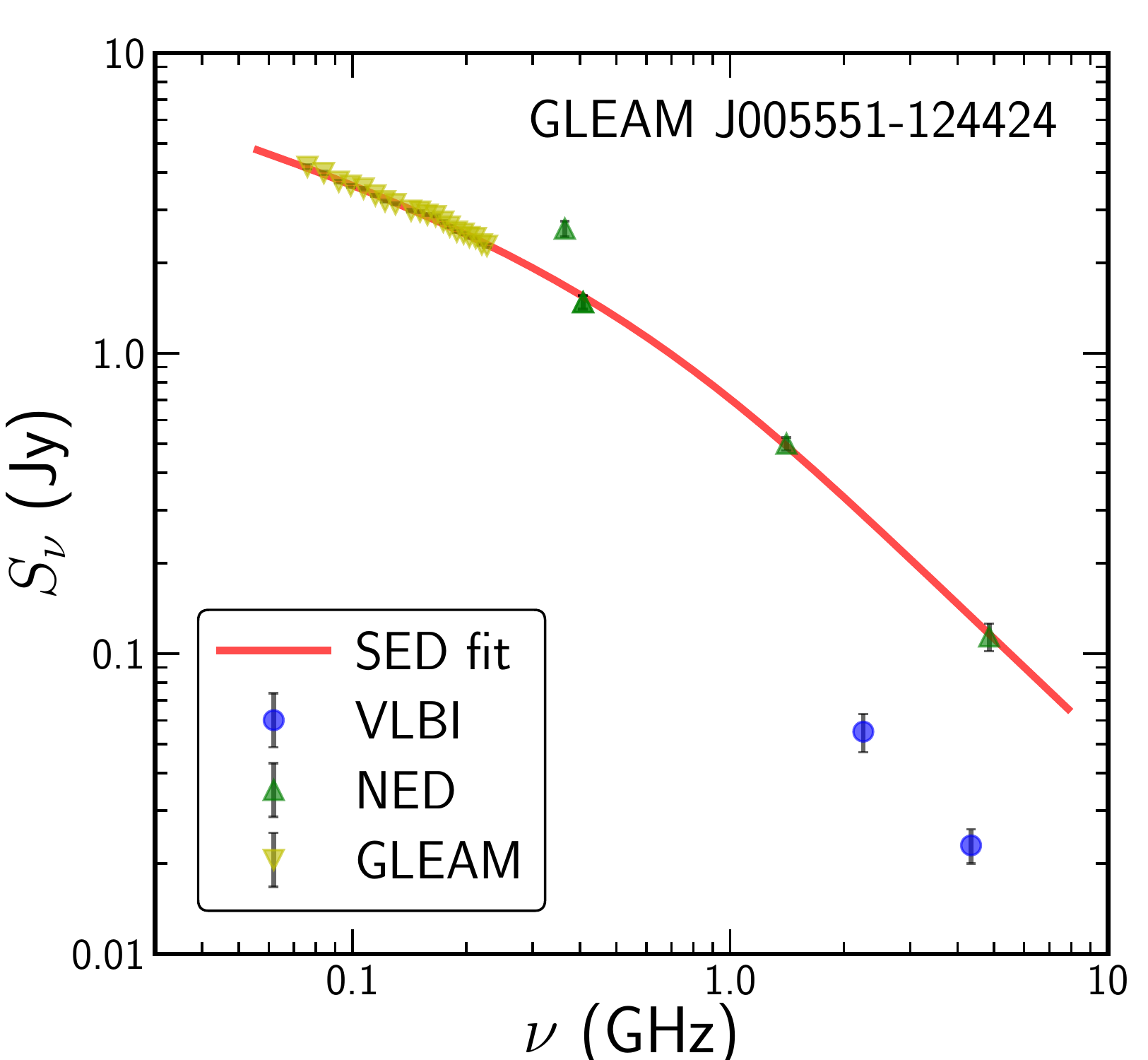}\\
\includegraphics[scale=0.26]{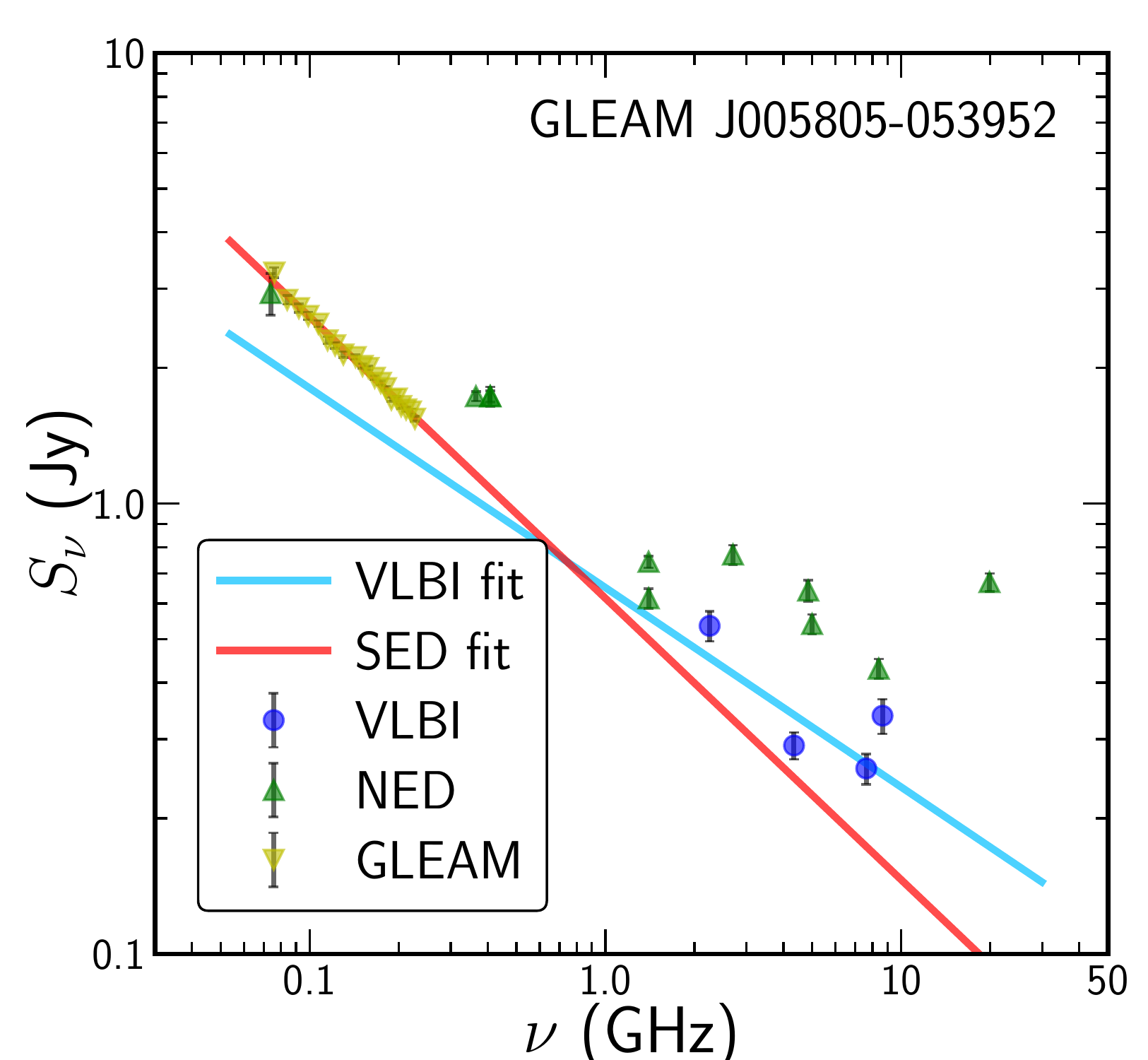}
\includegraphics[scale=0.26]{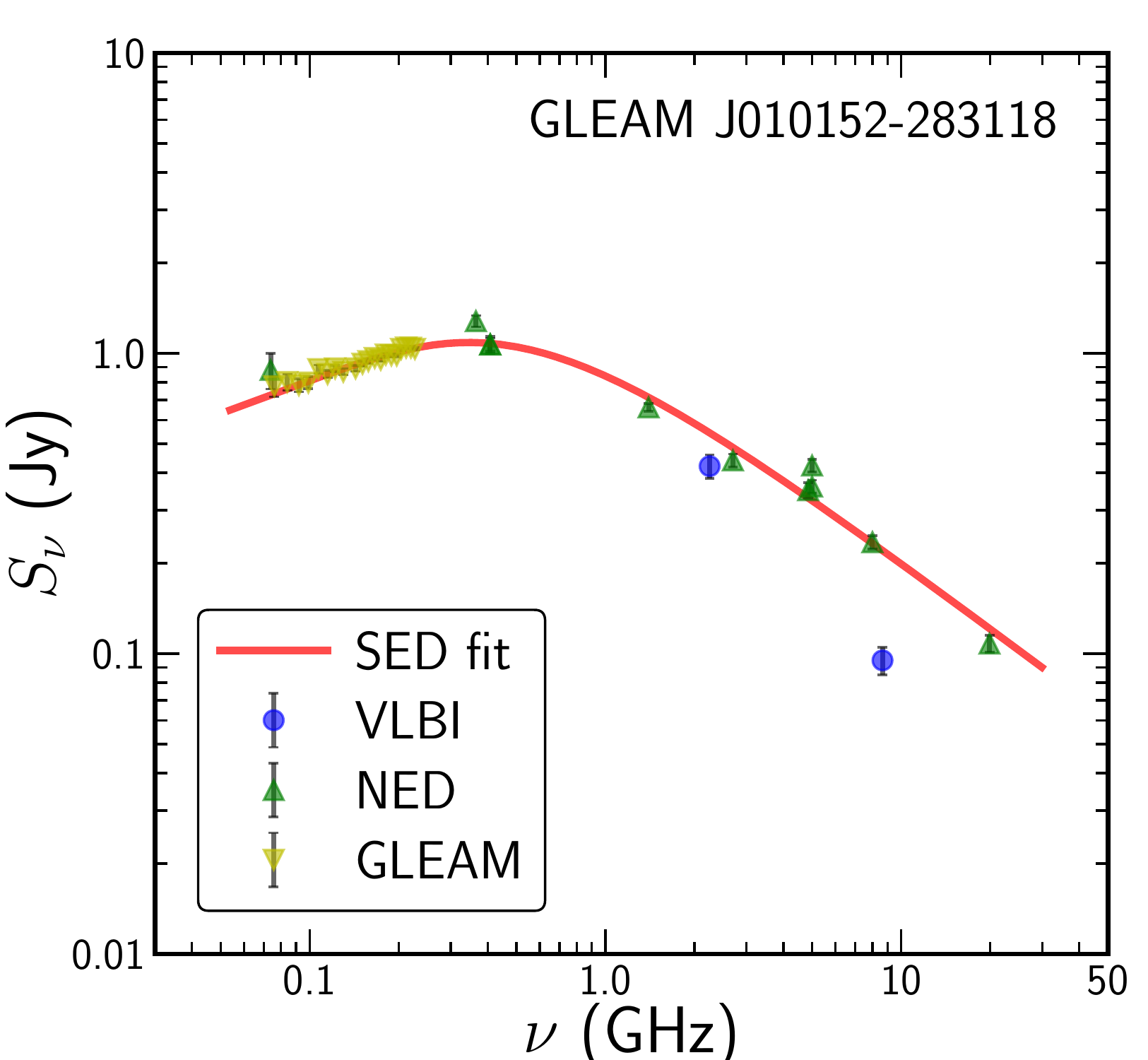}
\includegraphics[scale=0.26]{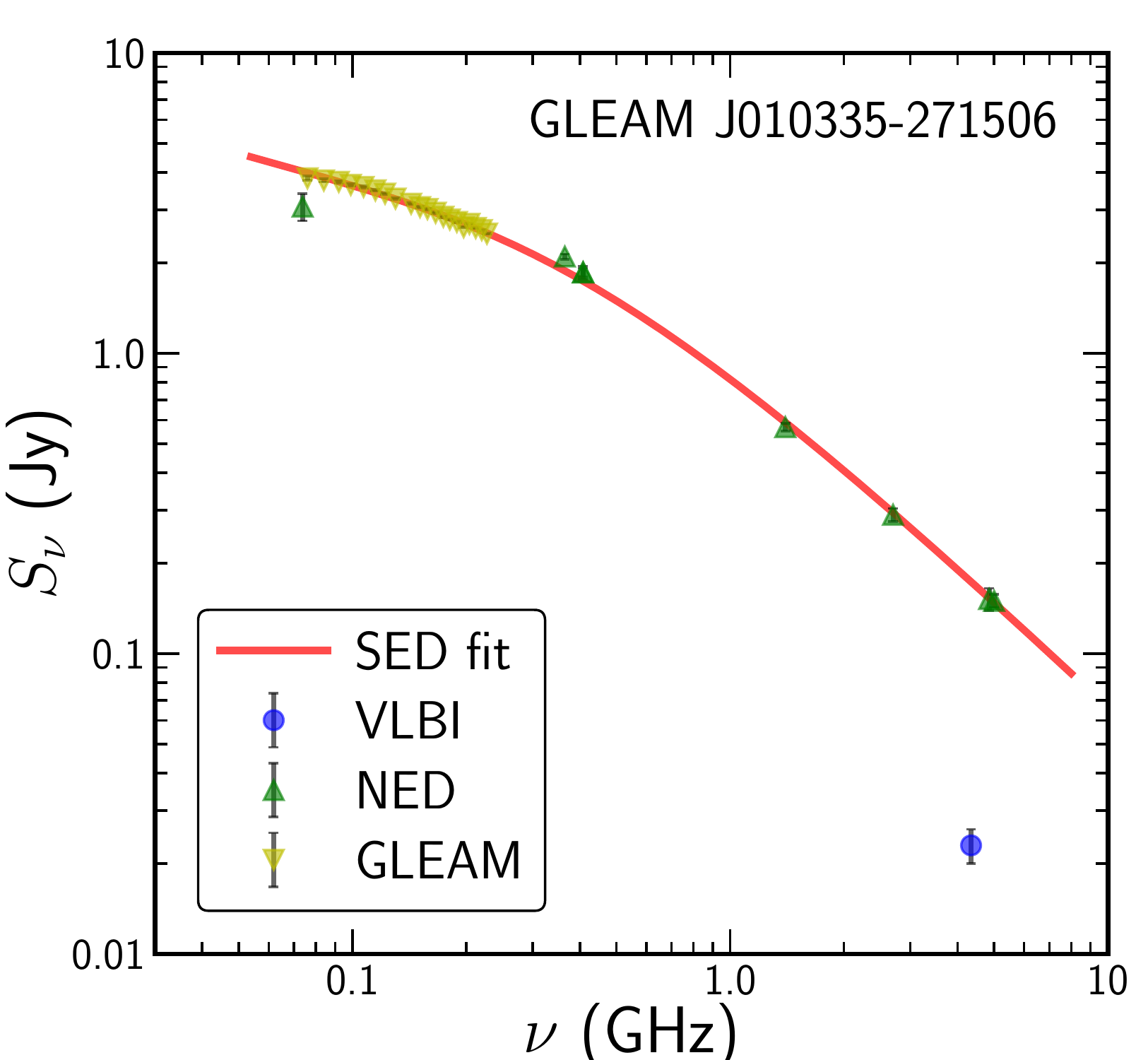}
\includegraphics[scale=0.26]{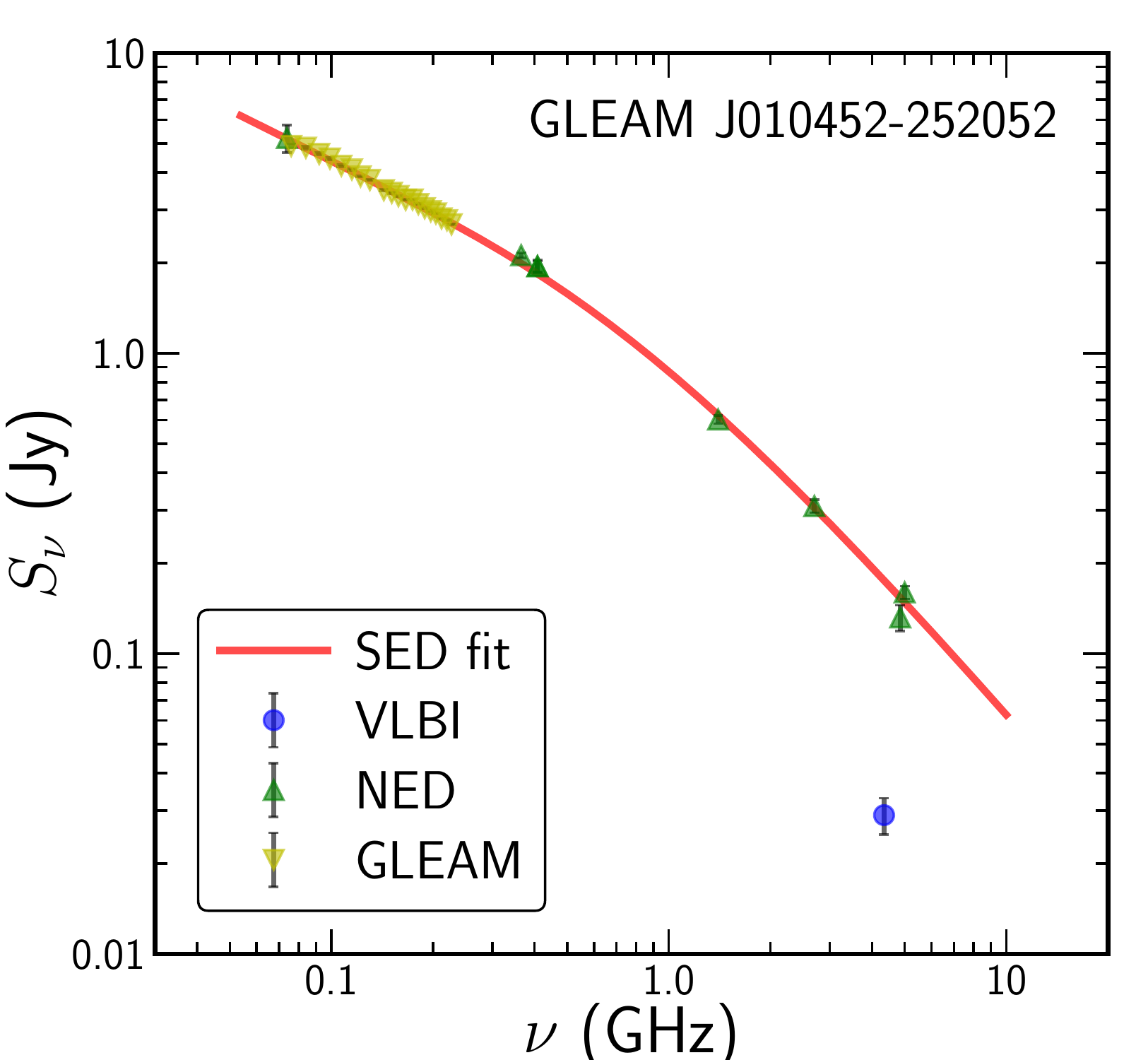}\\
\includegraphics[scale=0.26]{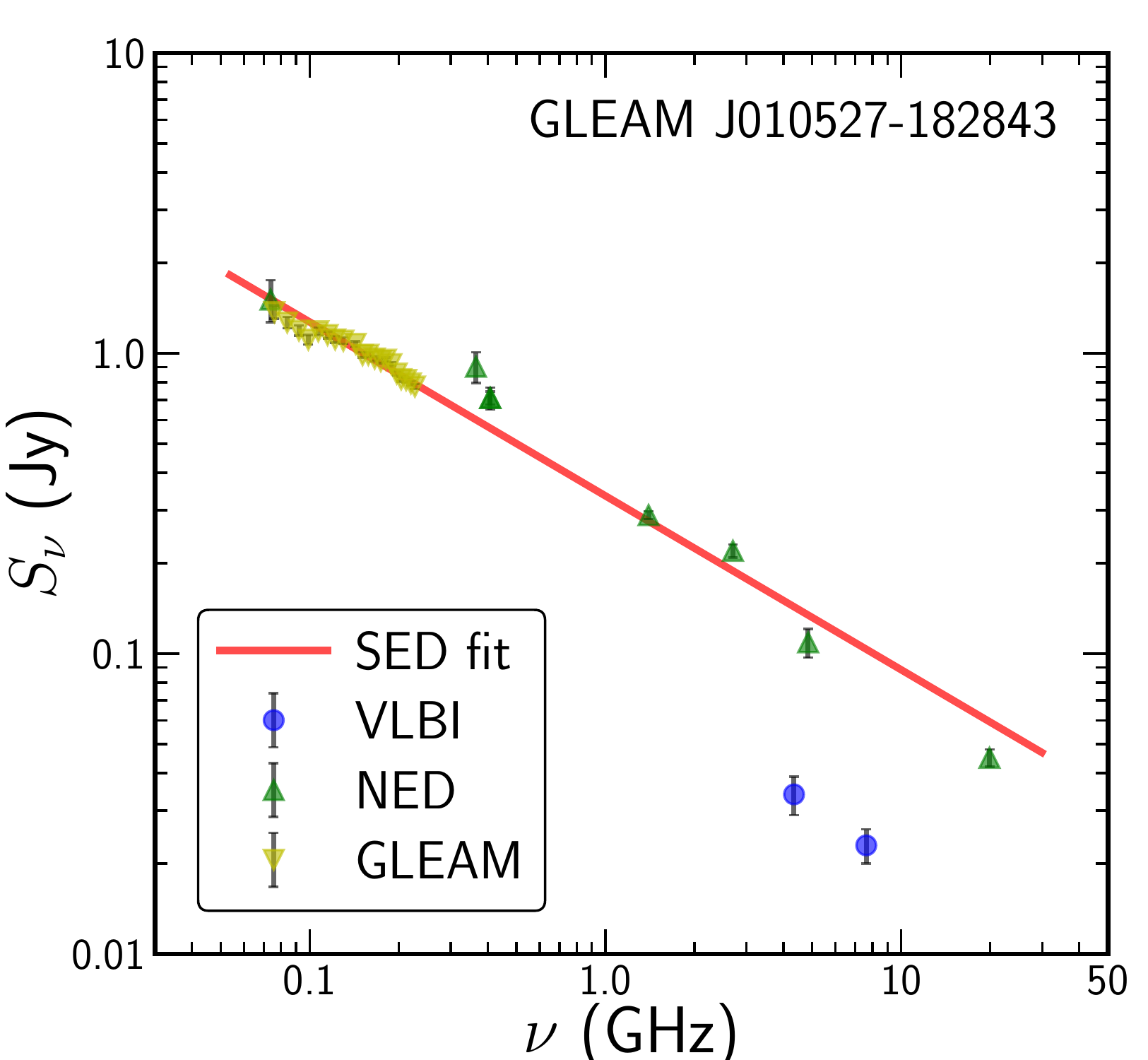}
\includegraphics[scale=0.26]{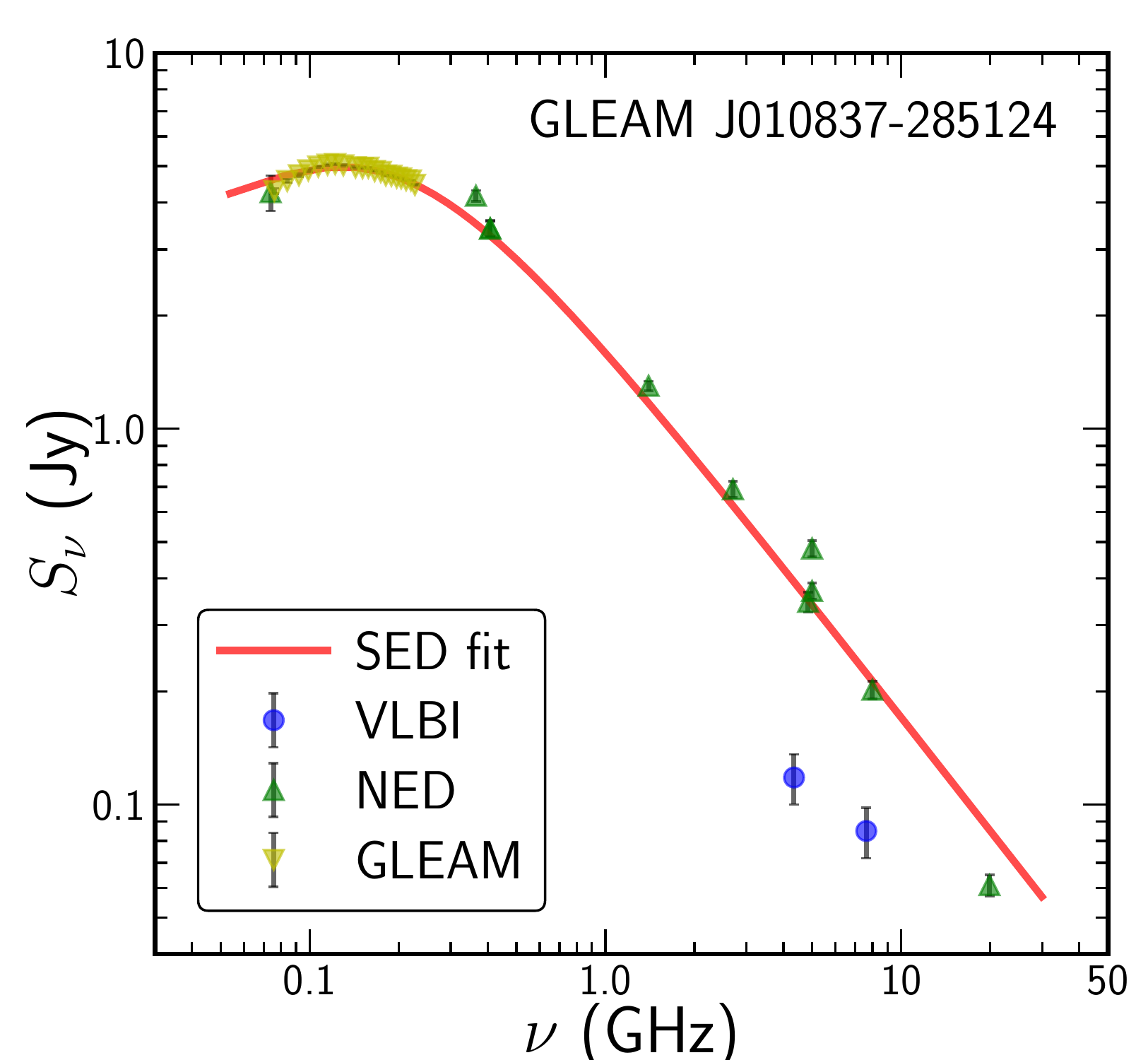}
\includegraphics[scale=0.26]{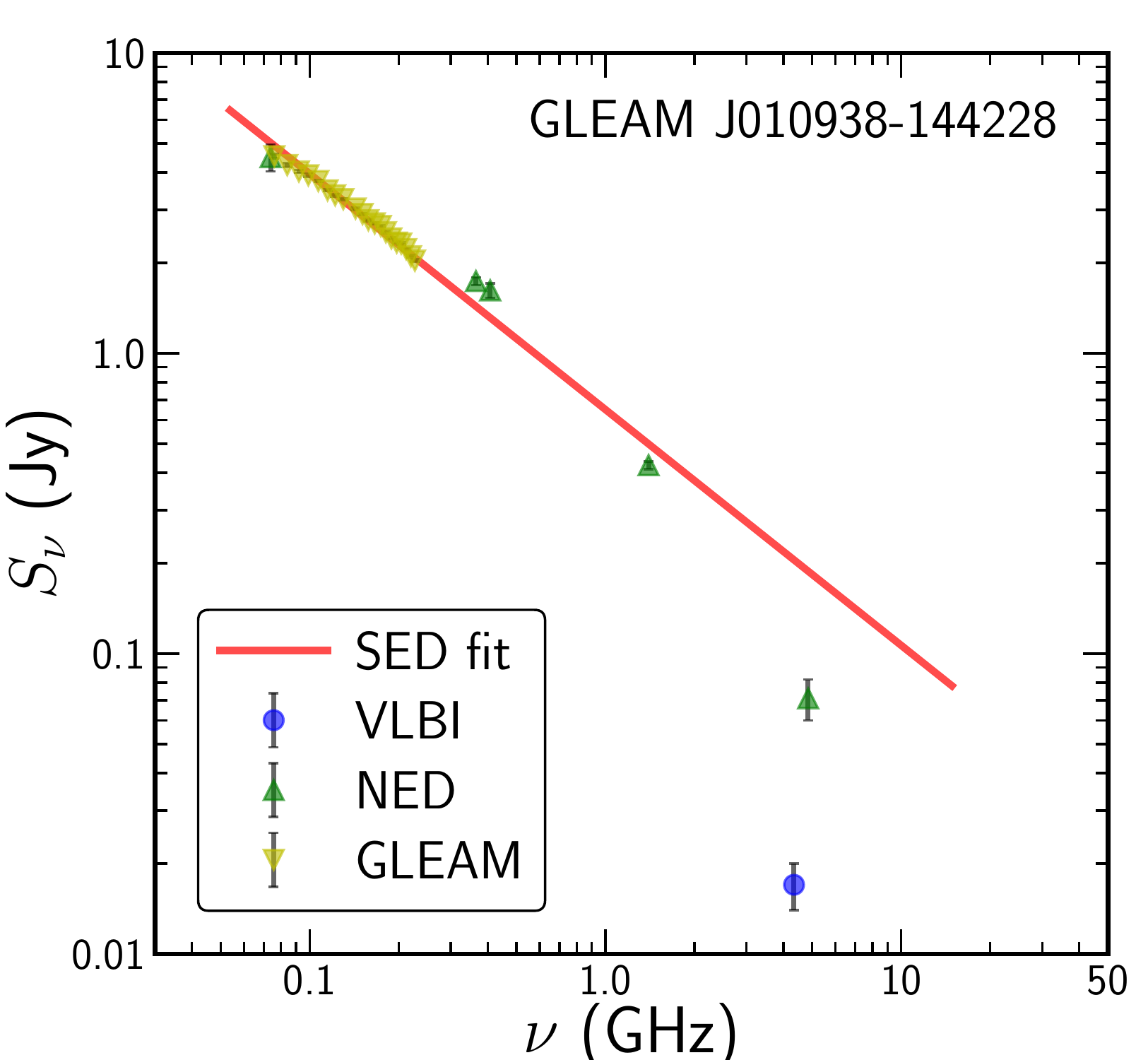}
\includegraphics[scale=0.26]{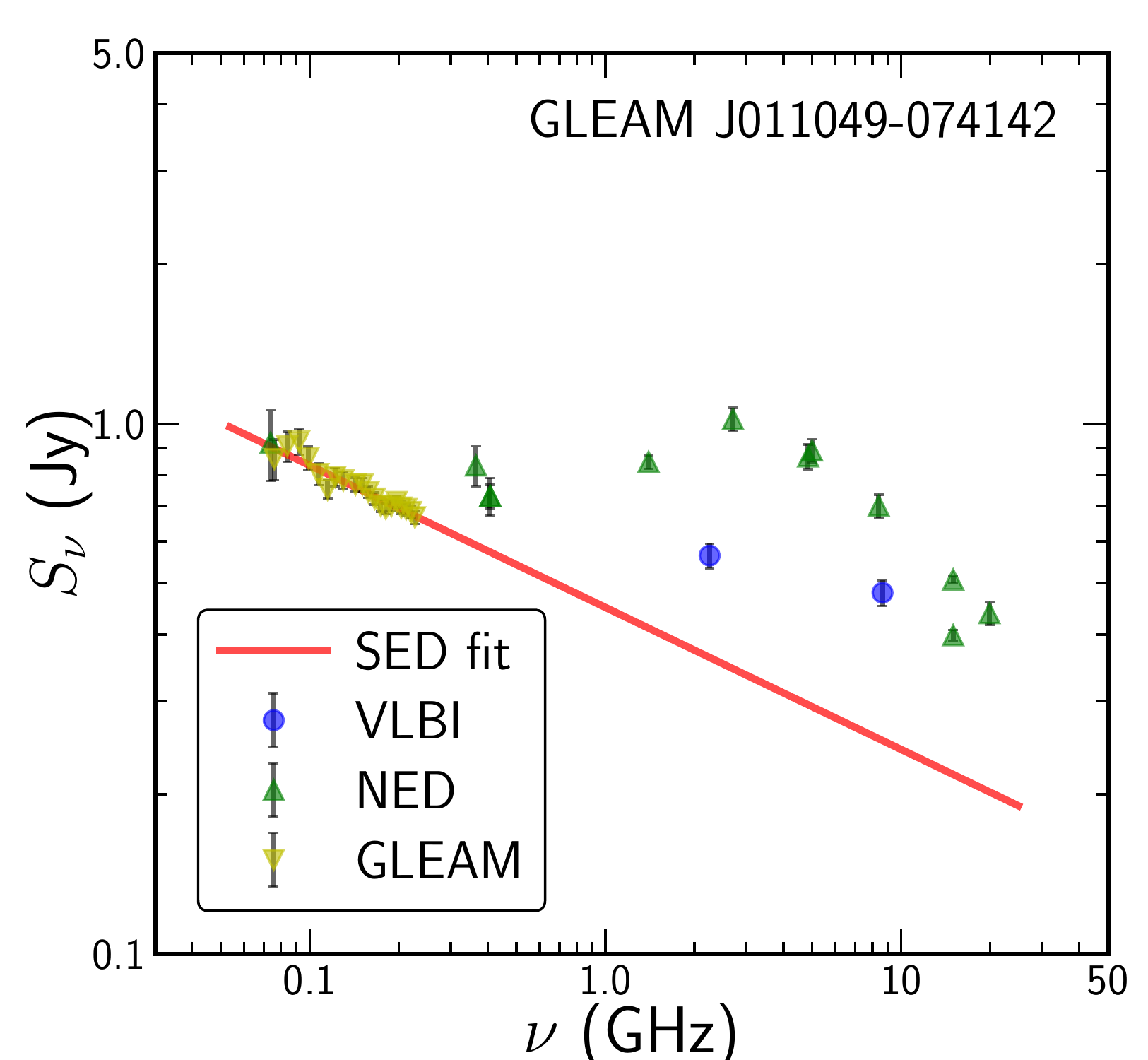}\\
\includegraphics[scale=0.26]{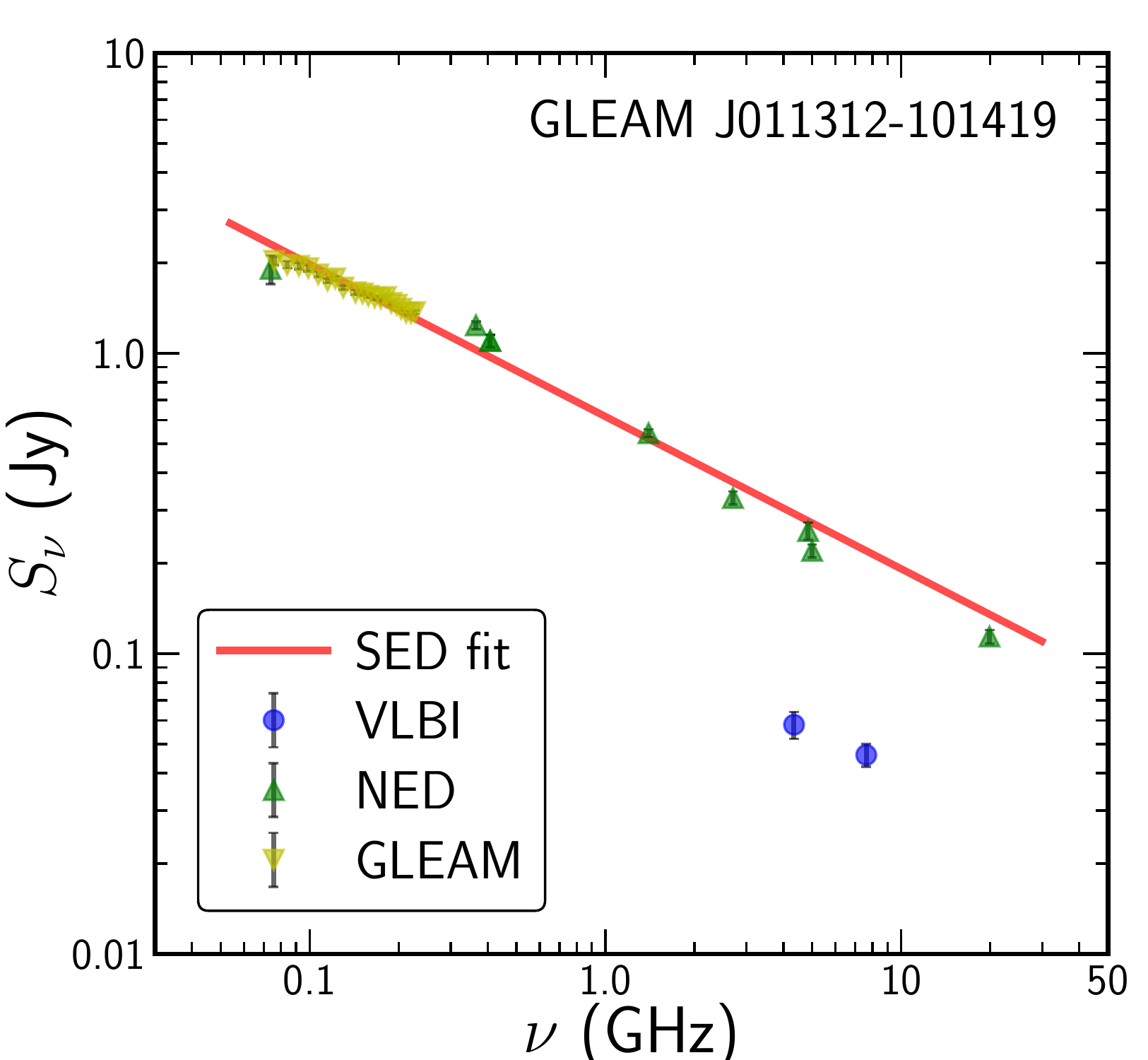}
\includegraphics[scale=0.26]{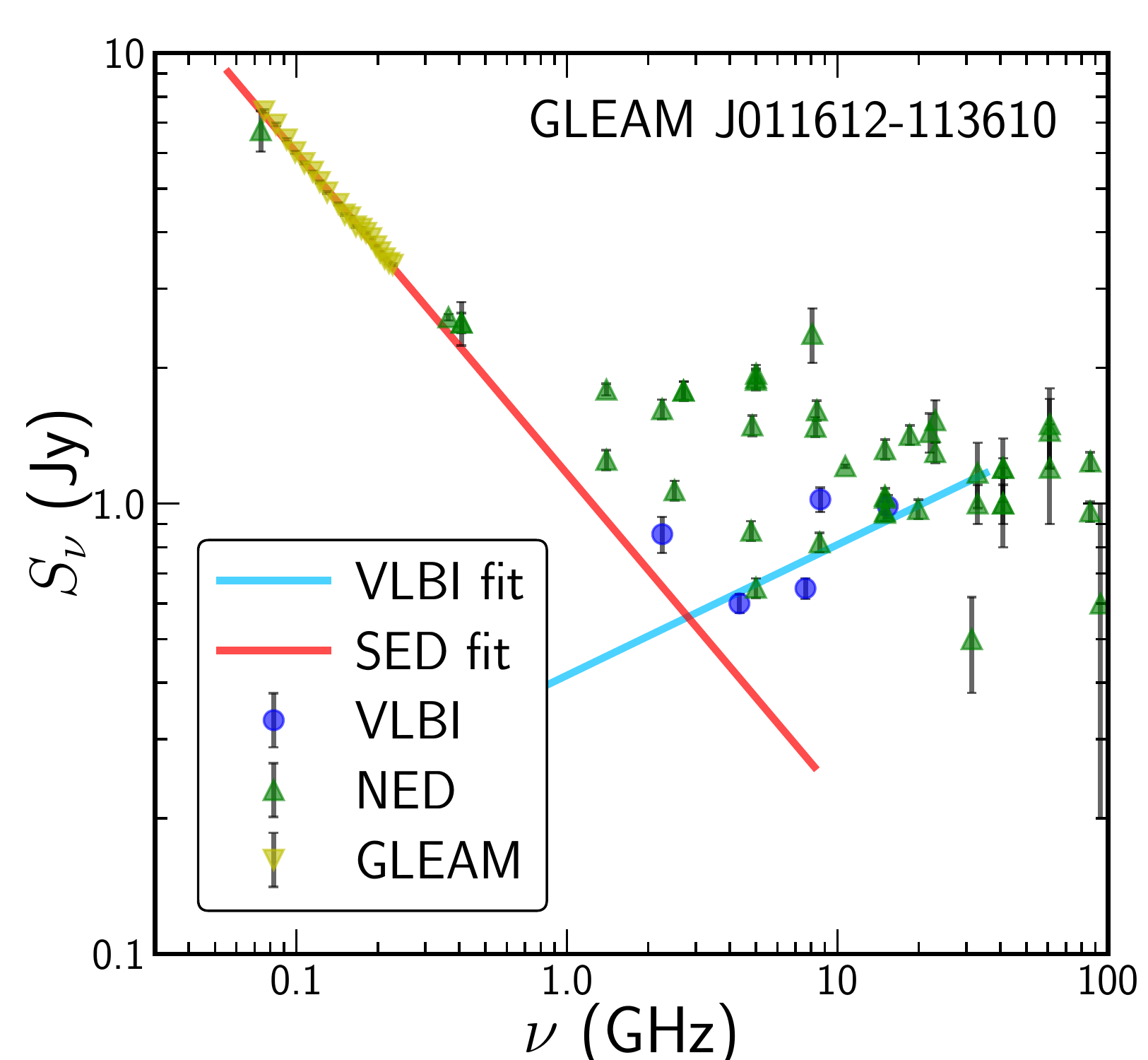}
\includegraphics[scale=0.26]{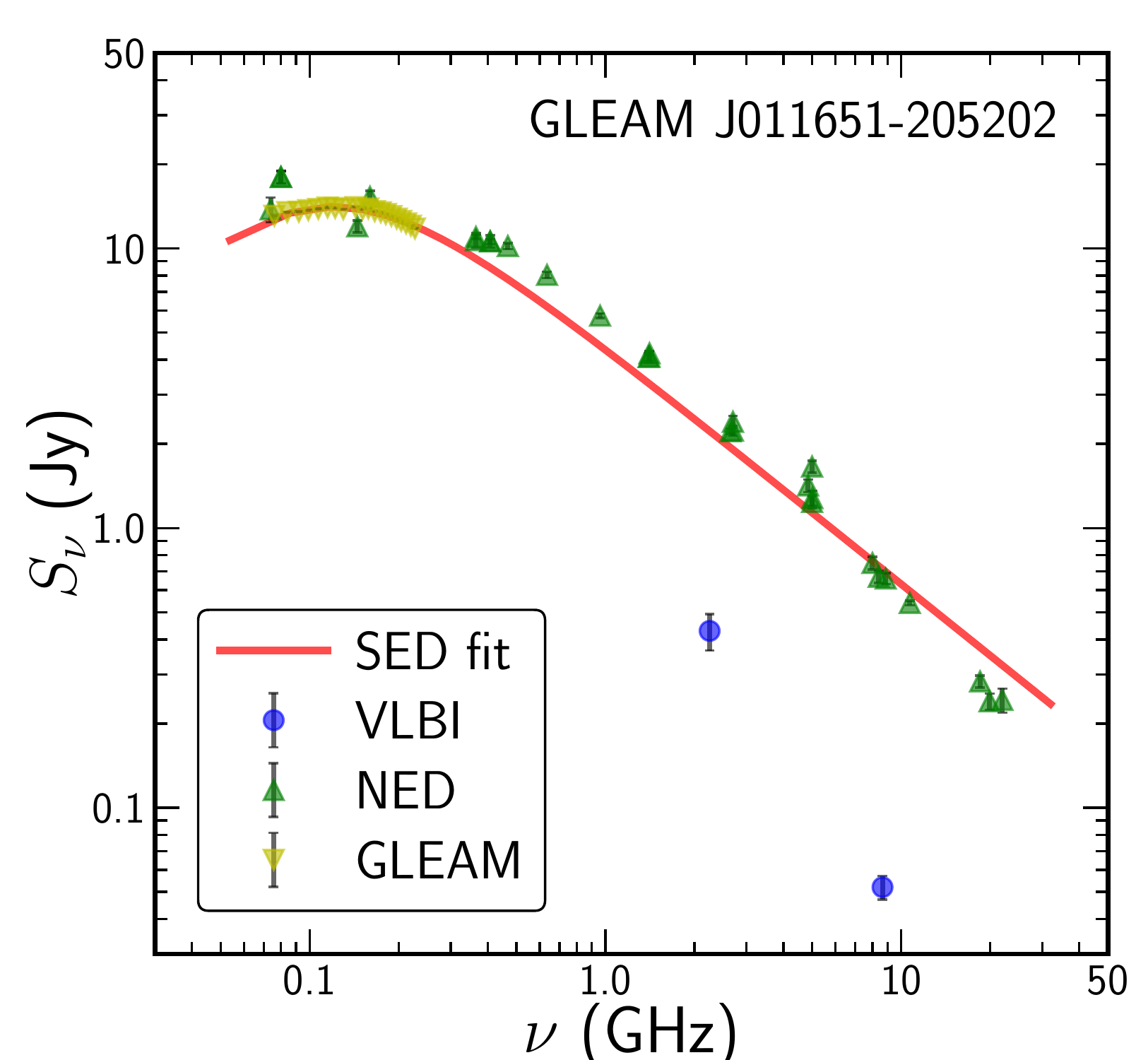}
\includegraphics[scale=0.26]{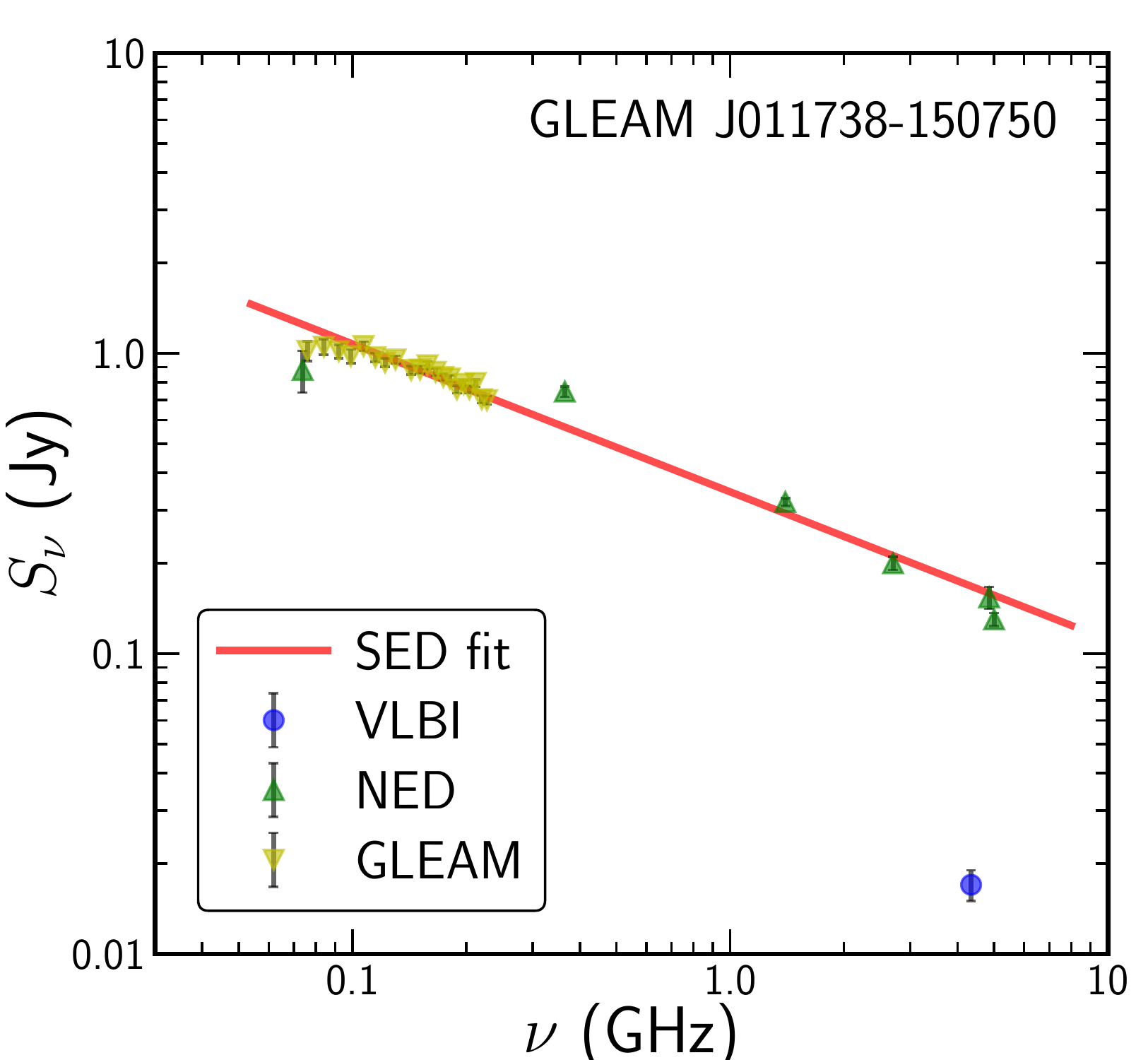}\\
\includegraphics[scale=0.26]{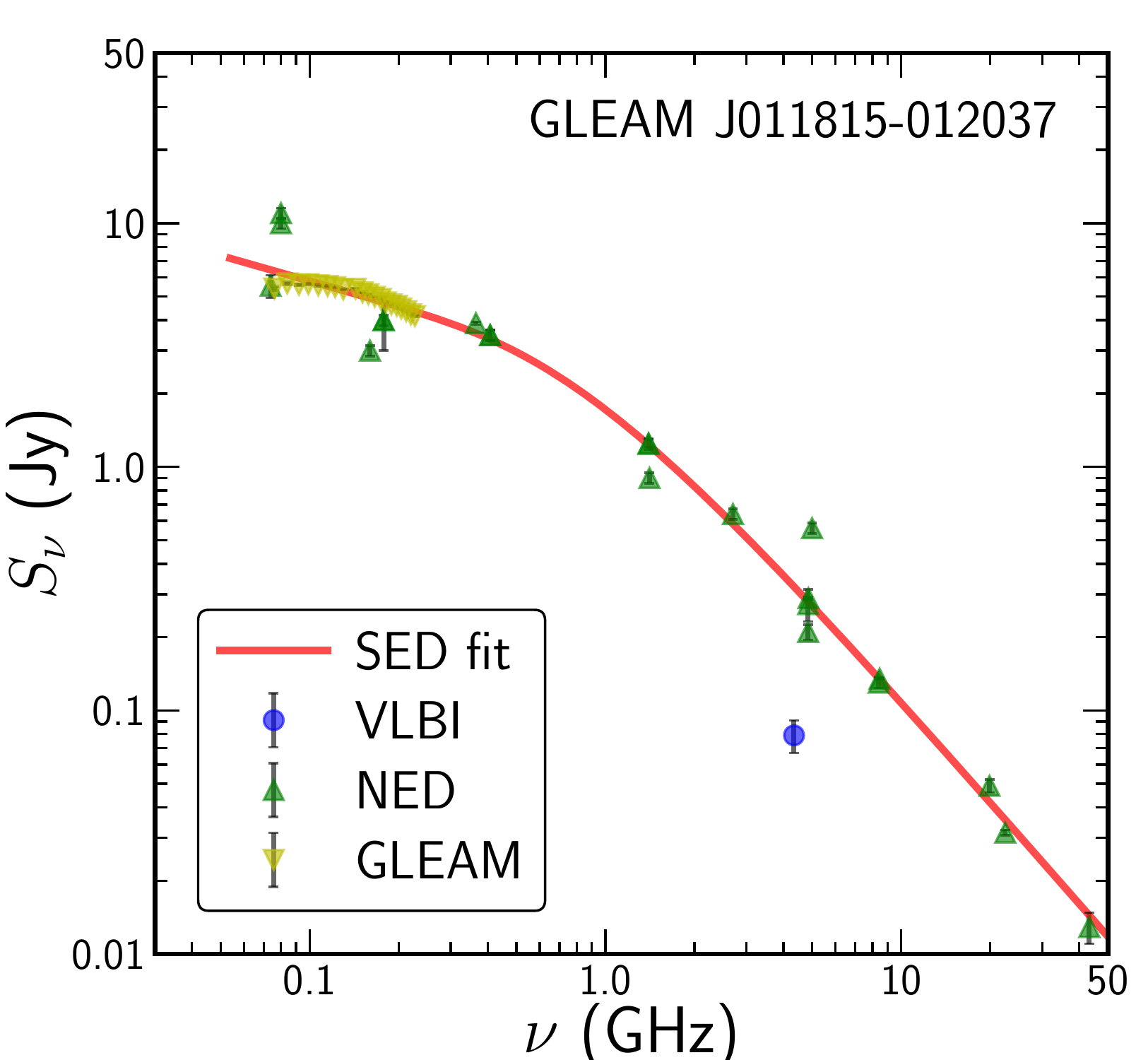}
\includegraphics[scale=0.26]{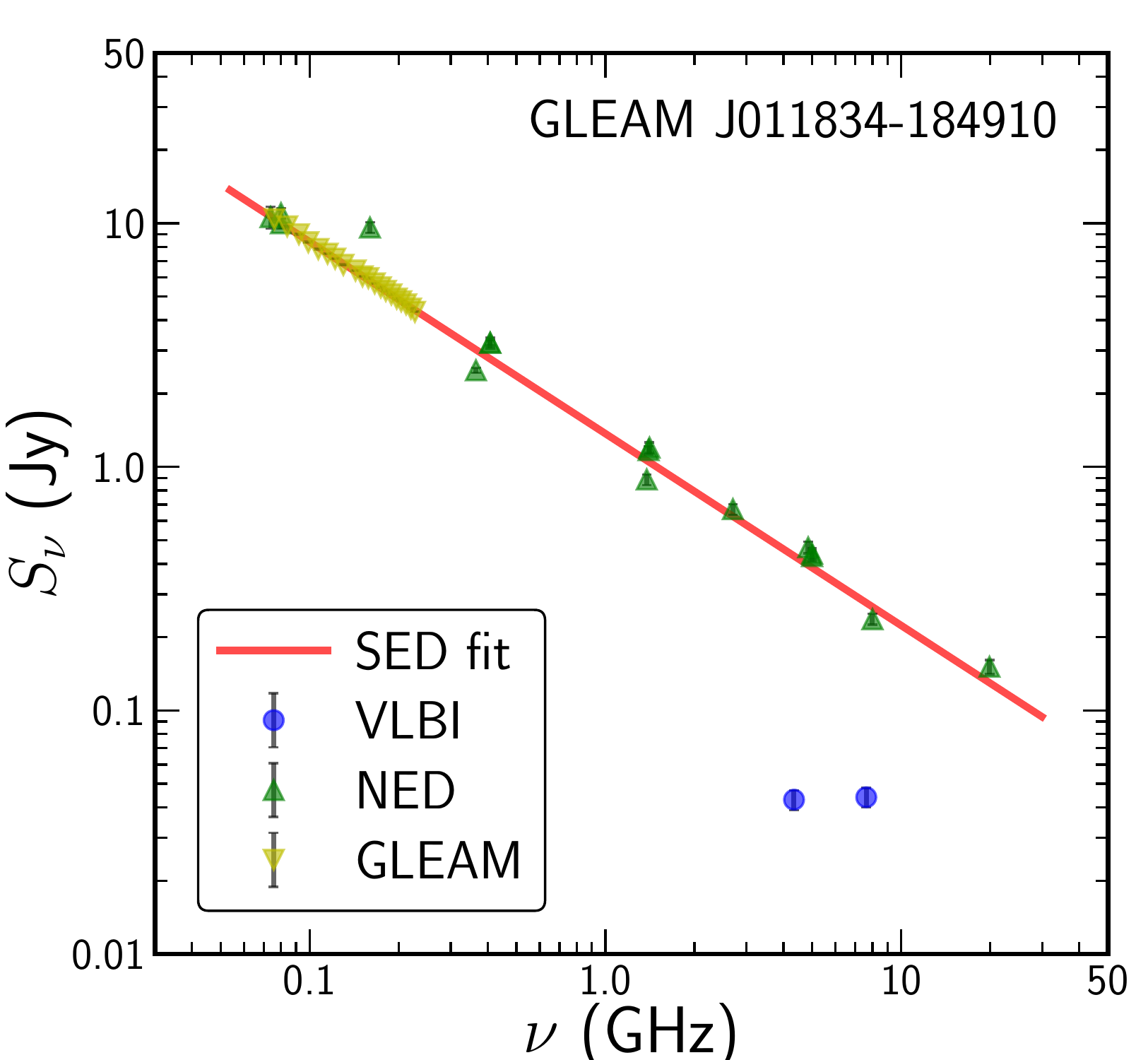}
\includegraphics[scale=0.26]{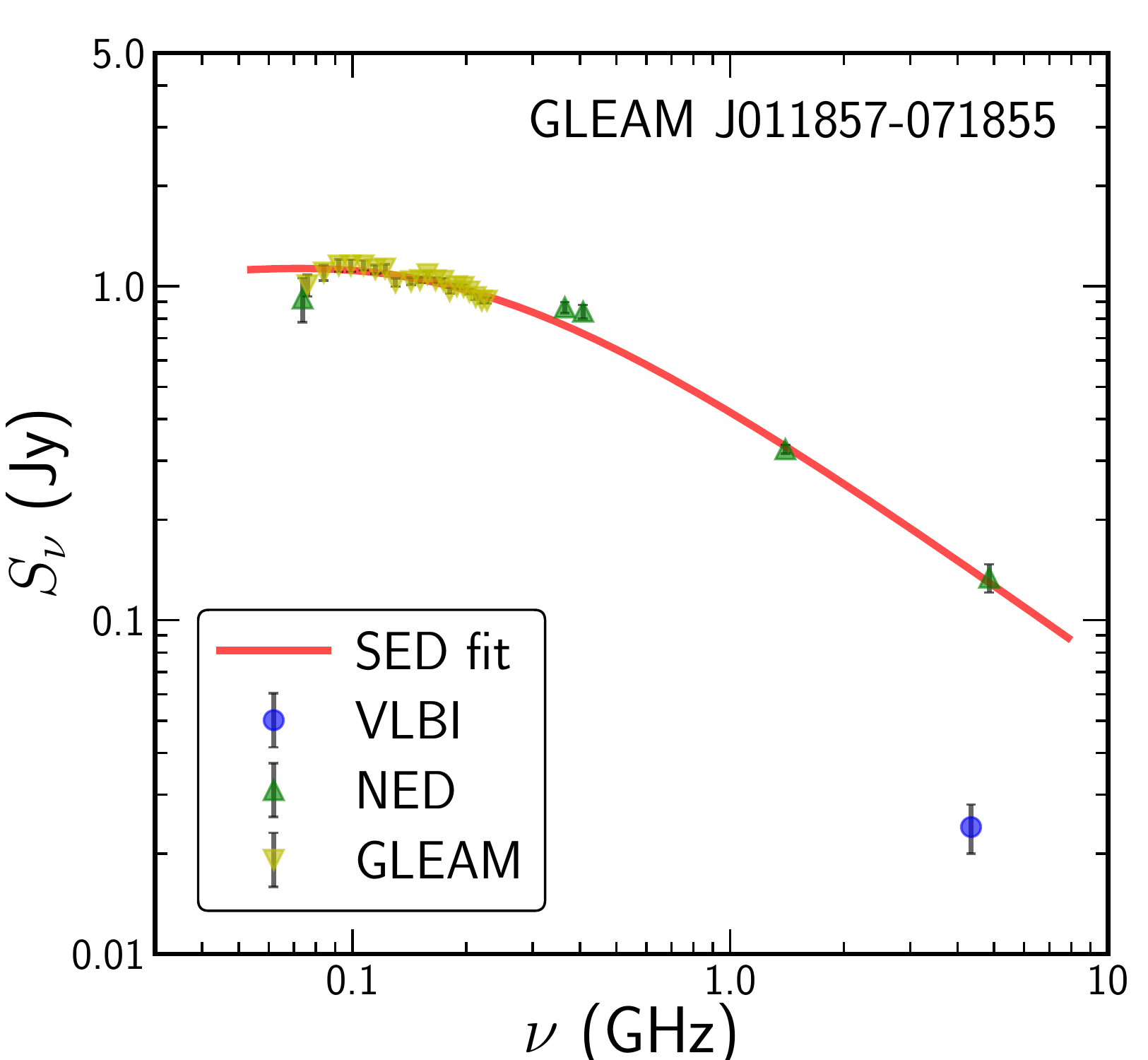}
\includegraphics[scale=0.26]{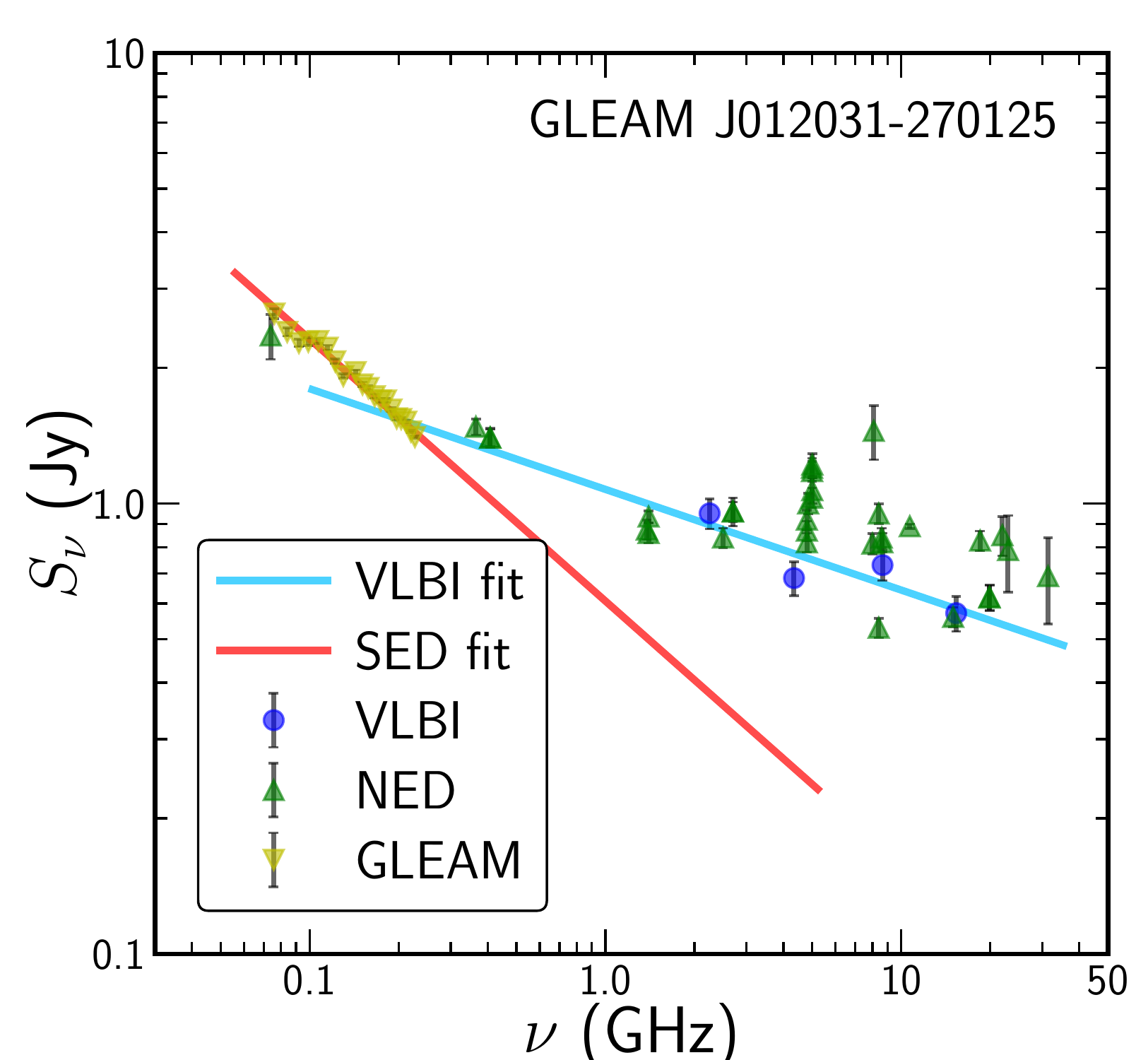}
\caption{\it - continued}
\end{figure*}
\addtocounter{figure}{-1}
\begin{figure*}
\centering
\includegraphics[scale=0.26]{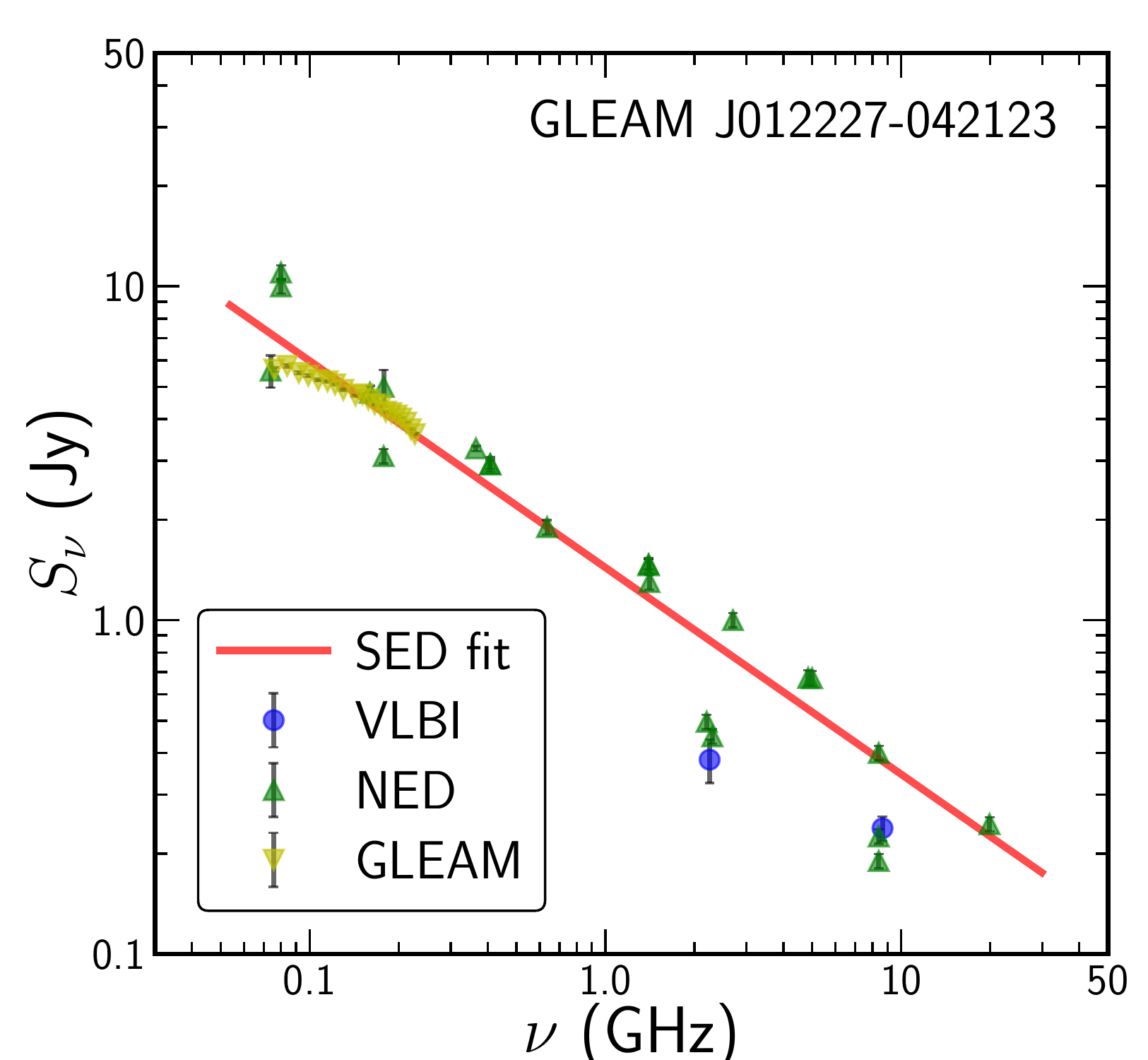}
\includegraphics[scale=0.26]{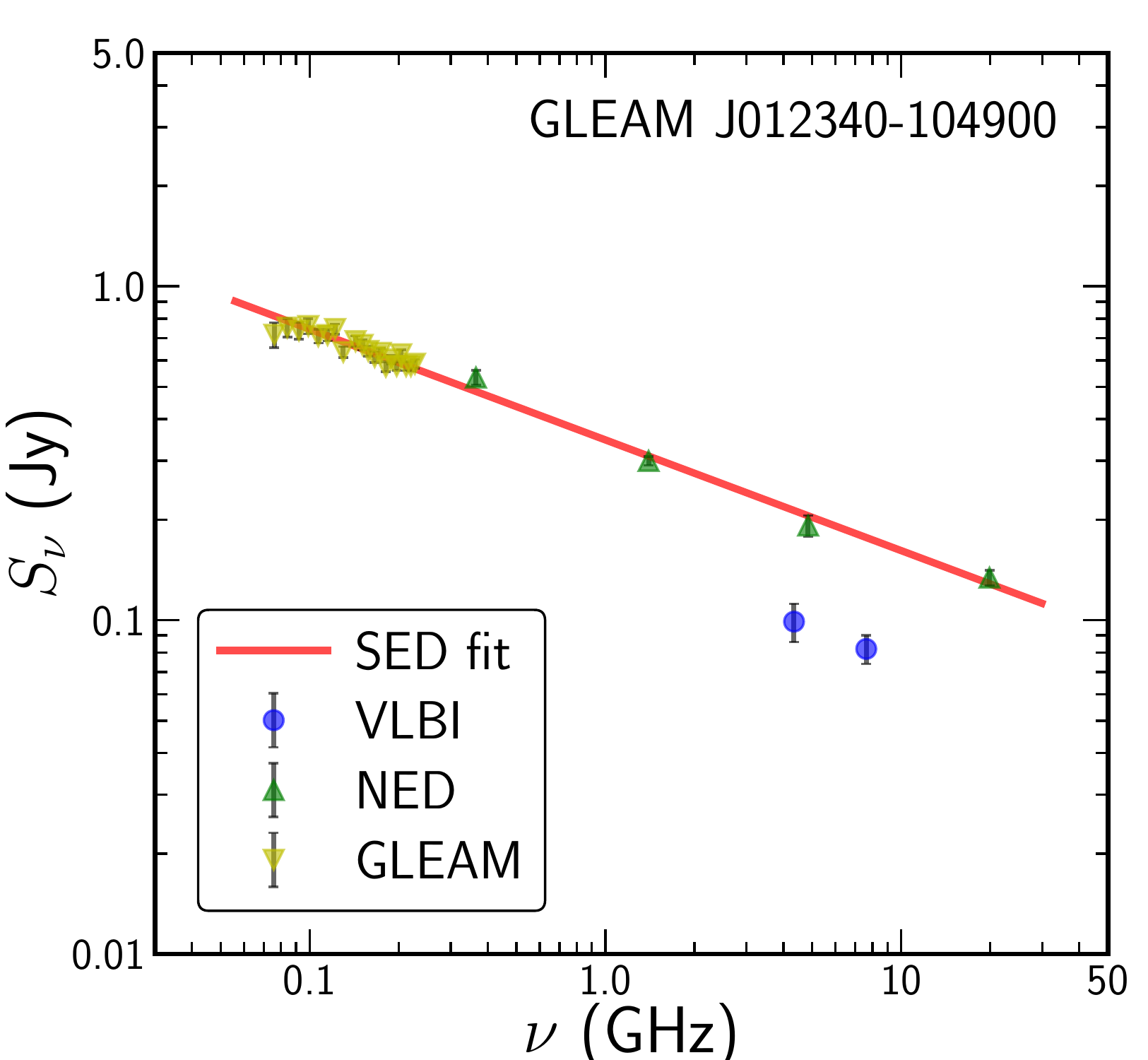}
\includegraphics[scale=0.26]{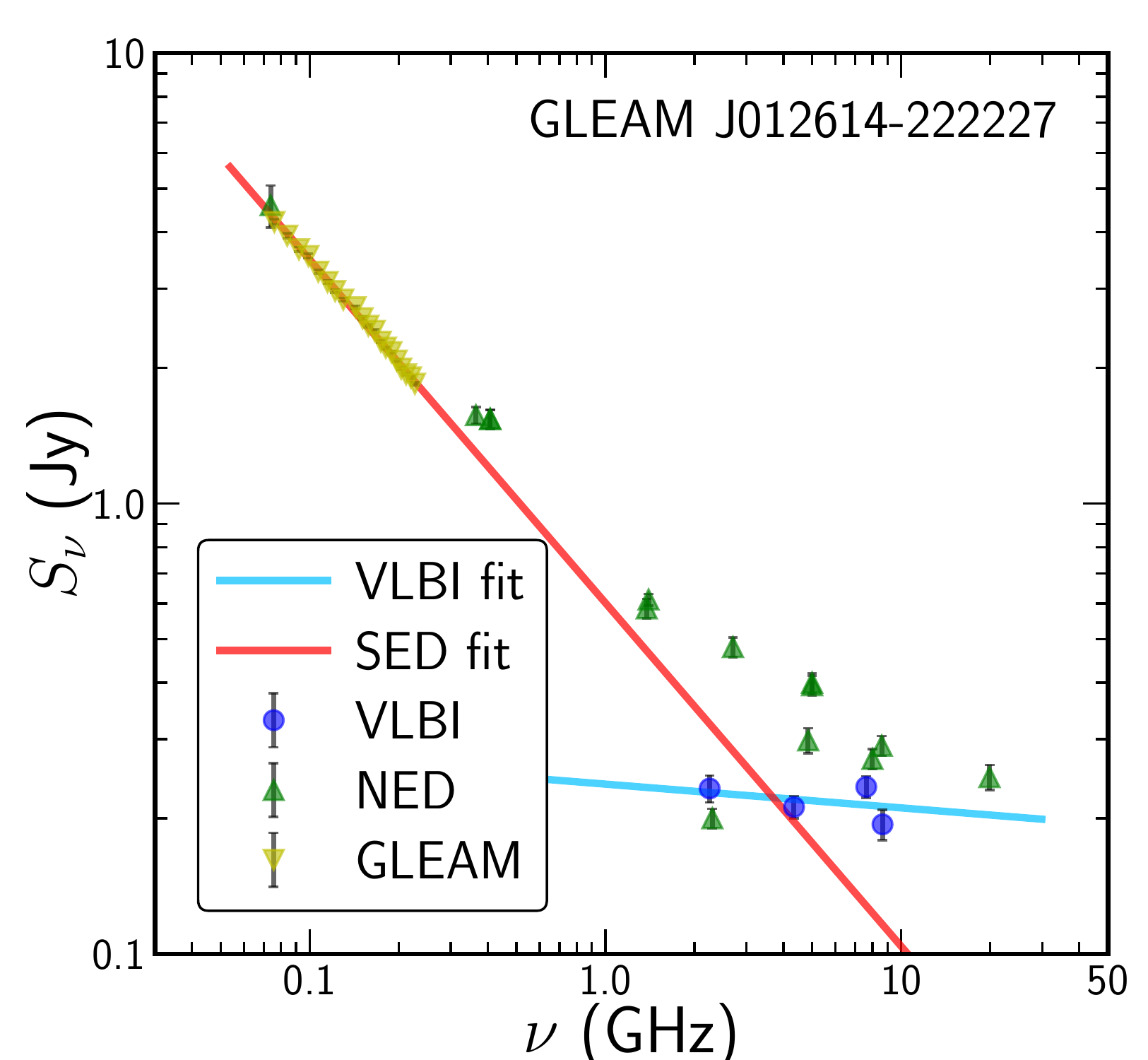}
\includegraphics[scale=0.26]{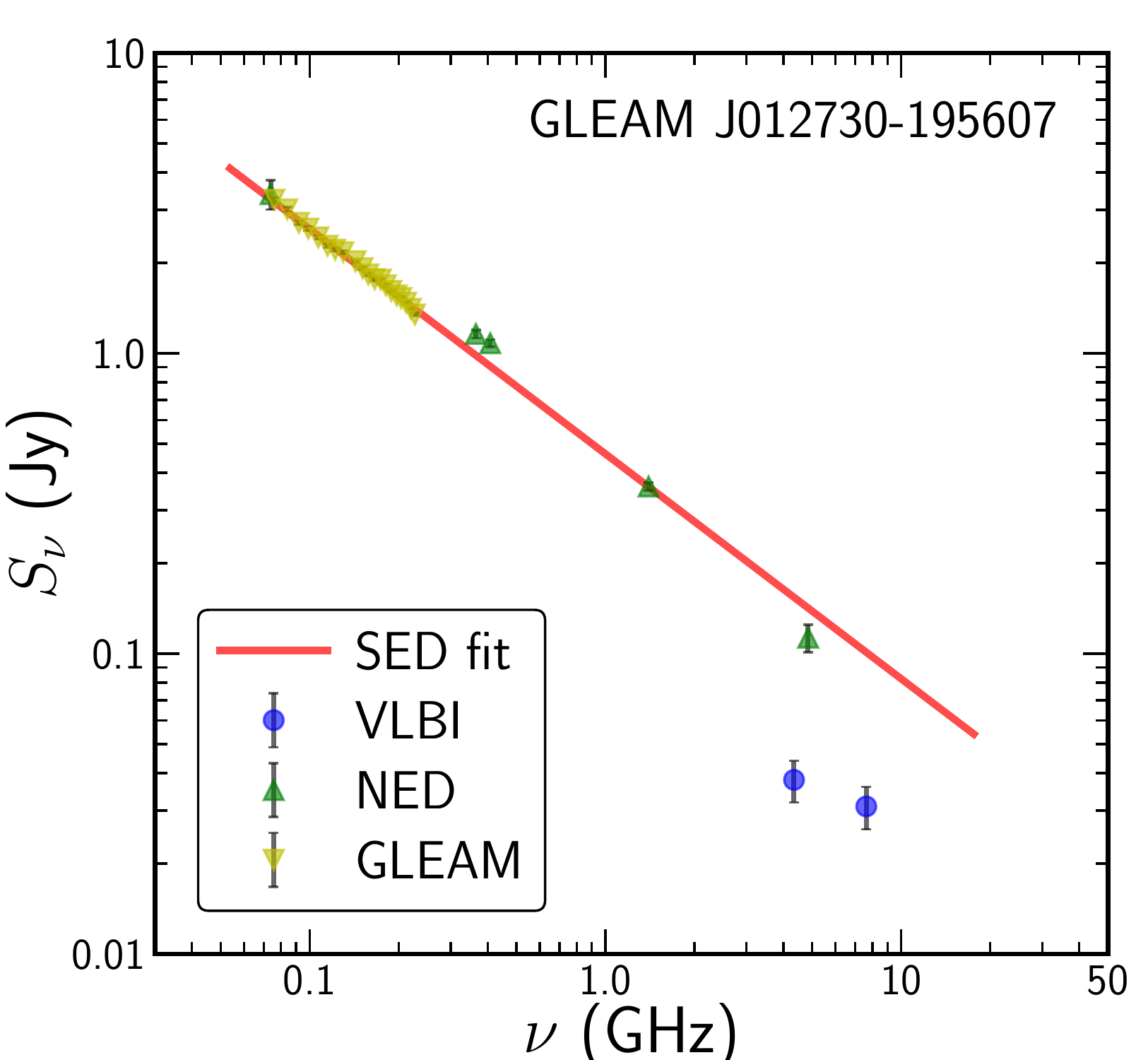}\\
\includegraphics[scale=0.26]{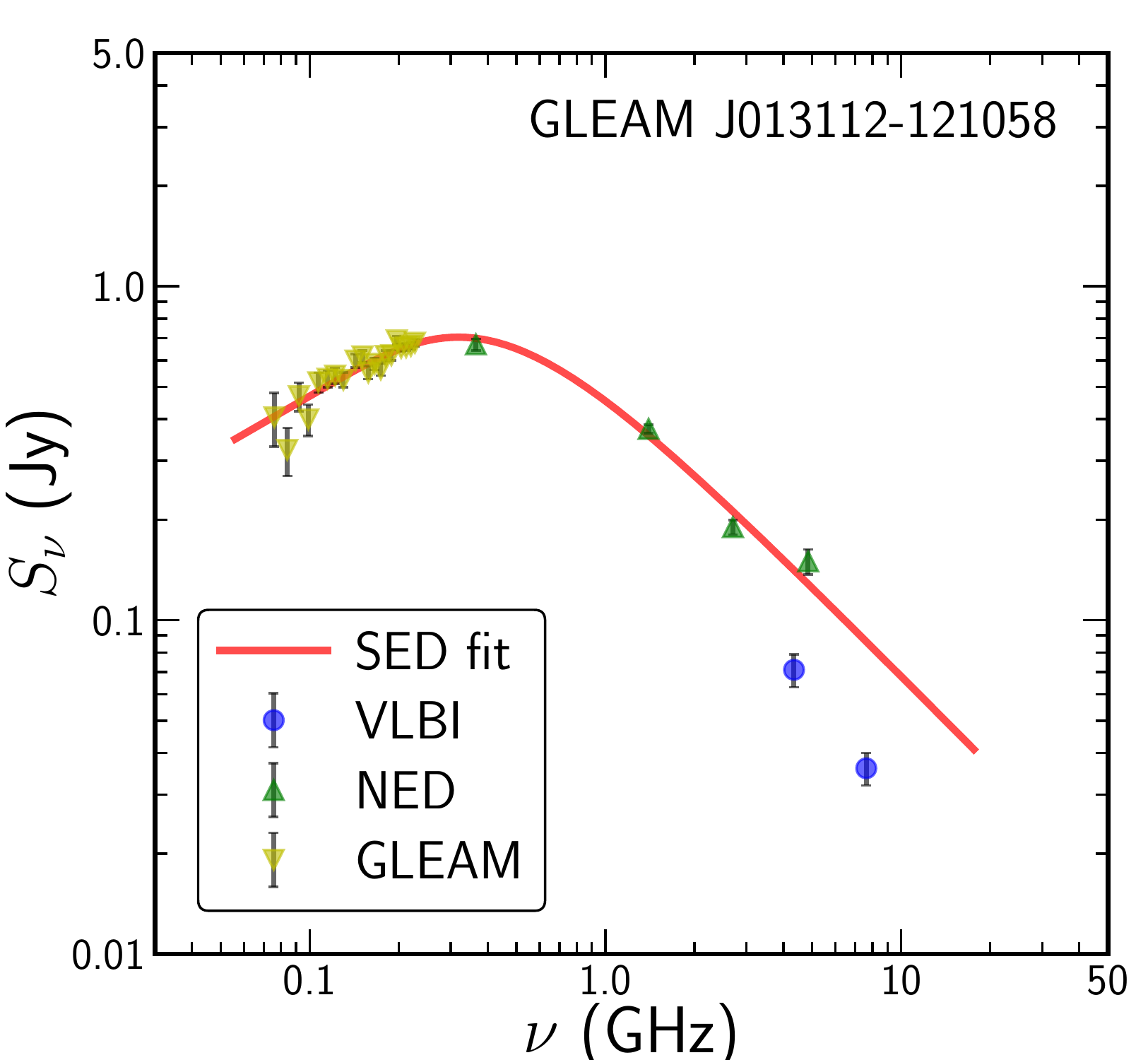}
\includegraphics[scale=0.26]{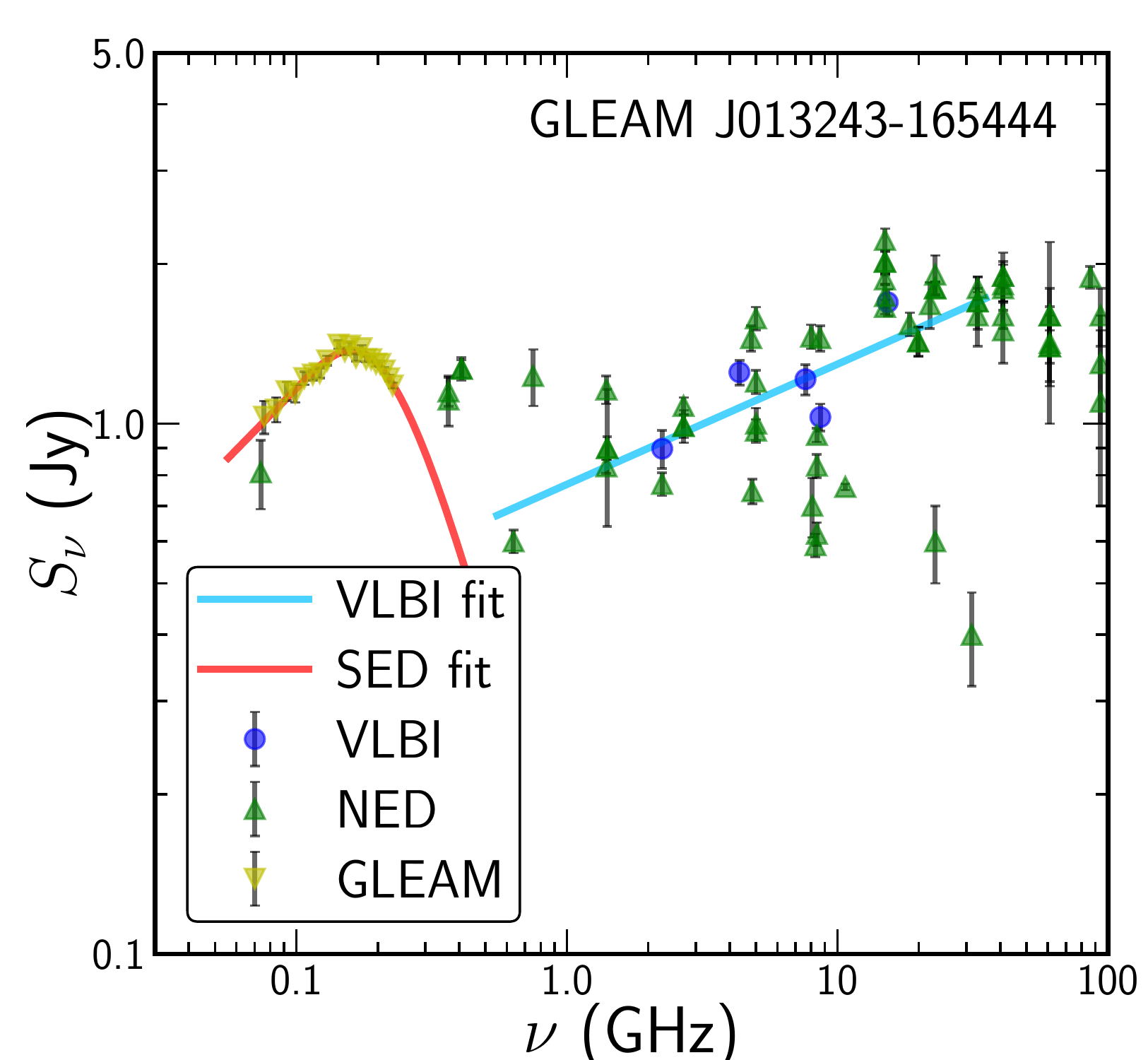}
\includegraphics[scale=0.26]{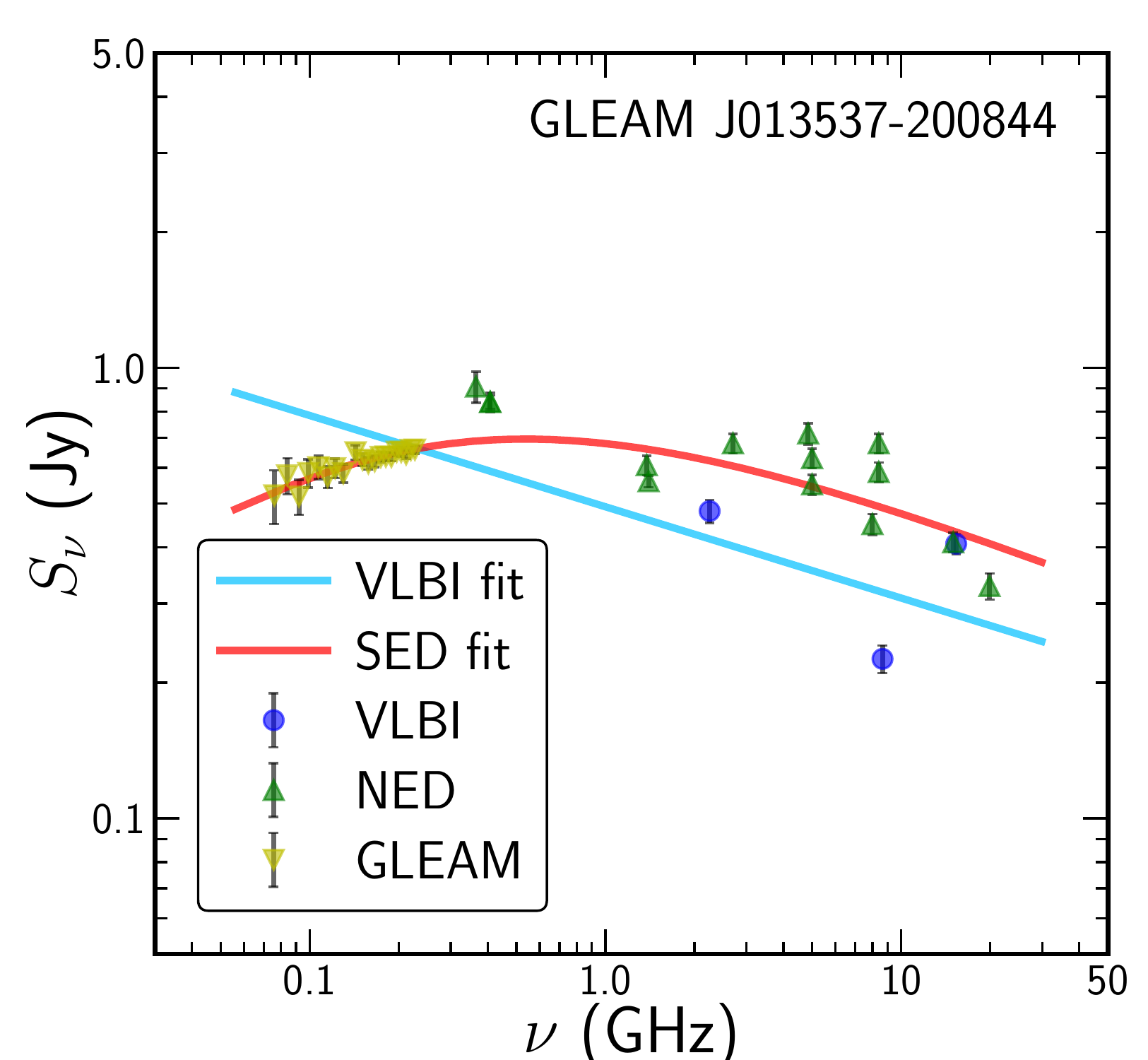}
\includegraphics[scale=0.26]{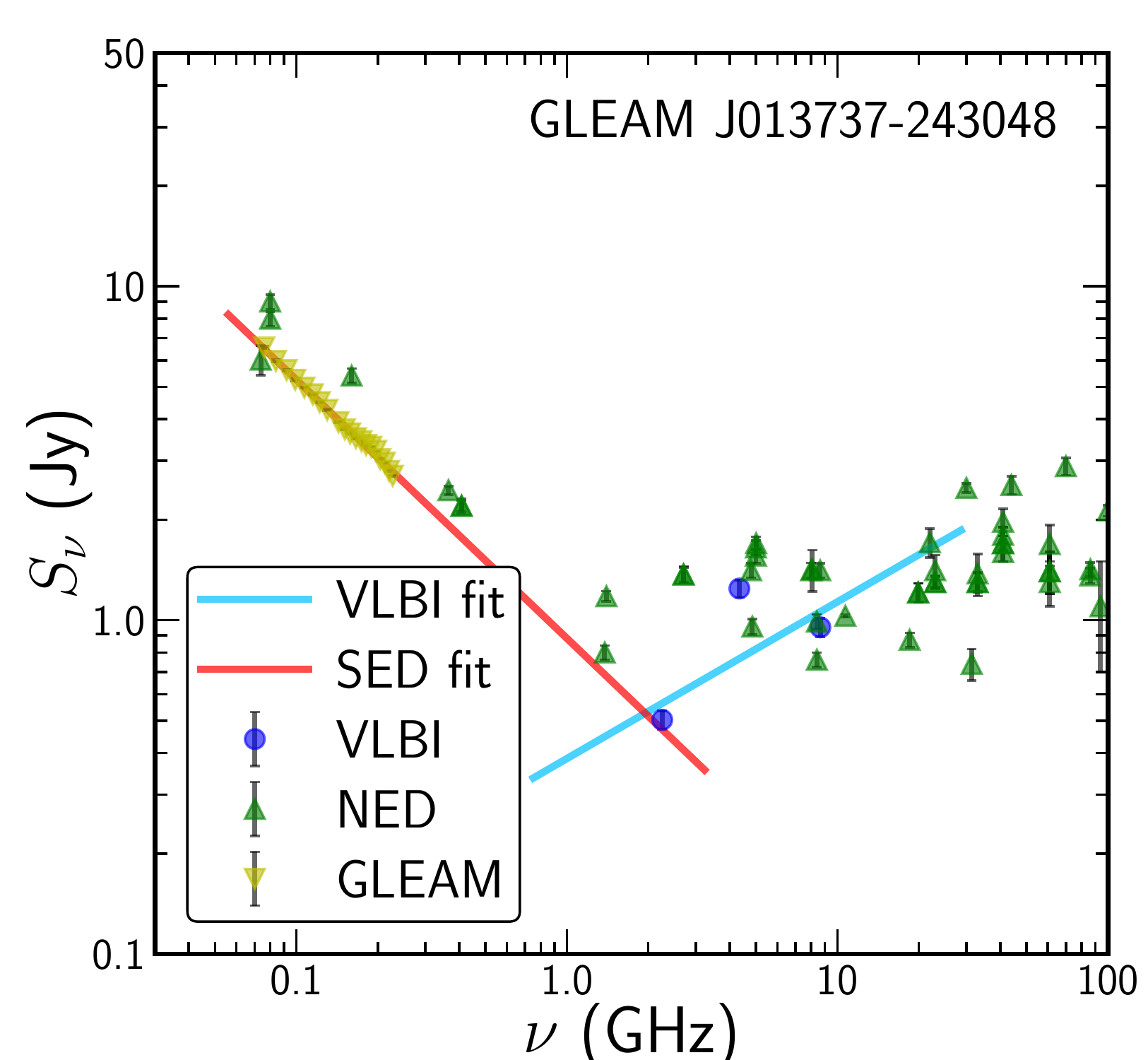}\\
\includegraphics[scale=0.26]{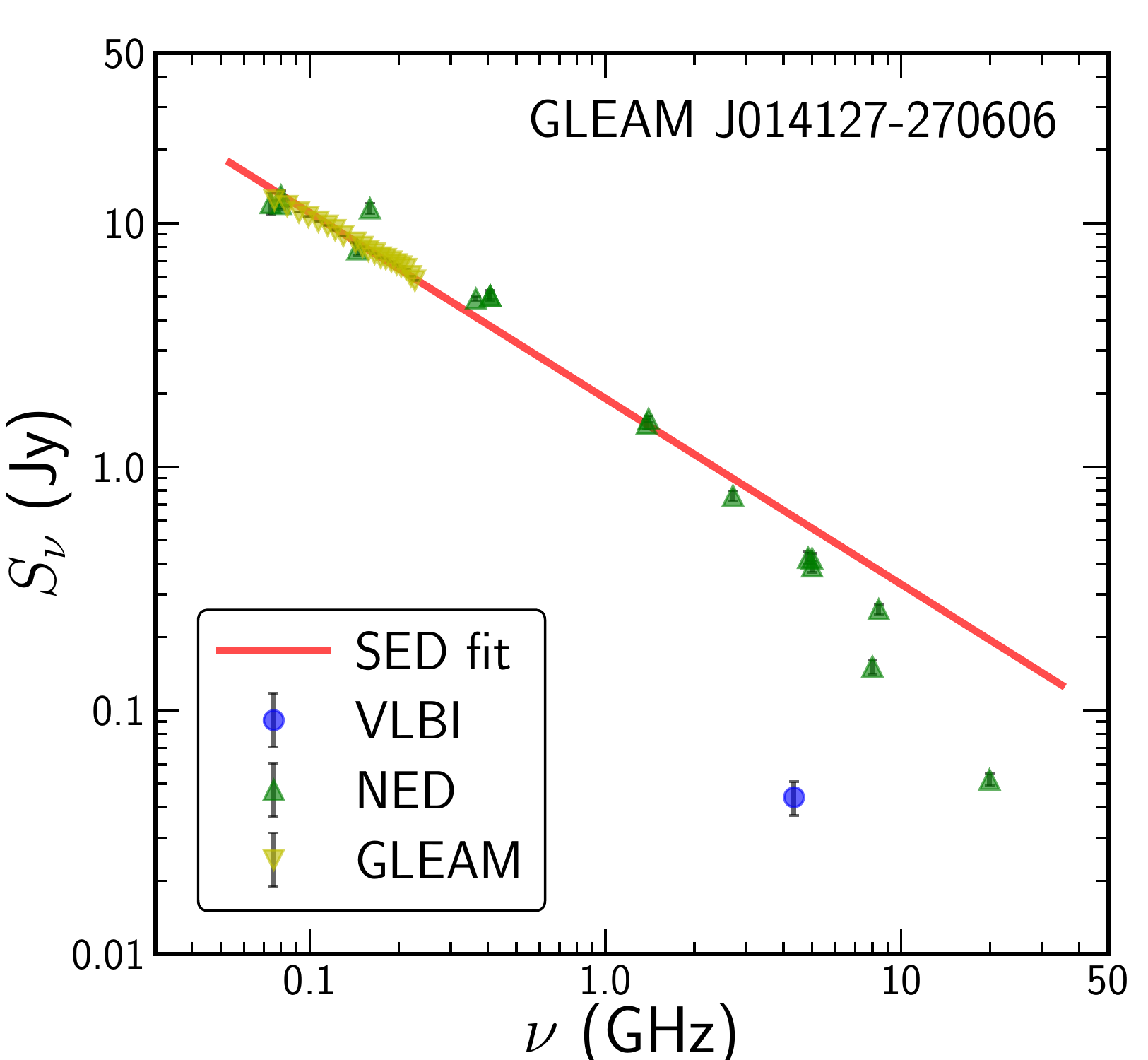}
\includegraphics[scale=0.26]{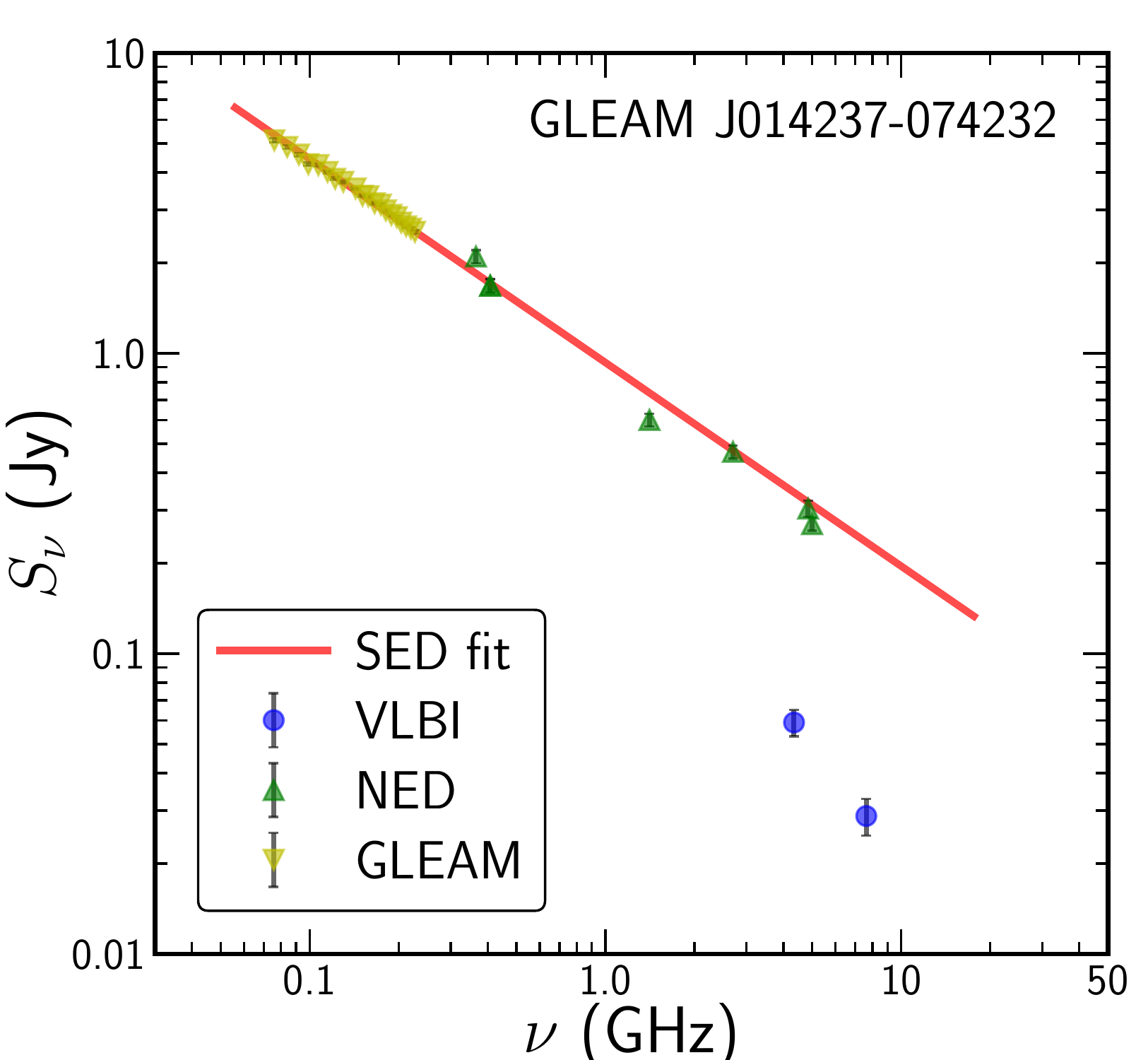}
\includegraphics[scale=0.26]{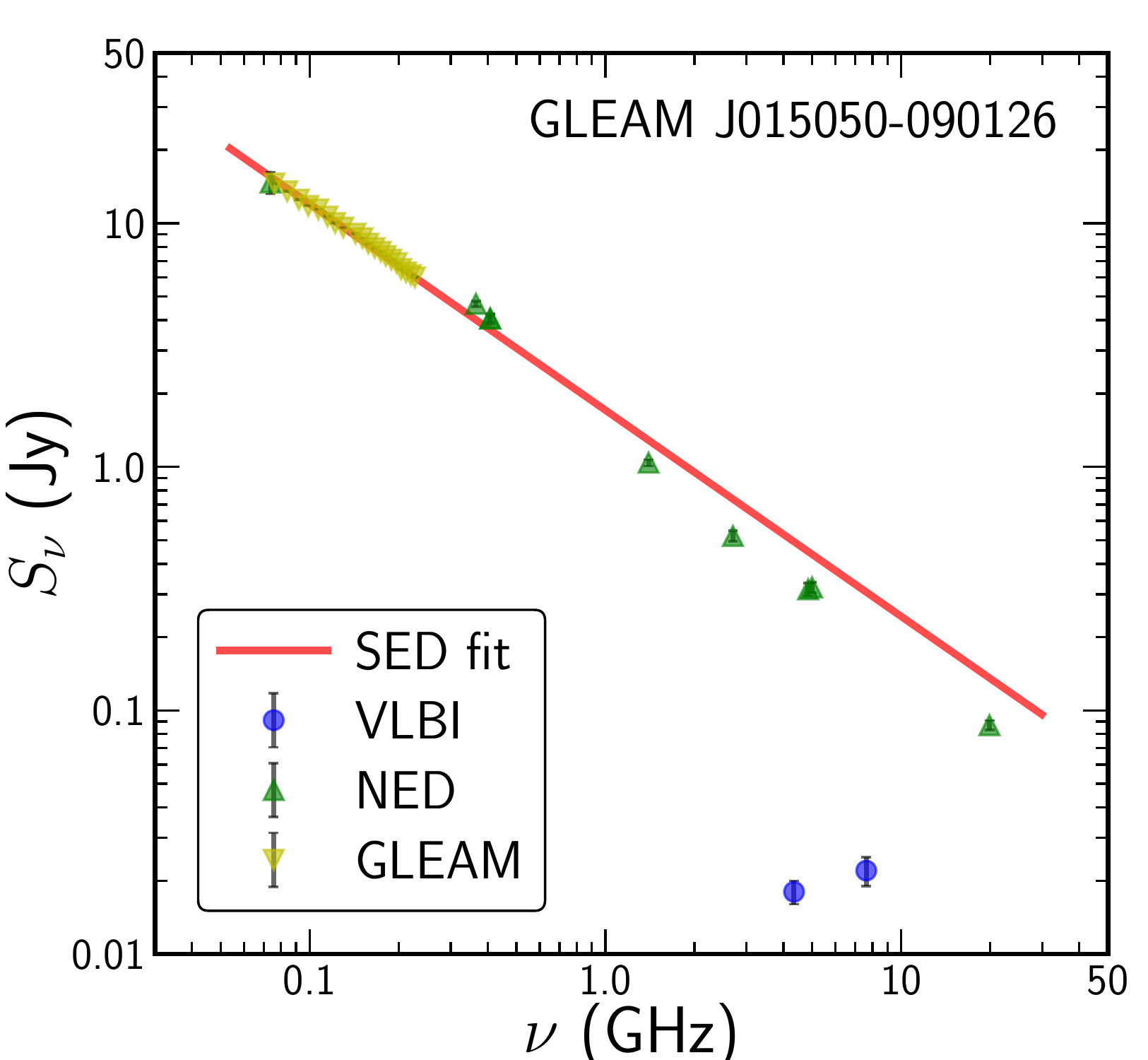}
\includegraphics[scale=0.26]{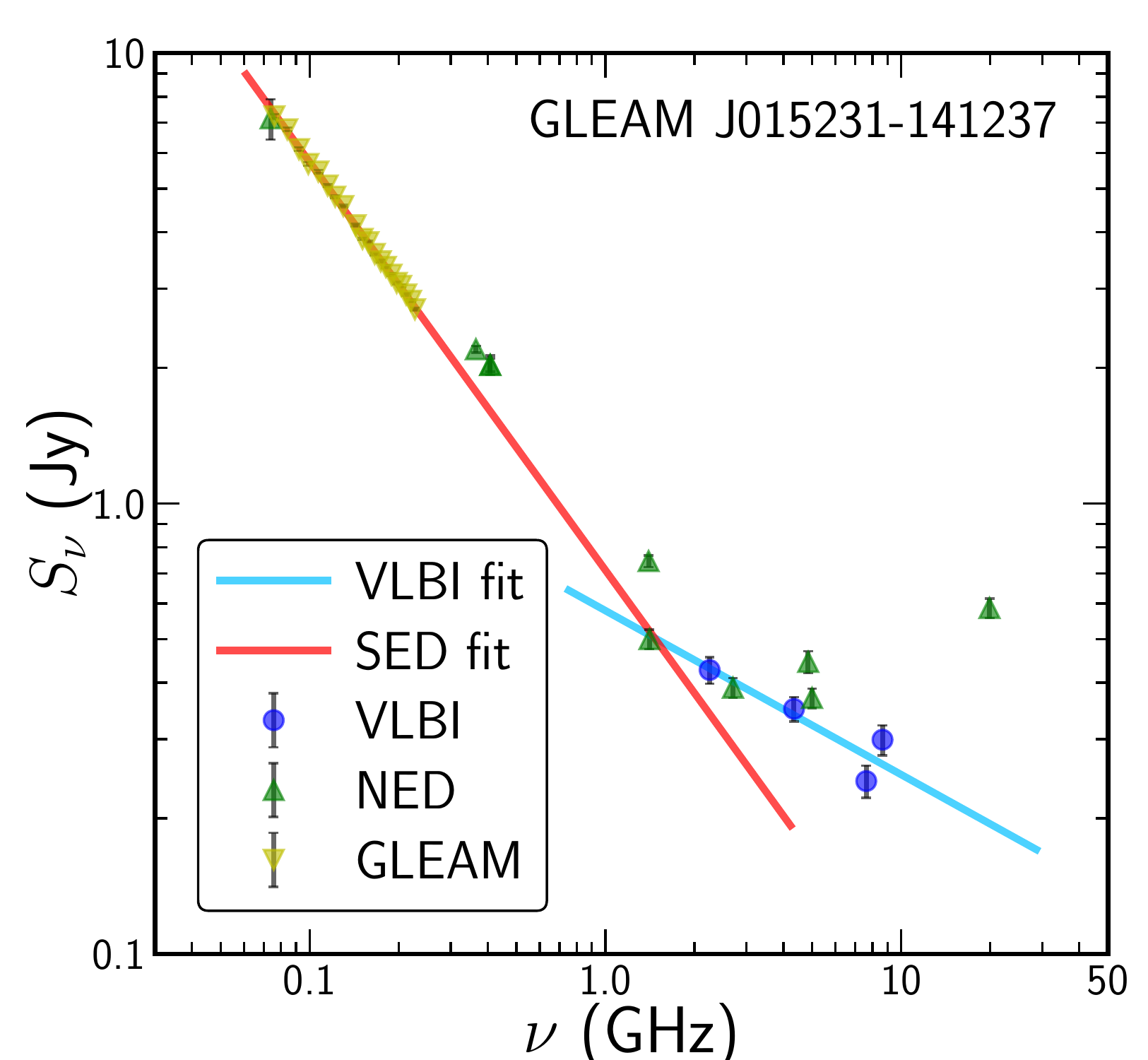}\\
\includegraphics[scale=0.26]{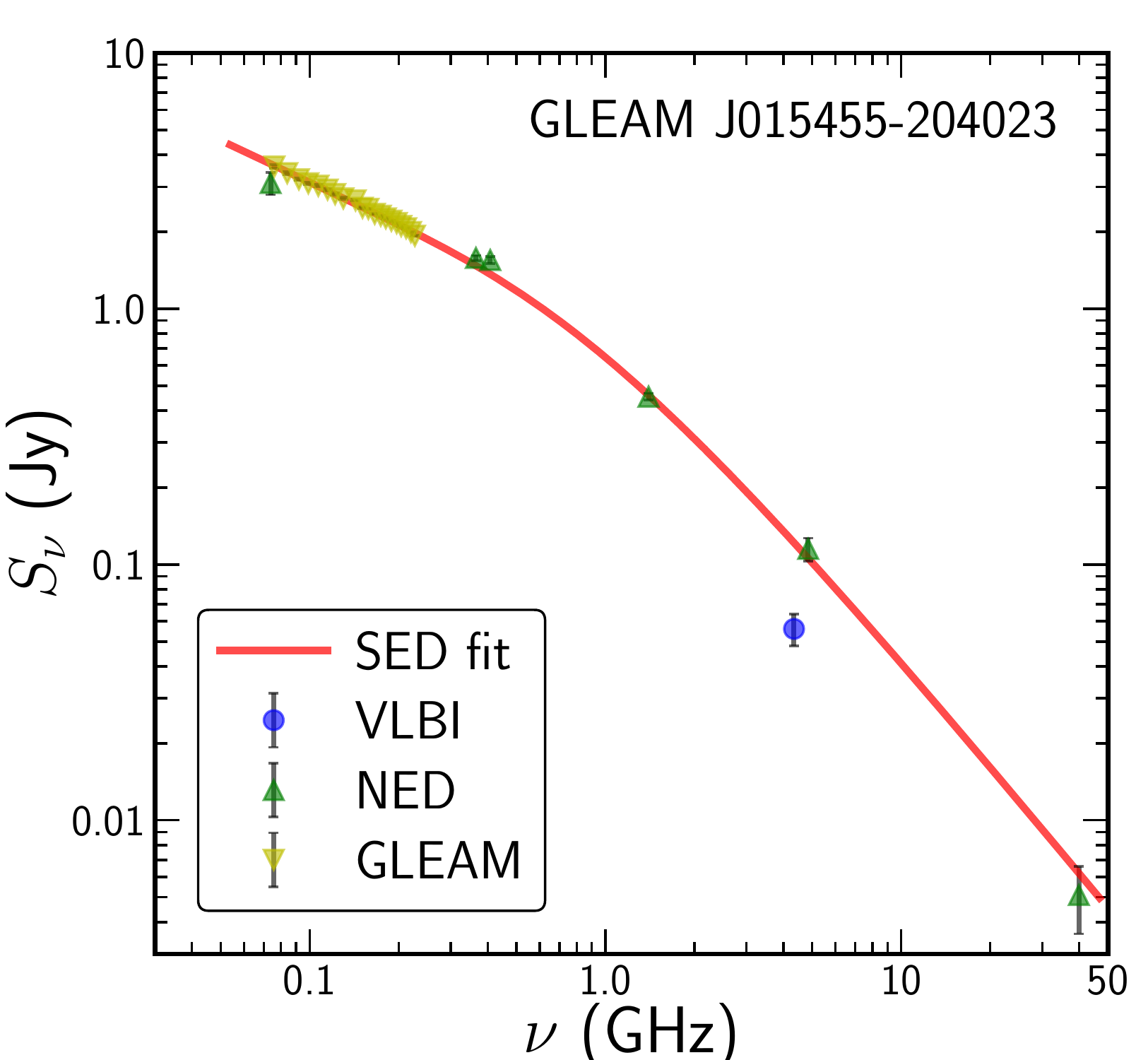}
\includegraphics[scale=0.26]{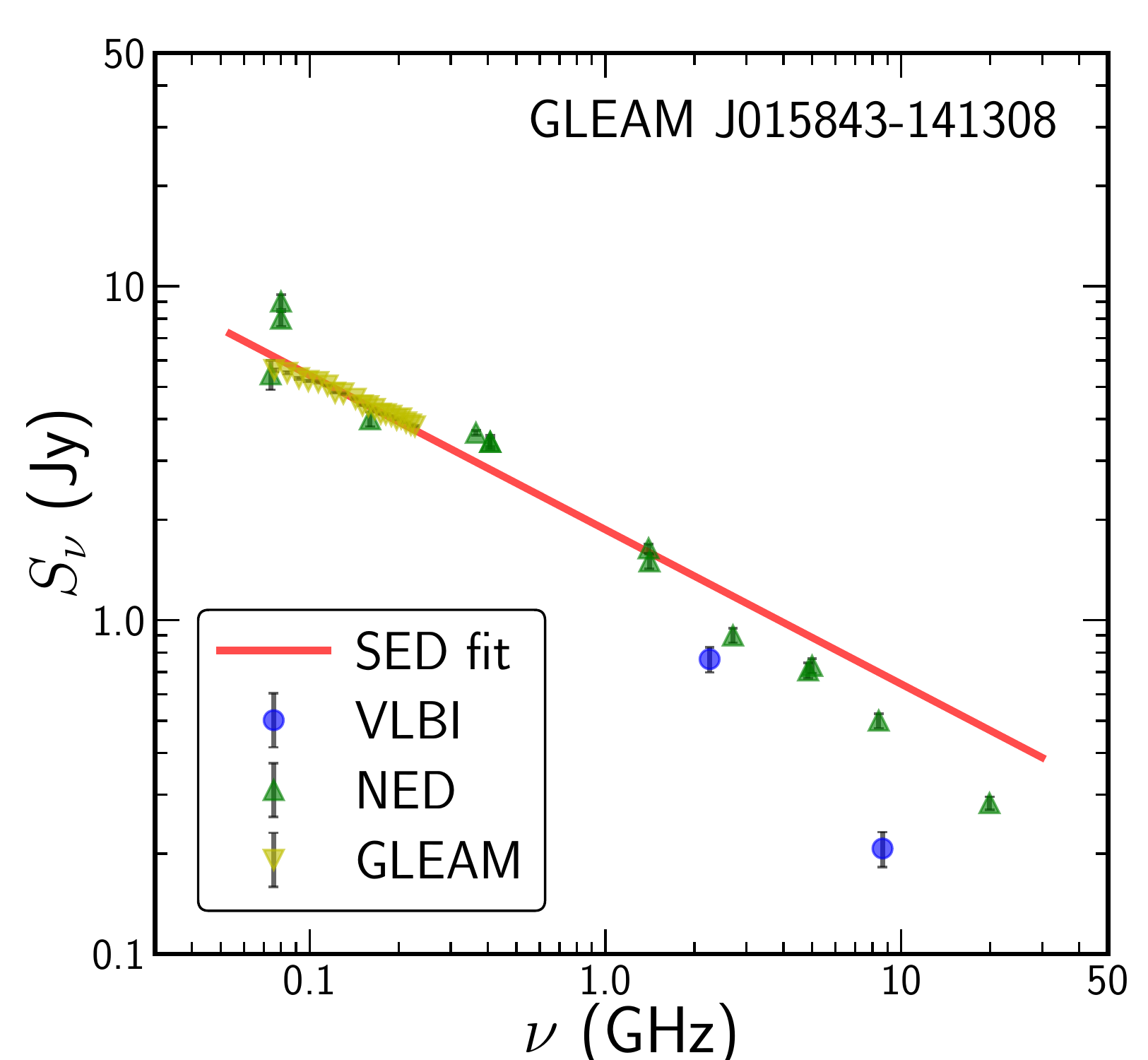}
\includegraphics[scale=0.26]{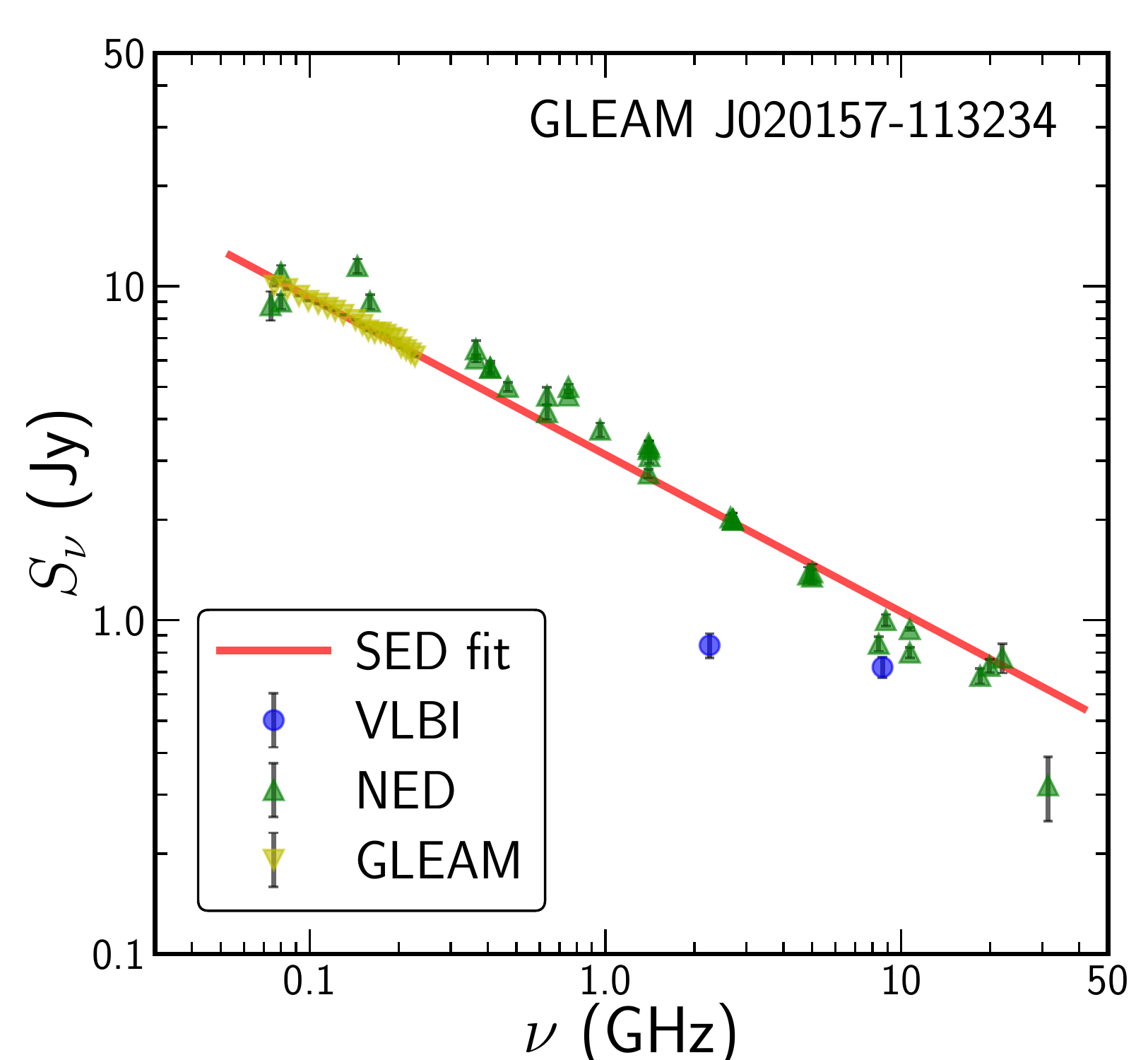}
\caption{\it - continued}
\end{figure*}



\label{lastpage}
\end{document}